\documentclass[tradiabstract]{aa}
\usepackage{graphicx} 
\usepackage{txfonts}
\usepackage{natbib}
\bibpunct{(}{)}{;}{a}{}{,}

\usepackage{xcolor}
\usepackage[normalem]{ulem}

\newcommand{\hl}[1]{{\color{black}{#1}}}
\newcommand{\strikeout}[1]{$\!\!$}

\newcommand{\thetav}{\theta_v}
\newcommand{\Yrec}{Y_{rec}}
\newcommand{\DeltaYrec}{\Delta Y_{rec}}
\newcommand{\thetarec}{\theta_{rec}}
\newcommand{\Deltathetarec}{\Delta\theta_{rec}}

\newcommand{\GContam}{\Gamma_{g}}

\begin{document}
   \title{A Comparison of Algorithms for the Construction of SZ Cluster Catalogues}


   \author{J.-B. Melin\inst{1}
          \and
          N. Aghanim\inst{2}
          \and
          M. Bartelmann\inst{3}
          \and
          J. G. Bartlett\inst{4,5,6}
          \and
          M. Betoule\inst{4}
          \and
          J. Bobin\inst{7}
          \and
          P. Carvalho\inst{8}
          \and
          G. Chon\inst{9}
          \and
          J.~Delabrouille\inst{4}
          \and
          J. M. Diego\inst{10}
          \and
          D. L. Harrison\inst{11}
          \and
          D. Herranz\inst{10}
          \and
          M. Hobson\inst{8}
          \and
          R. Kneissl\inst{12}
          \and
          A. N. Lasenby\inst{8,11}
          \and
          M.~Le~Jeune\inst{4}
          \and
          M. Lopez-Caniego\inst{10}
          \and
          P. Mazzotta\inst{13}
          \and
          G. M. Rocha\inst{6}
          \and
          B. M. Schaefer \inst{14}
          \and
          J.-L. Starck\inst{7}
          \and
          J. C. Waizmann\inst{3}
          \and
          D. Yvon\inst{1}
          }

   \institute{DSM/Irfu/SPP, CEA/Saclay, F-91191 Gif-sur-Yvette Cedex\\
              \email{jean-baptiste.melin@cea.fr}
         \and
         Institut d'Astrophysique Spatiale, CNRS et Universit\'e de Paris XI, B\^atiment 120-121, Centre Universitaire d'Orsay, 91400 Orsay CEDEX, France
         \and
          Institut fur Theoretische Astrophysik, Zentrum fur Astronomie der Universitat Heidelberg, Albert-Ueberle-Str.2, 69120 Heidelberg, Germany
        \and
         APC, 10 rue Alice Domon et L\'eonie Duquet, 75205 Paris Cedex 13, France
         \and
         Universit\'e Paris Diderot -- Paris 7, 75205 Paris Cedex 13, France
         \and
         Jet Propulsion Laboratory, California Institute of Technology, 4800 Oak Grove Drive, Pasadena, CA 91109, U.S.A.
         \and
         DSM/Irfu/SAp, CEA/Saclay, F-91191 Gif-sur-Yvette Cedex
         \and
         Astrophysics Group, Cavendish Laboratory, J.J. Thomson Avenue, Cambridge CB3 0HE
         \and
         Max-Planck-Institut f\"ur extraterrestrische Physik Giessenbachstr, 85748 Garching, Germany
         \and
         Instituto de F\'isica de Cantabria (CSIC-UC), Santander 39005, Spain
         \and
         Kavli Institute for Cosmology Cambridge and Institute of Astronomy, Madingley Road, Cambridge, CB3 OHA
         \and
         ALMA JAO, Av. El Golf 40 - Piso 18, Las Condes, Santiago, Chile \\
         ESO, Alonso de C\'ordova 3107, Vitacura, Casilla 19001, Santiago, Chile
         \and
         Universita' di Roma "Tor Vergata", Via della ricerca scientifica, 1, 00133 Roma, Italy
         \and
         Astronomisches Rechen-Institut, Monchhofstrasse 12-14, 69120 Heidelberg, Germany
             }

   \date{Received ; accepted }




  \abstract{We evaluate the construction methodology of an
      all--sky catalogue of galaxy clusters detected through the
      Sunyaev-Zel'dovich (SZ) effect. We
      perform an extensive comparison of twelve algorithms applied
      to the same detailed simulations of the millimeter and submillimeter sky based
      on a Planck-like case.  We present the results of this ``SZ
      Challenge" in terms of catalogue completeness, purity,
      astrometric and photometric reconstruction.  Our results provide
      a comparison of a representative sample of SZ detection
      algorithms and highlight important issues in their
      application. In our study case, we show
      that the exact expected number of clusters remains uncertain
      (about a thousand cluster candidates at $|b|>20$ deg with 90\%
      purity) and that
      it depends on the SZ model and on the detailed sky simulations,
      and on
      algorithmic implementation of the detection methods. We also estimate 
      the astrometric precision of the cluster candidates which is found of
      the order of $\sim 2$~arcmins on average, and the photometric
      uncertainty of order $\sim 30$\%, depending on flux.} 


   \maketitle
%

\section{Introduction}
Galaxy cluster catalogues have played a long--standing, vital role in
cosmology, providing important information on topics ranging from
cosmological parameters to galaxy formation~\citep{rosatietal:2002,
  voit:2005}.  In particular, recent X--ray cluster catalogs have
proved valuable in establishing the standard cosmological model (e.g., \cite{schuecker03}, \cite{vikhlinin09}).  The
science potential of large cluster surveys is strong: They are, for
instance, considered one of the central observational tools for
illuminating the nature of dark energy (e.g., the Dark Energy Task
Force Report~\citep{detf,FoMSWG}).  A suite of large cluster surveys
planned over the coming years in the optical/IR, X--ray and millimeter
bands will greatly extend the reach of cluster science by probing much
larger volumes to higher redshifts with vastly superior statistics and
control of systematics.

The Planck SZ cluster catalogue will be one of the important players
in this context. Surveying the entire sky in 9
millimeter/submillimeter bands with $\sim 5-10$~arcmin resolution over the
channels most sensitive to the cosmic microwave background (CMB)
anisotropies, the Planck satellite will find large numbers of clusters
through the Sunyaev--Zel'dovich (SZ) effect \citep{sz:1970, sz:1972,
  birk:1999, carlstrometal:2002}.  The advantages of this, much
anticipated, technique include efficient detection of distant clusters
and selection based on an observable expected to correlate tightly with
cluster mass~\citep{bartlett:2002, daSilvaetal:2004, motletal:2005,
  nagai:2006, bonaldietal:2007, shawetal:2008}.  An official mission
deliverable, the Planck SZ catalogue will be the first all--sky
cluster catalogue since the workhorse catalogs from the ROSAT All-Sky
Survey (RASS) \citep{trumper:1992}; in other words, Planck
will be the first all--sky cluster survey since the early 1990s!

Within the Planck Consortium, a considerable effort has been
  conducted for the scientific evaluation of the cluster catalogue
  construction methodology.  As part of this evaluation effort, we
  completed an extensive comparison of twelve
algorithms applied to detailed simulations of Planck data based
on the Planck Sky Model.  This study was 
dubbed ``The SZ Challenge'' and was carried out in two steps
using different SZ cluster models and cosmologies; these are referred
to as Versions 1 and 2 and more fully explained below.  We report the
findings of these initial studies in terms of catalogue completeness and
purity, as well as astrometric and photometric accuracy and precision.

The article is organized as follows: In the next section we detail our
sky simulations, including a brief description of
their basis, the Planck Sky Model (PSM, \cite{delabrouille2012}).  The following section then introduces the different
catalogue construction methodologies employed, before moving on to a
presentation of each of the twelve algorithms in the study. We
  present the results of the challenge in
  Section~\ref{sec:results}, followed by a comparative study of
  algorithmic performance.  We conclude with a discussion of both the
  limitations of this study and future directions.

\section{Simulations}
\label{sec:simulations}
\subsection{The Planck Sky Model}
\label{sec:PSM}
One of the important differences between the present work and previous
studies of Planck SZ capabilities is the detailed and rather
sophisticated simulation of millimeter and submillimeter sky emission
used here.  Our sky simulations are based on an early development
version of the Planck Sky Model (PSM,  \cite{delabrouille2012}), a flexible
software package developed within the Planck collaboration for
making predictions, simulations and constrained realisations of the
microwave sky.  The simulations used for this challenge are not
polarised (only temperature maps are useful for detecting clusters
using the thermal SZ effect).

The CMB sky is based on the best fit angular power spectrum model of
WMAP.  The CMB realisation is not constrained by actual observed CMB
multipoles, in contrast to the simulations used by
\citet{leachetal:2008}.

Diffuse Galactic emission is described by a four component
model of the interstellar medium comprising free-free, synchrotron,
thermal dust and spinning dust and is based on
\citet{mdetal:2008} (see \citet{mdetal:2009} for a review). The
predictions rely on a number of sky templates with
different angular resolutions.  In order to simulate the sky at Planck
resolution we have added small-scale fluctuations to some of the
templates to increase the fluctuation level as a
function of the local brightness and therefore reproduce the
non-Gaussian and non-uniform properties of the interstellar
emission. The procedure used to add small scales is presented in
\citet{mdetal:2007}.

Free-free emission is based on the model of \citet{dickinsonetal:2003}
assuming an electronic temperature of 7000~K.  The spatial structure
of the emission is modeled using an H$\alpha$ template corrected
for dust extinction.  The H$\alpha$ map is a combination of the
Southern H-Alpha Sky Survey Atlas (SHASSA, \citet{shassa}) and the
Wisconsin H-Alpha Mapper (WHAM, \citet{wham}), smoothed to obtain a
uniform angular resolution of 1$^{\circ}$.  Dust extinction is
inferred using the $E(B-V)$ all-sky map of \citet{sfd:1998}. As
mentioned earlier, small scales were added in both templates to match
the Planck resolution.  The free-free emission law is constant over
the sky, as it depends only on the electronic temperature, taken as a
constant here (see however \citet{Wakkeretal:2008} for a description
of high-velocity clouds not detected by the WHAM survey and hence not
included in our simulations).

Synchrotron emission is based on an extrapolation of the 408~MHz map
of \citet{haslametal:1982} from which an estimate of the free-free
emission was removed.  In any direction on the sky, the spectral
emission law of the synchrotron is assumed to follow a power law, $
T_b^{sync} \propto\nu^\beta$.  We use a pixel-dependent spectral index
$\beta$ derived from the ratio of the 408~MHz map and the estimate of
the synchrotron emission at 23~GHz in the {\sc WMAP} data obtained by
\citet{mdetal:2008}.

The thermal emission from interstellar dust is estimated using model 7
of \citet{fds:1999}. This model, fitted to the FIRAS data (7$^{\circ}$
resolution), assumes that each line of sight can be modeled by
the sum of the emission from two dust populations, one cold and one
hot. Each grain population is in thermal equilibrium with the
radiation field and thus has a grey-body spectrum, so that the total
dust emission is modelled as
\begin{eqnarray}
\label{eq:dustmodel}
I_\nu \propto \sum_{i=1}^2 f_i \nu^{\beta_i} B_\nu(T_i)
\end{eqnarray}
where $B_\nu(T_i)$ is the Planck function at temperature $T_i$.  In
model 7 the emissivity indices are $\beta_1=1.5$, $\beta_2=2.6$, and
$f_1=0.0309$ and $f_2 = 0.9691$.  Once these values are fixed, the
dust temperature of the two components is determined using only the
ratio of the observations at 100~$\mu$m and 240~$\mu$m.  For this
purpose, we use the 100/240~$\mu$m map ratio published by
\cite{fds:1999}.  Knowing the temperature and $\beta$ of each dust
component at a given position on the sky, we use the 100~$\mu$m
brightness at that position to scale the emission at any frequency
using Eq.~(\ref{eq:dustmodel}).

Spinning dust emission uses as a template the dust extinction map
$E(B-V)$, and uses an emission law uniform on the sky, based on the
model of \citet{drainelazarian:1998}, assuming a Warm Neutral Medium
(WNM).

We emphasise that the emission laws of both synchrotron and dust vary
across the sky. The spectral index of free-free and the emission law
of spinning dust, however, are taken as uniform on the sky.

Point sources are modeled with two main categories: radio and
infrared. In the present simulation, none of two is correlated
  with the SZ signal. Simulated radio sources are based on the NVSS
or SUMSS and GB6 or PMN catalogues. Measured fluxes at 1 and/or 4.85
GHz are extrapolated to 20 GHz using their measured SED when observed
at two frequencies.  Sources for which a flux measurement is available
at a single frequency have been randomly assigned to either the steep-
or to the flat-spectrum class in the proportions observationally
determined by \citet{fomalontetal:1991} for various flux intervals,
and assigned a spectral index randomly drawn from the corresponding
distribution.  Source counts at 5 and 20 GHz obtained in this way are
compared, for consistency, with observed counts, with the model by
\citet{toffolattietal:1998}, and with an updated version of the model
by \citet{dezottietal:2005}, allowing for a high-redshift decline of
the space density of both flat-spectrum quasars (FSQs) and
steep-spectrum sources (not only for FSQs as in the original
model). Further extrapolation at Planck frequencies has been made
allowing a change in SED above 20 GHz, assuming again a distribution
in flat and steep populations. For each of these two populations, the
spectral index is randomly drawn within a set of values compatible
with the typical average and dispersion.
Infrared sources are based on the {\sc IRAS} catalogue, and modeled
as dusty galaxies~\citep{serjeantharrison:2005}. IRAS coverage gaps
were filled by randomly adding sources with a flux distribution
consistent with the mean counts.  Fainter sources were assumed to be
mostly sub-millimeter bright galaxies, such as those detected by SCUBA
surveys. These were modelled following \citet{granatoetal:2004} and
assumed to be strongly clustered, with a comoving clustering radius
$r_0 \simeq 8\,\hbox{h}^{-1}\,$Mpc. Since such sources have a very
high areal density, they are not simulated individually but make up
the sub-mm background.

The SZ component is described in detail in the following section.

Component maps are produced at all Planck central frequencies.  They
are then co-added and smoothed with Gaussian beams as indicated in
Table~\ref{tab:freqchann_Planck}, extracted from the Planck Blue book.
We thus obtain a total of nine monochromatic sky maps.

Finally, inhomogeneous noise is simulated according to the pixel
hit count corresponding to a nominal 14-month mission\footnote{Note
  that since launch in May 2009, the observed Helium consumption for
  the Planck-HFI dilution cooler indicates that the instrument can
  operate for about 30 months.  A mission extension has been approved
  by ESA accordingly.} using the Level-S simulation tool
\citep{levelS}. The RMS noise level in the simulated maps is given in
Table~\ref{tab:freqchann_Planck} from the Planck Blue book. It is
  worth noting that the in-flight performances of Planck-HFI as
  reported by the Planck collaboration \citep{PEPVI}
  are better than the requirements.

\begin{table*}
  \caption{Characteristics of instrumental values taken from 
    {\sc Planck}'s Blue Book for a 14 month
    nominal mission.  $\sigma_{\rm
      pixel}$ refers to the standard deviation of the $1.7'$ (Healpix
    $n_{\rm side}$=2048) pixel noise maps at the considered
    frequency.}  \centering
  \begin{tabular}{c|ccccccccc}
    \hline
    \hline
    Channel & 30 GHz & 44 GHz & 70 GHz & 100 GHz & 143 GHz & 217 GHz & 353 GHz & 545 GHz & 857 GHz \\ \hline
    FWHM [arcmin]  & 33 & 24 & 14 &  10 & 7.1 & 5 & 5 & 5 & 5 \\
    $\sigma_{\rm pixel}$ [${\rm m}K_{\rm CMB}$] & 0.131 & 0.130 & 0.126 & 0.057 & 0.029 & 0.046 & 0.137 & 1.241 & 56.639 \\
    \end{tabular}
  \vspace{0.3 cm}
  \\
  \label{tab:freqchann_Planck}
\end{table*}

\subsection{The SZ component}

We simulate the SZ component using a semi-analytic approach based on
an analytic mass function $dN(M,z)/dM dz$.  After selecting
cosmological parameters ($h$, $\Omega_{\mathrm m}$, $\Omega_\Lambda$,
$\sigma_8$), the cluster distribution in the mass-redshift plane
$(M,z)$ is drawn from a Poisson law whose mean is given by the
mass function.  Clusters are spanning the mass range $5 \times 10^{13} M_\odot < M_{\rm vir} < 5  \times 10^{15} M_\odot$ and the redshift range $0 < z < 4$.
Cluster Galactic coordinates $(l,b)$ are then
uniformly drawn on the sphere. We compute the SZ signal for each dark
matter halo following two different models, producing two simulations
(v1 and v2) with different sets of cosmological parameters and mass
functions and SZ signals\footnote{We assume spherical symmetry for
  the individual SZ clusters and do not take into account any scatter
  in the distribution of pressure profiles.}.

\subsubsection{SZ Challenge v1}

For the first version of the SZ Challenge, we used $h=0.7$,
$\Omega_{\mathrm m}=0.3$, $\Omega_\Lambda=0.7$, $\sigma_8=0.85$ and
the Sheth-Tormen mass function \citet{st:1999}. We assume that the
clusters are isothermal and that the electron density profile is given
by the $\beta$-model, with $\beta=2/3$, and core radius scaling as
$M^{1/3}$. We truncate the model at the virial radius, $r_{\rm
    vir}$, and choose the core radius $r_{\rm c}=r_{\rm vir}/10$.  The
  virial radius here is defined according to the spherical collapse
  model. The temperature of each cluster is derived using a
mass-temperature given in \citet{colafrancescoetal:1997} with $T_{15}
= 7.75{\rm keV}$, consistent with the simulations of
\citet{ekeetal:1998}. For more details on this model, we refer the
reader to \S~5 of \citet{dezottietal:2005}.

\subsubsection{SZ Challenge v2}

The second version of the SZ Challenge was produced using a WMAP5 only
cosmology \citep{wmap5only} ($h=0.719$, $\Omega_{\mathrm m}=0.256$,
$\Omega_\Lambda=0.744$, $\sigma_8=0.798$). We used the Jenkins mass
function \citep{jenkinsetal:2001}.  The SZ emission is modeled using
the universal pressure profile derived from the X-ray REXCESS cluster
sample \citep{arnaudetal:2009} which predicts profile and
normalisation of SZ clusters given their mass and redshift. The
profile is well fitted by a generalized NFW profile that is much
steeper than the $\beta$-profile in the outskirts. 
Moreover, for a given mass, the normalisation of the SZ flux is
$\sim15\%$ lower than the normalisation of SZ Challenge v1. This
  profile was used as the baseline profile in the SZ early results
  from Planck \citep{PEPVIII, PEPIX, PEPX, PEPXI, PEPXII}. \\

Neither of the two sets of simulations (v1 and v2) contains point
sources within clusters. The effect of contamination by radio or
infrared point sources in clusters was therefore not studied here\footnote{The effect of radio sources ($\nu<217{\rm GHz}$) is to reduce the observed SZ signal at a given frequency while the effect of infrared sources ($\nu>217{\rm GHz}$) is to increase it. However, the extraction algorithms being multifrequency, their sensitivity to point sources is expected to be weaker than for single frequency extractions because of the different spectral dependence of point sources and SZ clusters.}. We neither include relativistic electronic populations within clusters. As for point sources, this effect was not studied here\footnote{The effect has been very recently studied within the Planck Collaboration: assuming a non relativistic spectrum for extracting clusters biases the flux low by about 10\% in direction of massive ($M_{\rm vir}>10^{15} M_\odot$) clusters (Planck Collaboration et al. 2013, in preparation).}.

\section{Methods and Algorithms}

The SZ Challenge was run as a blind test by providing the simulated
sky maps.  Participating teams, ten, were then asked to run
their algorithms, twelve in total on the simulated data and
supply a cluster catalogue with
\begin{enumerate}
\item $(\alpha, \delta)$: cluster sky coordinates
\item $\Yrec$: recovered total SZ flux, in terms of the
  integrated Compton--$Y$ parameter
\item $\DeltaYrec$: estimated flux error, i.e., the method's
  internal estimate of flux error
\item $\thetarec$: recovered cluster angular size, in terms of
  equivalent virial radius
\item $\Deltathetarec$: estimated size error (internal error).
\end{enumerate}

The different methods were divided into two classes:
direct methods that produce a cluster catalogue applying filters
directly to a set of frequency maps, and indirect methods that first
extract a thermal SZ map and then apply source finding algorithms.

In this classification, the 12 algorithms studied were:

\begin{itemize}
\renewcommand{\labelitemii}{$\bullet$}
\item Direct methods:
\begin{itemize}
\item MMF1: Matched Multi--Filters (MMF) as implemented by
  D. Harrison.
\item MMF2: MMF as described in~\cite{herranz:2002}.
\item MMF3: MMF as described in~\cite{melin:2006}.
\item MMF4: MMF as described in~\cite{schaefer:2006}.
\item PwS: Bayesian method PowellSnakes as described
  in~\cite{2009MNRAS.393..681C}. 
\end{itemize}
\item Indirect methods:
\begin{itemize}
\item BNP: Bayesian Non-parametric method as described
  in~\cite{diego:2002}, followed by SExtractor
  \citep{1996A&AS..117..393B} to detect clusters and perform
  photometry.
\item ILC1: All-sky Internal Linear combination (ILC) on needlet
coefficients\footnote{needlet coefficients are the equivalent of Fourier coefficients in the adopted spherical wavelet domain} (similar to the method used for CMB extraction in
\citet{delabrouille:2009}) to get an SZ map, followed by Matched
Filters on patches to extract clusters and perform photometry.
\item ILC2: Same SZ map \citet{delabrouille:2009}, but followed by 
SExtractor on patches instead
of a matched filter to extract the clusters. Photometry or flux measurement is however
done using Matched Filters at the position of the detected clusters. 
\item ILC3 developed by G. Chon and R. Kneissl: ILC in real space
and filtering in harmonic space to obtain an SZ map, followed by fitting
a cluster model.
\item ILC4 developed by J.-B. Melin: ILC on patches in Fourier
space to obtain SZ maps, followed by SExtractor to detect clusters and
Matched Multifilters to perform photometry.
\item ILC5 developed by D. Yvon: ILC on patches in wavelet
space to obtain SZ maps, followed by SExtractor to detect clusters and
perform photometry.
\item GMCA: Generalized Morphological Component Analysis as
described in~\cite{bobin:2008}, followed by wavelet filtering and
SExtractor to extract clusters and Matched Multifilters to perform
photometry.

\end{itemize}
\end{itemize}

All of the algorithms make use of the known frequency spectrum
of the SZ signal; attempts to detect clusters without this prior
knowledge perform significantly worse. A summary of the
characteristics of the codes as well as their treatment of the
point sources, foreground removal and masking is given in
Table~\ref{tab:methods}. Further details about the algorithms are
  given in the following subsections.

\begin{table*}
  \caption{Summary of the algorithms compared in the SZ
    Challenge. The first column shows the name of the method. The
      second column indicates when the code is using a prior on the SZ cluster shape. The uperscript $^a$
      indicate that the detection did not use a shape prior but that
      the computation of the SZ flux did. The third column
    gives the performance in terms of CPU hours needed to complete the
    analysis. The fourth and fifth column show whether the
    analysis was made using all-sky maps or projected patches, their
    area in square degrees and the number of patches. Methods ILC1 and
    ILC2 work with full sky maps for producing an SZ map (by ILC on
    needlet coefficients) and then work with 504 small patches for
    cluster detection by matched filtering (ILC1) or using SExtractor
    (ILC2). The sixth and seventh columns provide information
    about any specific method used for subtracting point sources
    (PS) and Galactic foregrounds (FG). Note that both the MMF
    and the ILC methods have a built-in way for subtracting both point
    sources and diffuse foregrounds, by treating them as additional
    noise (of astrophysical origin) correlated across the channels.
    Note, also, that the study is made only on clusters at
    $|b|>20$~degrees Galactic latitude for all methods.
    The eighth column summarizes the main characteritics of each algorithm in terms of yield at 90\% purity and photometric accuracy.}
    \centering
  \begin{tabular}{c|cccc|cc|c}
    \hline
    \hline
    Method 	& Shape matching & CPU      & Patches 	& Number of 	& PS subtraction  	& FG subtraction  & Main characteristics \\
    		& & Time (h) & (size deg.) 	& patches     	& method    	 & method 	\\
    \hline
    MMF1         & Yes & 50-60  	& $14.6^\circ \times 14.6^\circ$  	& 640   	& -- 					& --  	&	Best yield among MMFs		\\
                       &       &             &                                                              &                           &                                           &      &       Good photometry  \\
                       \hline
    MMF2        & Yes	&    31 	& $14.6^\circ \times 14.6^\circ$  	& 371     		& --  					& -- & 	Good yield			\\
                        &       &             &                                                              &                           &                                           &      &       Good photometry  \\
                        \hline
    MMF3         & Yes	& 5     	& $10^\circ \times 10^\circ$    		& 504     		 & mask $10\sigma$ PS  	& -- 	&	Good yield		 \\
                            &       &             &                                                              &                           &                                           &      &       Good photometry  \\
                            \hline
    MMF4         & Yes	&        	& full sky          					& --  			& --   					& --  	&	Poor yield (see Sec.~\ref{sec:mmf4}) \\
                                &       &             &                                                              &                           &                                           &      &       No photometry  \\
                                \hline
    PwS           & Yes	& 5.73  	& $14.6^\circ \times 14.6^\circ$  	& 2064 		& --   					 & -- 	&	Good yield		 \\
                                &       &             &                                                              &                           &                                           &      &       Good photometry  \\
                                \hline
    BNP          & No	&  15   	& $10^\circ \times 10^\circ$ 		& 512    		& MHW 				& Subtract 857	&	Median yield  \\
                              &       &             &                                                              &                           &                                           &      &       Poor photometry  \\ 
                              \hline
    ILC1         & Yes	& 2-3    	& see caption          				& (504)     		 & --  					& --  &		Good yield		 \\
                                    &       &             &                                                              &                           &                                           &      &       Good photometry  \\
                                    \hline
    ILC2         & No	& 2-3    	& see caption          				& (504)     		 & --  					& --  	&	Best yield among ILCs 		 \\
                                      &       &             &                                                              &                           &                                           &      &       Good photometry  \\
                              \hline
    ILC3         & Yes& 24    	& full sky            					& --  			 & --  					& Template fitting & 	Poor yield  \\
                         &       &             &                                                              &                           &                                           &      &       Good photometry  \\
                         \hline
    ILC4         & Yes$^a$	&  6    	& $10^\circ \times 10^\circ$             	& 504   		 & mask $10\sigma$ PS  	& --  &	Good yield			 \\
                             &       &             &                                                              &                           &                                           &      &       Good photometry  \\
                             \hline
    ILC5         & No &  0.2  	& $11^\circ \times 11^\circ$    		& 461   		 & SExt.  				& --  &	Poor yield (see Sec.~\ref{sec:ilc5})		 \\
                      &       &             &                                                              &                           &                                           &      &       Poor photometry  \\
                      \hline
    GMCA	  & No &  4		& $10^\circ \times 10^\circ$		& 504   		& --  					 & -- 	&	Median yield	          \\
                         &       &             &                                                              &                           &                                           &      &       Good photometry  \\    	
    \hline
  \end{tabular}
  \label{tab:methods}
\end{table*}

\subsection{Matched Multifilter Methods}

The multifrequency matched filter (MMF) enhances the contrast
(signal-to-noise) of objects of known shape and known spectral
emission law over a set of observations containing correlated
contamination signals.  It offers a practical way of extracting a
SZ clusters using multifrequency maps. The method makes use
of the universal thermal SZ effect frequency dependance (assuming
electrons in clusters are non-relativistic), and adopts a spatial
  (angular profile) template.  The filter rejects foregrounds using a
  linear combination of maps (which requires an estimate of the
  statistics of the contamination) and uses spatial filtering to
  suppress both foregrounds and noise (making use of prior knowledge
  of the cluster profile).  In all cases discussed here, the adopted
  template is identical to the simulated cluster profiles, except for MMF2 on SZ Challenge v2. The MMF
  has been studied extensively by \cite{herranz:2002} and
  \cite{melin:2006}.

Three of the MMF methods tested here work with projected flat patches
of the sky, and one method works directly on the pixelised sphere.

In the first case, the full-sky frequency maps are projected onto an
atlas of overlapping square flat regions. The filtering is then
implemented on sets of small patches comprising one patch for each
frequency channel. For each such region, the nine frequency maps are
processed with the MMF. A simple thresholding detection algorithm is
used to find the clusters and produce local catalogues.  The MMF
  is applied with varying cluster sizes to find the best detection for
  each cluster.  This provides an estimate of the angular size in
  addition to the central Compton parameter.  Each algorithm explored
  its own, but similar, range of angular scales; MMF3, for example,
  runs from $\thetav=2$ to 150~arcmins.  The catalogues extracted from
individual patches are then merged into a full-sky catalogue that
contains the position of the clusters, their estimated central Compton
parameter, the virial radius and an estimation of the error in the two
later quantities. The integrated Compton parameter is derived from the
value of the central Compton parameter and the radius of the cluster.

In the following, we give relevant details specific to each
implementation of the MMF.

\subsubsection{MMF1}

The performance of the MMF1 algorithm is sensitive to the accuracy of
the evaluation of the power spectra and cross-power spectra of the
non-SZ component of the input maps. The detection is performed in two
passes, the first detecting the highest signal-to-noise SZ clusters,
and the second detecting fainter clusters after the removal of the
contribution of the brightest ones from the power spectra estimated on
the maps.

The merging of the catalogues from distinct patches is implemented
with the option of discarding detections found in the smallest radius
bin. These detections essentially correspond to spatial profiles
indistinguishable from that of a point source. This option permits
better control of the contamination by point sources, as a
disproportionate fraction of the spurious detections occur in this
bin ; despite their different spectral signature, point sources
  can occasionally pass through the filter. Using this option
  of MMF1 reduces the contamination at a given threshold (see
  Section~\ref{sec:cat-cont} for definitions of this and other
  diagnostics of catalogue content) depending on the actual profile
of the SZ clusters.

\subsubsection{MMF2}

The MMF2 algorithm follows closely the method described in
~\citet{herranz:2002}.  The method is simple and quite robust,
although the performance depends on the model assumed for the radial
profile of the clusters. For this work, a truncated multiquadric
profile similar to a $\beta$-model has been used for SZ Challenge v1 and v2.
The profile is not a good match for the simulated profile in SZ Challenge v2.
This does not affect significantly the completeness and purity of the method (as shown in Section~\ref{sec:results}) but the extracted flux is biased with respect to the input.
The family of profiles used by the algorithm can however be
adjusted.

\subsubsection{MMF3}

The MMF3 SZ extraction algorithm is an all-sky extension of the
Matched Multifrequency Filter described in ~\citet{melin:2006}. 
  It has been used for the production of the early SZ cluster sample
  \citep{PEPVIII}. In the version used for the SZ challenge, auto and
cross power spectra used by the filter do not rely on prior
assumptions about the noise, but are directly estimated from the
data. They are thus adapted to the local instrumental noise and
astrophysical contamination.

\subsubsection{MMF4}
\label{sec:mmf4}

The spherical matched and scale adaptive filters \citep{schaefer:2006}
are generalisations of the filters proposed by \citet{herranz:2002}
for spherical coordinates.  Just like their counterparts they can be
derived from an optimisation problem and maximise the signal to noise
ratio while being linear in the signal (matched filter) and being
sensitive to the size of the object. The algorithm interfaces to the
common HEALPIX package and treats the entire celestial sphere in
one pass.

The most important drawback is the strong Galactic contamination - the
filter was not optimised to deal with a Galactic cut like many
of the other algorithms, although it is in principle possible to
include that extension. The large noise contribution due to the
Galaxy is the principal reason why the performance of the filter
suffers in comparison to the approach of discarding a large fraction
of the sky.

\subsection{Bayesian Methods}

\subsubsection{PowellSnakes}

PowellSnakes (PwS) (\citet{2009MNRAS.393..681C}) is a
fast multi-frequency Bayesian detection algorithm. It analyses
flat sky patches using the ratio
\begin{equation}
\rho \equiv \frac{\Pr(H_{1}|\vec{d})}{\Pr(H_{0}|\vec{d})},
\label{eq:PwSBI_PostProbsRatio}
\end{equation}
where $H_1$ is the detection hypothesis, `\emph{There is a source}'
and $H_0$ the null hypothesis `\emph{Only background is present}'
\citep{jaynes}. Applying Bayes theorem to the above formula one gets 
\begin{equation}
\rho
=\frac{\Pr(\vec{d}|H_{1})\Pr(H_{1})}{\Pr(\vec{d}|
H_{0})\Pr(H_{0})}
=\frac{\mathcal{Z}_1}{\mathcal{Z}_0}\frac{\Pr(H_{1})}{\Pr(H_{0})},
\label{eq:PwSBI_ModelRatio}
\end{equation}

where

\begin{equation}
\mathcal{Z} =
\int{\mathcal{L}(\vec{\Theta})\,\pi(\vec{\Theta})}\,d^D\vec{\Theta}, \label{eq:PwSBI_EvidDef}
\end{equation}
is the evidence, $\mathcal{L}(\vec{\Theta})$ is the likelihood,
$\pi(\vec{\Theta})$ is the prior and $\vec{\Theta}$ a vector
representing the parameter set.

An SZ parameterised template profile of the clusters
$\vec{s}(\vec{X},\vec{A}) \equiv \xi~\tau(\vec{a},~x - X,~y - Y)$, is
assumed known and fairly representative of the majority of the
clusters according to the resolution and \emph{S/N} ratio of the
instrumental setup, where $\tau(\ldots)$ is the general shape of the
objects (beta or Arnaud et al. profile) and $\vec{a}_j$ a vector which contains the parameters
controlling the geometry of one specific element (core/scale radius, parameters of the beta or Arnaud et al. profile). 

The algorithm may be operated on either `\textit{Frequentist mode}'
where the detection step closely resembles a multi-frequency
multi-scale `\textit{Matched Filter}' or `\textit{Bayesian mode}'
where the posterior distributions are computed resorting to a simple
`\textit{Nested Sampling}' algorithm \citep{ferozhobson:2008}.

The acceptance/rejection threshold may be defined either by using
`\textit{Decision theory}' where the expected loss criterion
is minimised or by imposing a pre-defined contamination ratio. In
  the case of a loss criterion, the symmetric criterion --- `\emph{An
    undetected cluster is as bad as spurious cluster}' --- is used.

\subsubsection{BNP}

This method is described in detail in ~\citet{diego:2002}. It is based
on the maximization of the Bayesian probability of having an SZ
cluster given the multifrequency data with no assumptions about the
shape nor size of the clusters. The method devides the sky
into multiple patches of about 100 sq. degrees and performs a
basic cleaning of the Galactic components (by subtracting the properly
weighted 857 GHz map from the channels of interest for SZ) and point
sources (using a Mexican Hat Wavelet). The cleaned maps are combined
in the Bayesian estimator and the output map of Compton parameters is
derived. SExtractor is applied to the map of reconstructed Compton
parameter to detect objects above a given threshold and compute their
flux. The thresholds are based on the background or noise
  estimated by SExtractor.  In order to compute the purity, different
  signal-to-noise ratios (ranging from 3 to 10) are used to compute
  the flux. The method assumes a power spectrum for the cluster
population although this is not critical.

The main advantage of the method resides in its robustness (almost no
assumptions) and its ability to reconstruct both extended and compact
clusters. The main limitation is the relatively poor reconstruction of
the total flux of the cluster as compared to matched filters.

\subsection{Internal Linear Combination Methods}

The internal linear combination (ILC) is a simple method for
extracting one single component of interest out of multifrequency
observations.  It has been widely used for CMB estimation on WMAP data
\citep{2003ApJS..148...97B,2003PhRvD..68l3523T,2004ApJ...612..633E,2007ApJ...660..959P,delabrouille:2009,2009PhRvD..79b3003K,2009arXiv0903.3634S}.
  A general description of the method can be found in
  \citet{2009LNP...665..159D}.

The general idea behind the ILC is to form a linear combination of all
available observations which has unit response to the component of
interest, while minimizing the total variance of the output map. This
method assumes that all observations $y_i(p)$, for channel $i$ and
pixel $p$, can be written as the sum of one single template of
interest scaled by some coefficient $a_i$, and of unspecified
contaminants which comprise noise and foregrounds, i.e. 
\begin{equation}
	y_i(p) = a_i s(p) + n_i(p)
\end{equation}
where $y_i(p)$ is the observed map in channel $i$, $s(p)$ is the
template of interest (here, the SZ map), and $n_i(p)$ comprise the
contribution of both all astrophysical foregrounds (CMB, galactic
emission, point sources...) and of instrumental noise. This equation
can be recast as:
\begin{equation}
	\vec{y}(p) = \vec{a} s(p) + \vec{n}(p)
\end{equation}
 To first order, linear combinations of the inputs of the form $\sum_i
 w_i y_i(p)$ guarantee unit response to the component of interest
 provided that the constraint $\sum_i w_i a_i = $1 is satisfied (there
 are, however, restrictions and higher order effects, which are
 discussed in detail in the appendix of \citet{delabrouille:2009} and
 in \citet{2010MNRAS.401.1602D}). It can be shown straightforwardly
 \citep{2004ApJ...612..633E,2009LNP...665..159D} that the linear
 weights which minimize the variance of the output map are:
 \begin{equation}\label{eq:weights}
 	\vec{w} = \frac{\vec{a}^t \widehat \tens{R}^{-1}}{\vec{a}^t
          \widehat \tens{R}^{-1} \vec{a}} 
\end{equation}
where $\widehat{R}$ is the empirical covariance matrix of the observations.
What distinguishes the different ILC implementations is essentially
the domains over which the above solution is implemented. 

\subsubsection{ILC1}

The needlet ILC method works in two steps. First, an SZ map is
produced by internal linear combination, with a needlet space
implementation similar to that of \citet{delabrouille:2009}.  The use
of spherical needlets permits the ILC filter to adapt to local
conditions in both direct (pixel) space and harmonic space. Input maps
include all simulated Planck maps, as well as an external template of
emission at 100 microns \citep{sfd:1998}, which helps subtracting dust emission.  The
cluster catalogue is then obtained by matched filtering on small
patches extracted from the needlet ILC SZ full-sky map, as
described in ~\citet{melin:2006}.

\subsubsection{ILC2}

The ILC2 approach relies on the same processing for photometry,
cluster size, and signal-to-noise estimates as in ILC1, but in
this case the detection of cluster candidates is made using SExtractor
on a Wiener-filtered version of the ILC1 map.

While the extraction of the ILC map works on full sky maps, using the
needlet framework to perform localized filtering, here again the
detection of cluster candidates, and the estimation of size and flux,
are performed on small patches (obtained by gnomonic
projection).

\subsubsection{ILC3}

In this method, a filter in the harmonic domain is applied to
construct a series of maps that are sensitive to the range of cluster
scales. We used a Mexican-hat filter constructed from two
  Gaussians, one with 1/4 the width of the other.  A list of cluster
candidates is compiled using a peak finding algorithm, which
  searches for enhanced signal levels in the individual map by
 fitting cluster model parameter. We employed the
  $\beta$-model profile with $\beta=2/3$ convolved with the Planck
  beams. The catalogue produced then includes as parameters the
cluster location, the flux, and the size estimate. The errors on
  these parameters are incorporated as given by the likelihoods of
the fit.

Additional improvements to the method can be achieved by using more
optimal foreground estimators (but probably with slower
convergence). This is useful especially for the fainter clusters, or
those confused to a high degree by source contamination.

\subsubsection{ILC4}

The ILC4 method is a standard ILC in Fourier space performed on
the square patches. It is implemented independently in annuli in
wave number (modulus of the Fourier mode) by applying weights,
  according to Equation~\ref{eq:weights}, this time in the Fourier
  domain. The cluster detection is performed using SExtractor
on the reconstructed SZ map.  The flux estimation is performed on the
original multifrequency maps (small patches) using the MMF at the
position of the SExtractor detections.

\subsubsection{ILC5}
\label{sec:ilc5}
This algorithm is designed to work on local noisy multichannel
maps in the wavelet domain. The representation of galaxy clusters in
an appropriate biorthogonal wavelet basis is expected to be sparse
compared to the contributions of other astrophysical components.
  This should ease the subsequent SZ separation using the ILC method
at each wavelet scale. The reverse wavelet transform is then applied
to estimate local SZ-maps. The latter are convolved using a Gaussian
beam of 5 arcmin FWHM to reduce the noise prior to cluster detection using SExtractor with
a threshold fixed classically to a multiple of the rms noise (signal-to-noise ratio ranging from 3 to 6). The
brightest IR galaxies and radiosources are masked to reduce
contamination. Finally, multiple detections due to the overlap of
local maps are removed.  Multiscale ILC proved to be more efficient
than regular ILC to remove large angular scale contamination on local
map simulations.
However, the cluster catalogue appears comparatively more
contaminated which may be due to an imperfect cleaning of multiple
detections. Also, the known point sources were only masked before
detection with SExtractor. Masking the observed sky maps earlier could
improve the component separation using Multiscale ILC. Doing so
may prevent a few very bright pixels biasing the
ILC parameter estimation but in turn raises the question of data
interpolation across masked regions. The other implementations of the ILC do mask point sources
before combining the maps and are thus not subject to this bias.

\subsection{GMCA}

Generalized morphological component analysis is a blind source
separation method devised for separating sources from instantaneous
linear mixtures using the model given by: ${\bf Y} = {\bf AS} +
{\bf n}$. The components ${\bf S}$ are assumed to be sparsely represented
(i.e. have a few significant samples in a specific basis) in a
so-called sparse representation ${\bf \Phi}$ (typically
wavelets). Assuming that the components have a sparse representation
in the wavelet domain is equivalent to assuming that most components
have a certain spatial regularity. These components and their spectral
signatures are then recovered by minimizing the number of significant
coefficients in ${\bf \Phi}$~:
\begin{eqnarray}
\mbox{min}_{\{{{\bf S},{\bf S}}\}} \lambda \|{\bf S}{\bf \Phi}^T\| +
\frac{1}{2}\|{\bf Y} - {\bf AS}\|_2^2 
\end{eqnarray}
where $||\ldots||_2$ is the $L_2$ (Euclidean) norm.
In \citet{bobin:2008}, it was shown that sparsity enhances the
diversity between the components thus improving the separation
quality. The spectral signatures of CMB and SZ are assumed to be
known. The spectral signature of the free-free component is
approximately known up to a multiplicative constant (power law with
fixed spectral index).

Hence, GMCA furnishes a noisy SZ map in which we want to detect the SZ
clusters.  This is done in three steps:

\begin{itemize}
\item wavelet denoising
\item SExtractor to extract the clusters from the
  noise-free SZ map, and finally
\item a maximum likehood to get the flux of the detected sources.
\end{itemize}

\section{Results}
\label{sec:results}
We evaluated each extracted catalogue in terms of catalogue
  content and photometric recovery based on comparisons between the
extracted catalogues and the simulated input SZ cluster catalogue.
For this purpose, we cut the input catalogue at $Y>5\times
10^{-4}$~arcmin$^2$, well below the theoretical Planck detection
limit (see below), and restrict ourselves to the high latitude sky at
$|b|>20$ degs to reduce contamination by galactic foregrounds.
We then cross-match the candidate cluster in a given catalogue
to a corresponding input cluster.  Each match results in a true
  detection, while candidates without a match are labeled as {\em
  false detections}.  In a second step, we compare the extracted
properties, namely SZ Compton parameter and size, of the true
detections to the input cluster properties.

Angular proximity was the only association criterion used for the
matching.  Specifically, we matched an extracted cluster to an input
cluster if their separation on the sky $\theta <
\theta_{max}=f(\thetav)$, a function of the {\em true} angular virial
radius of the (input) cluster.  The function $f(\thetav)$ varied over
three domains: $f(\thetav) = 5$~arcmins for $\thetav< 5$~arcmins;
$f(\thetav) = \thetav$ for 5~arcmins~$<\thetav< 20$~arcmins; and
$f(\thetav) = 20$~arcmins for $\thetav> 20$~arcmins.

We first focus on the catalogue {\em completeness} and {\em
  purity}, both of which we define immediately below. We then
test the accuracy of the recovered flux and size estimates, as well as
each algorithm's ability to internally evaluate the uncertainties on
these photometric quantities. Since many fewer codes ran on the
  Challenge v2, we focus mainly on Challenge v1.  We include results
  for those codes that did run on Challenge 2 to gauge the influence
  of the underlying cluster model used for the simulation.

\subsection{Catalogue Content}
\label{sec:cat-cont}

A useful {\em global} diagnostic is the curve of {\em yield} versus
{\em global purity} for a given catalogue (see e.g.~\citet{pires:2006}).
The former is simply the total number of clusters detected and the
latter we define as $1-\GContam$, where $\GContam$ is the global
contamination rate:
\begin{equation}
\GContam \equiv \frac{{\rm total\ number\ of\ false\ detections}}{{\rm
    total\ number\ of\ detections}} 
\end{equation}
The yield curve is parametrized by the effective detection threshold
of the catalogue construction algorithm.  It is a global diagnostic
because it gauges the total content of a catalogue, rather than its
content as a function of flux or other measurable quantities.

\begin{figure*}
\centering
\includegraphics[scale=0.45]{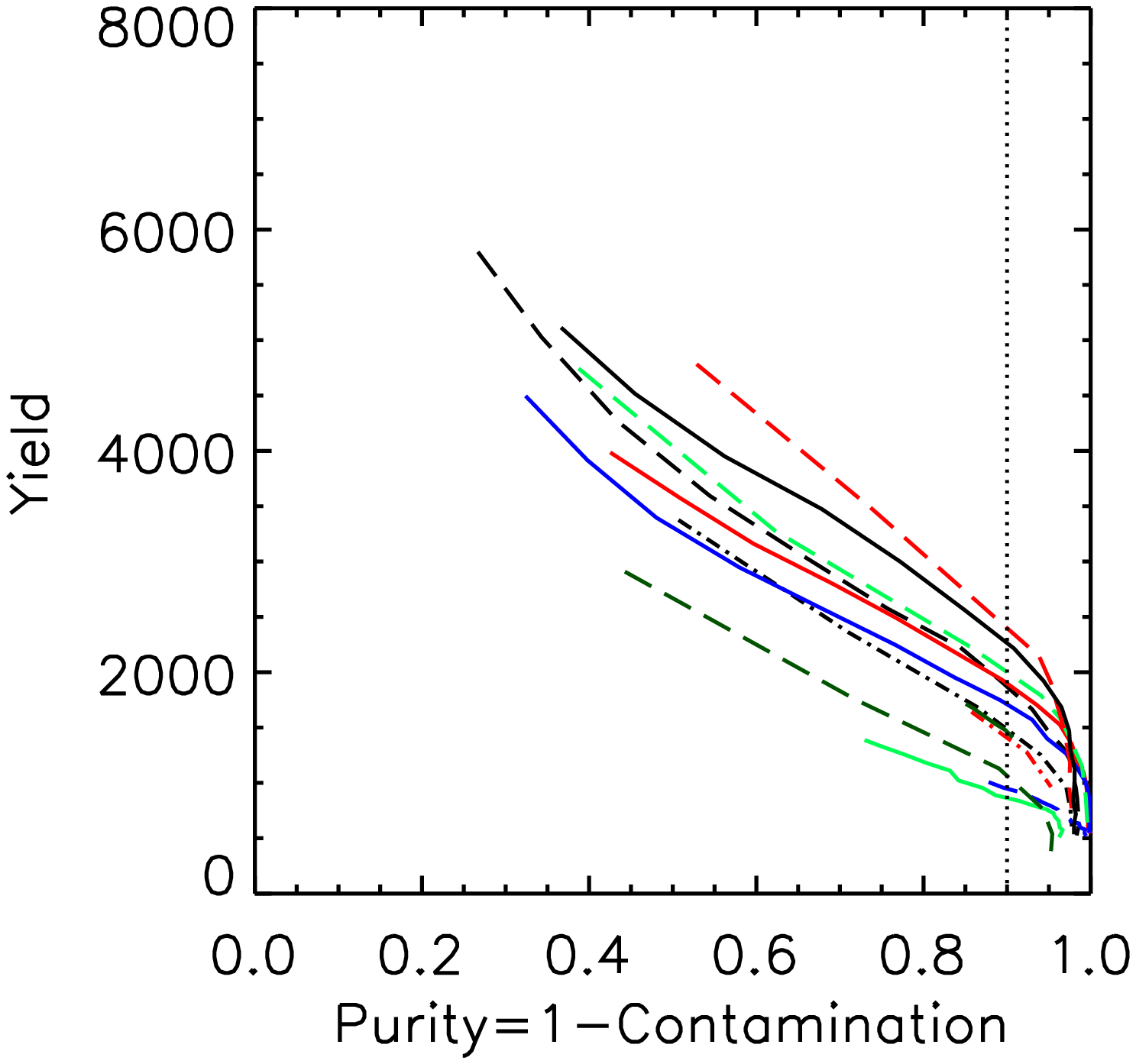}
\includegraphics[scale=0.35]{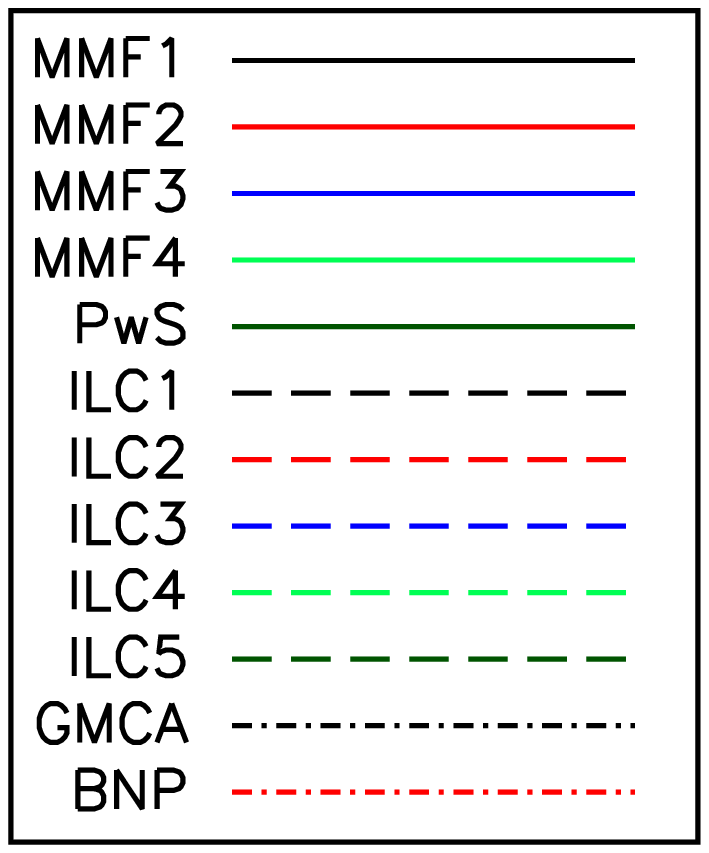}
\includegraphics[scale=0.45]{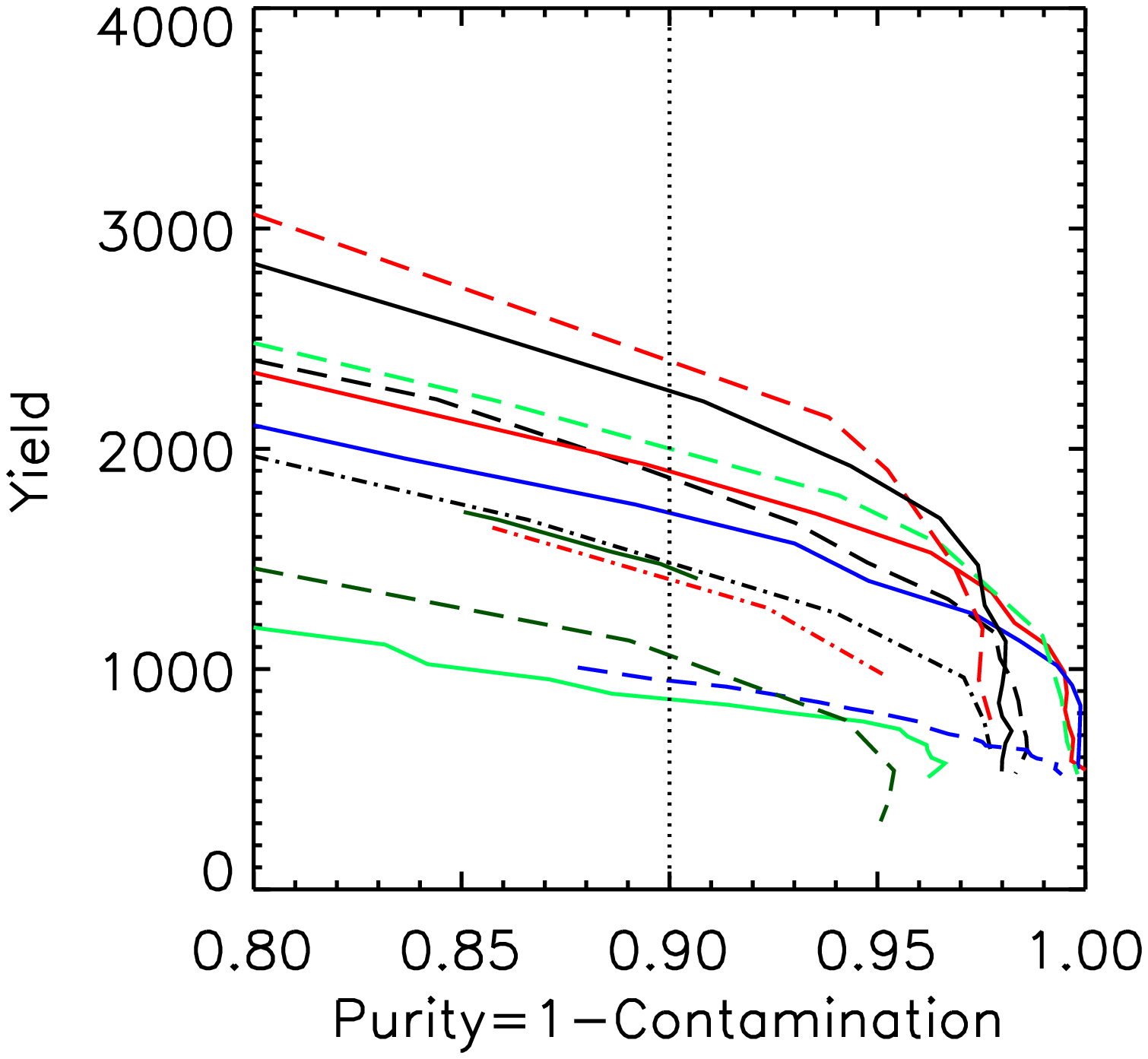}
\caption{For SZ Challenge v1: Yield as a function of global
  purity. The right handside panel is a zoom on the high--purity
  region.  Each curve is parameterized by the detection threshold of
  the corresponding algorithm.  As discussed in the text, the overall
  value of the yields should be considered with caution, due to
  remaining modeling uncertainties (see text).  We focus instead on
  relative yield between algorithms as a measure of performance.}
\label{fig:YieldVSPurity}
\end{figure*}

\begin{figure*}
\centering
\includegraphics[scale=0.45]{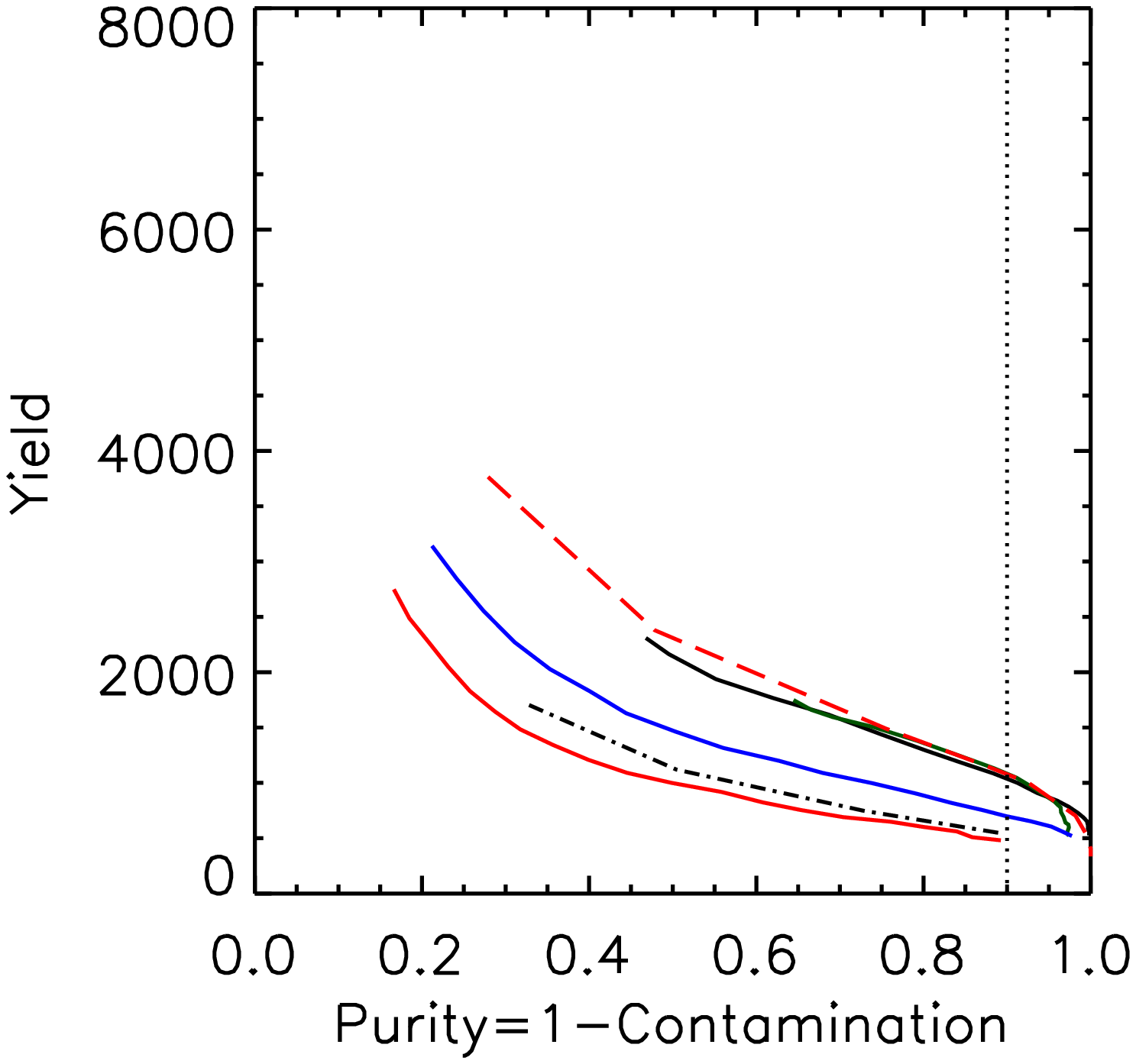}
\includegraphics[scale=0.35]{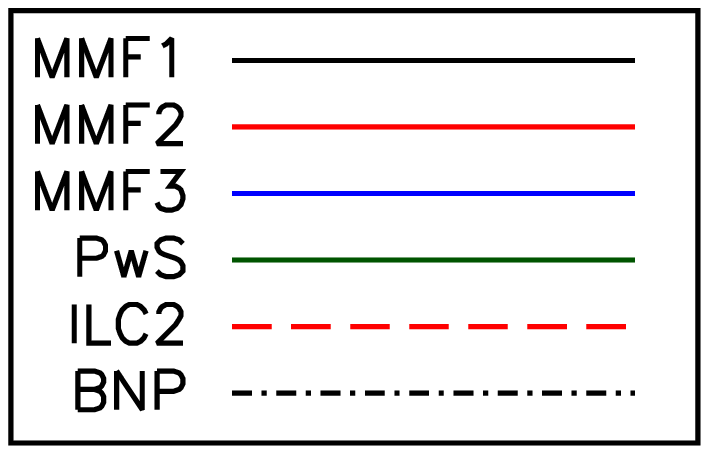}
\includegraphics[scale=0.45]{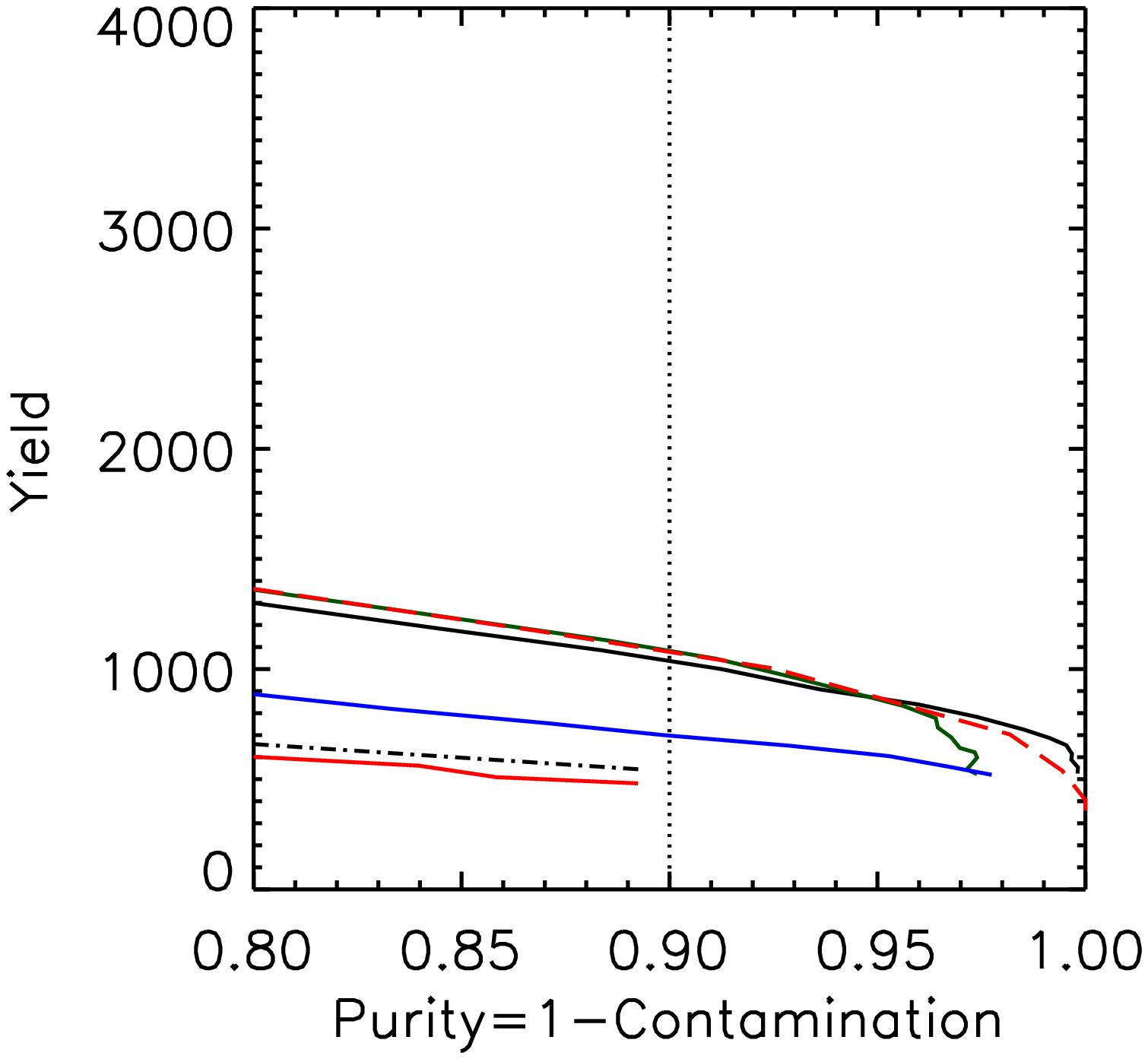}
\caption{For SZ Challenge v2: Yield as a function of global
    purity.  The same comment applies concerning the overall yield
  values; in particular, the cluster model changed significantly
  between the versions v1 and v2 of the Challenge resulting in
  lower overall yields here.  Fewer codes participated in the
  Challenge v2 (see text).}
 \label{fig:YieldVSPurity_run2}
\end{figure*}

Figure~\ref{fig:YieldVSPurity} compares the yield curves of
  outputs of all the
algorithms in the SZ Challenge v1, and
Figure~\ref{fig:YieldVSPurity_run2} those for the SZ Challenge v2.
Increasing the detection threshold moves a catalogue along its curve to
higher purity and lower yield.  Algorithms increase in performance
towards the upper right--hand corner, i.e., both high yield and high
purity.

\begin{figure*}
\centering
\includegraphics[scale=0.6,
  angle=90]{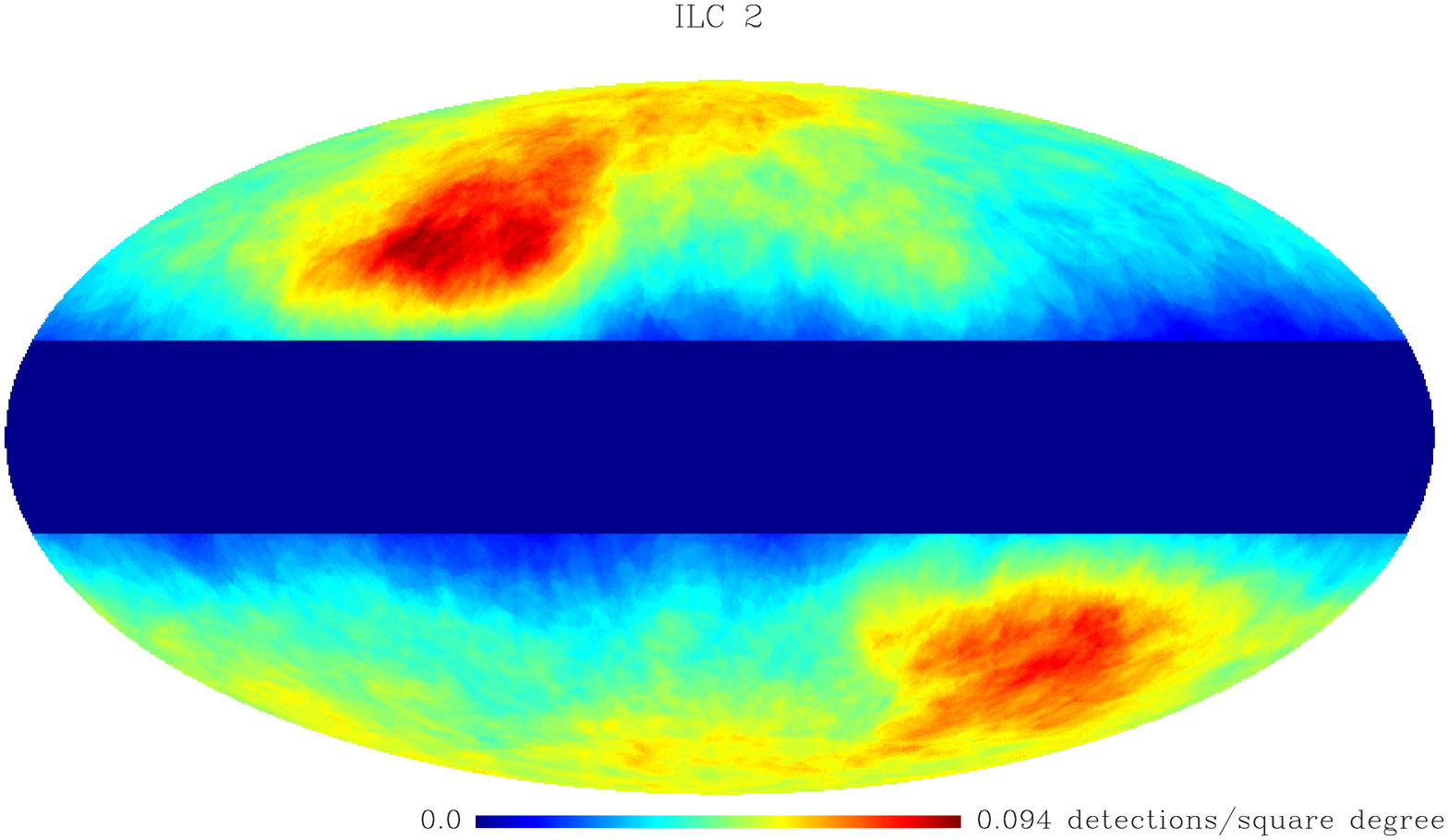} 
\includegraphics[scale=0.6, angle=90]{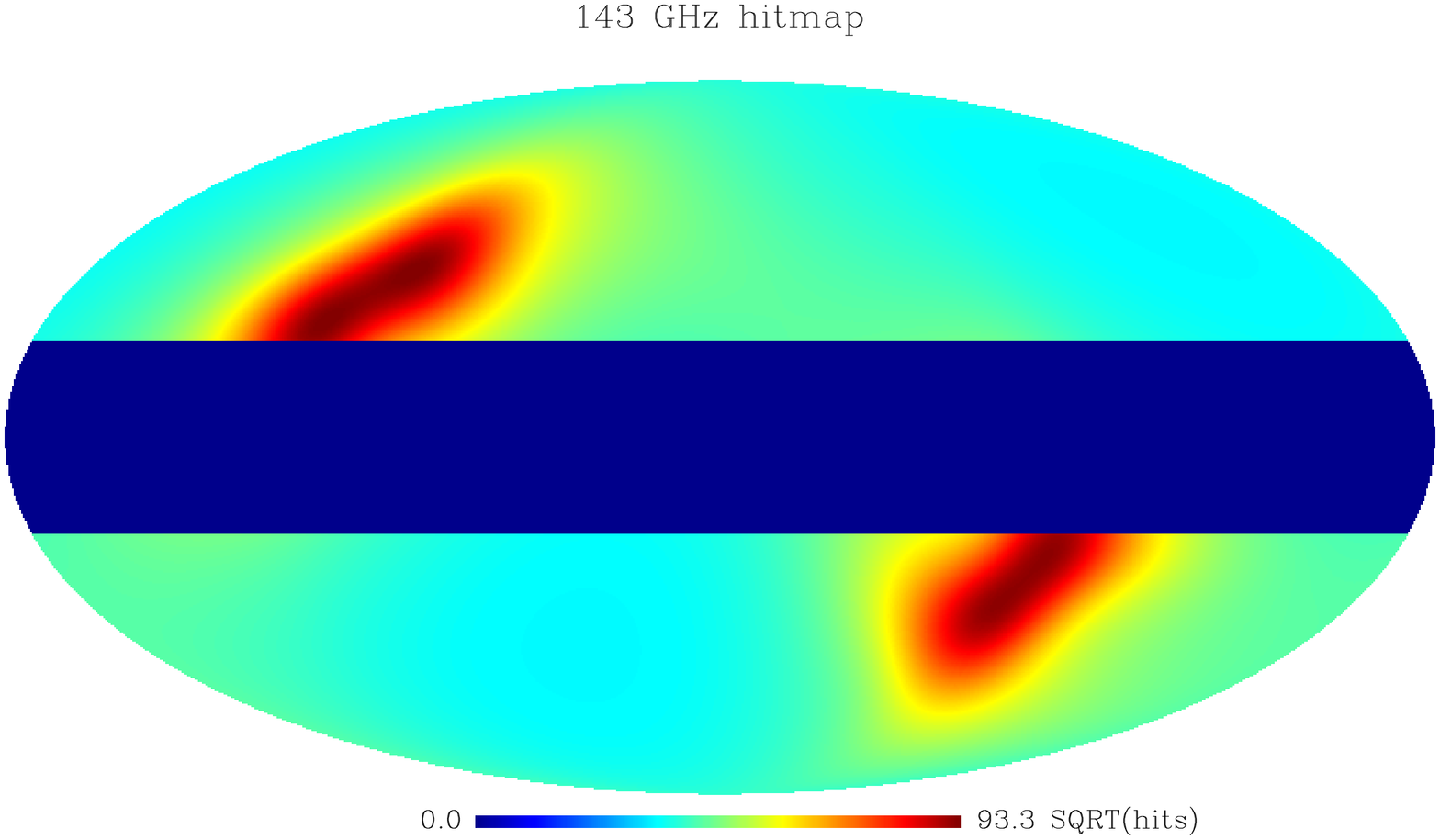}
\caption{Illustrated for ILC2 on the SZ Challenge v1, the
    detection density (top panel) is compared to the pixel hit count
  for the map at 143~GHz (bottom panel).  The noise in the simulated
  Planck maps scales as $1/\sqrt{hits}$. At high Galactic latitudes,
  the detection density clearly anticorrelates with map noise.  Both
  maps are smoothed on a scale of 20 degrees, and the Galactic plane
  ($|b|<20$ deg) is masked. }
 \label{fig:detdens}
\end{figure*}

As to be expected, algorithms that locally estimate the noise
(both instrumental and astrophysical), i.e. on local patches of a few
square degrees, perform much better than those that employ a global
noise estimate, such as MMF4 and ILC3.  For those methods with local
noise estimation, we note that their effective survey depth appears to
anticorrelate with the {\em instrumental} noise, indicating that
astrophysical confusion is effectively removed.  This can be seen in
Figure~\ref{fig:detdens}, which compares the density of detected SZ
sources (top panel) to the pixel hit count (bottom panel). The
  result illustrated with one single method, ILC2 run on the SZ
  challenge v1, holds for the other algorithms. The cluster detection
  limit appears to be primarily modulated by the instrumental noise at
  high Galactic latitude, as opposed to foreground emission.

Less expected, perhaps, is the fact that all algorithms tend to
  miss nearby clusters.  These are extended objects, and
  although they have large total SZ flux, these clusters are
``resolved--out'' -- missed because of their low surface brightness.
This is an extreme example of resolution effects expected in the case
 of SZ detection in relatively low resolution experiments like
  Planck. It is not related to the foreground removing efficiency since the effect
  can also be mimicked in simulations including only instrumental white noise. Apart from the few clusters that are fully resolved, many
have angular sizes comparable to the effective beam, and this leads to
a non--trivial selection function \citep{white2003, melin:2005,
  melin:2006}.

  We emphasize that the numerical values of the yields depend on
  the cosmological model, on the foreground model and on the cluster
  model used in the simulations. They must be considered with caution
  because of the inherent modeling uncertainties. As for the
  foreground model, the templates used to model Galactic components in
  the PSM were chosen so that they are reasonably representative of the
  complexity of the diffuse galactic emission. Thanks to many new
  observations in particular in the IR and submm domain
  (\cite{Lagache2007}, \cite{Viero2009}, \cite{Hall2010},
  \cite{Amblard2011}, \cite{PEPXVIII}), the models of
  point sources have evolved very much between the beginning of the SZ
  challenge and the publication of these results. These updates were
  not taken into account in the PSM when the study was performed.
  Moreover, the cluster model in challenge v1 was based on the
  isothermal $\beta$-model, while v2 employed a modified NFW pressure
  profile favoured by X--ray determinations of the gas pressure \citep{arnaudetal:2009} with a normalization of the $Y-M$ relation lower by
  $\sim 15\%$ than in v1. Finally, $\sigma_8$ changed from $0.85$ in
  challenge v1 to $0.796$ in v2 which strongly influences the total
  cluster yield. 

The more peaked profile actually improves detection efficiency, while
the lower normalization reduces the predicted yield.  Along with the
much lower value of $\sigma_8$, the net result is that the yields in
challenge v2 are noticeably lower than in v1 as seen in Figs.
\ref{fig:YieldVSPurity} and \ref{fig:YieldVSPurity_run2}. We thus only
discuss, in this study, the relative yields of the codes as a gauge of
performance treating the absolute value of the yield with caution.

Focusing on the relative merit of the algorithms, we see that
Figures~\ref{fig:YieldVSPurity} and \ref{fig:YieldVSPurity_run2}
display large dispersion in the yield at a given purity.  This
  reflects of course the intrinsic performance of the algorithms, but
  also for the detection methods that share similar underlying
  algorithms, e.g. MMF and ILC, the dispersion in the yield reflects
  the differences in implementations (e.g. noise estimation,
  de--blending, etc).

\begin{figure*}
\centering
\includegraphics[scale=0.75]{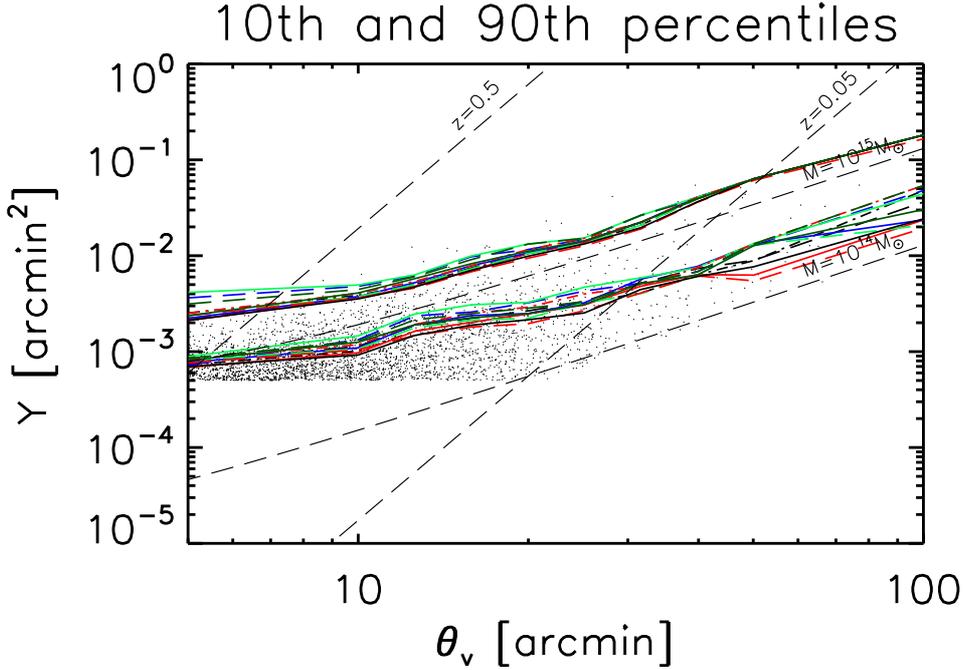}
\caption{Selection curves for each algorithm in the true $Y$ -- true
  $\theta_v$ plane for SZ Challenge v1.  The lower set of curves indicate the 90th
  percentiles, i.e., the curve above which lie 90\% of the detected
  clusters; the upper set corresponds to the 10th percentile. The
    color codes are as in Figure~\ref{fig:YieldVSPurity}. We see that
  the Planck selection function depends not just on flux, but also on
  cluster angular size.  Many clusters are at least marginally
  resolved by Planck, leading to these size effects in the selection
  fuction.  The dashed lines show contours of fixed mass and redshift,
  as indicated, while the cloud of points shows the distribution of
  the input catalogue(in fact a subsample of 1 in 4 randomly selected
  input clusters).  We see that small variations in selection curves
  generate significant yield changes.  }
 \label{fig:YVStv}
\end{figure*}

Somewhat deceptively, these yield variations correspond to only minor
differences in detection threshold, as illustrated for the
  SZ Challenge v1 in Figure~\ref{fig:YVStv}. This figure traces the
curves above which 90\% and 10\% (lower set and upper set
  respectively) of the clusters detected by each method lie in the
{\em true} $Y$ -- {\em true} $\theta_{\rm v}$ plane.  As already
mentioned, many clusters are marginally resolved (sizes at least
comparable to the beam), which means that detection efficiency depends
not only on flux, but also on size\footnote{This was shown on
  real data in \cite{PEPVIII}}.

The algorithms all have similar curves in this plane. This means that the
differences in yield are due to only small variations of the selection
curve since completeness is expected to be monotonic.  The black points represent a random sample of 1/4 of the input
clusters and show where the bulk of the catalogue lies.  These small
variations have important consequences for cosmological interpretation
of the counts and hence must be properly quantified.

 We compute for all the methods and in the SZ challenges v1 and v2
  the {\em completeness} defined as the ratio {\em [true detections
      (recovered clusters)/simulated clusters]} over bins of true
  (simulated) flux $Y$. We find it varies from 80 to 98\% at $Y_{\rm
    lim}=10^{-2}$ arcmin$^2$ for the direct methods (based on frequency
  maps). The {\em completness} is of order of 80\%
  at the same $Y_{\rm lim}$ for the indirect methods (based on
  detections in SZ maps). We note a slight increase of the completeness
  from the SZ Challenge v1 to v2. We also estimate the contamination
  of the output catalogues defined as the ratio {\em [false
      detections/total detections]}. This is evaluated as as a
  function of recovered flux. The average contamination of the output catalogues ranges between 6 and
  13\% both in the case of the challenge 1 and 2. However, the purity
  with respect to the $Y$ bins differ significantly from method to
  method. As a general trend, the lowest $Y$ bin, i.e. the smallet
  recovered fluxes centered around $10^{-3}$ arcmin$^2$, is more
  contaminated ($\sim$75\% on average) in the case of the indirect method
  than in the case of direct methods ($\sim$50\% on average).
  
\subsection{Photometric and Astrometric Recovery}
Cluster characterization is a separate issue from detection.  It
involves determination of angular positions as well as photometry.
Since Planck will marginally resolve many clusters, photometry here
means both flux $Y_{rec}$ and characteristic size measurements.
Moreover, each method should provide an estimate of the errors on
these quantities for each object in the catalog.

\begin{figure*}
\centering
\includegraphics[scale=0.7]{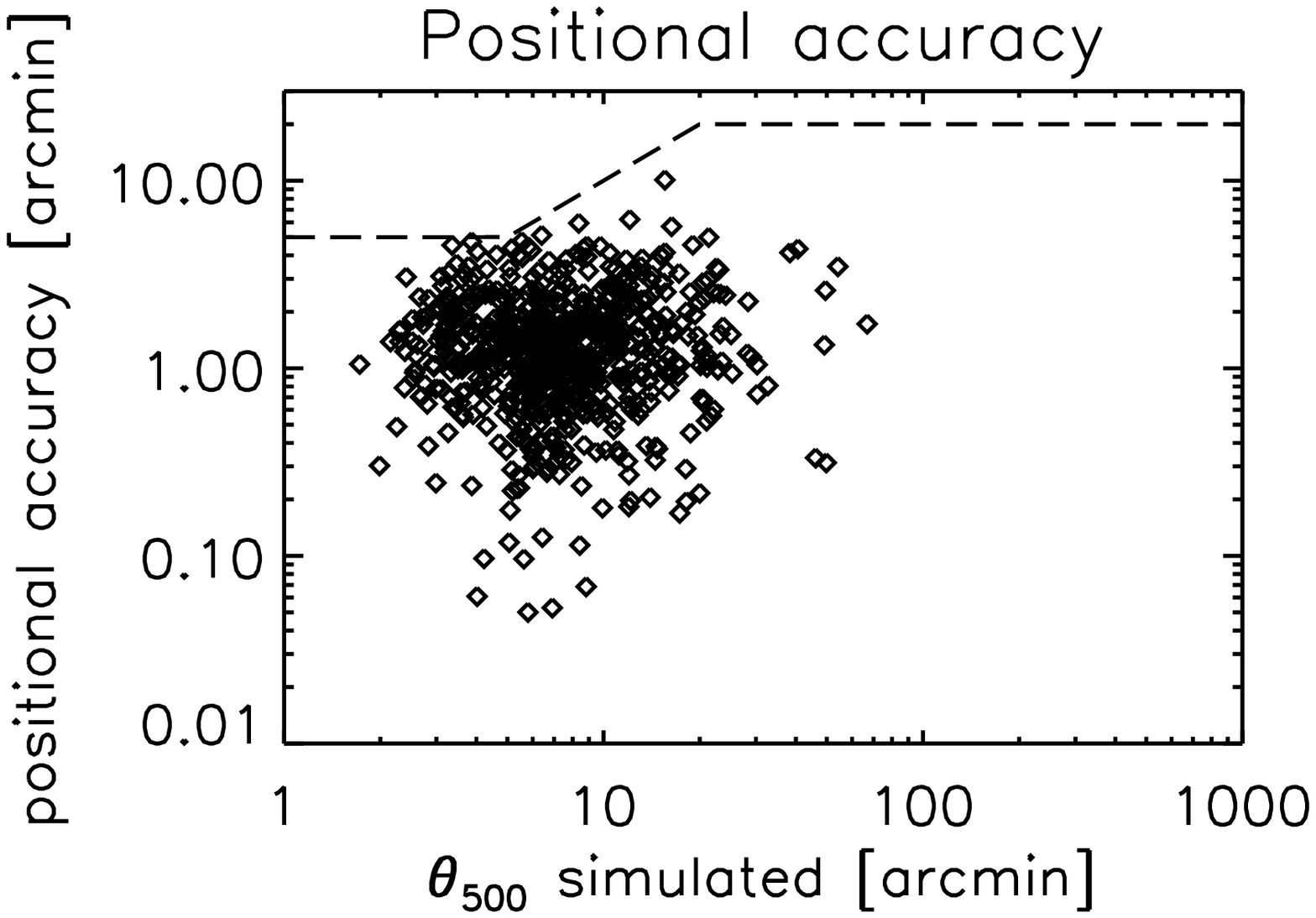}
\caption{Positional accuracy of the recovered clusters illustrated for
  MMF3 in the SZ challenge v2.}
 \label{fig:pos}
\end{figure*}

We illustrate in Fig. \ref{fig:pos}, a scatter diagram of
positional offset for MMF3
as a function of {\em true} cluster size, $\theta_{\rm v}$.  On average, all the
  algorithms perform similarly and recover cluster position to $\sim
  2$~arcmins with a large scatter.  In addition, we see
  that it is more difficult to accurately determine the position of
  intrinsically extended clusters, as shown by the fact that the
  cloud of points is elongated and inclined.  

\begin{figure*}
\centering
\includegraphics[scale=0.7]{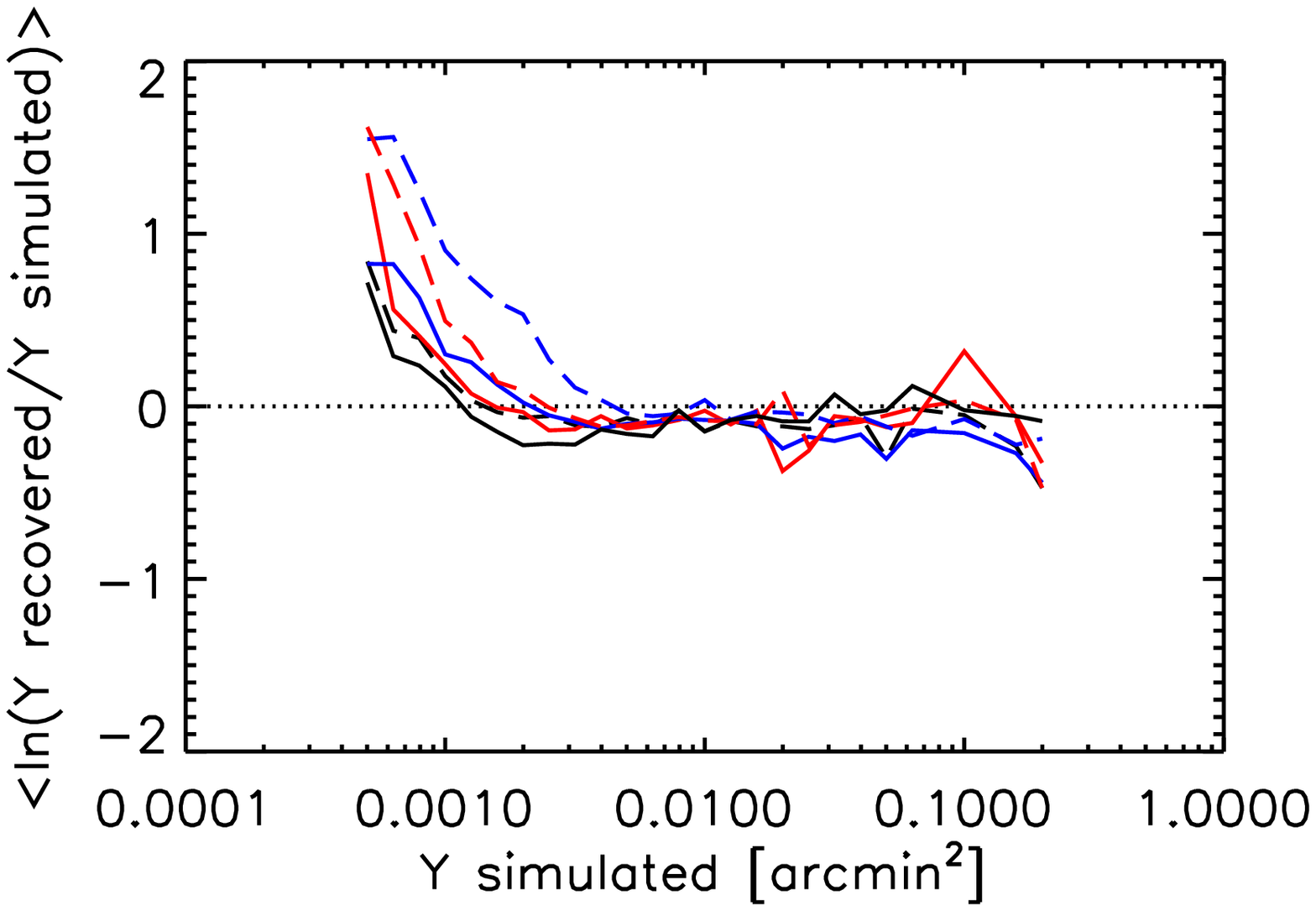}
\includegraphics[scale=0.5,clip=true,trim=0 100 0 0]{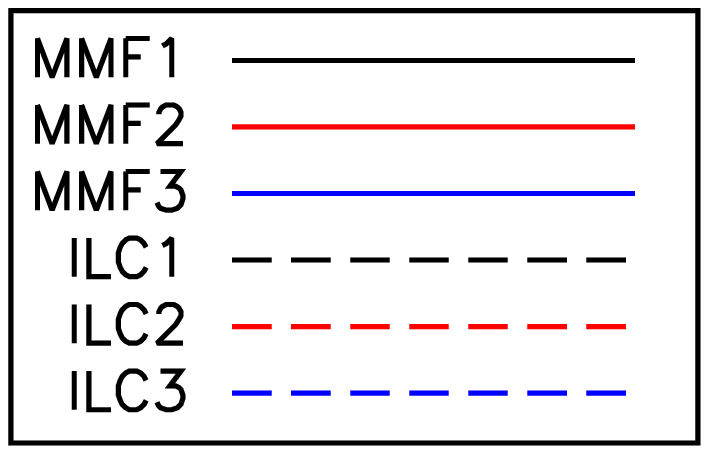}
\caption{Flux recovery bias.  The Figure shows for a subset of methods
  the average recovered $Y_{\mathrm rec}$, normalized to the true
  input $Y$, as a function of $Y$. At the bright end, most codes
  extract an unbiased estimate of cluster flux, while the expected
  Malmquist bias appears at the faint-end, just below $Y\sim 2\times
  10^{-3}$~arcmin$^2$.}
 \label{fig:YvsYrecov}
\end{figure*}

\begin{figure*}
\centering
\includegraphics[scale=0.7]{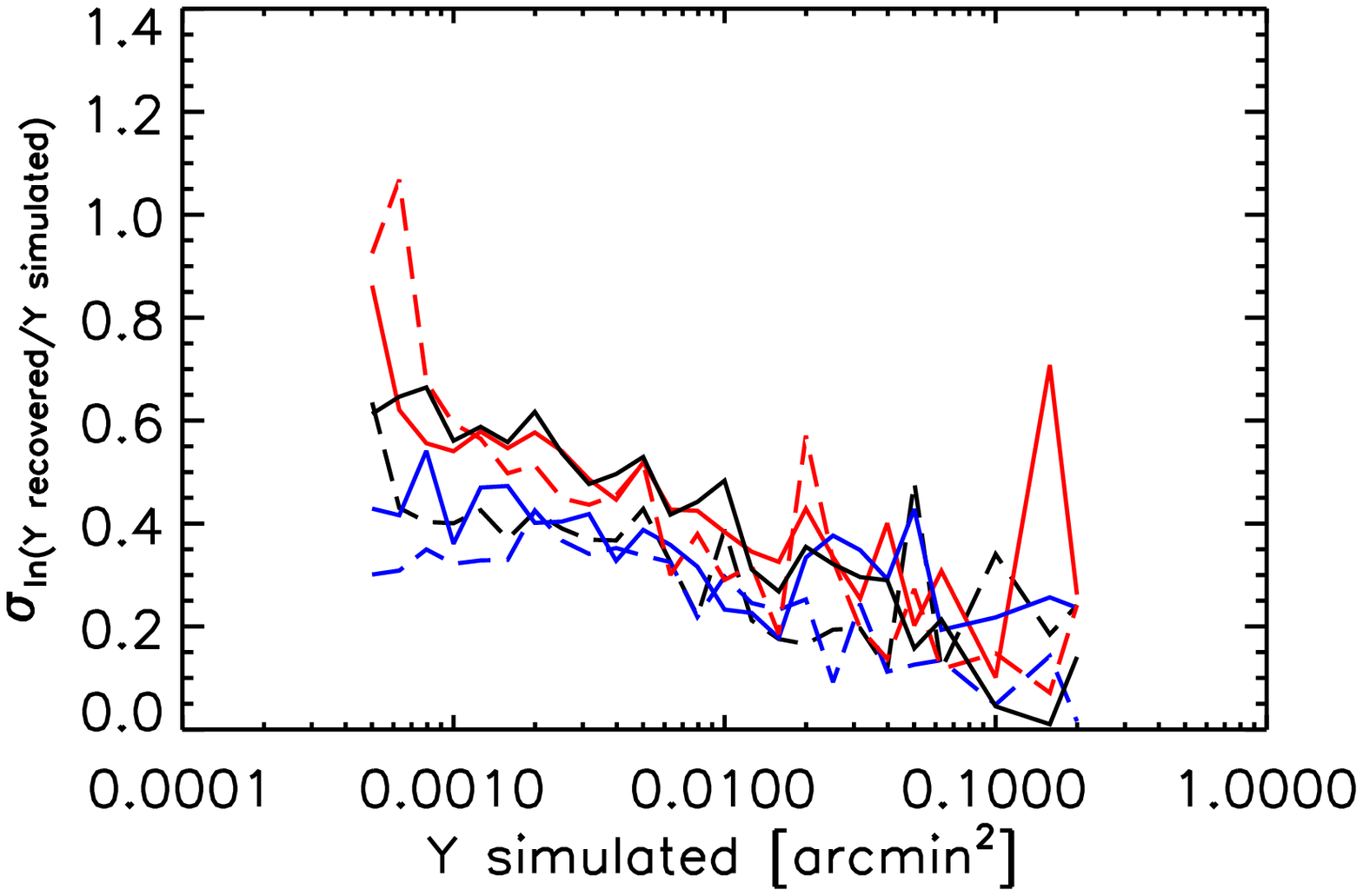}
\includegraphics[scale=0.5,clip=true,trim=0 100 0 0]{Figs/cy_bias_sigma_legend_v5.eps}
\caption{Flux recovery uncertainty for the subset of methods given in
  Figure~\ref{fig:YvsYrecov}.  The Figure shows the dispersion in
  measured flux about the true input $Y$ flux as a function of the
  true input flux. The best algorithms are affected by $\sim 30$\%
  dispersion.  }
 \label{fig:sigYvsY}
\end{figure*}

Concerning photometry, we show in Figure~\ref{fig:YvsYrecov} and
\ref{fig:sigYvsY} the mean recovered SZ flux, $Y_\mathrm{rec}$,
normalized to the true (simulated) flux, $Y$, as a function of the
latter. Only true detections are used in this comparison. We
  illustrate our results for a subset of methods
   namely MMF1, MMF2, MMF3, ILC1, ILC2, ILC3. Methods that
  filter the maps such as ILC4 and GMCA, or that use the SExtractor
  photometry such as BNP and ILC5 exhibit significant bias in flux
  recovery at the bright end. At the faint end, we see the appearance
of Malmquist bias as an upturn in the measured flux.  The importance
of this bias varies from algorithm to algorithm.

Figure~\ref{fig:sigYvsY} gives the dispersion in the recovered flux
$\sigma_{Y_\mathrm{rec}}$ as a function of true $Y$ (once again, only
involving true detections). Here, we see that some codes perform
significantly better than others.  Those that adjust an SZ profile to
each cluster outperform by a large margin those that do not.
Photometry based on SExtractor, for example, fares much worse than the
MMF codes. Even among the best performing codes, however, the
intrinsic photometric dispersion is of order 30\%.  This is important,
because we expect SZ flux to tightly correlate with cluster mass, with
a scatter as low as $\sim 10\%$ as indicated by both numerical
simulations \citep{dasilva:2004, motl:2005, kravtsov:2006} and recent
X--ray data \citep{arnaud:2007, nagai:2007}.  The SZ flux hence should
offer a good mass proxy.  What we see from this figure, however, is
that the observational scatter will dominate the intrinsic scatter of
this mass proxy and needs to be properly accounted for in the
cosmological analyses.

We have attributed the origin of the photometric scatter to
difficulty in determining cluster size. Although methods
  adjusting a profile to the SZ are able to estimate the size of many
  clusters, they do so with significant dispersion. Furthermore, this
  issue arises specifically for Planck-like resolution because a
  large number of clusters are only marginally resolved. Imposing the
  cluster size, for example from external data, such as X--ray or optical
  observations or higher resolution SZ measurements, would
  significantly reduce the observational scatter.


\section{Discussion and conclusion}

In the present study, we compare different codes and algorithms
  to detect SZ galaxy clusters from multi-wavelength experiments using 
Plank's instrumental characteristics. These methods
may be usefully divided into {\it direct methods} (four
matched--filter approaches and PowellSnakes) using individual
frequency maps, and {\it indirect methods} (five ILC methods, GMCA and
BNP) that first construct an SZ map in which they subsequently
search for clusters.

As already emphasized, the global yield values of all methods must be
considered with caution because of inherent modeling uncertainties of
the sky simulations and cluster models used, and to the underlying
cosmological model. Therefore, we focus on relative yields as a gauge
of performance of the algorithms. It is worth noting that results of a
direct or indirect method significantly vary (within factors of as
much as three) with the details of their implementation, with clear
impact on the survey yield as demonstrated in
Figures~\ref{fig:YieldVSPurity} and \ref{fig:YieldVSPurity_run2}.
Using the PSM simulations and including th noise as described in
\ref{sec:PSM}, we would expect of order of 1000-2000 clusters at a
purity of $\sim 90$\% with $|b|>20$ deg. This number depends on the
extraction method used and may vary with a more detailed modeling of
the sky.  The cluster yield can be increased by accepting a higher
contamination rate and calling for extensive follow-up to eliminate
false detections a posteriori.
 
The indirect methods seem to offer greater opportunity for
optimization with a larger number of tuning parameters. They are also
less model dependent for the SZ map construction and the cluster
  detection. Although, they can be coupled with matched filters for
  the SZ flux measurement.  In turn, the direct methods are linear,
easy to implement and robust.  One of their advantages relies in
  the fact that they can be optimized to detect objects of a given
  shape (SZ profile) and and a given spectral energy distribution -
  SED - (SZ spectrum). This characteristic of the direct method is
  particularly important for Planck-like multi-frequency surveys with
  moderate resolution. Indeed, due to lack of resolution, spurious
  sources (galactic features or point sources) may be detected by the
  spatial filter and in that case the frequency coverage and the
  spectral matching is the best strategy to monitor the false
  detections. Due to their robustness and easy implementation
  direct methods are more adapted to run in pipelines.\footnote{
      MMF1, MM2, MMF3 and PwS were all implemented in Planck's
      pipelines. Furthermore, MMF3 was used to extract the Planck
      clusters published in the ESZ sample.} The situation is quite different
for high resolution, arcminute-scale, SZ experiments such as ACT and SPT 
where the filtering of one unique low frequency
map (where the SZ signal is negative) is sufficiently efficient to unable 
cluster detection. In
these cases though, extended clusters are not well recovered as they 
suffer more from the CMB contamination and thus from the filtering.

The comparison of different codes and cluster detection methods exhibits
selection effects and catalogue uncertainties, neither of which depend, 
for example, on the actual cluster physics model. This is shown in the 
selection curves Fig.~\ref{fig:YVStv}. In a Planck-like case, with 
moderate resolution, clusters do not appear as
point sources, but many are resolved or have sizes comparable to the 
effective SZ beam. In view of this, the use of an adapted spatial filter
to optimally model the SZ profile provides a significant improvement
in the detection yield and in the photometry. Clusters being
partially resolved  leads to non--trivial
selection criteria that depend both on flux and true angular size, as
demonstrated by the fact that the curves in Fig.~\ref{fig:YVStv} are not
horizontal lines. 
In that respect, the use of X-ray information from the ROSAT All-Sky Survey (RASS) cluster
catalogues \citep{bohringer:2000, bohringer:2004,piffaretti2011} or from optical catalogues, e.g. in the 
SDSS area \citep{koester2007}, will be of particular value as they
will give us a handle on both flux and size of the clusters detected by Planck 
and, even more importantly, understand better the completeness in studying those which are
missed \citep{chamballu2010}.
As already stated, the exact position of each selection curve in Fig.~\ref{fig:YVStv}
depends on the algorithm and small variations in the position of this curve
produce significant changes in catalogue yield.  Most of the differences
bewteen catalogues occur at small flux and size, where the bulk of the
cluster population resides.  These objects are for the majority 
low mass, intermediate redshift clusters. Moreover, detection becomes
progressively less efficient for large objects; this is intrinsic and
hence shared by all algorithms.

Concerning individual cluster measurements we find that the 
astrometry is recovered to $\sim 2$~arcmins on average and photometry to
$\sim 30$\% for the best-behaving algorithms.  The positional error is not a problem for X-ray/optical follow-ups because Planck is expected to detect massive clusters
which will be easy to find in the XMM/Chandra/1 to 4meter class optical telescope fields of view. The photometric error in our Planck-like case
is dominated by the difficulty in accurately determining the cluster extent/size.  This is another
consequence of the fact that many clusters are marginally resolved by
Planck: large enough that their angular extent matters, but small
enough that we have difficulty fixing their true size.  One way of
reducing the photometric error is thus using external constraints on cluster
size.  One again ancillary data from RASS or optical cluster catalogues will
help in this regard, at least at low redshift ($z<0.3-0.5$) ; at higher redshift, we
will rely on follow--up observations if we want to reduce photometric
uncertainties.

The comparison of an ensemble of cluster extraction methods in the case of a
multi-frequency moderate resolution experiment shows that the optimization of the
cluster detection in terms of yield and purity, but also in terms of positional
accuracy and photometry, is very sensitive to the implementation of the code.
The global or local treatment of the noise estimate or the cleaning from point sources
are the two main causes of difference. However and most importantly, the use of as 
realistic as possible SZ profile (as opposed to model independent profile) to filter 
out the signal or to measure the fluxes
is a key aspect of cluster detection techniques in our context. In that respect, 
using external information from SZ observations or from other wavelength will
significantly help in improving the measurement of the cluster properties and in
turn optimize the catalogue yields and their selection function.

\section*{Acknowledgments}
\thanks{{The authors acknowledge the use of the Planck Sky Model
    (PSM), developed by the Component Separation Working Group (WG2)
    of the Planck Collaboration. We acknowledge also the use of the
    HEALPix package \citep{2005ApJ...622..759G}. We further thank
      M. White and S. White for helpful comments and suggestions.}

\bibliographystyle{aa}
\bibliography{SZChallenge_v15_with_appendix}

\begin{thebibliography}{85}
\expandafter\ifx\csname natexlab\endcsname\relax\def\natexlab#1{#1}\fi

\bibitem[{{Albrecht} {et~al.}(2009){Albrecht}, {Amendola}, {Bernstein},
  {Clowe}, {Eisenstein}, {Guzzo}, {Hirata}, {Huterer}, {Kirshner}, {Kolb}, \&
  {Nichol}}]{FoMSWG}
{Albrecht}, A., {Amendola}, L., {Bernstein}, G., {et~al.} 2009, ArXiv e-prints

\bibitem[{{Albrecht} {et~al.}(2006){Albrecht}, {Bernstein}, {Cahn}, {Freedman},
  {Hewitt}, {Hu}, {Huth}, {Kamionkowski}, {Kolb}, {Knox}, {Mather}, {Staggs},
  \& {Suntzeff}}]{detf}
{Albrecht}, A., {Bernstein}, G., {Cahn}, R., {et~al.} 2006, ArXiv Astrophysics
  e-prints

\bibitem[{Amblard {et~al.}(2011)Amblard, Cooray, Serra, \&
  et~al.}]{Amblard2011}
Amblard, A., Cooray, A., Serra, P., \& et~al. 2011, Nature, 1

\bibitem[{{Arnaud} {et~al.}(2007){Arnaud}, {Pointecouteau}, \&
  {Pratt}}]{arnaud:2007}
{Arnaud}, M., {Pointecouteau}, E., \& {Pratt}, G.~W. 2007, \aap, 474, L37

\bibitem[{{Arnaud} {et~al.}(2010){Arnaud}, {Pratt}, {Piffaretti},
  {B{\"o}hringer}, {Croston}, \& {Pointecouteau}}]{arnaudetal:2009}
{Arnaud}, M., {Pratt}, G.~W., {Piffaretti}, R., {et~al.} 2010, \aap, 517, A92

\bibitem[{{Bartlett}(2002)}]{bartlett:2002}
{Bartlett}, J.~G. 2002, in Astronomical Society of the Pacific Conference
  Series, Vol. 268, Tracing Cosmic Evolution with Galaxy Clusters, ed.
  {S.~Borgani, M.~Mezzetti, \& R.~Valdarnini}, 101--+

\bibitem[{{Bennett} {et~al.}(2003){Bennett}, {Hill}, {Hinshaw}, {Nolta},
  {Odegard}, {Page}, {Spergel}, {Weiland}, {Wright}, {Halpern}, {Jarosik},
  {Kogut}, {Limon}, {Meyer}, {Tucker}, \& {Wollack}}]{2003ApJS..148...97B}
{Bennett}, C.~L., {Hill}, R.~S., {Hinshaw}, G., {et~al.} 2003, \apjs, 148, 97

\bibitem[{{Bertin} \& {Arnouts}(1996)}]{1996A&AS..117..393B}
{Bertin}, E. \& {Arnouts}, S. 1996, \aaps, 117, 393

\bibitem[{{Birkinshaw}(1999)}]{birk:1999}
{Birkinshaw}, M. 1999, \physrep, 310, 97

\bibitem[{{Bobin} {et~al.}(2008){Bobin}, {Moudden}, {Starck}, {Fadili}, \&
  {Aghanim}}]{bobin:2008}
{Bobin}, J., {Moudden}, Y., {Starck}, J., {Fadili}, J., \& {Aghanim}, N. 2008,
  Statistical Methodology, 5, 307

\bibitem[{{B{\"o}hringer} {et~al.}(2004){B{\"o}hringer}, {Schuecker}, {Guzzo},
  {Collins}, {Voges}, {Cruddace}, {Ortiz-Gil}, {Chincarini}, {De Grandi},
  {Edge}, {MacGillivray}, {Neumann}, {Schindler}, \& {Shaver}}]{bohringer:2004}
{B{\"o}hringer}, H., {Schuecker}, P., {Guzzo}, L., {et~al.} 2004, \aap, 425,
  367

\bibitem[{{B{\"o}hringer} {et~al.}(2000){B{\"o}hringer}, {Voges}, {Huchra},
  {McLean}, {Giacconi}, {Rosati}, {Burg}, {Mader}, {Schuecker}, {Simi{\c c}},
  {Komossa}, {Reiprich}, {Retzlaff}, \& {Tr{\"u}mper}}]{bohringer:2000}
{B{\"o}hringer}, H., {Voges}, W., {Huchra}, J.~P., {et~al.} 2000, \apjs, 129,
  435

\bibitem[{{Bonaldi} {et~al.}(2007){Bonaldi}, {Tormen}, {Dolag}, \&
  {Moscardini}}]{bonaldietal:2007}
{Bonaldi}, A., {Tormen}, G., {Dolag}, K., \& {Moscardini}, L. 2007, \mnras,
  378, 1248

\bibitem[{{Carlstrom} {et~al.}(2002){Carlstrom}, {Holder}, \&
  {Reese}}]{carlstrometal:2002}
{Carlstrom}, J.~E., {Holder}, G.~P., \& {Reese}, E.~D. 2002, \araa, 40, 643

\bibitem[{{Carvalho} {et~al.}(2009){Carvalho}, {Rocha}, \&
  {Hobson}}]{2009MNRAS.393..681C}
{Carvalho}, P., {Rocha}, G., \& {Hobson}, M.~P. 2009, \mnras, 393, 681

\bibitem[{{Chamballu} {et~al.}(2010){Chamballu}, {Bartlett}, \&
  {Melin}}]{chamballu2010}
{Chamballu}, A., {Bartlett}, J.~G., \& {Melin}, J.~. 2010, ArXiv e-prints

\bibitem[{{Colafrancesco} {et~al.}(1997){Colafrancesco}, {Mazzotta},
  {Rephaeli}, \& {Vittorio}}]{colafrancescoetal:1997}
{Colafrancesco}, S., {Mazzotta}, P., {Rephaeli}, Y., \& {Vittorio}, N. 1997,
  \apj, 479, 1

\bibitem[{{da Silva} {et~al.}(2004{\natexlab{a}}){da Silva}, {Kay}, {Liddle},
  \& {Thomas}}]{daSilvaetal:2004}
{da Silva}, A.~C., {Kay}, S.~T., {Liddle}, A.~R., \& {Thomas}, P.~A.
  2004{\natexlab{a}}, \mnras, 348, 1401

\bibitem[{{da Silva} {et~al.}(2004{\natexlab{b}}){da Silva}, {Kay}, {Liddle},
  \& {Thomas}}]{dasilva:2004}
{da Silva}, A.~C., {Kay}, S.~T., {Liddle}, A.~R., \& {Thomas}, P.~A.
  2004{\natexlab{b}}, \mnras, 348, 1401

\bibitem[{{de Zotti} {et~al.}(2005){de Zotti}, {Ricci}, {Mesa}, {Silva},
  {Mazzotta}, {Toffolatti}, \& {Gonz{\'a}lez-Nuevo}}]{dezottietal:2005}
{de Zotti}, G., {Ricci}, R., {Mesa}, D., {et~al.} 2005, \aap, 431, 893

\bibitem[{{Delabrouille} {et~al.}(2012){Delabrouille}, {Betoule}, {Melin},
  {Miville-Desch{\^e}nes}, {Gonzalez-Nuevo}, {Le Jeune}, {Castex}, {de Zotti},
  {Basak}, {Ashdown}, {Aumont}, {Baccigalupi}, {Banday}, {Bernard}, {Bouchet},
  {Clements}, {da Silva}, {Dickinson}, {Dodu}, {Dolag}, {Elsner}, {Fauvet},
  {Fa{\"y}}, {Giardino}, {Leach}, {Lesgourgues}, {Liguori}, {Macias-Perez},
  {Massardi}, {Matarrese}, {Mazzotta}, {Montier}, {Mottet}, {Paladini},
  {Partridge}, {Piffaretti}, {Prezeau}, {Prunet}, {Ricciardi}, {Roman},
  {Schaefer}, \& {Toffolatti}}]{delabrouille2012}
{Delabrouille}, J., {Betoule}, M., {Melin}, J.-B., {et~al.} 2012, ArXiv
  e-prints

\bibitem[{{Delabrouille} \& {Cardoso}(2009)}]{2009LNP...665..159D}
{Delabrouille}, J. \& {Cardoso}, J. 2009, in Lecture Notes in Physics, Berlin
  Springer Verlag, Vol. 665, Lecture Notes in Physics, Berlin Springer Verlag,
  ed. {V.~J.~Martinez, E.~Saar, E.~M.~Gonzales, \& M.~J.~Pons-Borderia },
  159--+

\bibitem[{{Delabrouille} {et~al.}(2009){Delabrouille}, {Cardoso}, {Le Jeune},
  {Betoule}, {Fay}, \& {Guilloux}}]{delabrouille:2009}
{Delabrouille}, J., {Cardoso}, J., {Le Jeune}, M., {et~al.} 2009, \aap, 493,
  835

\bibitem[{{Dick} {et~al.}(2010){Dick}, {Remazeilles}, \&
  {Delabrouille}}]{2010MNRAS.401.1602D}
{Dick}, J., {Remazeilles}, M., \& {Delabrouille}, J. 2010, \mnras, 401, 1602

\bibitem[{{Dickinson} {et~al.}(2003){Dickinson}, {Davies}, \&
  {Davis}}]{dickinsonetal:2003}
{Dickinson}, C., {Davies}, R.~D., \& {Davis}, R.~J. 2003, \mnras, 341, 369

\bibitem[{{Diego} {et~al.}(2002){Diego}, {Vielva},
  {Mart{\'{\i}}nez-Gonz{\'a}lez}, {Silk}, \& {Sanz}}]{diego:2002}
{Diego}, J.~M., {Vielva}, P., {Mart{\'{\i}}nez-Gonz{\'a}lez}, E., {Silk}, J.,
  \& {Sanz}, J.~L. 2002, \mnras, 336, 1351

\bibitem[{{Draine} \& {Lazarian}(1998)}]{drainelazarian:1998}
{Draine}, B.~T. \& {Lazarian}, A. 1998, \apjl, 494, L19+

\bibitem[{{Dunkley} {et~al.}(2009){Dunkley}, {Komatsu}, {Nolta}, {Spergel},
  {Larson}, {Hinshaw}, {Page}, {Bennett}, {Gold}, {Jarosik}, {Weiland},
  {Halpern}, {Hill}, {Kogut}, {Limon}, {Meyer}, {Tucker}, {Wollack}, \&
  {Wright}}]{wmap5only}
{Dunkley}, J., {Komatsu}, E., {Nolta}, M.~R., {et~al.} 2009, \apjs, 180, 306

\bibitem[{{Eke} {et~al.}(1998){Eke}, {Navarro}, \& {Frenk}}]{ekeetal:1998}
{Eke}, V.~R., {Navarro}, J.~F., \& {Frenk}, C.~S. 1998, \apj, 503, 569

\bibitem[{{Eriksen} {et~al.}(2004){Eriksen}, {Banday}, {G{\'o}rski}, \&
  {Lilje}}]{2004ApJ...612..633E}
{Eriksen}, H.~K., {Banday}, A.~J., {G{\'o}rski}, K.~M., \& {Lilje}, P.~B. 2004,
  \apj, 612, 633

\bibitem[{{Feroz} \& {Hobson}(2008)}]{ferozhobson:2008}
{Feroz}, F. \& {Hobson}, M.~P. 2008, \mnras, 384, 449

\bibitem[{{Finkbeiner} {et~al.}(1999){Finkbeiner}, {Davis}, \&
  {Schlegel}}]{fds:1999}
{Finkbeiner}, D.~P., {Davis}, M., \& {Schlegel}, D.~J. 1999, \apj, 524, 867

\bibitem[{{Fomalont} {et~al.}(1991){Fomalont}, {Windhorst}, {Kristian}, \&
  {Kellerman}}]{fomalontetal:1991}
{Fomalont}, E.~B., {Windhorst}, R.~A., {Kristian}, J.~A., \& {Kellerman}, K.~I.
  1991, \aj, 102, 1258

\bibitem[{{Gaustad} {et~al.}(2001){Gaustad}, {McCullough}, {Rosing}, \& {Van
  Buren}}]{shassa}
{Gaustad}, J.~E., {McCullough}, P.~R., {Rosing}, W., \& {Van Buren}, D. 2001,
  \pasp, 113, 1326

\bibitem[{{G{\'o}rski} {et~al.}(2005){G{\'o}rski}, {Hivon}, {Banday},
  {Wandelt}, {Hansen}, {Reinecke}, \& {Bartelmann}}]{2005ApJ...622..759G}
{G{\'o}rski}, K.~M., {Hivon}, E., {Banday}, A.~J., {et~al.} 2005, \apj, 622,
  759

\bibitem[{{Granato} {et~al.}(2004){Granato}, {De Zotti}, {Silva}, {Bressan}, \&
  {Danese}}]{granatoetal:2004}
{Granato}, G.~L., {De Zotti}, G., {Silva}, L., {Bressan}, A., \& {Danese}, L.
  2004, \apj, 600, 580

\bibitem[{{Haffner} {et~al.}(2003){Haffner}, {Reynolds}, {Tufte}, {Madsen},
  {Jaehnig}, \& {Percival}}]{wham}
{Haffner}, L.~M., {Reynolds}, R.~J., {Tufte}, S.~L., {et~al.} 2003, \apjs, 149,
  405

\bibitem[{Hall {et~al.}(2010)Hall, Keisler, Knox, \& et~al.}]{Hall2010}
Hall, N.~R., Keisler, R., Knox, L., \& et~al. 2010, The Astrophysical Journal,
  718, 632

\bibitem[{{Haslam} {et~al.}(1982){Haslam}, {Salter}, {Stoffel}, \&
  {Wilson}}]{haslametal:1982}
{Haslam}, C.~G.~T., {Salter}, C.~J., {Stoffel}, H., \& {Wilson}, W.~E. 1982,
  \aaps, 47, 1

\bibitem[{{Herranz} {et~al.}(2002){Herranz}, {Sanz}, {Hobson}, {Barreiro},
  {Diego}, {Mart{\'{\i}}nez-Gonz{\'a}lez}, \& {Lasenby}}]{herranz:2002}
{Herranz}, D., {Sanz}, J.~L., {Hobson}, M.~P., {et~al.} 2002, \mnras, 336, 1057

\bibitem[{Jaynes(2003)}]{jaynes}
Jaynes, E.~T. 2003, {Probability Theory: the logic of science}, ed. {G. Larry
  Bretthorst} (Cambridge University Press)

\bibitem[{{Jenkins} {et~al.}(2001){Jenkins}, {Frenk}, {White}, {Colberg},
  {Cole}, {Evrard}, {Couchman}, \& {Yoshida}}]{jenkinsetal:2001}
{Jenkins}, A., {Frenk}, C.~S., {White}, S.~D.~M., {et~al.} 2001, \mnras, 321,
  372

\bibitem[{{Kim} {et~al.}(2009){Kim}, {Naselsky}, \&
  {Christensen}}]{2009PhRvD..79b3003K}
{Kim}, J., {Naselsky}, P., \& {Christensen}, P.~R. 2009, \prd, 79, 023003

\bibitem[{{Koester} {et~al.}(2007){Koester}, {McKay}, {Annis}, {Wechsler},
  {Evrard}, {Bleem}, {Becker}, {Johnston}, {Sheldon}, {Nichol}, {Miller},
  {Scranton}, {Bahcall}, {Barentine}, {Brewington}, {Brinkmann}, {Harvanek},
  {Kleinman}, {Krzesinski}, {Long}, {Nitta}, {Schneider}, {Sneddin}, {Voges},
  \& {York}}]{koester2007}
{Koester}, B.~P., {McKay}, T.~A., {Annis}, J., {et~al.} 2007, \apj, 660, 239

\bibitem[{{Kravtsov} {et~al.}(2006){Kravtsov}, {Vikhlinin}, \&
  {Nagai}}]{kravtsov:2006}
{Kravtsov}, A.~V., {Vikhlinin}, A., \& {Nagai}, D. 2006, \apj, 650, 128

\bibitem[{Lagache {et~al.}(2007)Lagache, Bavouzet, Fernandez-Conde, Ponthieu,
  Rodet, Dole, Miville-Desch\^{e}nes, \& Puget}]{Lagache2007}
Lagache, G., Bavouzet, N., Fernandez-Conde, N., {et~al.} 2007, The
  Astrophysical Journal, 665, L89

\bibitem[{{Leach} {et~al.}(2008){Leach}, {Cardoso}, {Baccigalupi}, {Barreiro},
  {Betoule}, {Bobin}, {Bonaldi}, {Delabrouille}, {de Zotti}, {Dickinson},
  {Eriksen}, {Gonz{\'a}lez-Nuevo}, {Hansen}, {Herranz}, {Le Jeune},
  {L{\'o}pez-Caniego}, {Mart{\'{\i}}nez-Gonz{\'a}lez}, {Massardi}, {Melin},
  {Miville-Desch{\^e}nes}, {Patanchon}, {Prunet}, {Ricciardi}, {Salerno},
  {Sanz}, {Starck}, {Stivoli}, {Stolyarov}, {Stompor}, \&
  {Vielva}}]{leachetal:2008}
{Leach}, S.~M., {Cardoso}, J., {Baccigalupi}, C., {et~al.} 2008, \aap, 491, 597

\bibitem[{{Melin} {et~al.}(2005){Melin}, {Bartlett}, \&
  {Delabrouille}}]{melin:2005}
{Melin}, J., {Bartlett}, J.~G., \& {Delabrouille}, J. 2005, \aap, 429, 417

\bibitem[{{Melin} {et~al.}(2006){Melin}, {Bartlett}, \&
  {Delabrouille}}]{melin:2006}
{Melin}, J.-B., {Bartlett}, J.~G., \& {Delabrouille}, J. 2006, \aap, 459, 341

\bibitem[{{Miville-Desch{\^e}nes}(2009)}]{mdetal:2009}
{Miville-Desch{\^e}nes}, M. 2009, ArXiv e-prints

\bibitem[{{Miville-Desch{\^e}nes} {et~al.}(2007){Miville-Desch{\^e}nes},
  {Lagache}, {Boulanger}, \& {Puget}}]{mdetal:2007}
{Miville-Desch{\^e}nes}, M., {Lagache}, G., {Boulanger}, F., \& {Puget}, J.
  2007, \aap, 469, 595

\bibitem[{{Miville-Desch{\^e}nes} {et~al.}(2008){Miville-Desch{\^e}nes},
  {Ysard}, {Lavabre}, {Ponthieu}, {Mac{\'{\i}}as-P{\'e}rez}, {Aumont}, \&
  {Bernard}}]{mdetal:2008}
{Miville-Desch{\^e}nes}, M., {Ysard}, N., {Lavabre}, A., {et~al.} 2008, \aap,
  490, 1093

\bibitem[{{Motl} {et~al.}(2005{\natexlab{a}}){Motl}, {Hallman}, {Burns}, \&
  {Norman}}]{motletal:2005}
{Motl}, P.~M., {Hallman}, E.~J., {Burns}, J.~O., \& {Norman}, M.~L.
  2005{\natexlab{a}}, \apjl, 623, L63

\bibitem[{{Motl} {et~al.}(2005{\natexlab{b}}){Motl}, {Hallman}, {Burns}, \&
  {Norman}}]{motl:2005}
{Motl}, P.~M., {Hallman}, E.~J., {Burns}, J.~O., \& {Norman}, M.~L.
  2005{\natexlab{b}}, \apjl, 623, L63

\bibitem[{{Nagai}(2006)}]{nagai:2006}
{Nagai}, D. 2006, \apj, 650, 538

\bibitem[{{Nagai} {et~al.}(2007){Nagai}, {Kravtsov}, \&
  {Vikhlinin}}]{nagai:2007}
{Nagai}, D., {Kravtsov}, A.~V., \& {Vikhlinin}, A. 2007, \apj, 668, 1

\bibitem[{{Park} {et~al.}(2007){Park}, {Park}, \& {Gott}}]{2007ApJ...660..959P}
{Park}, C., {Park}, C., \& {Gott}, III, J.~R. 2007, \apj, 660, 959

\bibitem[{{Piffaretti} {et~al.}(2011){Piffaretti}, {Arnaud}, {Pratt},
  {Pointecouteau}, \& {Melin}}]{piffaretti2011}
{Piffaretti}, R., {Arnaud}, M., {Pratt}, G.~W., {Pointecouteau}, E., \&
  {Melin}, J.-B. 2011, \aap, 534, A109

\bibitem[{{Pires} {et~al.}(2006){Pires}, {Juin}, {Yvon}, {Moudden}, {Anthoine},
  \& {Pierpaoli}}]{pires:2006}
{Pires}, S., {Juin}, J.~B., {Yvon}, D., {et~al.} 2006, \aap, 455, 741

\bibitem[{{Planck Collaboration} {et~al.}(2011{\natexlab{a}}){Planck
  Collaboration}, {Ade}, {Aghanim}, {Ansari}, {Arnaud}, {Ashdown}, {Aumont},
  {Banday}, {Bartelmann}, {Bartlett}, {Battaner}, {Benabed}, {Beno{\^i}t},
  {Bernard}, {Bersanelli}, {Bock}, {Bond}, {Borrill}, {Bouchet}, {Boulanger},
  {Bradshaw}, {Bucher}, {Cardoso}, {Castex}, {Catalano}, {Challinor},
  {Chamballu}, {Chary}, {Chen}, {Chiang}, {Church}, {Clements}, {Colley},
  {Colombi}, {Couchot}, {Coulais}, {Cressiot}, {Crill}, {Crook}, {de
  Bernardis}, {Delabrouille}, {Delouis}, {D{\'e}sert}, {Dolag}, {Dole},
  {Dor{\'e}}, {Douspis}, {Dunkley}, {Efstathiou}, {Filliard}, {Forni},
  {Fosalba}, {Ganga}, {Giard}, {Girard}, {Giraud-H{\'e}raud}, {Gispert},
  {G{\'o}rski}, {Gratton}, {Griffin}, {Guyot}, {Haissinski}, {Harrison},
  {Helou}, {Henrot-Versill{\'e}}, {Hern{\'a}ndez-Monteagudo}, {Hildebrandt},
  {Hills}, {Hivon}, {Hobson}, {Holmes}, {Huffenberger}, {Jaffe}, {Jones},
  {Kaplan}, {Kneissl}, {Knox}, {Kunz}, {Lagache}, {Lamarre}, {Lange},
  {Lasenby}, {Lavabre}, {Lawrence}, {Le Jeune}, {Leroy}, {Lesgourgues},
  {Mac{\'{\i}}as-P{\'e}rez}, {MacTavish}, {Maffei}, {Mandolesi}, {Mann},
  {Marleau}, {Marshall}, {Masi}, {Matsumura}, {McAuley}, {McGehee}, {Melin},
  {Mercier}, {Mitra}, {Miville-Desch{\^e}nes}, {Moneti}, {Montier}, {Mortlock},
  {Murphy}, {Nati}, {Netterfield}, {N{\o}rgaard-Nielsen}, {North}, {Noviello},
  {Novikov}, {Osborne}, {Pajot}, {Patanchon}, {Peacocke}, {Pearson},
  {Perdereau}, {Perotto}, {Piacentini}, {Piat}, {Plaszczynski},
  {Pointecouteau}, {Ponthieu}, {Pr{\'e}zeau}, {Prunet}, {Puget}, {Reach},
  {Remazeilles}, {Renault}, {Riazuelo}, {Ristorcelli}, {Rocha}, {Rosset},
  {Roudier}, {Rowan-Robinson}, {Rusholme}, {Saha}, {Santos}, {Savini},
  {Schaefer}, {Shellard}, {Spencer}, {Starck}, {Stolyarov}, {Stompor},
  {Sudiwala}, {Sunyaev}, {Sutton}, {Sygnet}, {Tauber}, {Thum}, {Torre},
  {Touze}, {Tristram}, {van Leeuwen}, {Vibert}, {Vibert}, {Wade}, {Wandelt},
  {White}, {Wiesemeyer}, {Woodcraft}, {Yurchenko}, {Yvon}, \&
  {Zacchei}}]{PEPVI}
{Planck Collaboration}, {Ade}, P.~A.~R., {Aghanim}, N., {et~al.}
  2011{\natexlab{a}}, \aap, 536, A6

\bibitem[{{Planck Collaboration} {et~al.}(2011{\natexlab{b}}){Planck
  Collaboration}, {Ade}, {Aghanim}, {Arnaud}, {Ashdown}, {Aumont},
  {Baccigalupi}, {Balbi}, {Banday}, {Barreiro}, \& et~al.}]{PEPVIII}
{Planck Collaboration}, {Ade}, P.~A.~R., {Aghanim}, N., {et~al.}
  2011{\natexlab{b}}, \aap, 536, A8

\bibitem[{{Planck Collaboration} {et~al.}(2011{\natexlab{c}}){Planck
  Collaboration}, {Ade}, {Aghanim}, {Arnaud}, {Ashdown}, {Aumont},
  {Baccigalupi}, {Balbi}, {Banday}, {Barreiro}, \& et~al.}]{PEPXI}
{Planck Collaboration}, {Ade}, P.~A.~R., {Aghanim}, N., {et~al.}
  2011{\natexlab{c}}, \aap, 536, A11

\bibitem[{{Planck Collaboration} {et~al.}(2011{\natexlab{d}}){Planck
  Collaboration}, {Ade}, {Aghanim}, {Arnaud}, {Ashdown}, {Aumont},
  {Baccigalupi}, {Balbi}, {Banday}, {Barreiro}, \& et~al.}]{PEPXVIII}
{Planck Collaboration}, {Ade}, P.~A.~R., {Aghanim}, N., {et~al.}
  2011{\natexlab{d}}, \aap, 536, A18

\bibitem[{{Planck Collaboration} {et~al.}(2011{\natexlab{e}}){Planck
  Collaboration}, {Aghanim}, {Arnaud}, {Ashdown}, {Aumont}, {Baccigalupi},
  {Balbi}, {Banday}, {Barreiro}, {Bartelmann}, {Bartlett}, {Battaner},
  {Benabed}, {Beno{\^i}t}, {Bernard}, {Bersanelli}, {Bhatia}, {Bock},
  {Bonaldi}, {Bond}, {Borrill}, {Bouchet}, {Brown}, {Bucher}, {Burigana},
  {Cabella}, {Cardoso}, {Catalano}, {Cay{\'o}n}, {Challinor}, {Chamballu},
  {Chary}, {Chiang}, {Chiang}, {Chon}, {Christensen}, {Churazov}, {Clements},
  {Colafrancesco}, {Colombi}, {Couchot}, {Coulais}, {Crill}, {Cuttaia}, {da
  Silva}, {Dahle}, {Danese}, {de Bernardis}, {de Gasperis}, {de Rosa}, {de
  Zotti}, {Delabrouille}, {Delouis}, {D{\'e}sert}, {Diego}, {Dolag},
  {Donzelli}, {Dor{\'e}}, {D{\"o}rl}, {Douspis}, {Dupac}, {Efstathiou},
  {En{\ss}lin}, {Finelli}, {Flores-Cacho}, {Forni}, {Frailis}, {Franceschi},
  {Fromenteau}, {Galeotta}, {Ganga}, {G{\'e}nova-Santos}, {Giard}, {Giardino},
  {Giraud-H{\'e}raud}, {Gonz{\'a}lez-Nuevo}, {G{\'o}rski}, {Gratton},
  {Gregorio}, {Gruppuso}, {Harrison}, {Henrot-Versill{\'e}},
  {Hern{\'a}ndez-Monteagudo}, {Herranz}, {Hildebrandt}, {Hivon}, {Hobson},
  {Holmes}, {Hovest}, {Hoyland}, {Huffenberger}, {Jaffe}, {Jones}, {Juvela},
  {Keih{\"a}nen}, {Keskitalo}, {Kisner}, {Kneissl}, {Knox}, {Kurki-Suonio},
  {Lagache}, {Lamarre}, {Lasenby}, {Laureijs}, {Lawrence}, {Leach}, {Leonardi},
  {Linden-V{\o}rnle}, {L{\'o}pez-Caniego}, {Lubin}, {Mac{\'{\i}}as-P{\'e}rez},
  {MacTavish}, {Maffei}, {Maino}, {Mandolesi}, {Mann}, {Maris}, {Marleau},
  {Mart{\'{\i}}nez-Gonz{\'a}lez}, {Masi}, {Matarrese}, {Matthai}, {Mazzotta},
  {Melchiorri}, {Melin}, {Mendes}, {Mennella}, {Mitra},
  {Miville-Desch{\^e}nes}, {Moneti}, {Montier}, {Morgante}, {Mortlock},
  {Munshi}, {Murphy}, {Naselsky}, {Natoli}, {Netterfield},
  {N{\o}rgaard-Nielsen}, {Noviello}, {Novikov}, {Novikov}, {Osborne}, {Pajot},
  {Pasian}, {Patanchon}, {Perdereau}, {Perotto}, {Perrotta}, {Piacentini},
  {Piat}, {Pierpaoli}, {Piffaretti}, {Plaszczynski}, {Pointecouteau},
  {Polenta}, {Ponthieu}, {Poutanen}, {Pratt}, {Pr{\'e}zeau}, {Prunet}, {Puget},
  {Rebolo}, {Reinecke}, {Renault}, {Ricciardi}, {Riller}, {Ristorcelli},
  {Rocha}, {Rosset}, {Rubi{\~n}o-Mart{\'{\i}}n}, {Rusholme}, {Sandri},
  {Santos}, {Schaefer}, {Scott}, {Seiffert}, {Smoot}, {Starck}, {Stivoli},
  {Stolyarov}, {Sunyaev}, {Sygnet}, {Tauber}, {Terenzi}, {Toffolatti},
  {Tomasi}, {Tristram}, {Tuovinen}, {Valenziano}, {Vibert}, {Vielva}, {Villa},
  {Vittorio}, {Wandelt}, {White}, {White}, {Yvon}, {Zacchei}, \&
  {Zonca}}]{PEPX}
{Planck Collaboration}, {Aghanim}, N., {Arnaud}, M., {et~al.}
  2011{\natexlab{e}}, \aap, 536, A10

\bibitem[{{Planck Collaboration} {et~al.}(2011{\natexlab{f}}){Planck
  Collaboration}, {Aghanim}, {Arnaud}, {Ashdown}, {Aumont}, {Baccigalupi},
  {Balbi}, {Banday}, {Barreiro}, {Bartelmann}, \& et~al.}]{PEPIX}
{Planck Collaboration}, {Aghanim}, N., {Arnaud}, M., {et~al.}
  2011{\natexlab{f}}, \aap, 536, A9

\bibitem[{{Planck Collaboration} {et~al.}(2011{\natexlab{g}}){Planck
  Collaboration}, {Aghanim}, {Arnaud}, {Ashdown}, {Aumont}, {Baccigalupi},
  {Balbi}, {Banday}, {Barreiro}, {Bartelmann}, \& et~al.}]{PEPXII}
{Planck Collaboration}, {Aghanim}, N., {Arnaud}, M., {et~al.}
  2011{\natexlab{g}}, \aap, 536, A12

\bibitem[{{Reinecke} {et~al.}(2006){Reinecke}, {Dolag}, {Hell}, {Bartelmann},
  \& {En{\ss}lin}}]{levelS}
{Reinecke}, M., {Dolag}, K., {Hell}, R., {Bartelmann}, M., \& {En{\ss}lin},
  T.~A. 2006, \aap, 445, 373

\bibitem[{{Rosati} {et~al.}(2002){Rosati}, {Borgani}, \&
  {Norman}}]{rosatietal:2002}
{Rosati}, P., {Borgani}, S., \& {Norman}, C. 2002, \araa, 40, 539

\bibitem[{{Samal} {et~al.}(2009){Samal}, {Saha}, {Delabrouille}, {Prunet},
  {Jain}, \& {Souradeep}}]{2009arXiv0903.3634S}
{Samal}, P.~K., {Saha}, R., {Delabrouille}, J., {et~al.} 2009, ArXiv e-prints

\bibitem[{{Sch{\"a}fer} {et~al.}(2006){Sch{\"a}fer}, {Pfrommer}, {Hell}, \&
  {Bartelmann}}]{schaefer:2006}
{Sch{\"a}fer}, B.~M., {Pfrommer}, C., {Hell}, R.~M., \& {Bartelmann}, M. 2006,
  \mnras, 370, 1713

\bibitem[{{Schlegel} {et~al.}(1998){Schlegel}, {Finkbeiner}, \&
  {Davis}}]{sfd:1998}
{Schlegel}, D.~J., {Finkbeiner}, D.~P., \& {Davis}, M. 1998, \apj, 500, 525

\bibitem[{{Schuecker} {et~al.}(2003){Schuecker}, {B{\"o}hringer}, {Collins}, \&
  {Guzzo}}]{schuecker03}
{Schuecker}, P., {B{\"o}hringer}, H., {Collins}, C.~A., \& {Guzzo}, L. 2003,
  \aap, 398, 867

\bibitem[{{Serjeant} \& {Harrison}(2005)}]{serjeantharrison:2005}
{Serjeant}, S. \& {Harrison}, D. 2005, \mnras, 356, 192

\bibitem[{{Shaw} {et~al.}(2008){Shaw}, {Holder}, \& {Bode}}]{shawetal:2008}
{Shaw}, L.~D., {Holder}, G.~P., \& {Bode}, P. 2008, \apj, 686, 206

\bibitem[{{Sheth} \& {Tormen}(1999)}]{st:1999}
{Sheth}, R.~K. \& {Tormen}, G. 1999, \mnras, 308, 119

\bibitem[{{Sunyaev} \& {Zeldovich}(1970)}]{sz:1970}
{Sunyaev}, R.~A. \& {Zeldovich}, Y.~B. 1970, Comments on Astrophysics and Space
  Physics, 2, 66

\bibitem[{{Sunyaev} \& {Zeldovich}(1972)}]{sz:1972}
{Sunyaev}, R.~A. \& {Zeldovich}, Y.~B. 1972, Comments on Astrophysics and Space
  Physics, 4, 173

\bibitem[{{Tegmark} {et~al.}(2003){Tegmark}, {de Oliveira-Costa}, \&
  {Hamilton}}]{2003PhRvD..68l3523T}
{Tegmark}, M., {de Oliveira-Costa}, A., \& {Hamilton}, A.~J. 2003, \prd, 68,
  123523

\bibitem[{{Toffolatti} {et~al.}(1998){Toffolatti}, {Argueso Gomez}, {de Zotti},
  {Mazzei}, {Franceschini}, {Danese}, \& {Burigana}}]{toffolattietal:1998}
{Toffolatti}, L., {Argueso Gomez}, F., {de Zotti}, G., {et~al.} 1998, \mnras,
  297, 117

\bibitem[{{Truemper}(1992)}]{trumper:1992}
{Truemper}, J. 1992, \qjras, 33, 165

\bibitem[{Viero {et~al.}(2009)Viero, Ade, Bock, \& et~al.}]{Viero2009}
Viero, M.~P., Ade, P. a.~R., Bock, J.~J., \& et~al. 2009, The Astrophysical
  Journal, 707, 1766

\bibitem[{{Vikhlinin} {et~al.}(2009){Vikhlinin}, {Kravtsov}, {Burenin},
  {Ebeling}, {Forman}, {Hornstrup}, {Jones}, {Murray}, {Nagai}, {Quintana}, \&
  {Voevodkin}}]{vikhlinin09}
{Vikhlinin}, A., {Kravtsov}, A.~V., {Burenin}, R.~A., {et~al.} 2009, \apj, 692,
  1060

\bibitem[{{Voit}(2005)}]{voit:2005}
{Voit}, G.~M. 2005, Reviews of Modern Physics, 77, 207

\bibitem[{{Wakker} {et~al.}(2008){Wakker}, {York}, {Wilhelm}, {Barentine},
  {Richter}, {Beers}, {Ivezi{\'c}}, \& {Howk}}]{Wakkeretal:2008}
{Wakker}, B.~P., {York}, D.~G., {Wilhelm}, R., {et~al.} 2008, \apj, 672, 298

\bibitem[{{White}(2003)}]{white2003}
{White}, M. 2003, \apj, 597, 650

\end{thebibliography}

\appendix

\section{Comparison plots for Challenge SZ v1}
\label{app:complotv1}

The plots are obtained at purity=0.9 (except when the catalog does \hl{not permit this value to be reached}).

\clearpage

\begin{table}[htbp]
\begin{center}
\begin{tabular}{cc}
\includegraphics[scale=0.45]{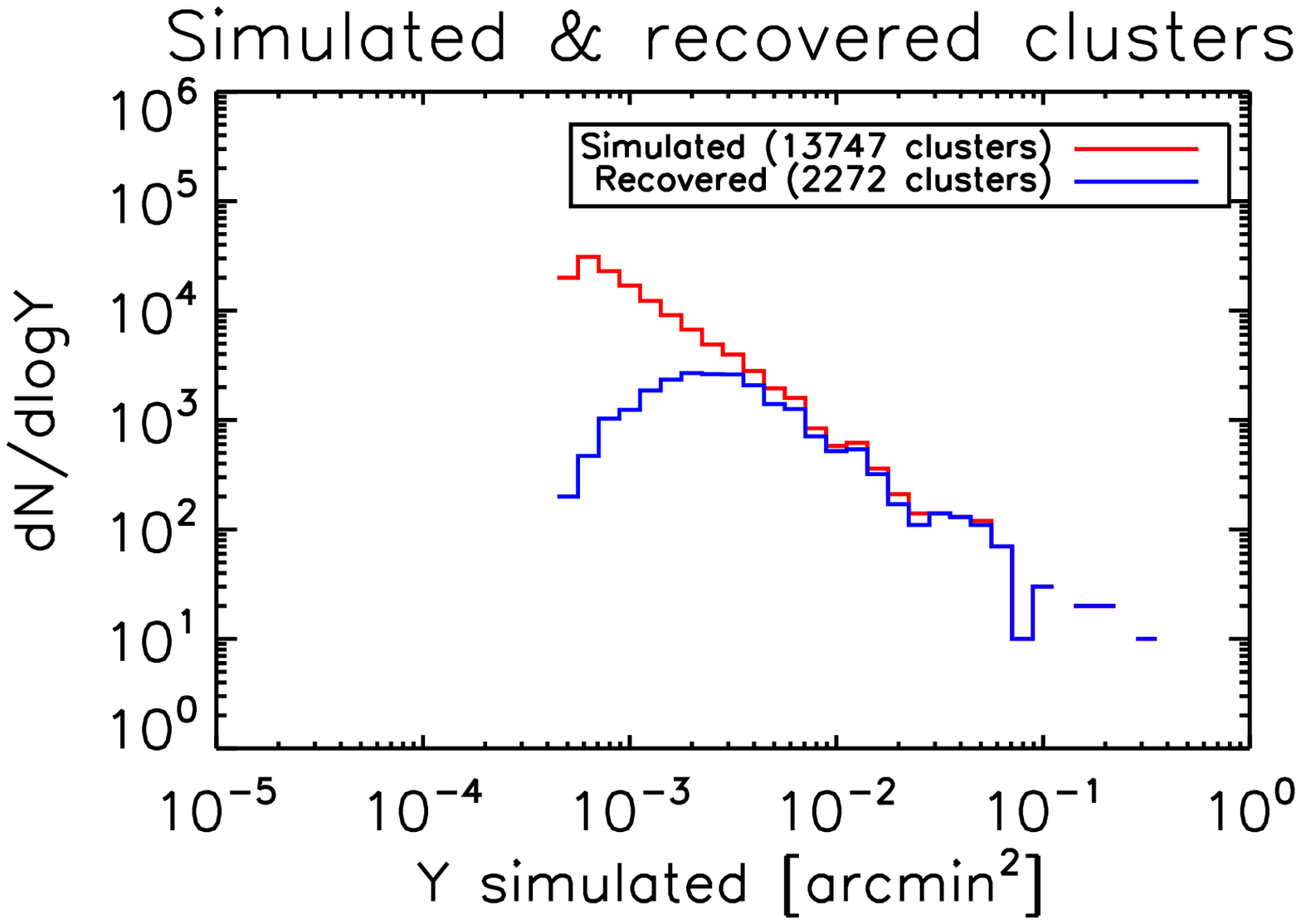}  &
\includegraphics[scale=0.45]{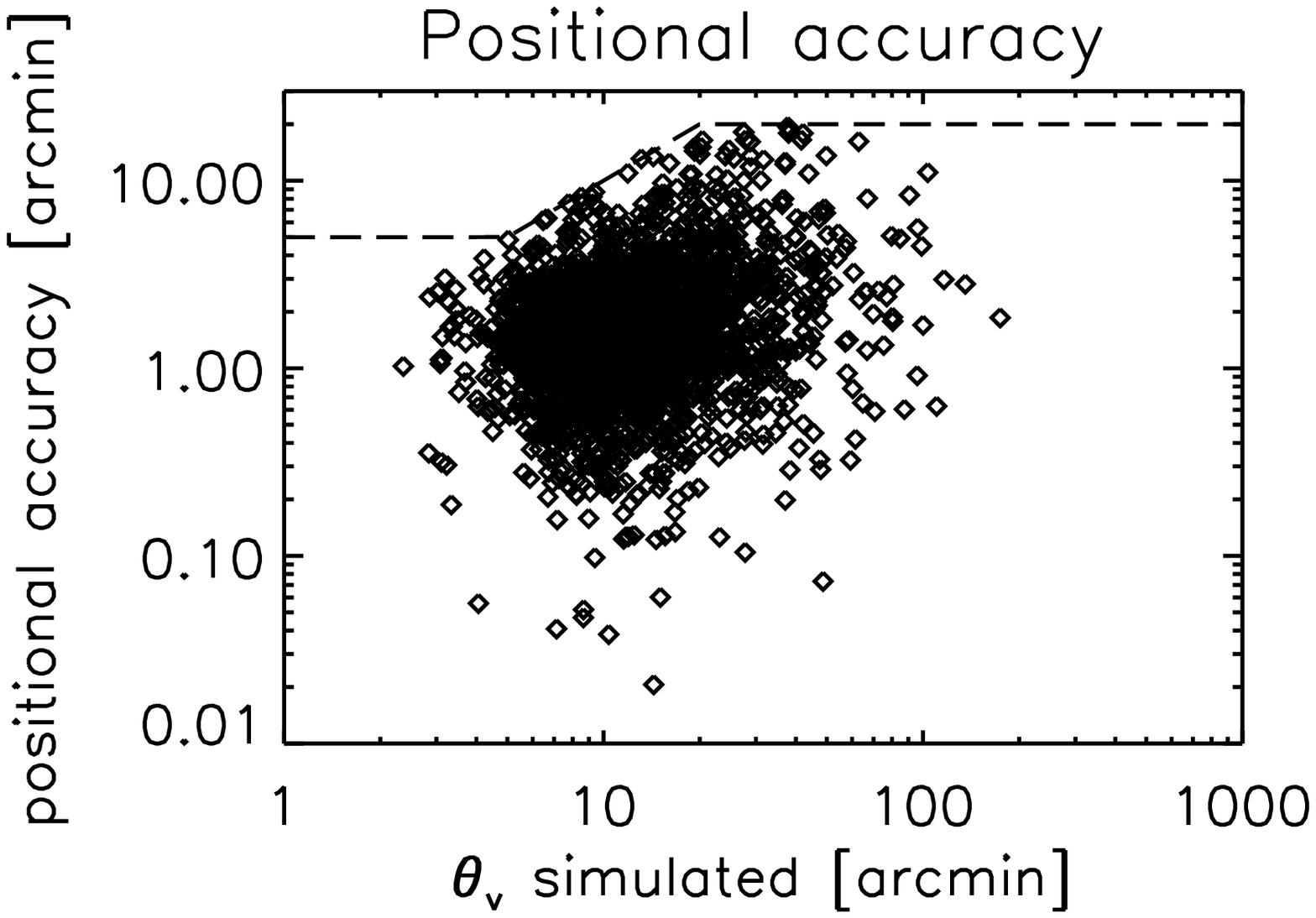} \\
\includegraphics[scale=0.45]{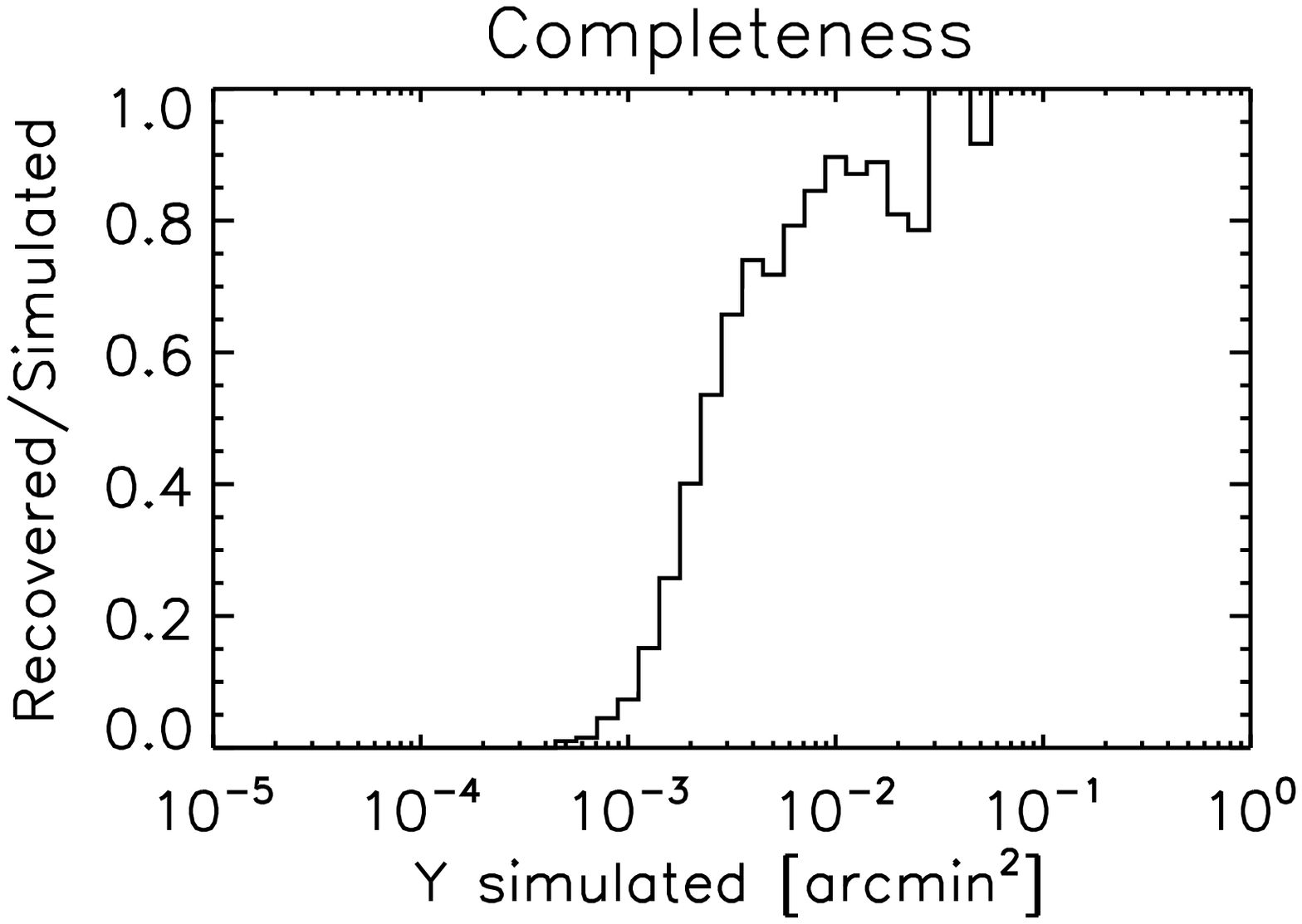}  &
\includegraphics[scale=0.45]{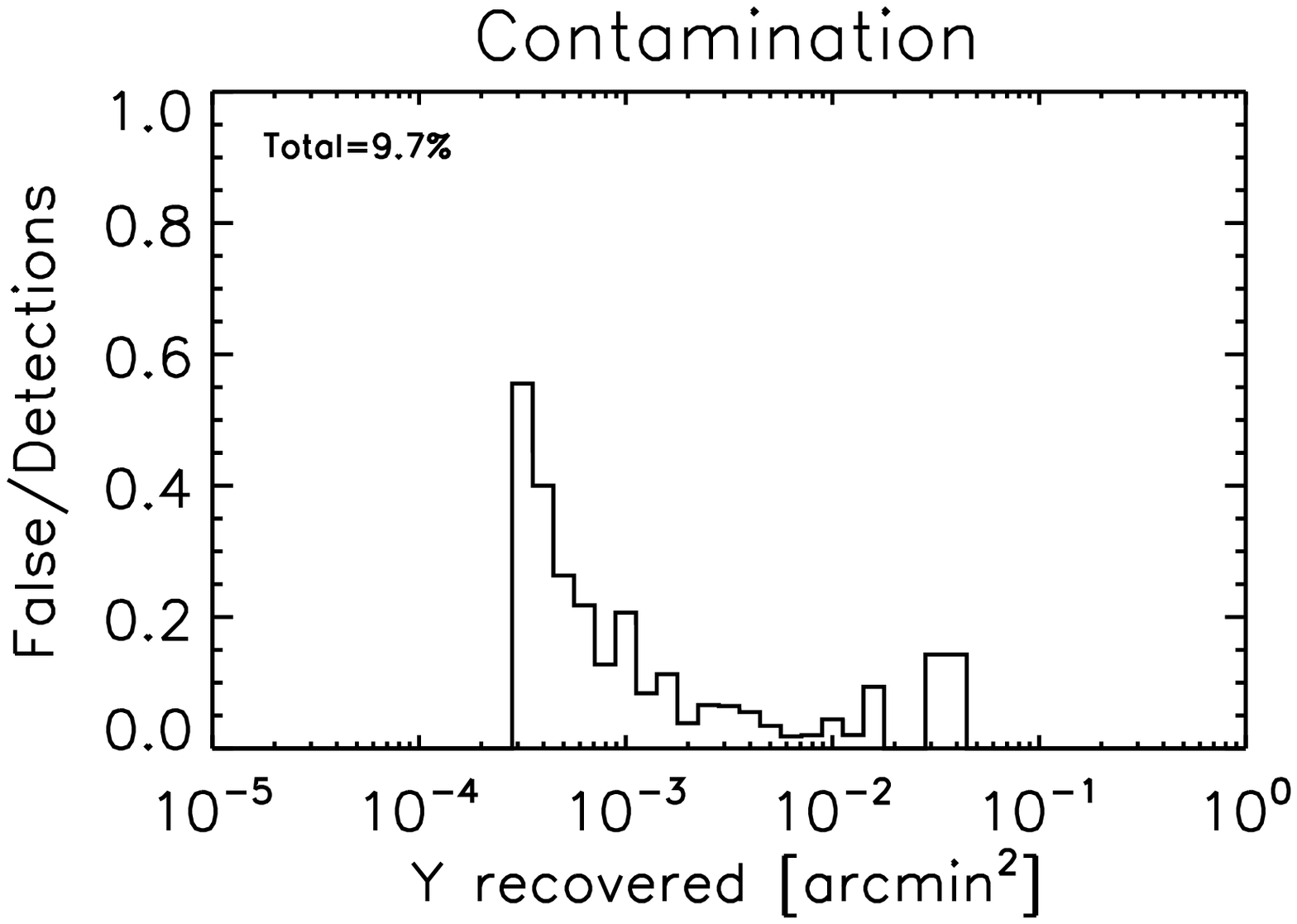} \\
\includegraphics[scale=0.45]{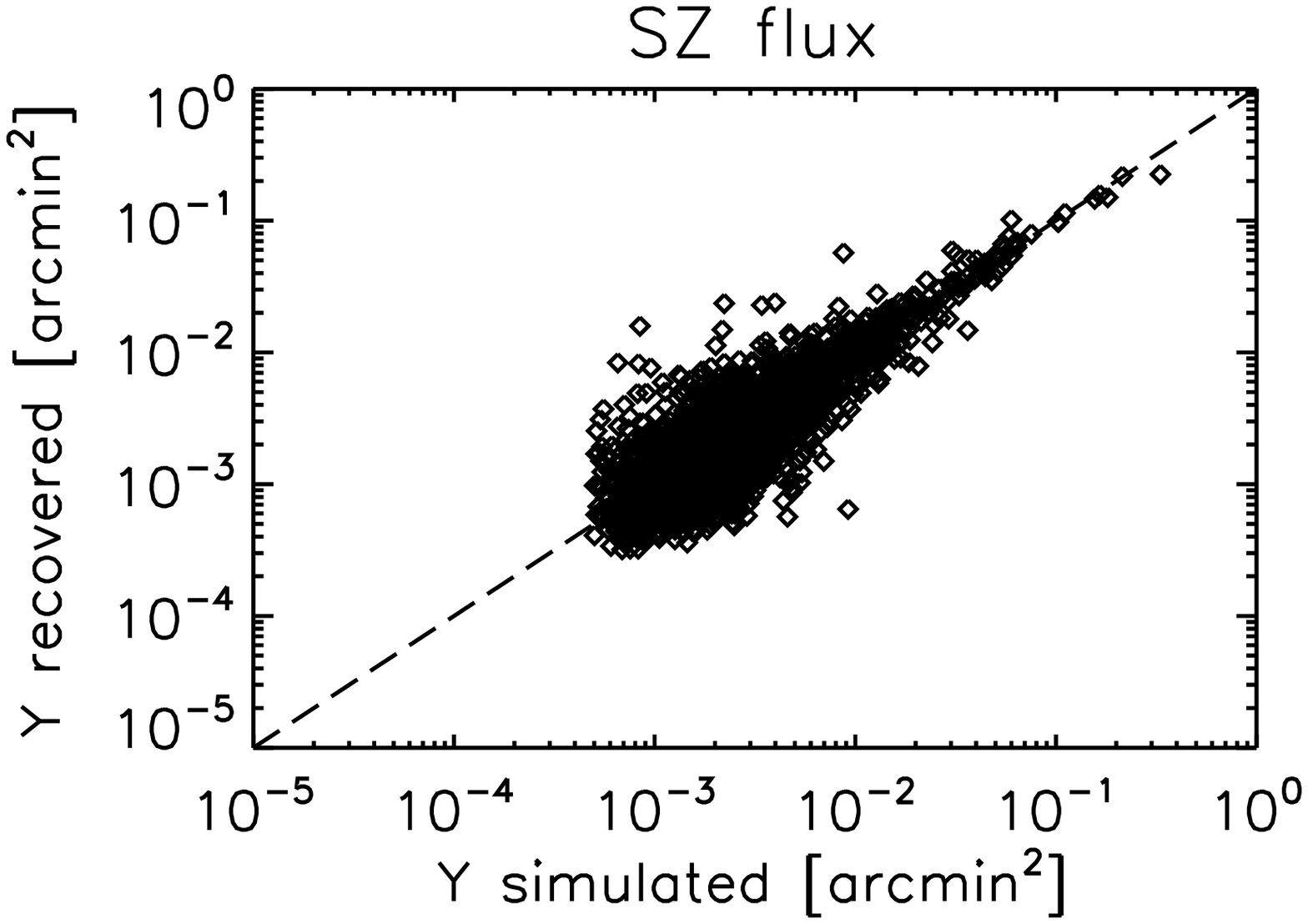}  &
\includegraphics[scale=0.45]{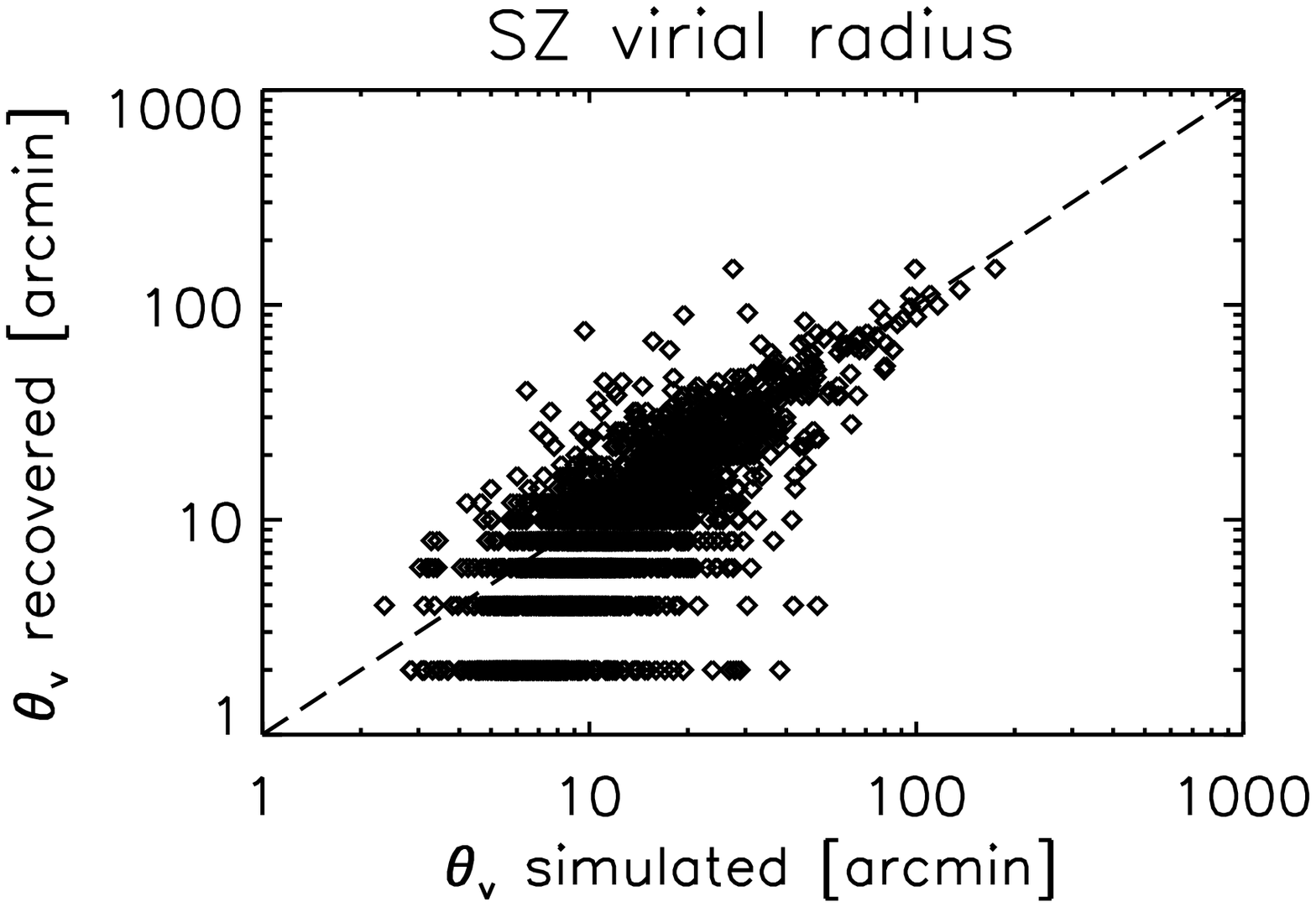} \\
\includegraphics[scale=0.45]{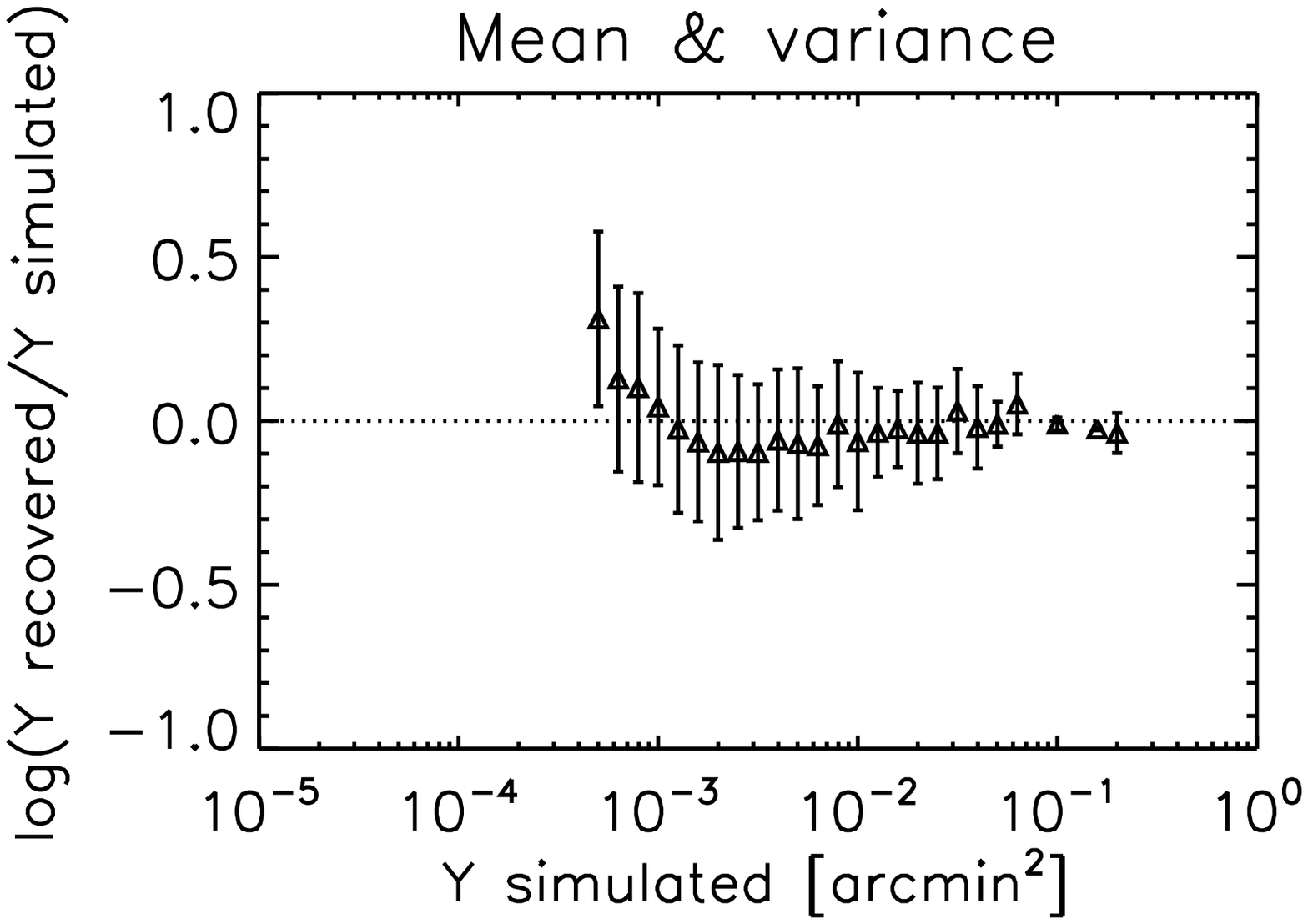} &
\includegraphics[scale=0.45]{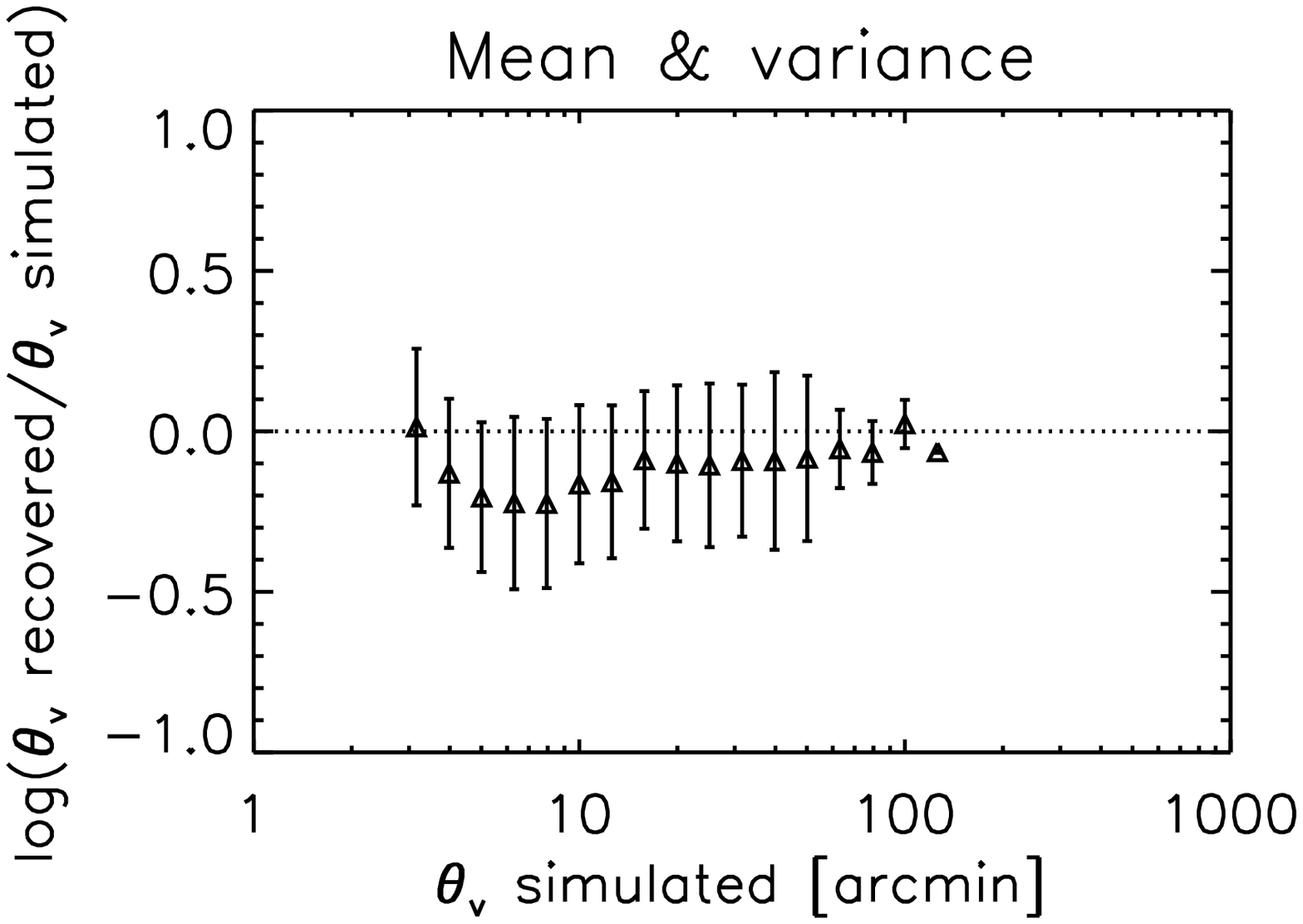} \\
\end{tabular}
\caption{{\bf MMF1}}
\end{center}
\end{table}

\clearpage

\begin{table}[htbp]
\begin{center}
\begin{tabular}{cc}
\includegraphics[scale=0.45]{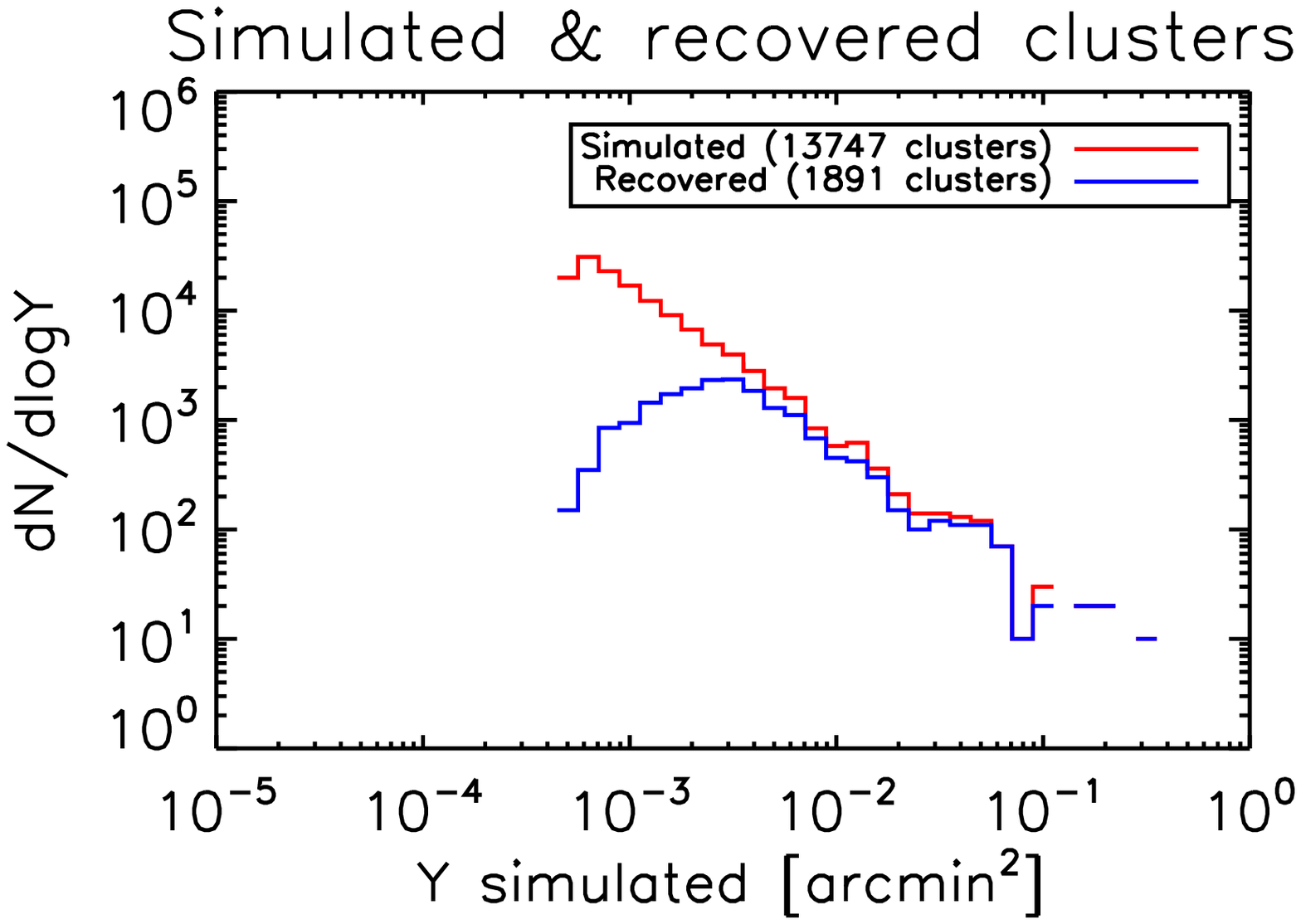}  &
\includegraphics[scale=0.45]{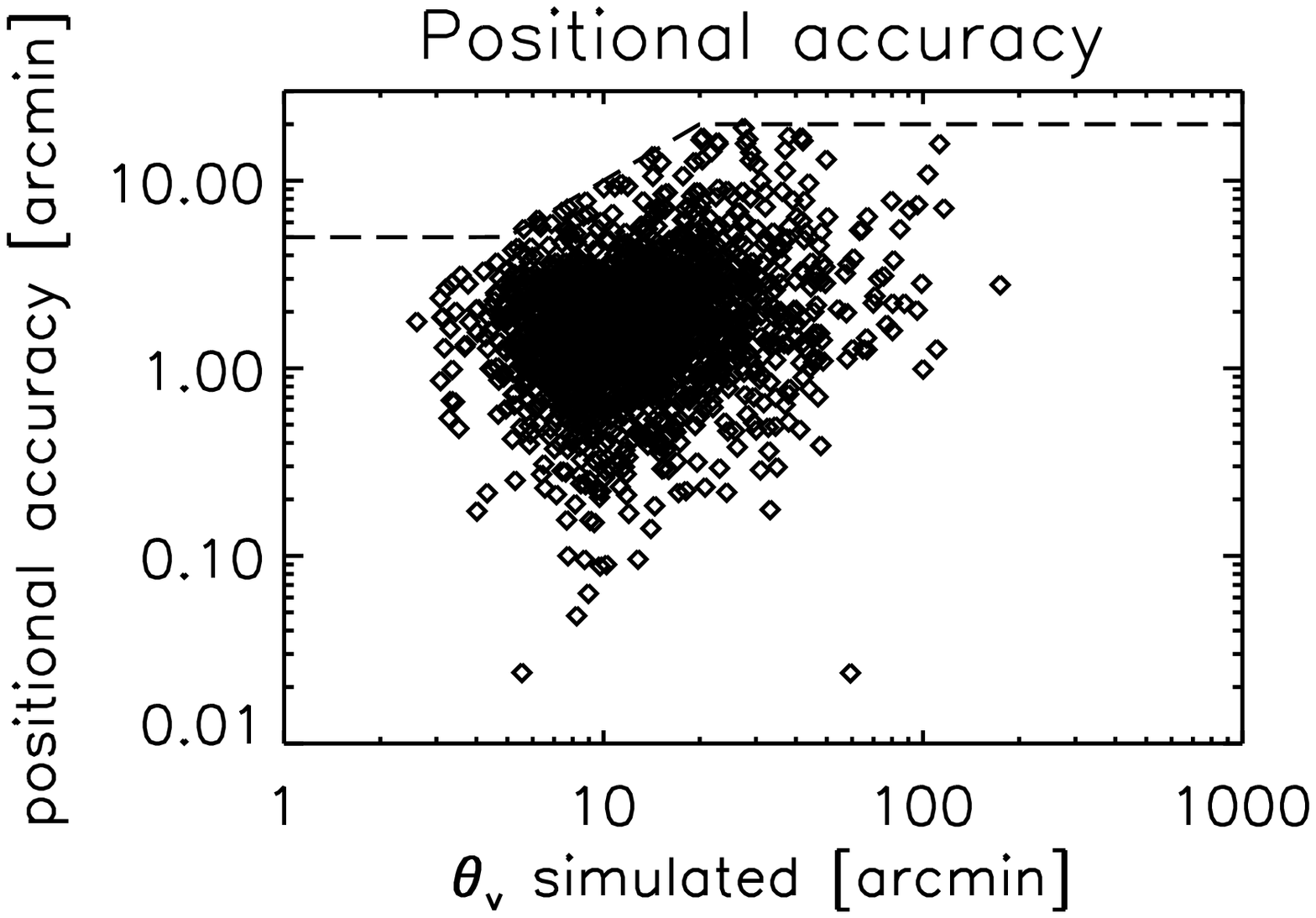} \\
\includegraphics[scale=0.45]{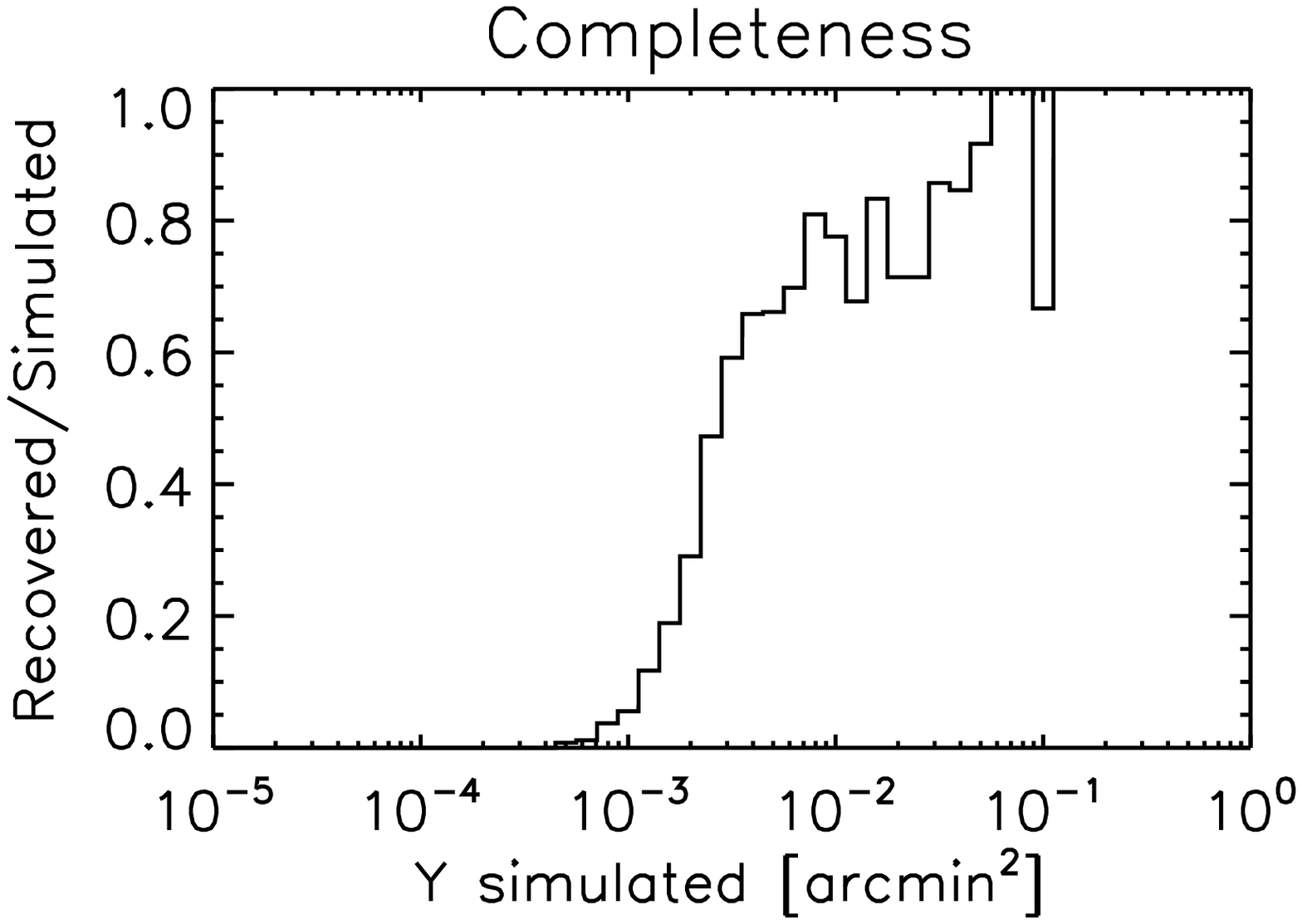}  &
\includegraphics[scale=0.45]{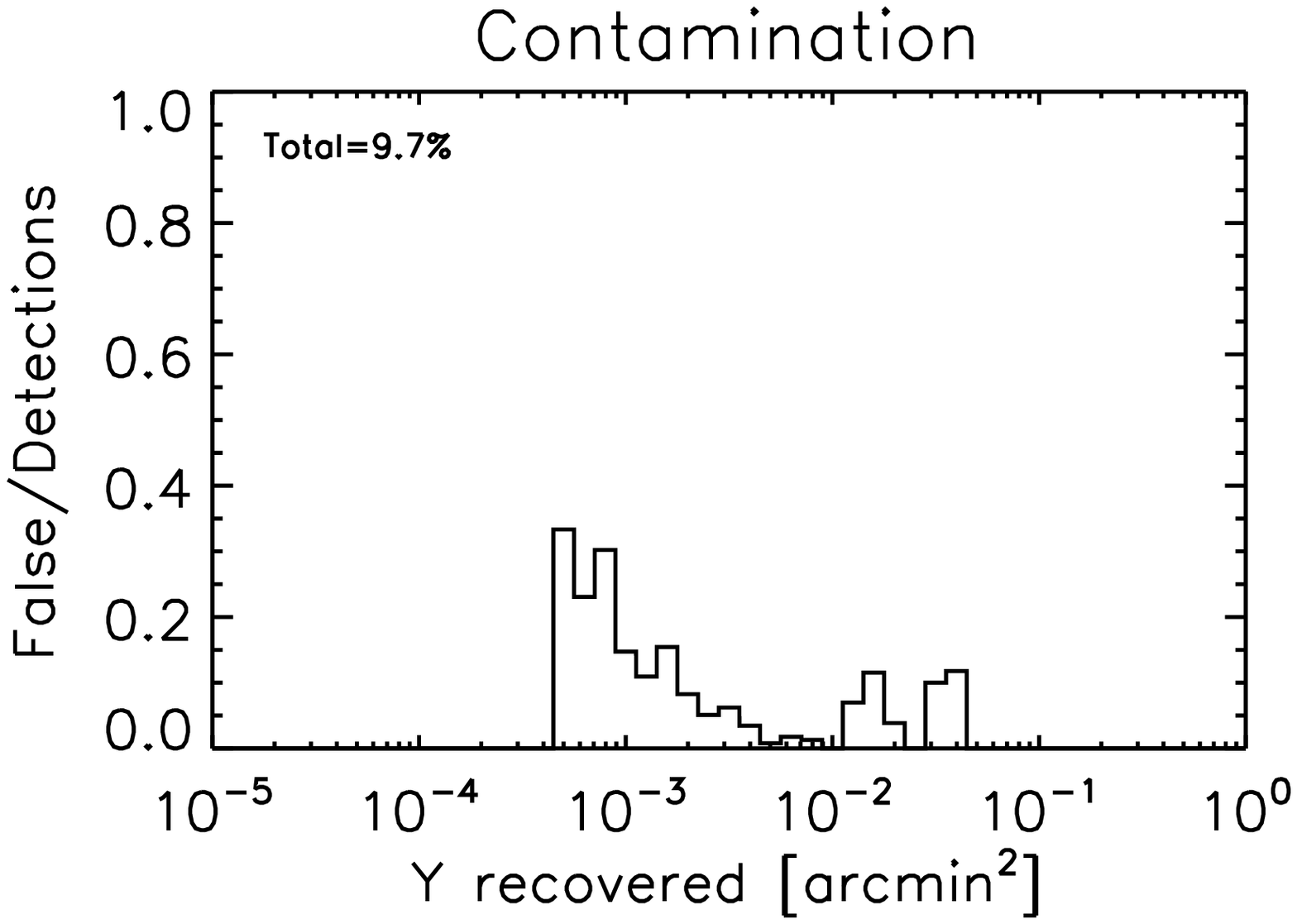} \\
\includegraphics[scale=0.45]{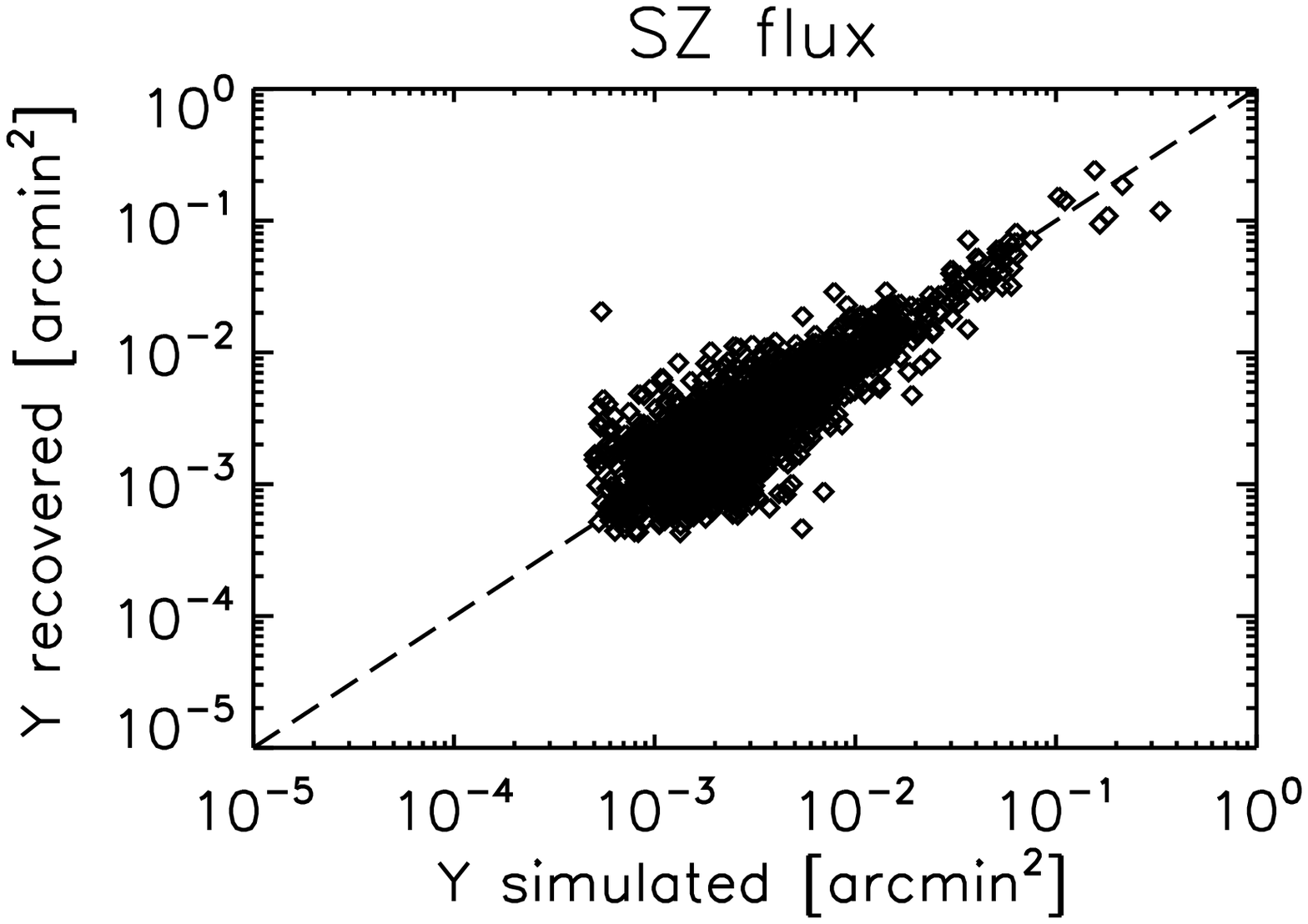}  &
\includegraphics[scale=0.45]{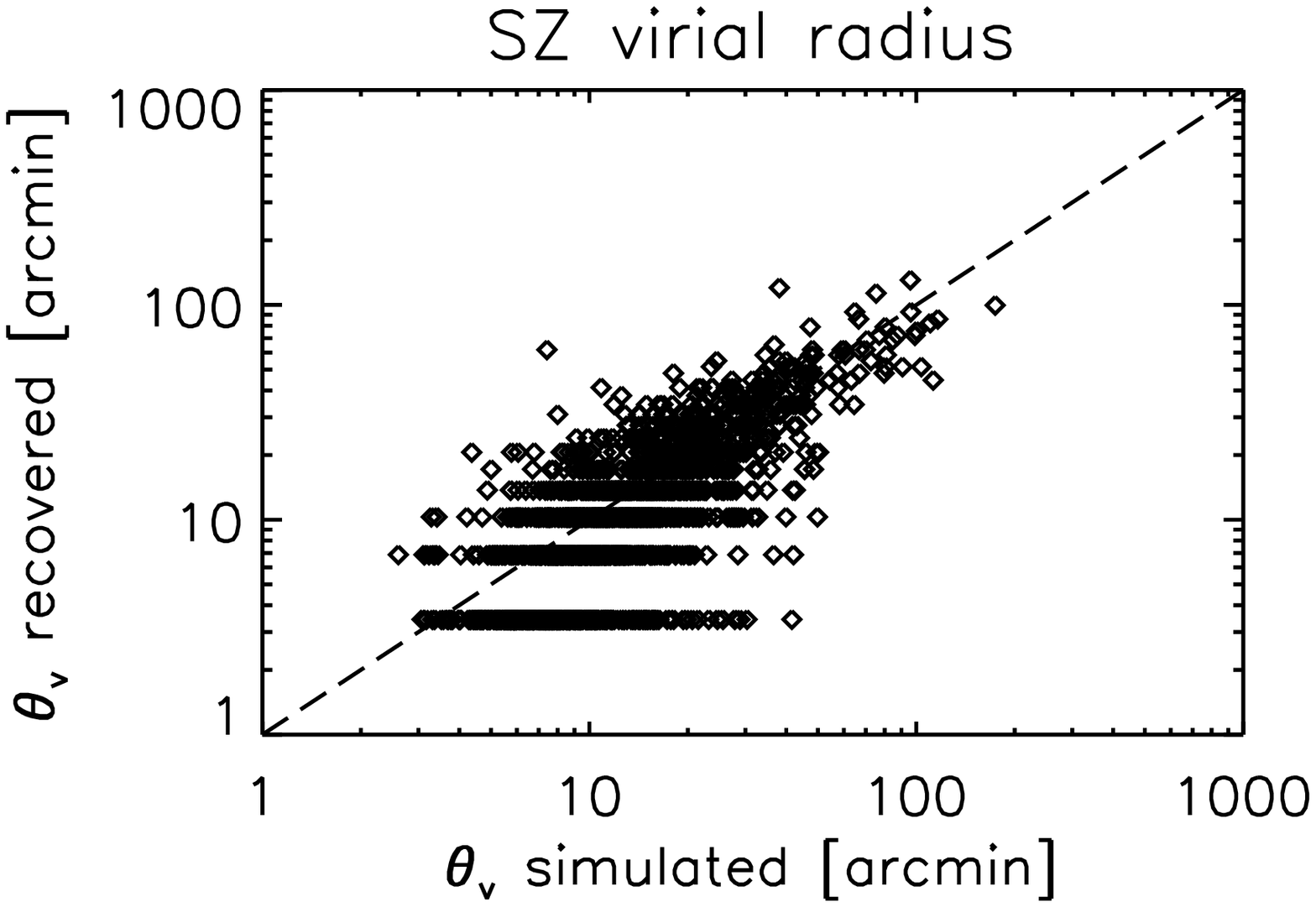} \\
\includegraphics[scale=0.45]{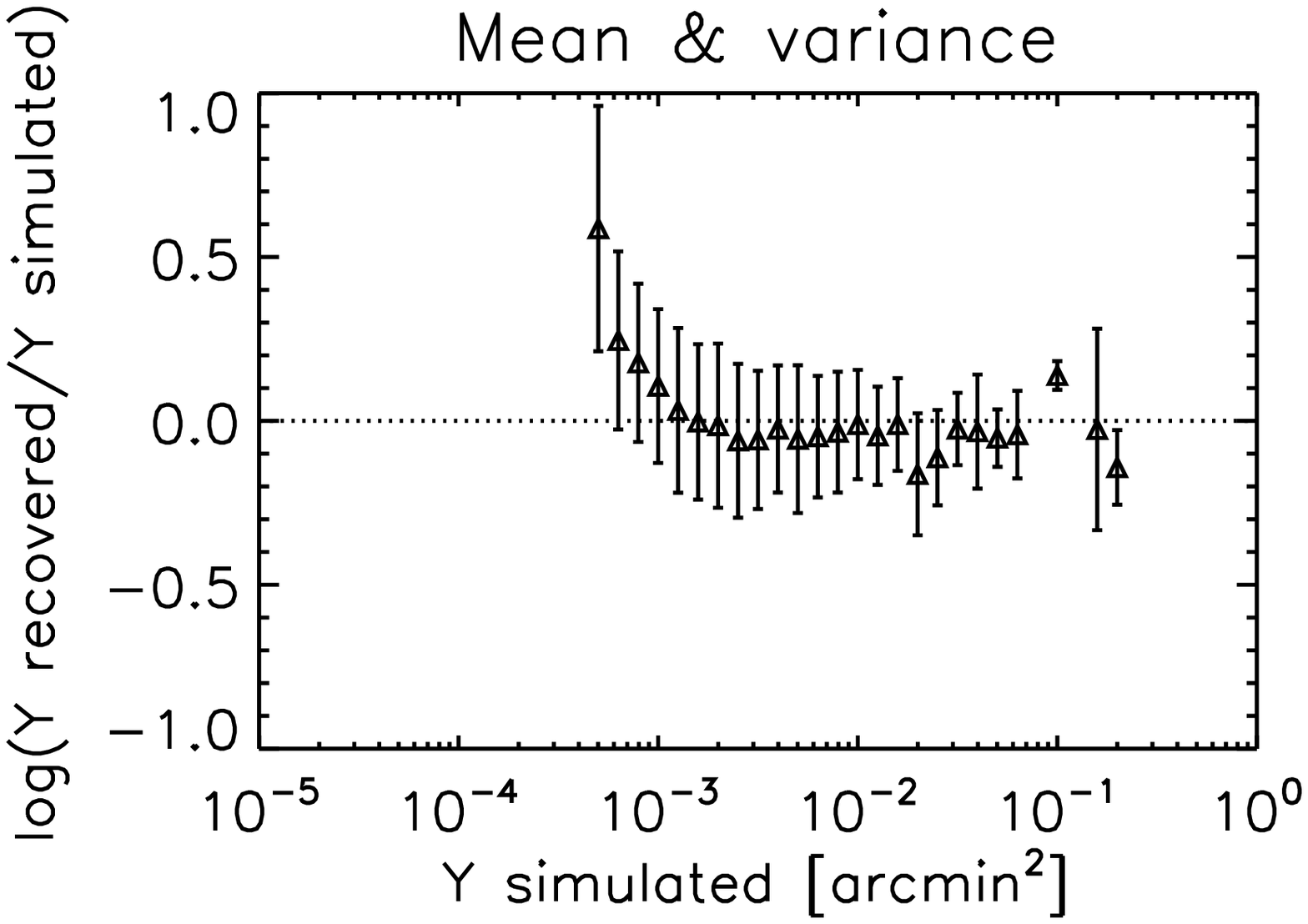} &
\includegraphics[scale=0.45]{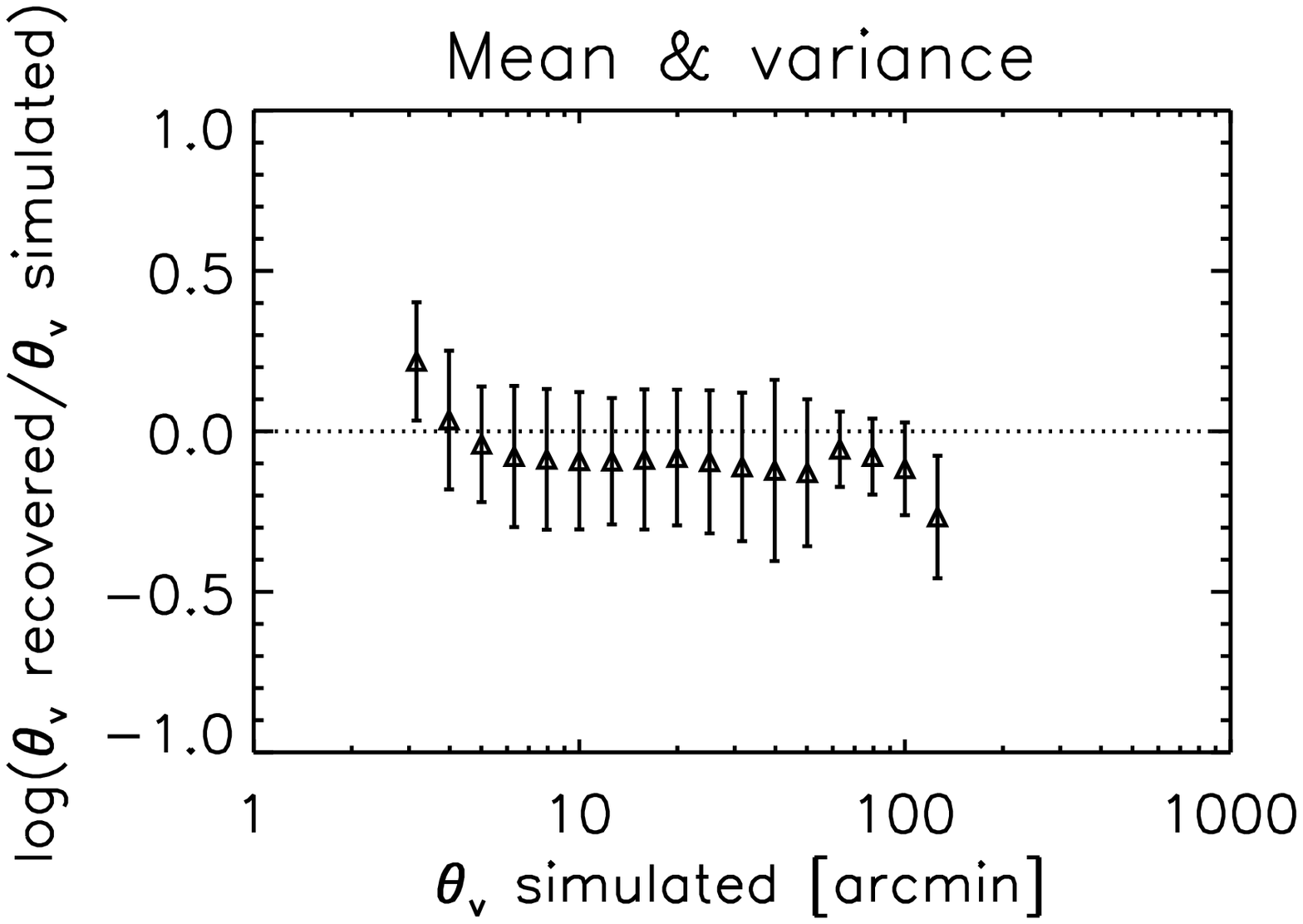} \\
\end{tabular}
\caption{{\bf MMF2}}
\end{center}
\end{table}

\clearpage

\begin{table}[htbp]
\begin{center}
\begin{tabular}{cc}
\includegraphics[scale=0.45]{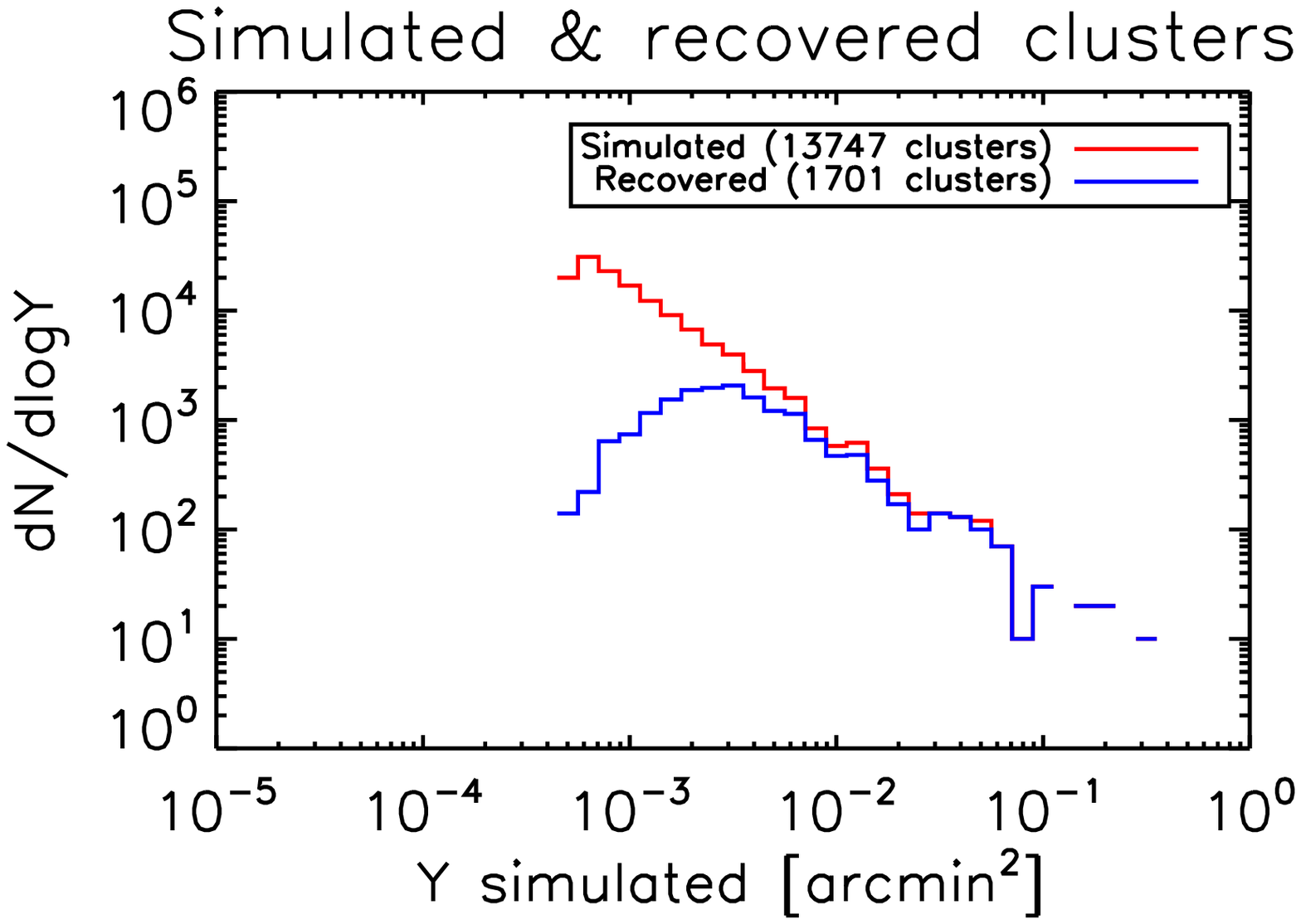}  &
\includegraphics[scale=0.45]{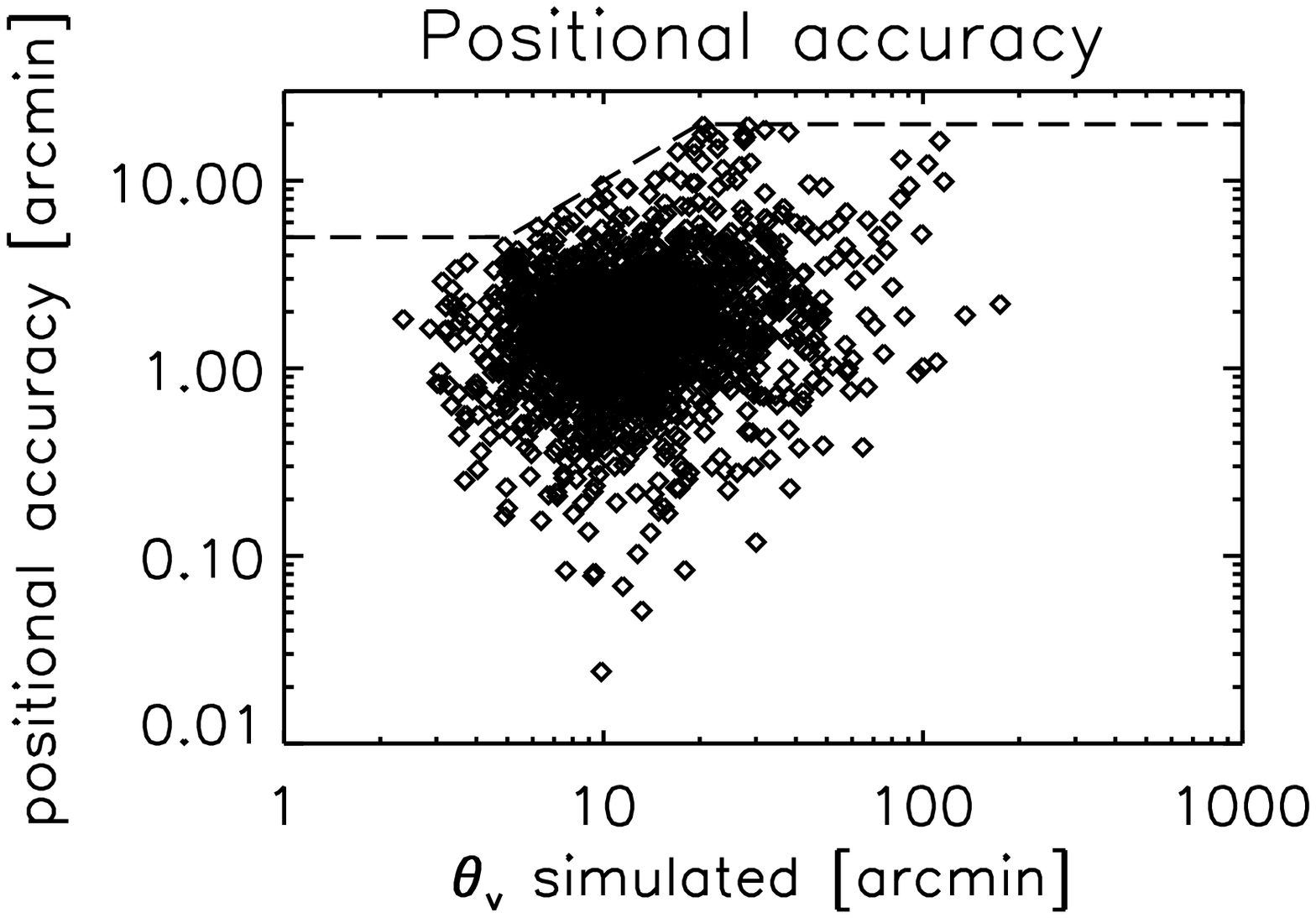} \\
\includegraphics[scale=0.45]{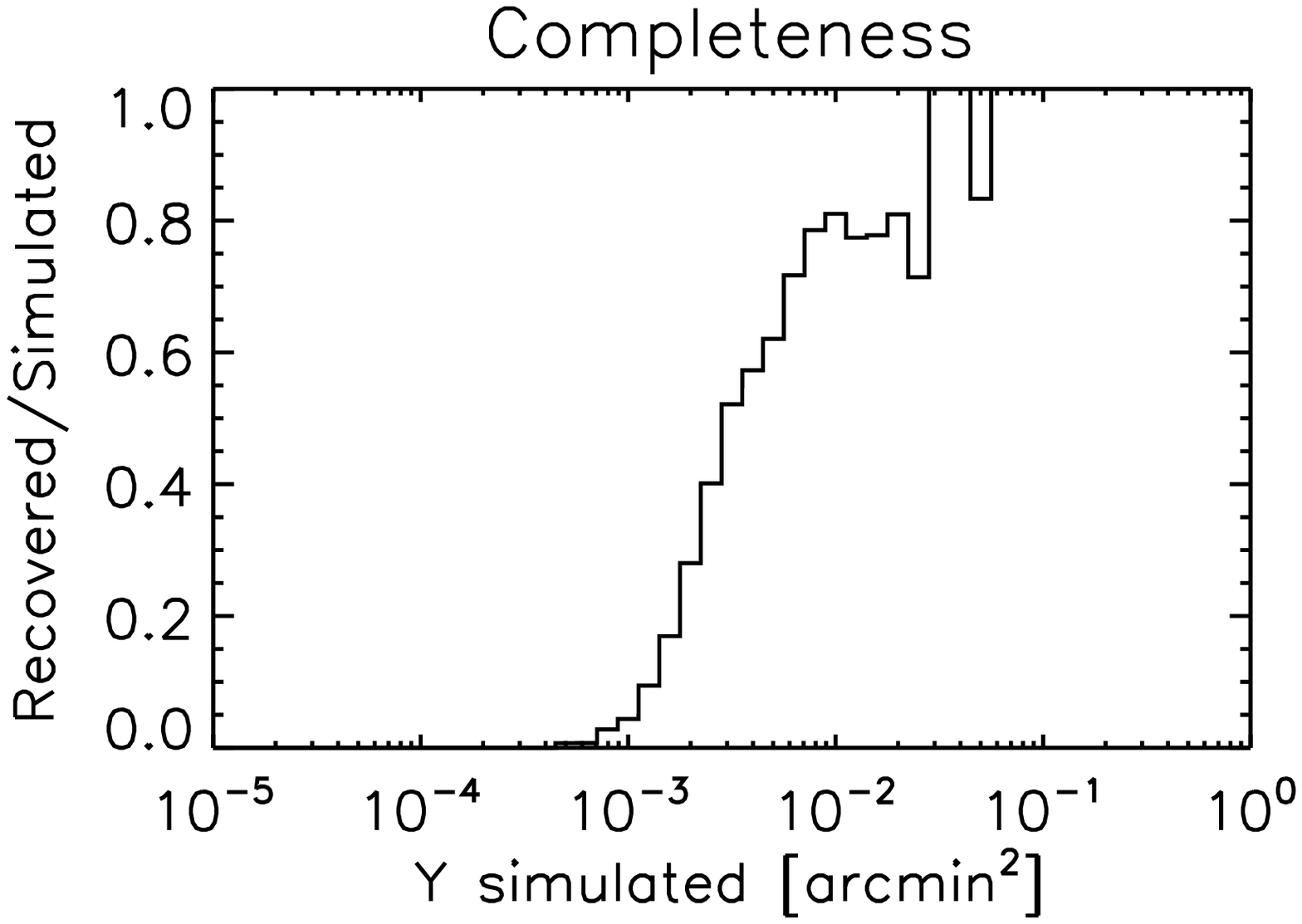}  &
\includegraphics[scale=0.45]{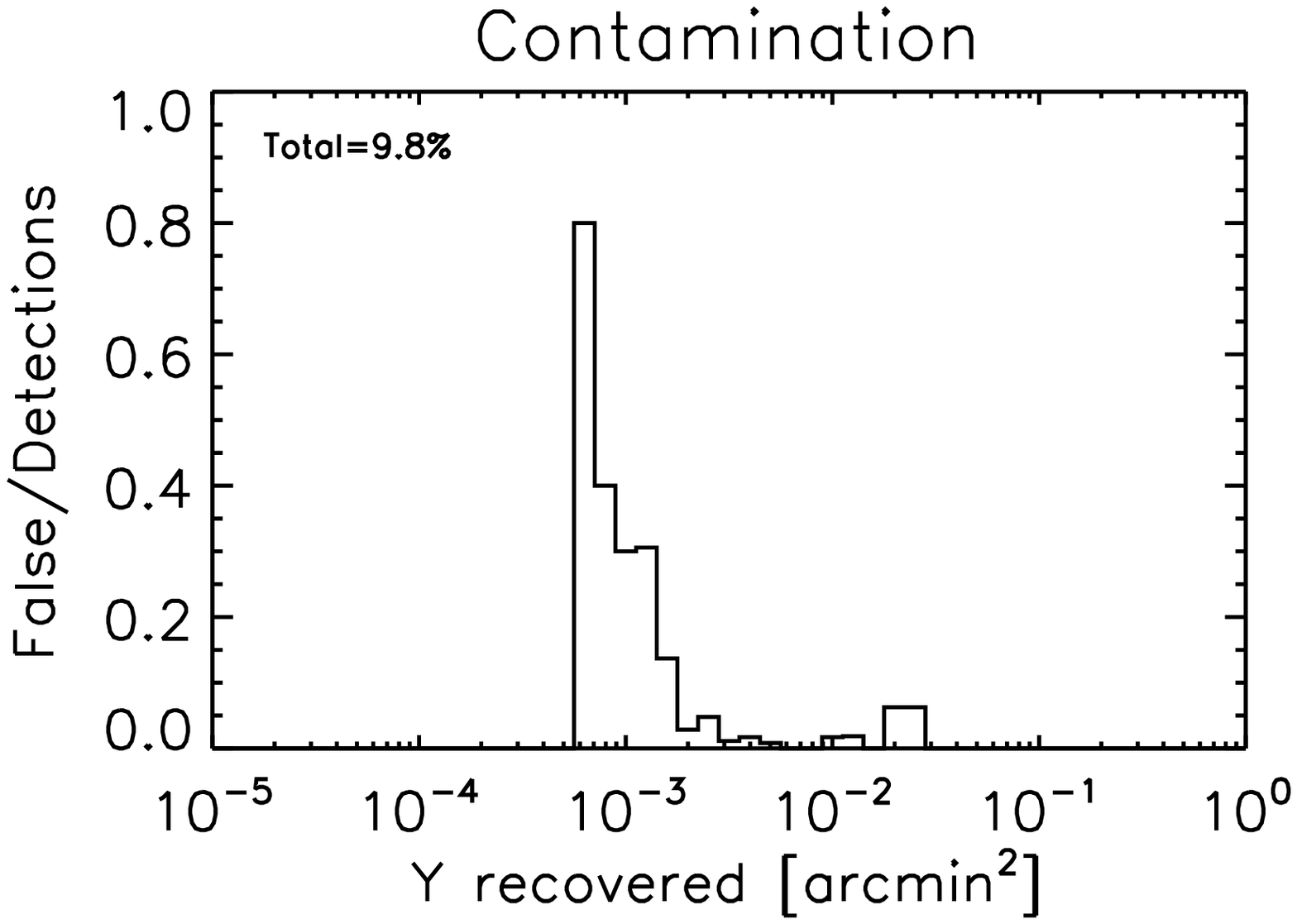} \\
\includegraphics[scale=0.45]{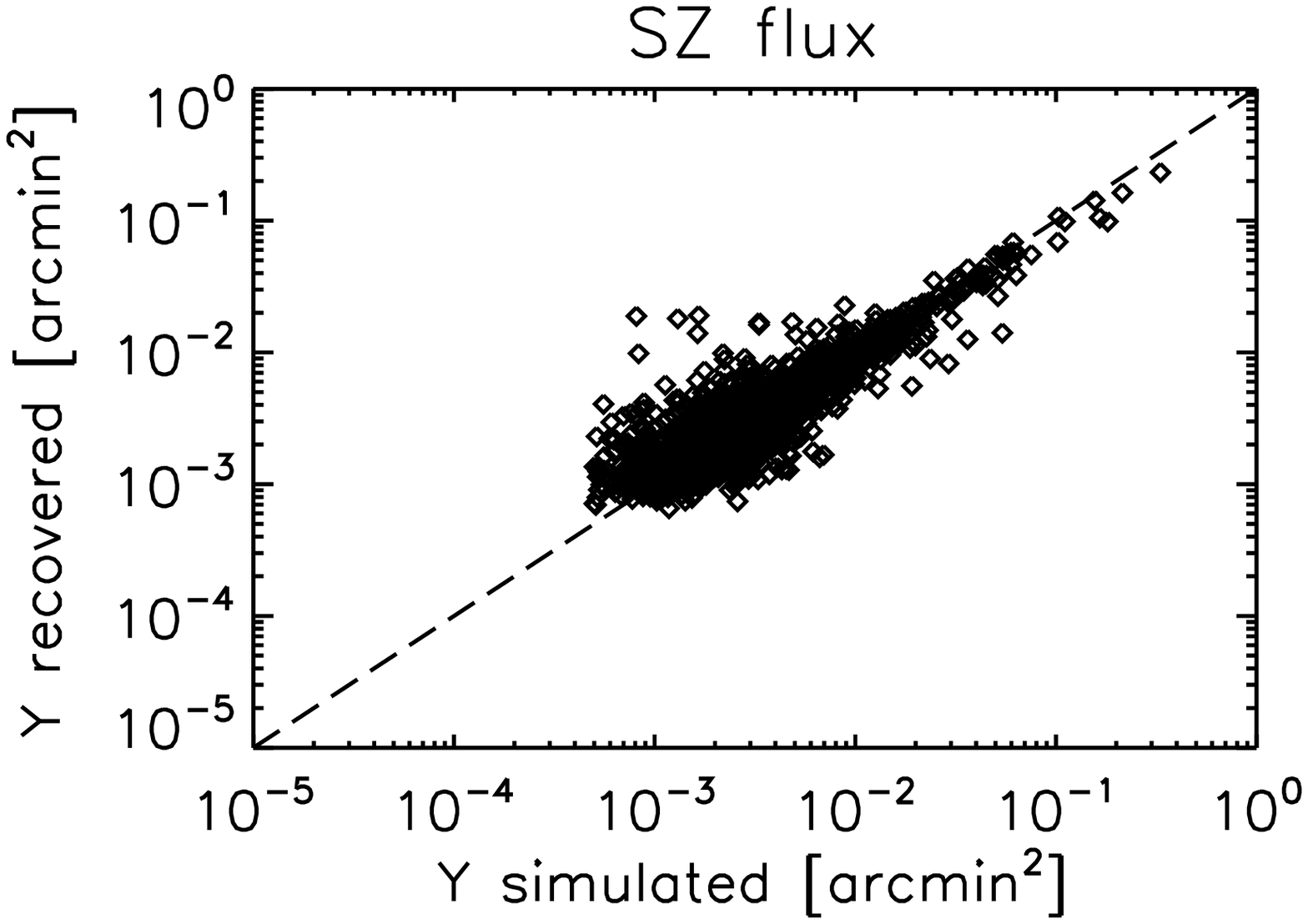}  &
\includegraphics[scale=0.45]{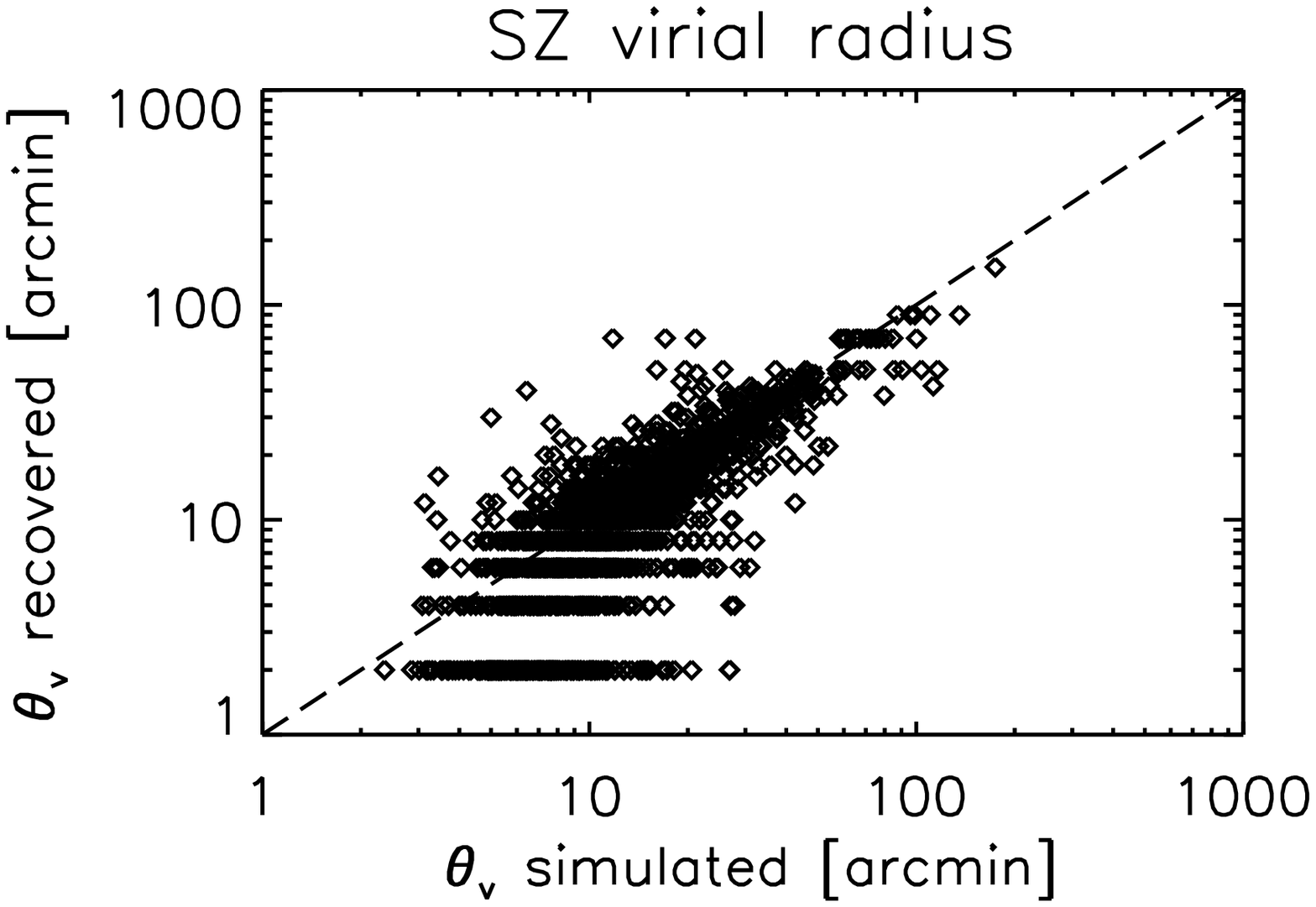} \\
\includegraphics[scale=0.45]{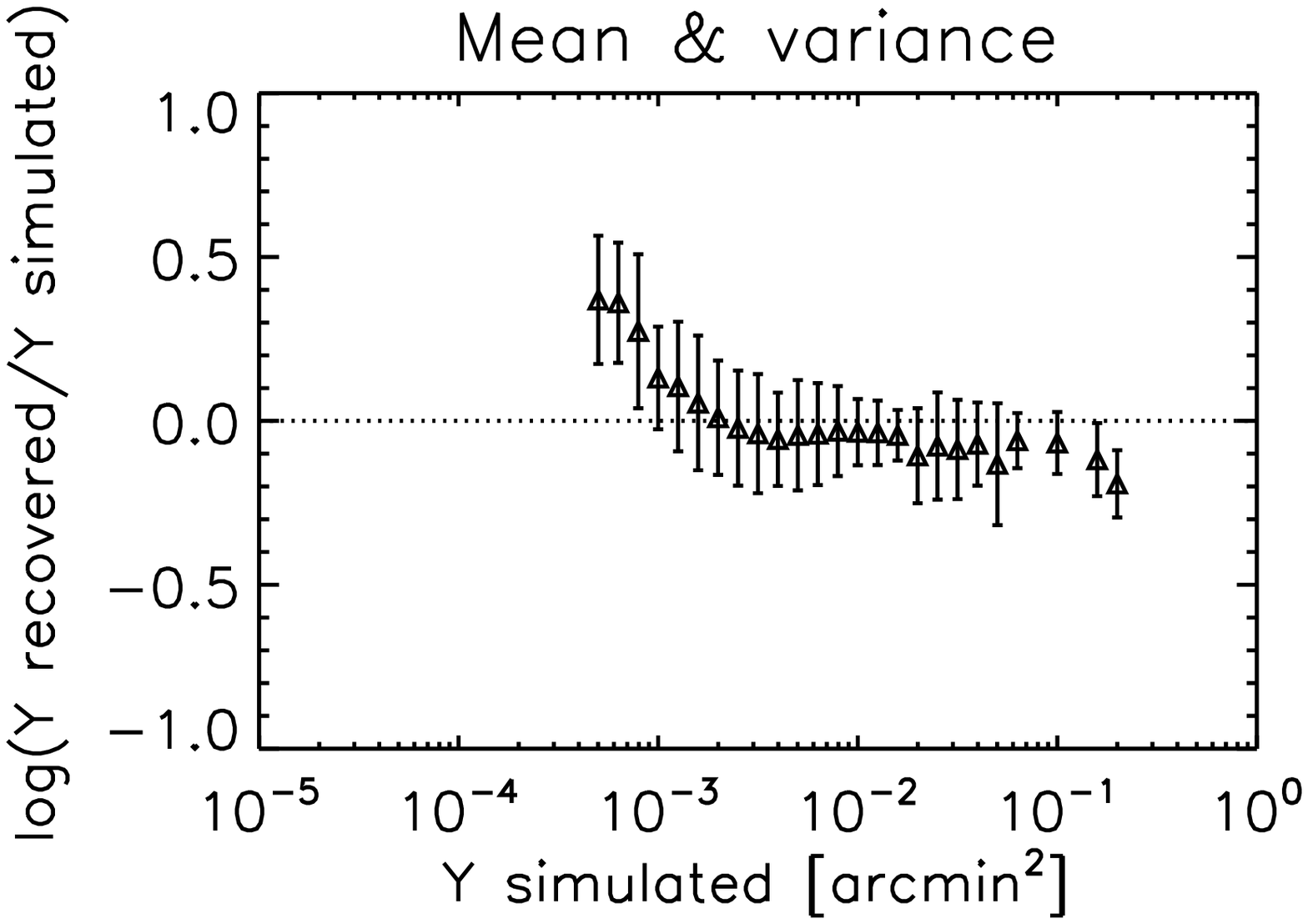}  &
\includegraphics[scale=0.45]{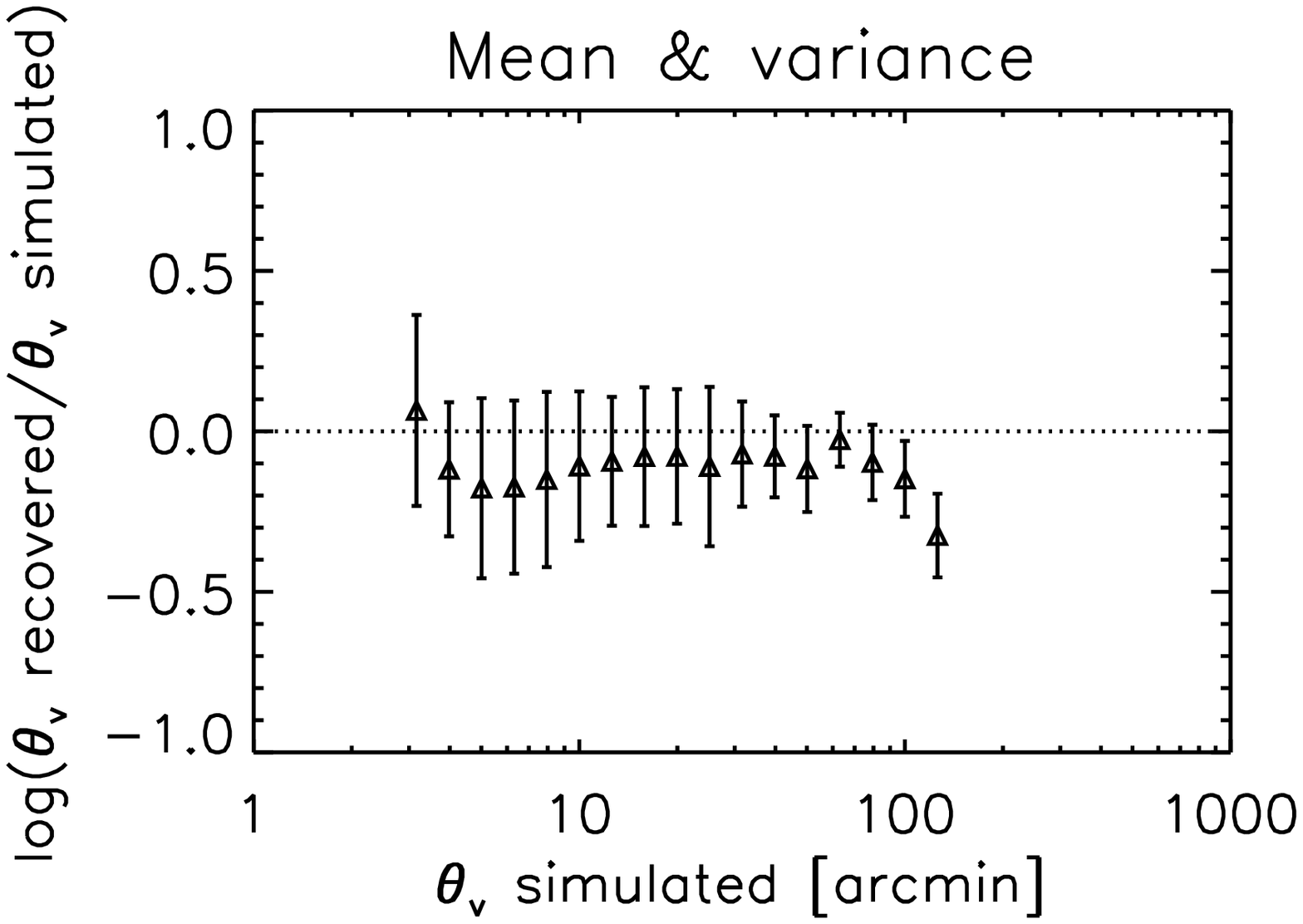} \\
\end{tabular}
\caption{{\bf MMF3}}
\end{center}
\end{table}

\clearpage

\begin{table}[htbp]
\begin{center}
\begin{tabular}{cc}
\includegraphics[scale=0.45]{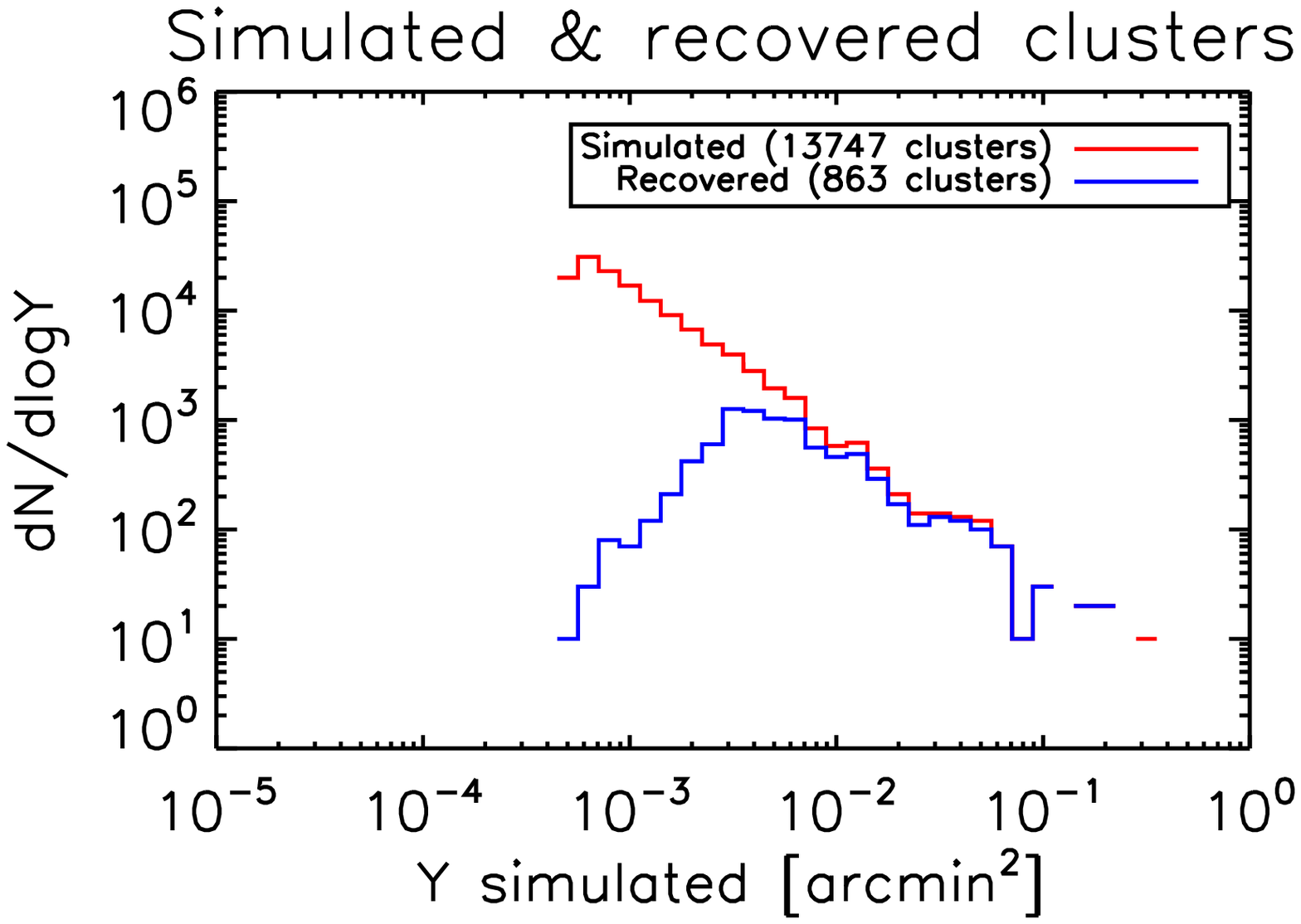}  & \\
\includegraphics[scale=0.45]{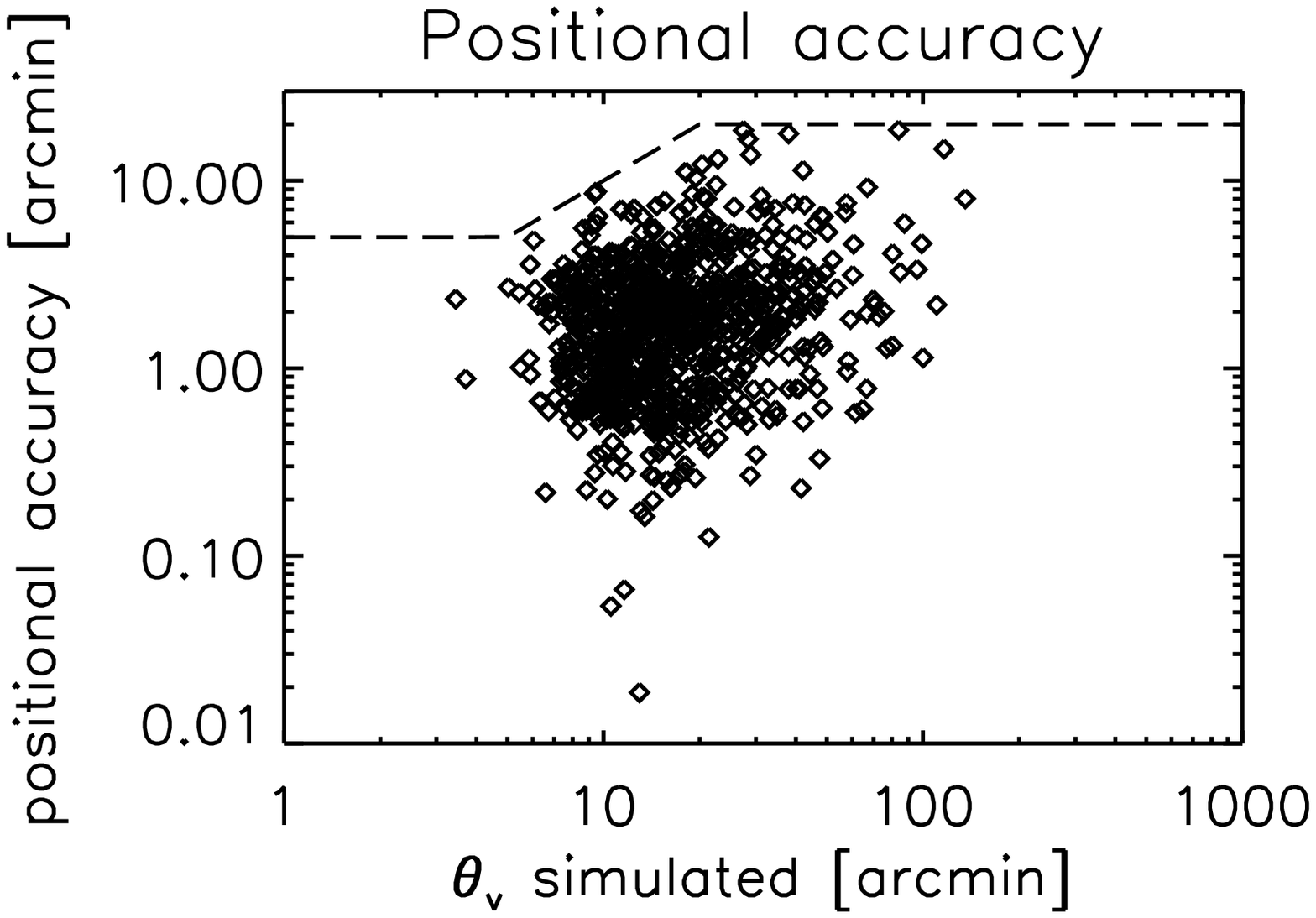} \\
\includegraphics[scale=0.45]{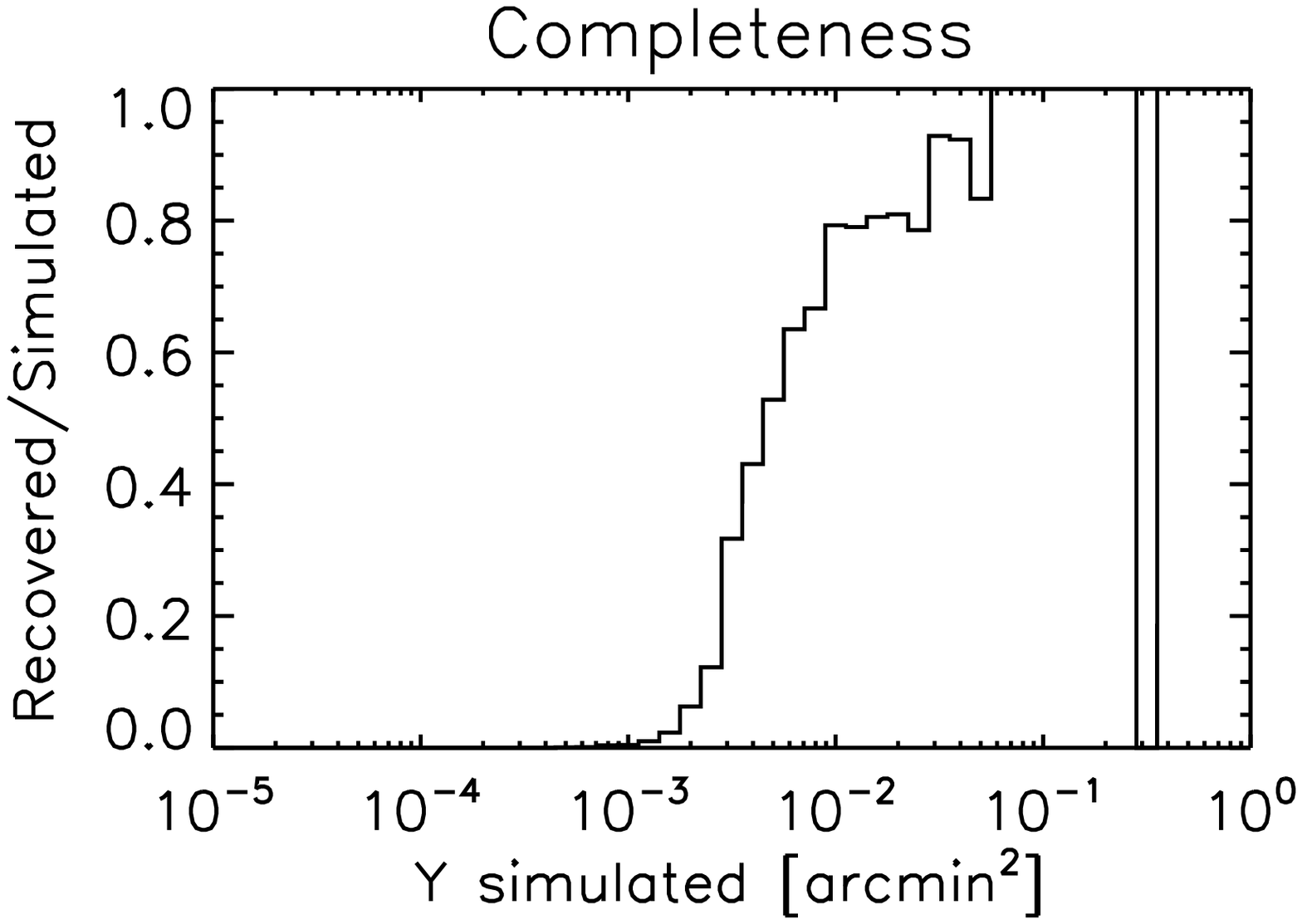}  & \\
\end{tabular}
\caption{{\bf MMF4} (photometry has not been computed for this algorithm.}
\end{center}
\end{table}

\clearpage

\begin{table}[htbp]
\begin{center}
\begin{tabular}{cc}
\includegraphics[scale=0.45]{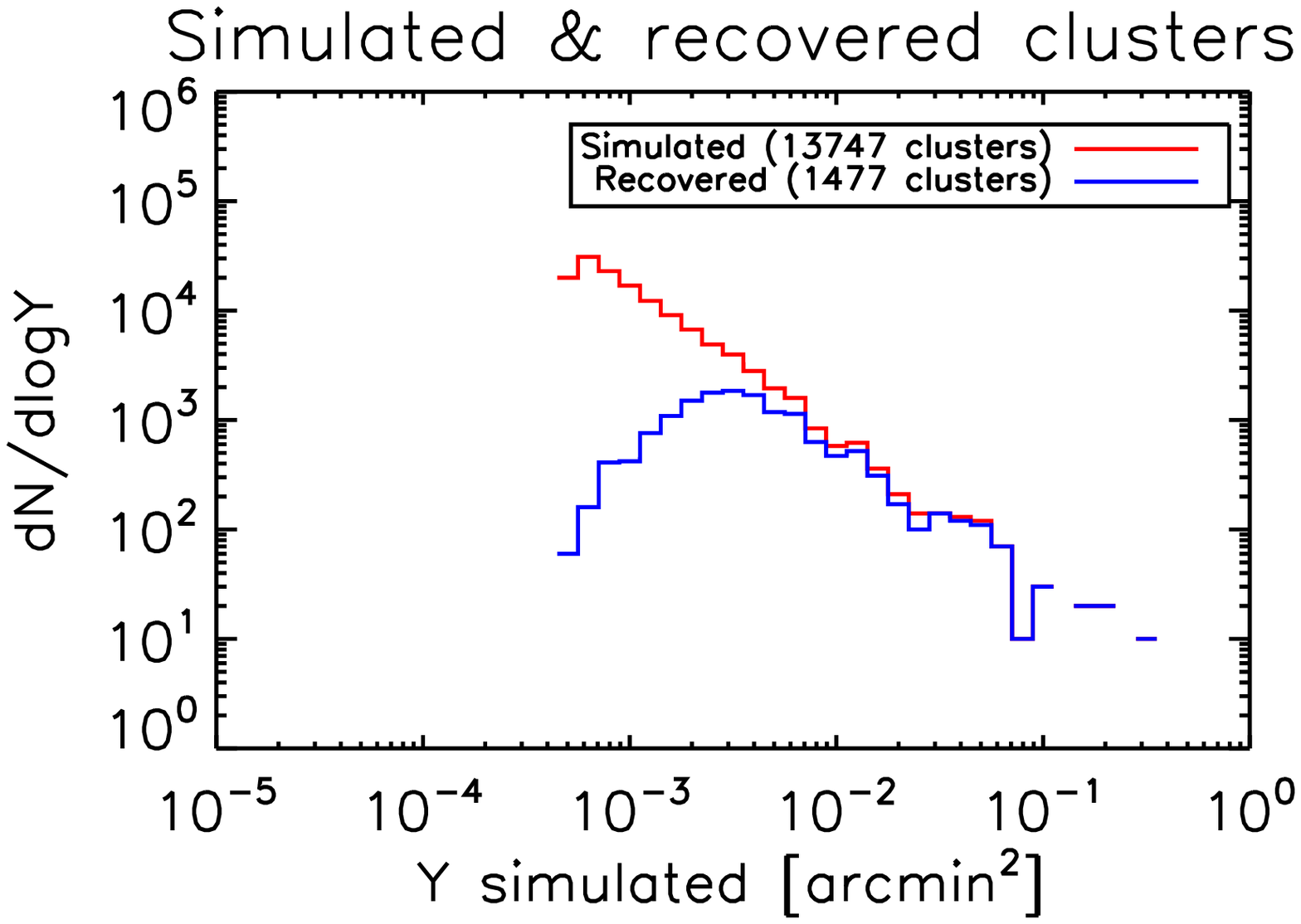}  &
\includegraphics[scale=0.45]{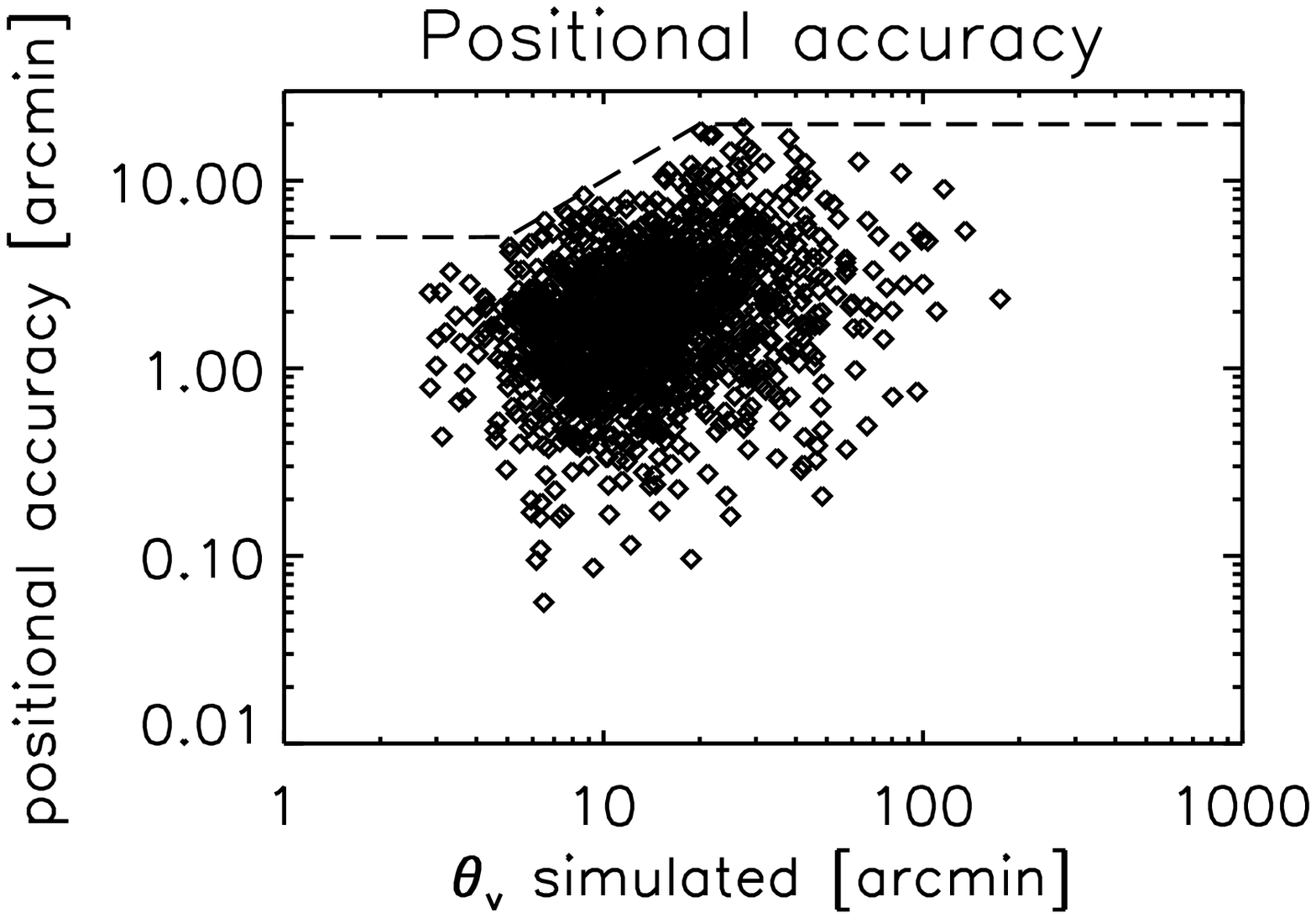} \\
\includegraphics[scale=0.45]{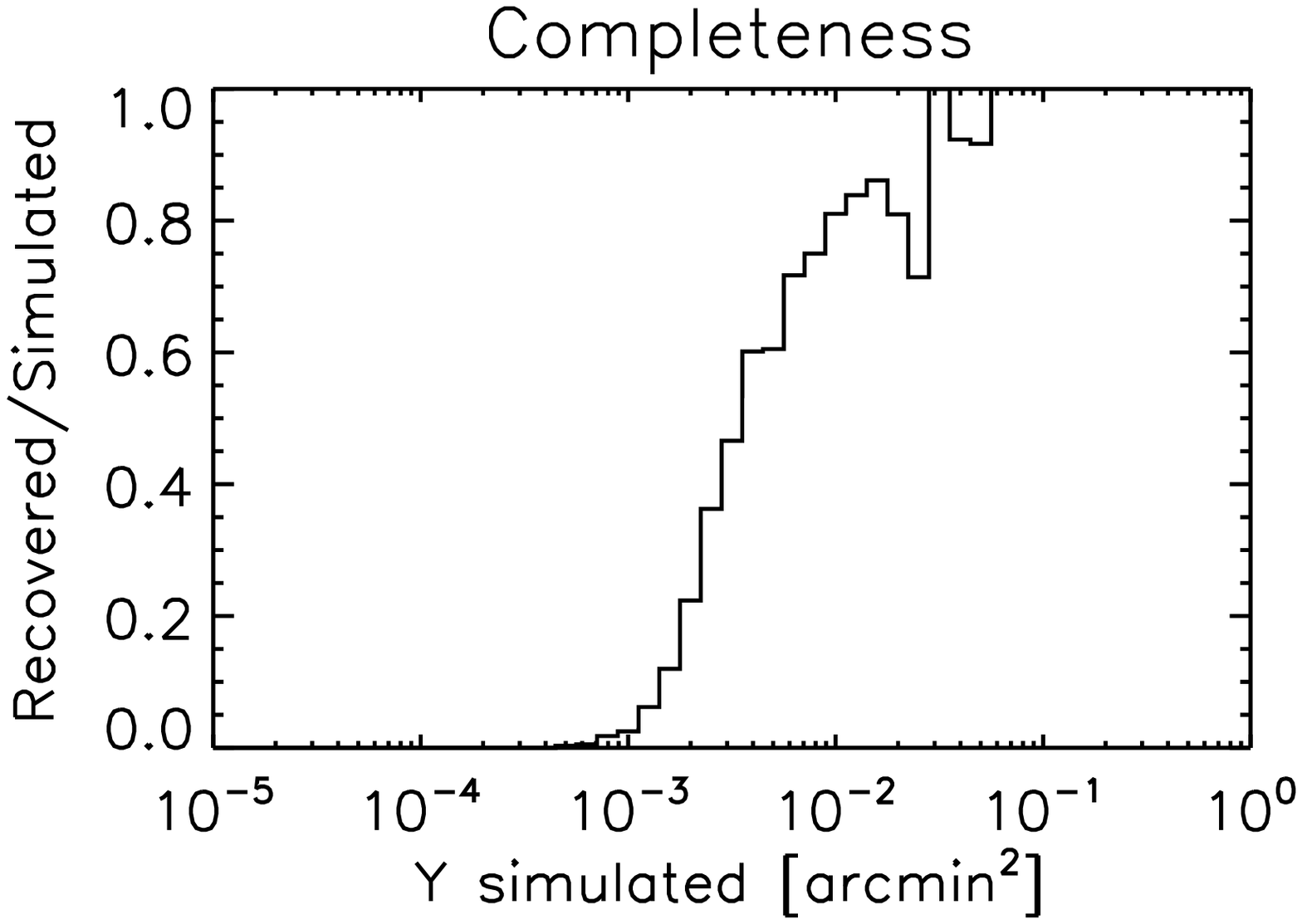}  &
\includegraphics[scale=0.45]{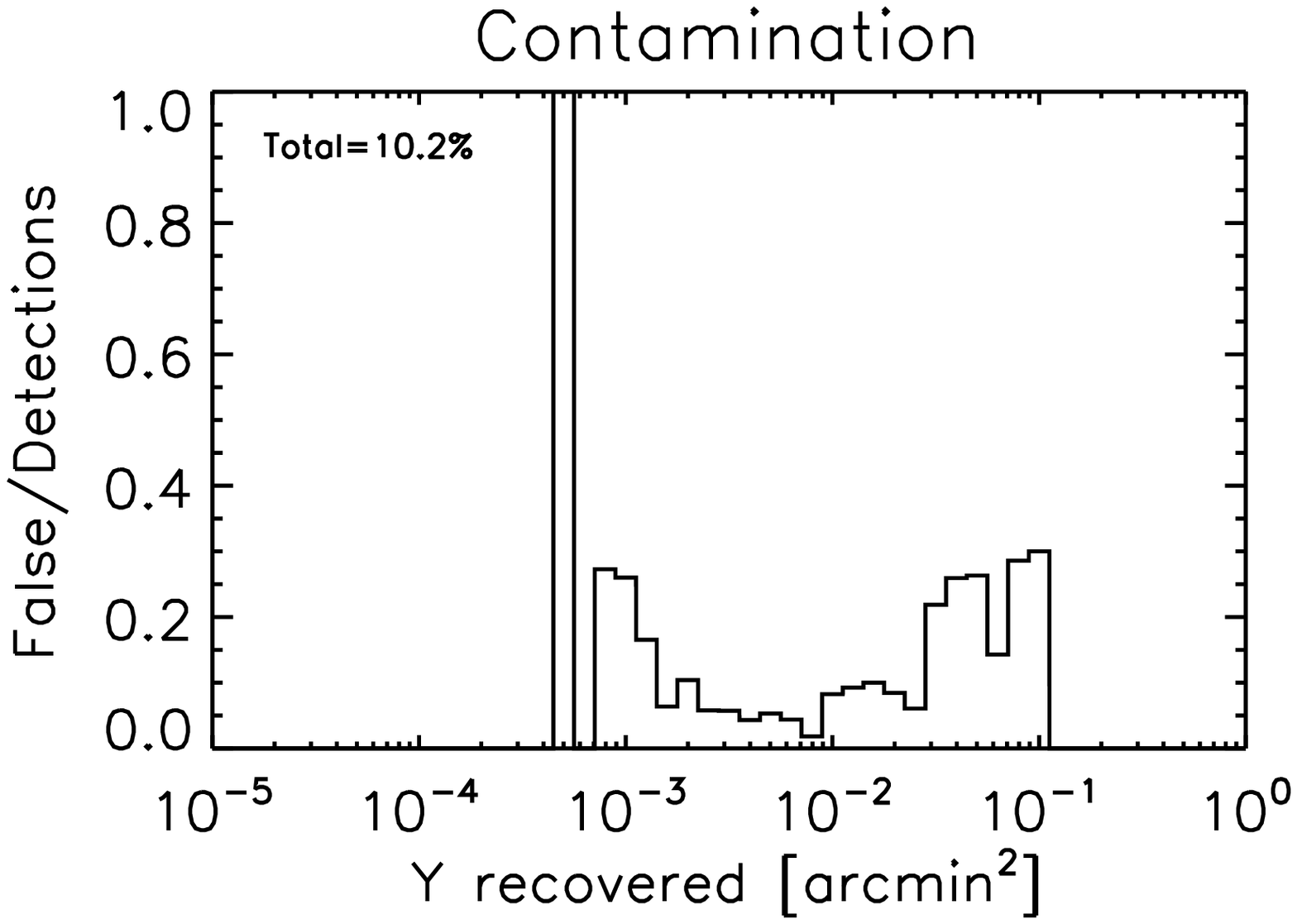} \\
\includegraphics[scale=0.45]{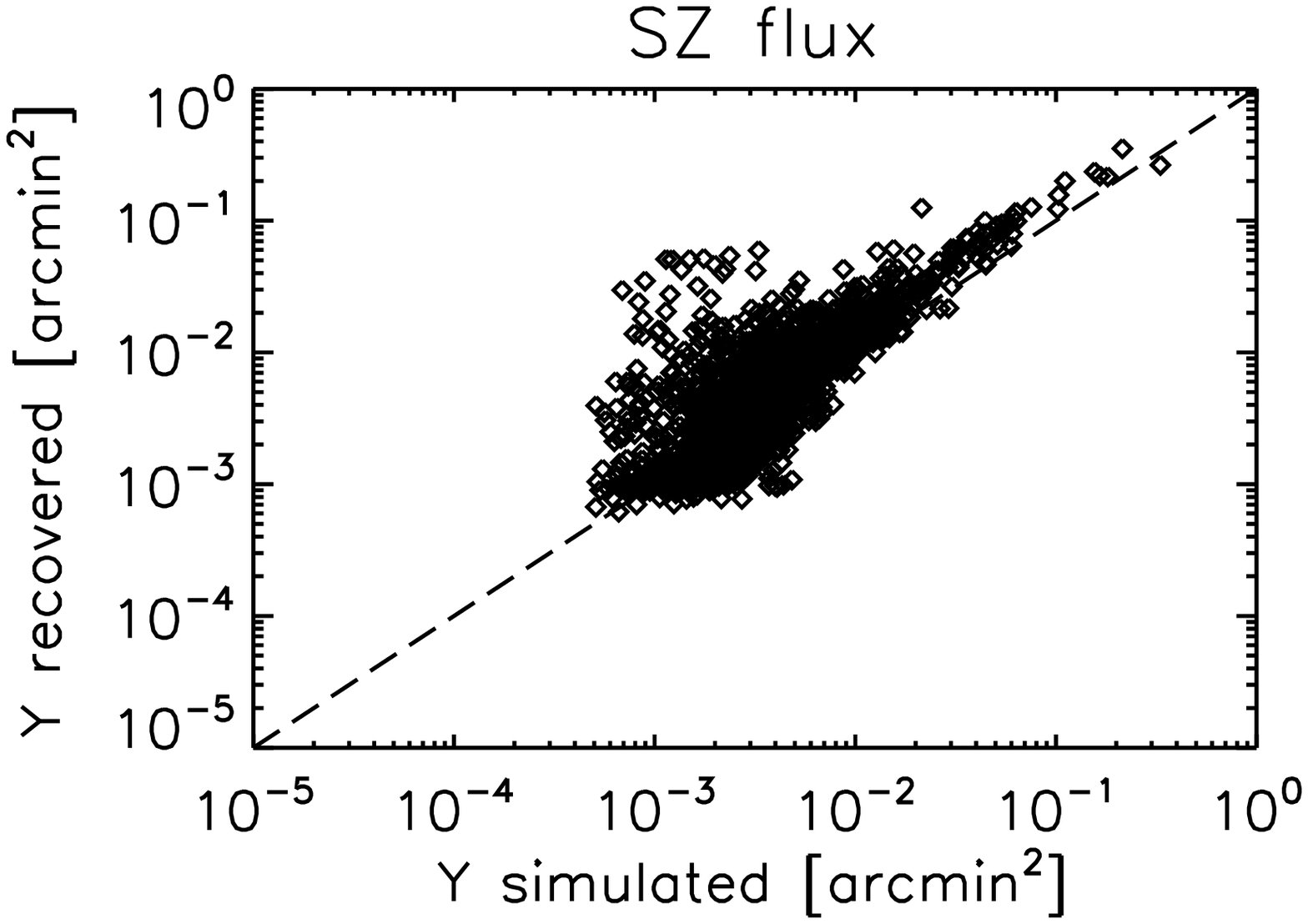}  &
\includegraphics[scale=0.45]{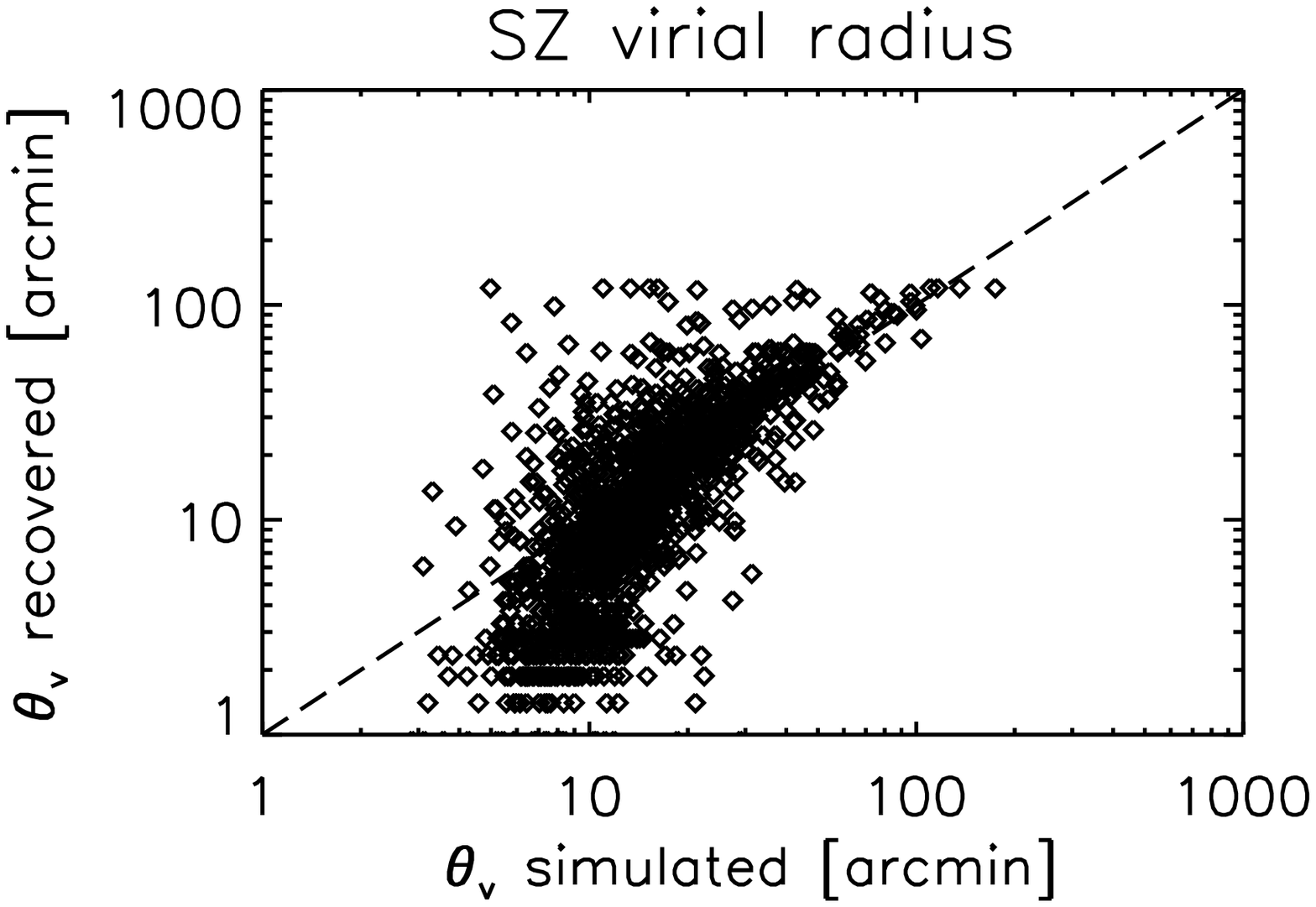} \\
\includegraphics[scale=0.45]{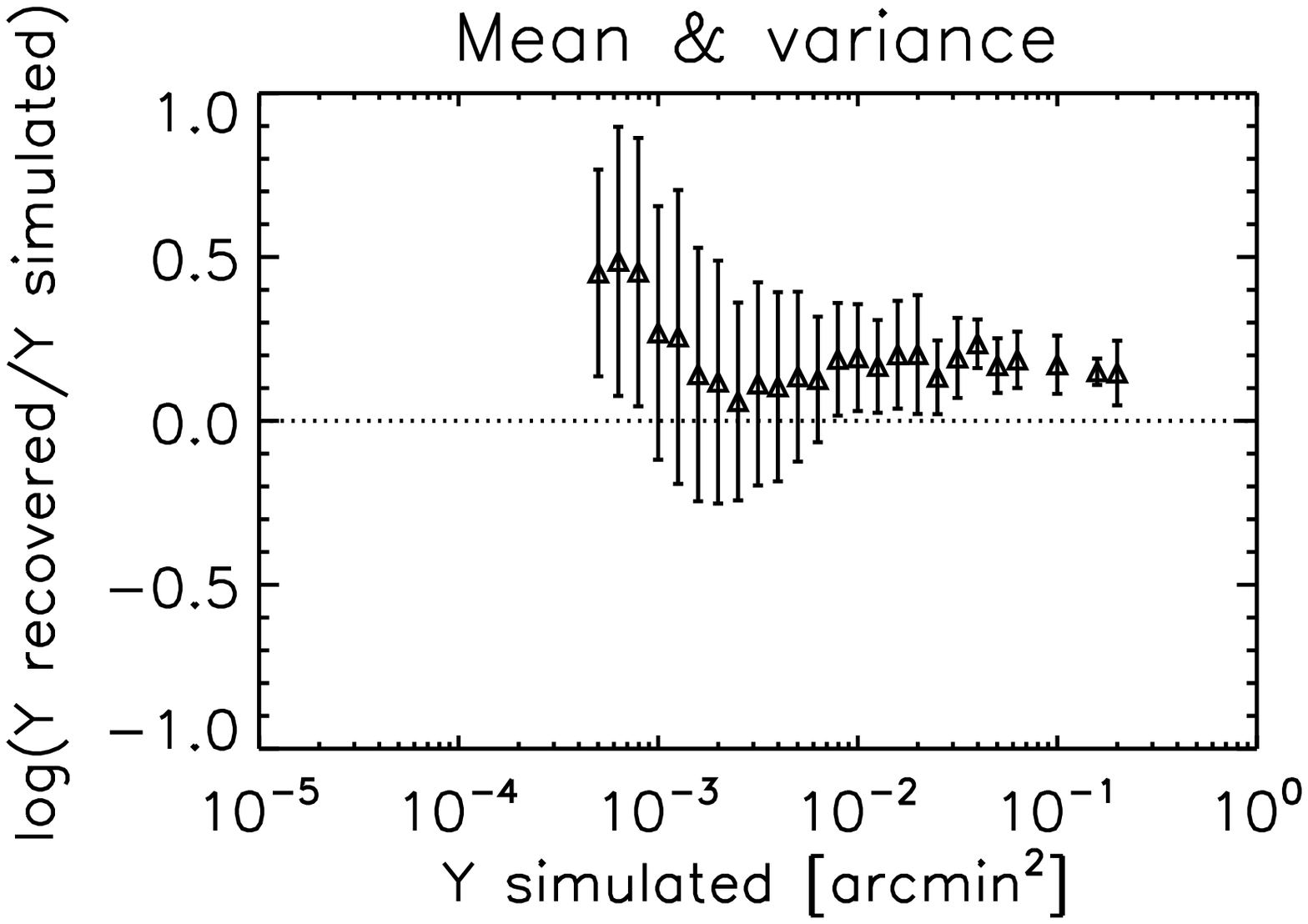}  &
\includegraphics[scale=0.45]{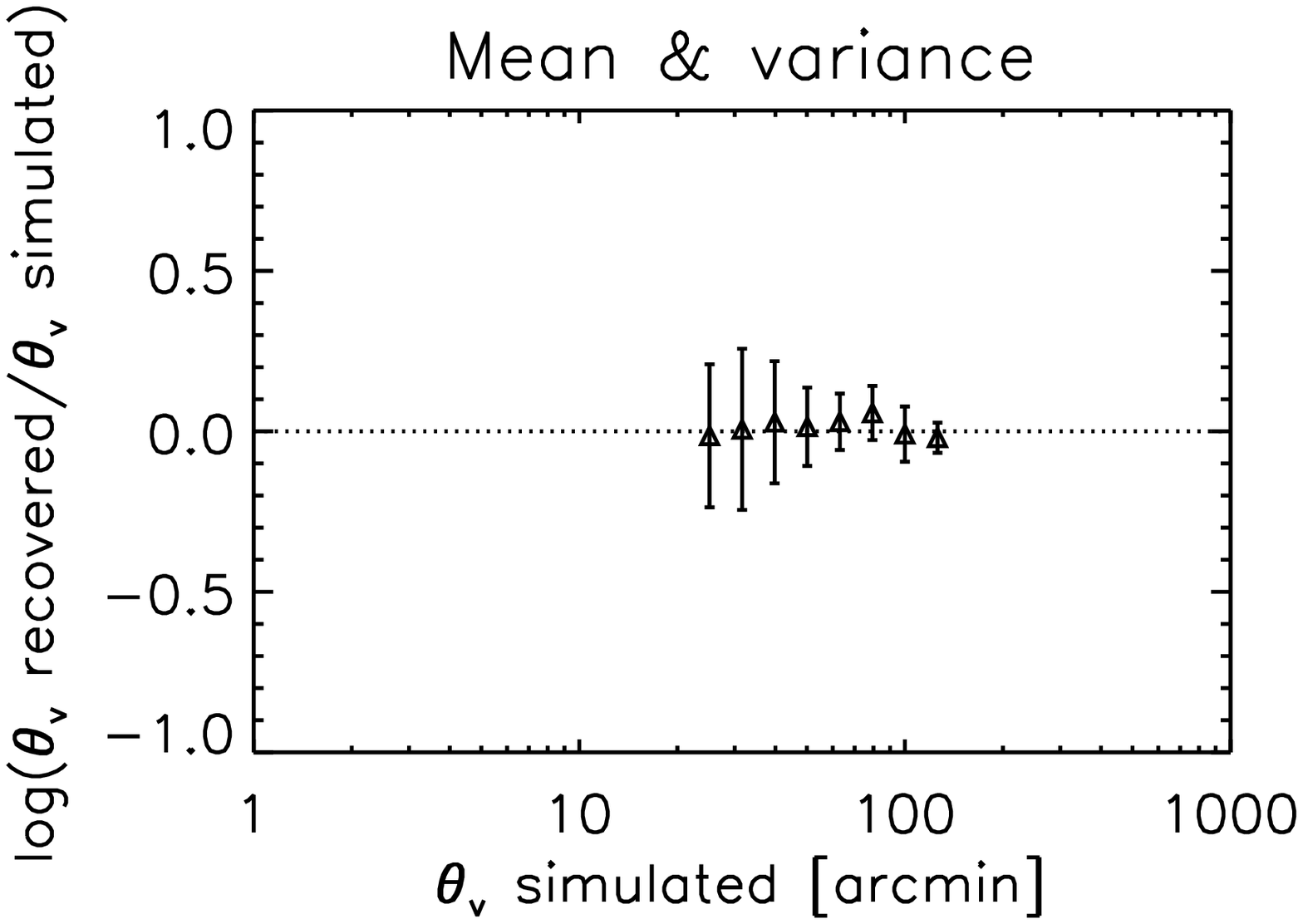} \\
\end{tabular}
\caption{{\bf PS}}
\end{center}
\end{table}

\clearpage

\begin{table}[htbp]
\begin{center}
\begin{tabular}{cc}
\includegraphics[scale=0.45]{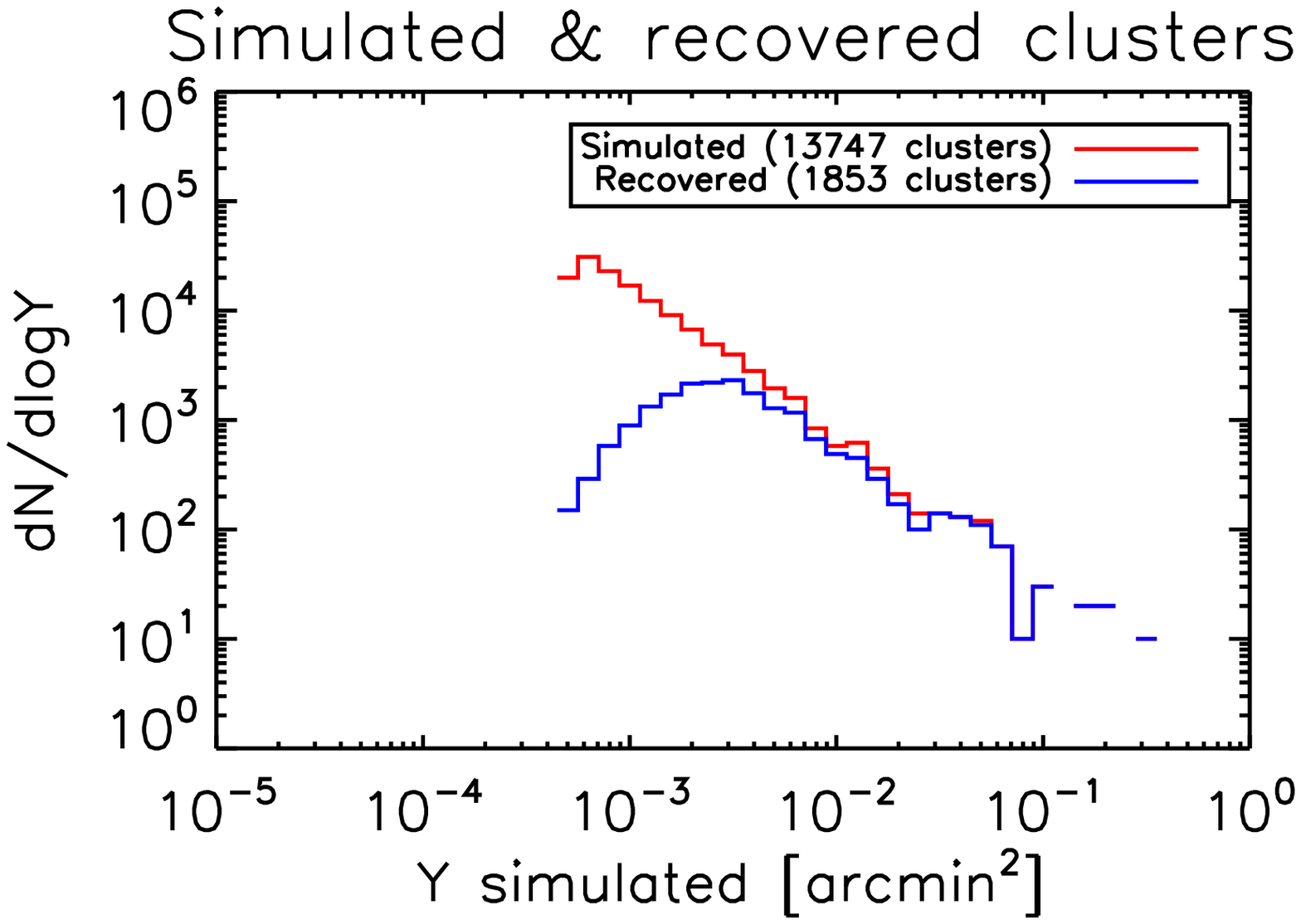}  &
\includegraphics[scale=0.45]{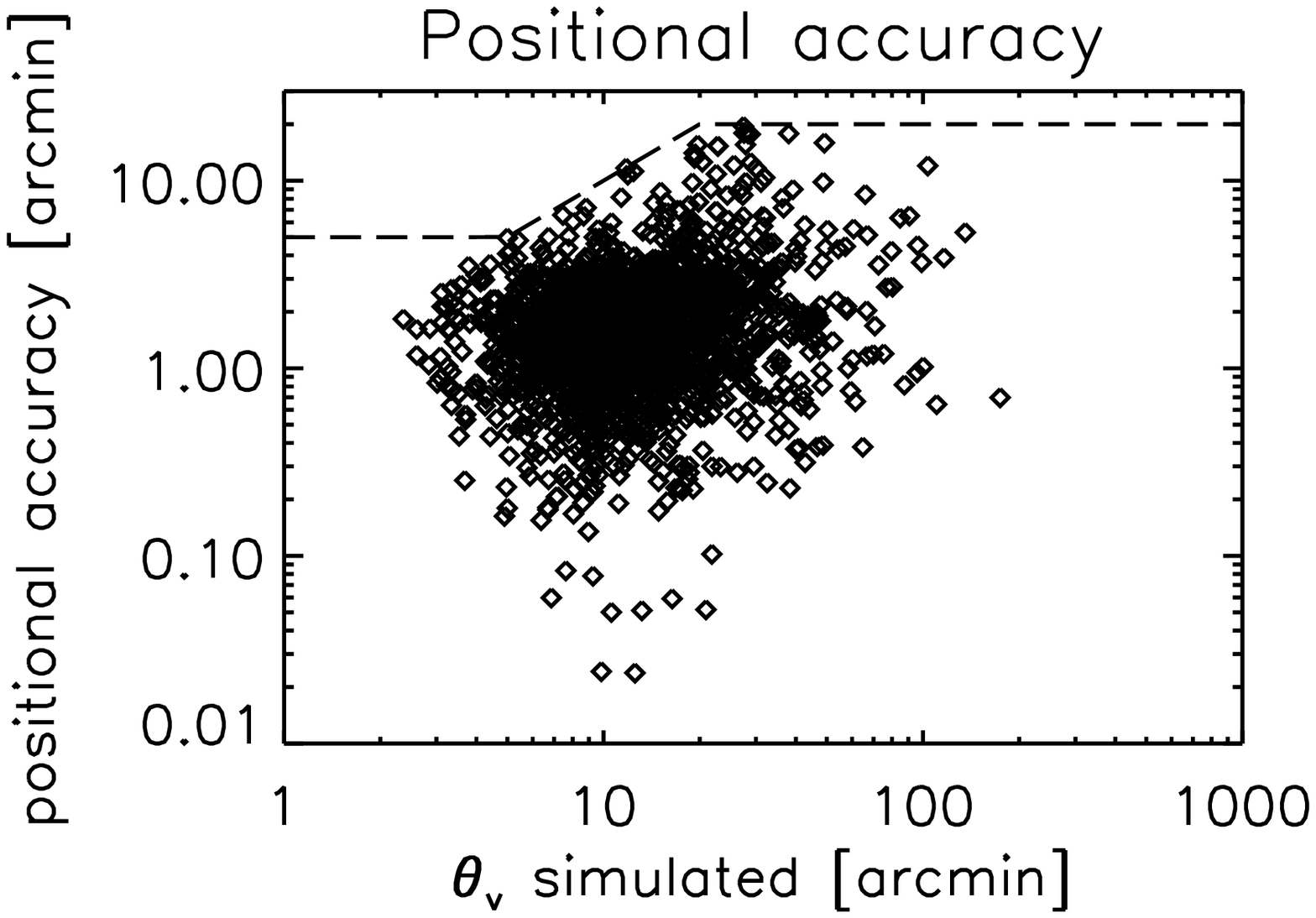} \\
\includegraphics[scale=0.45]{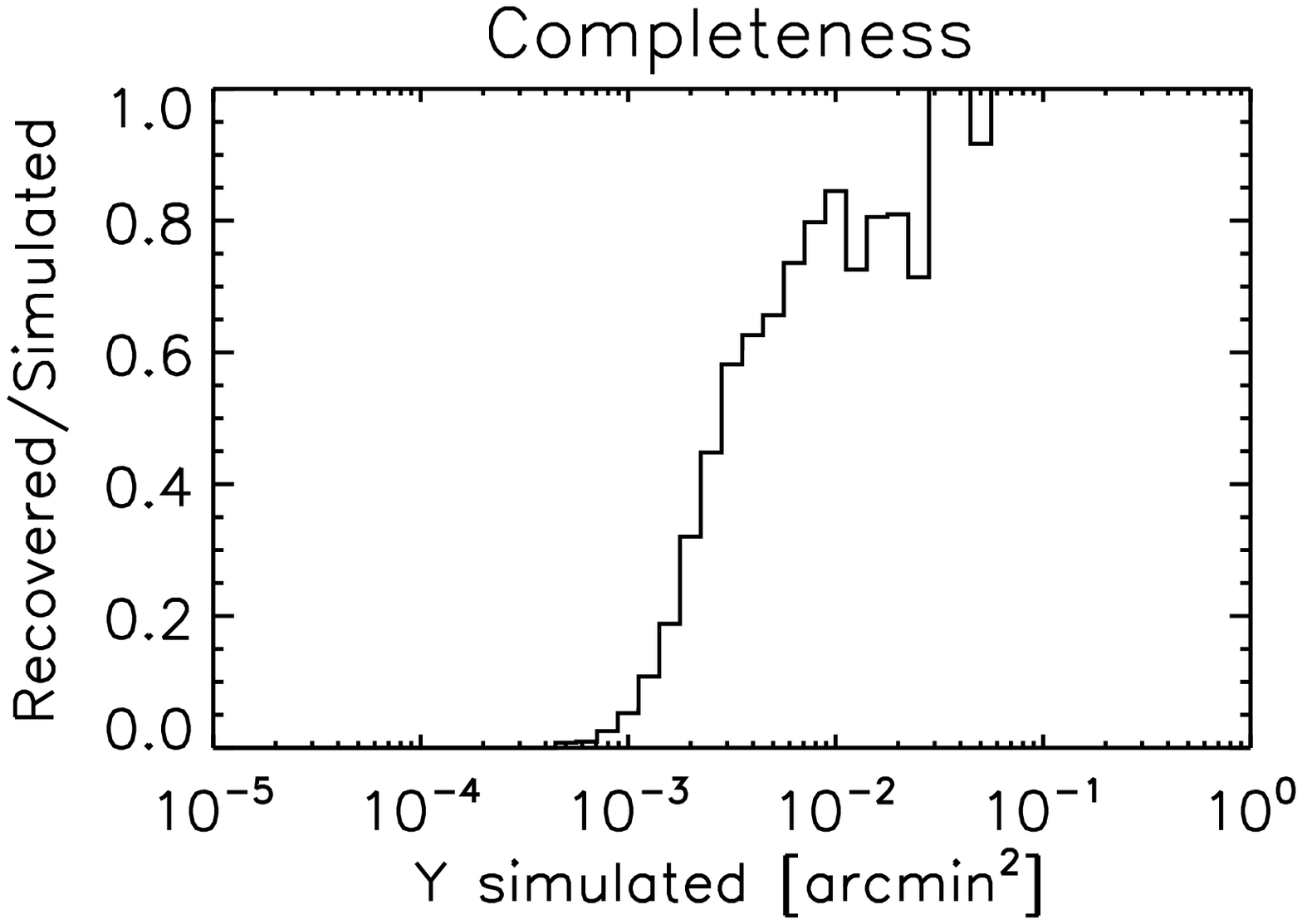}  &
\includegraphics[scale=0.45]{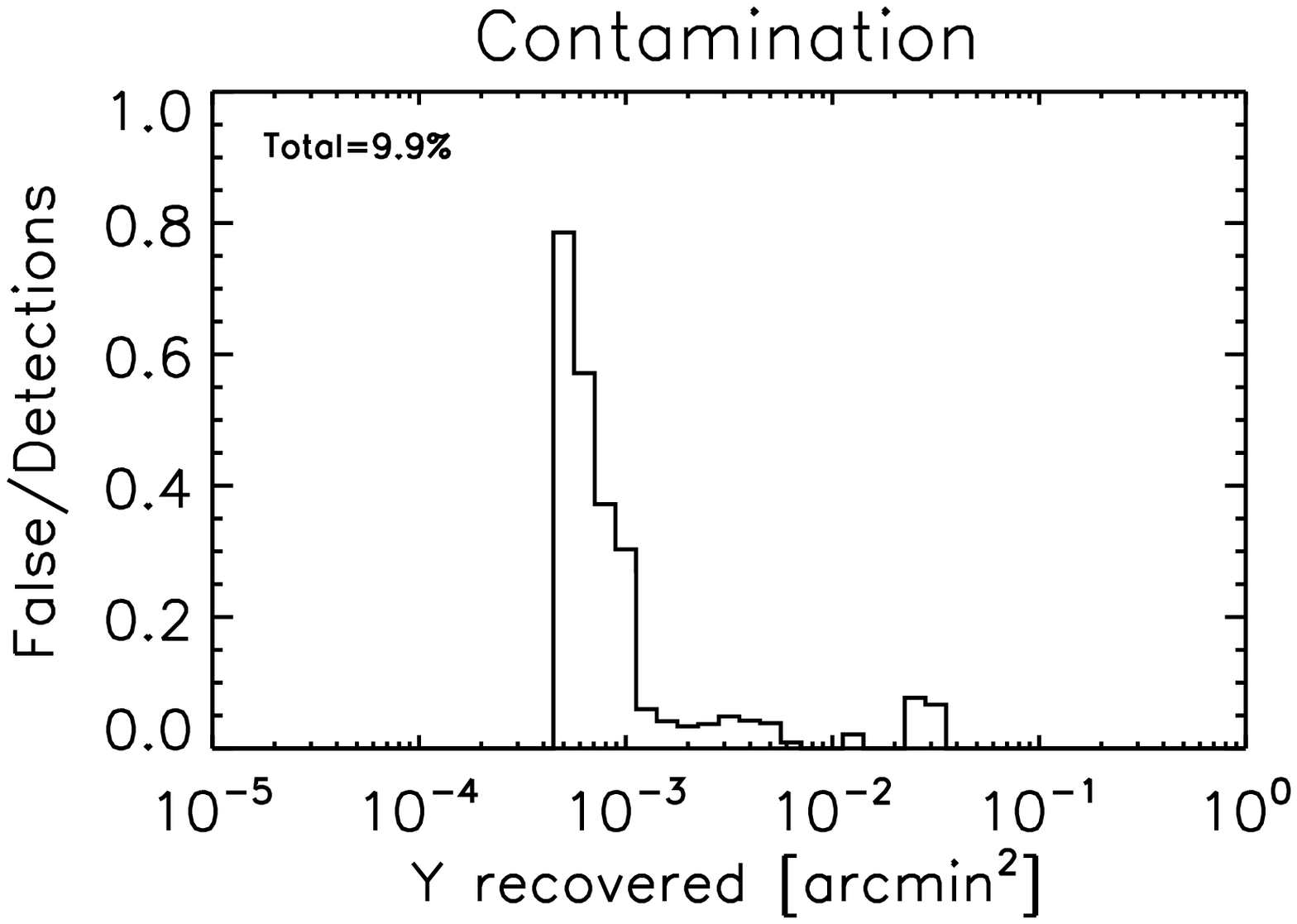} \\
\includegraphics[scale=0.45]{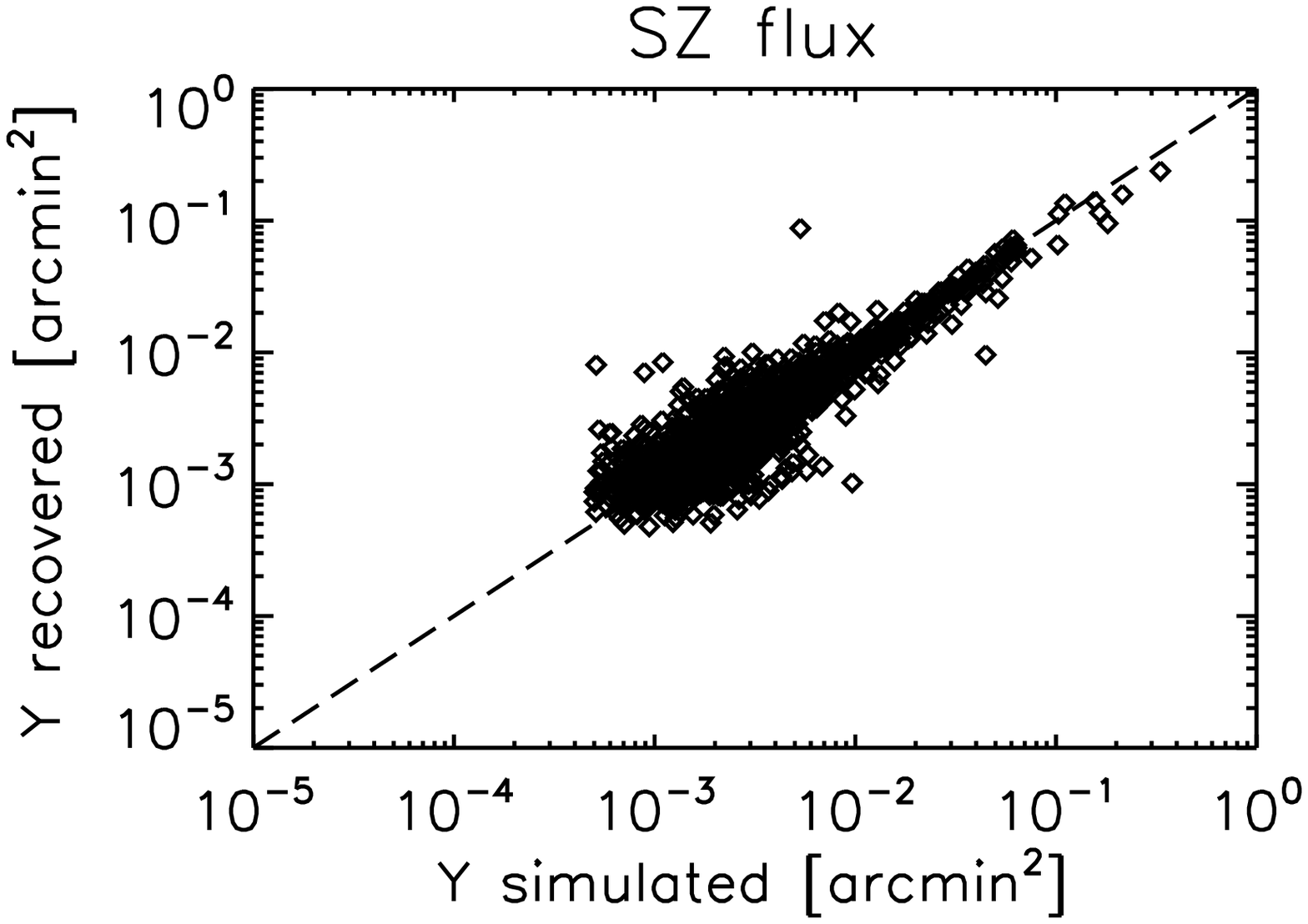}  &
\includegraphics[scale=0.45]{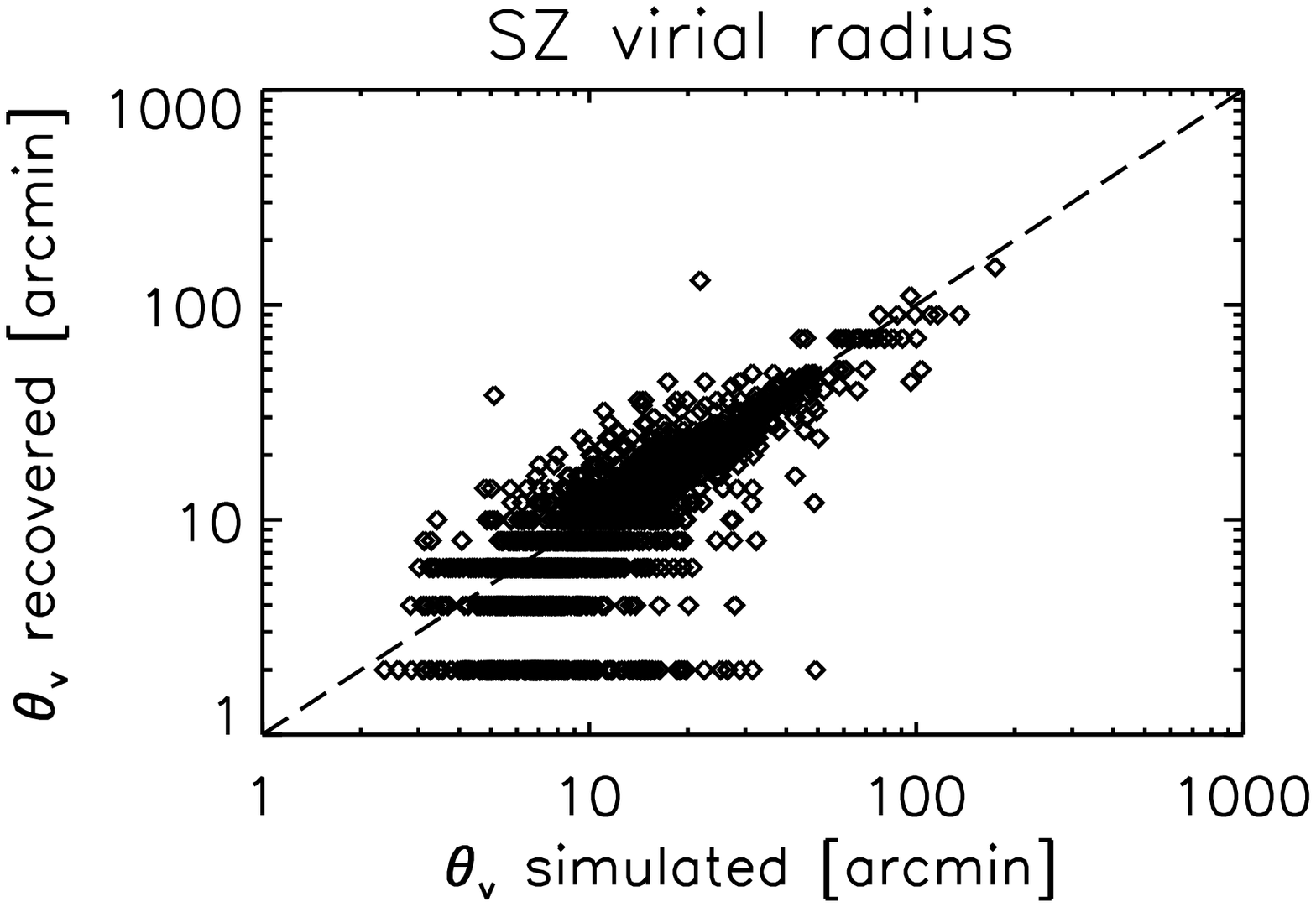} \\
\includegraphics[scale=0.45]{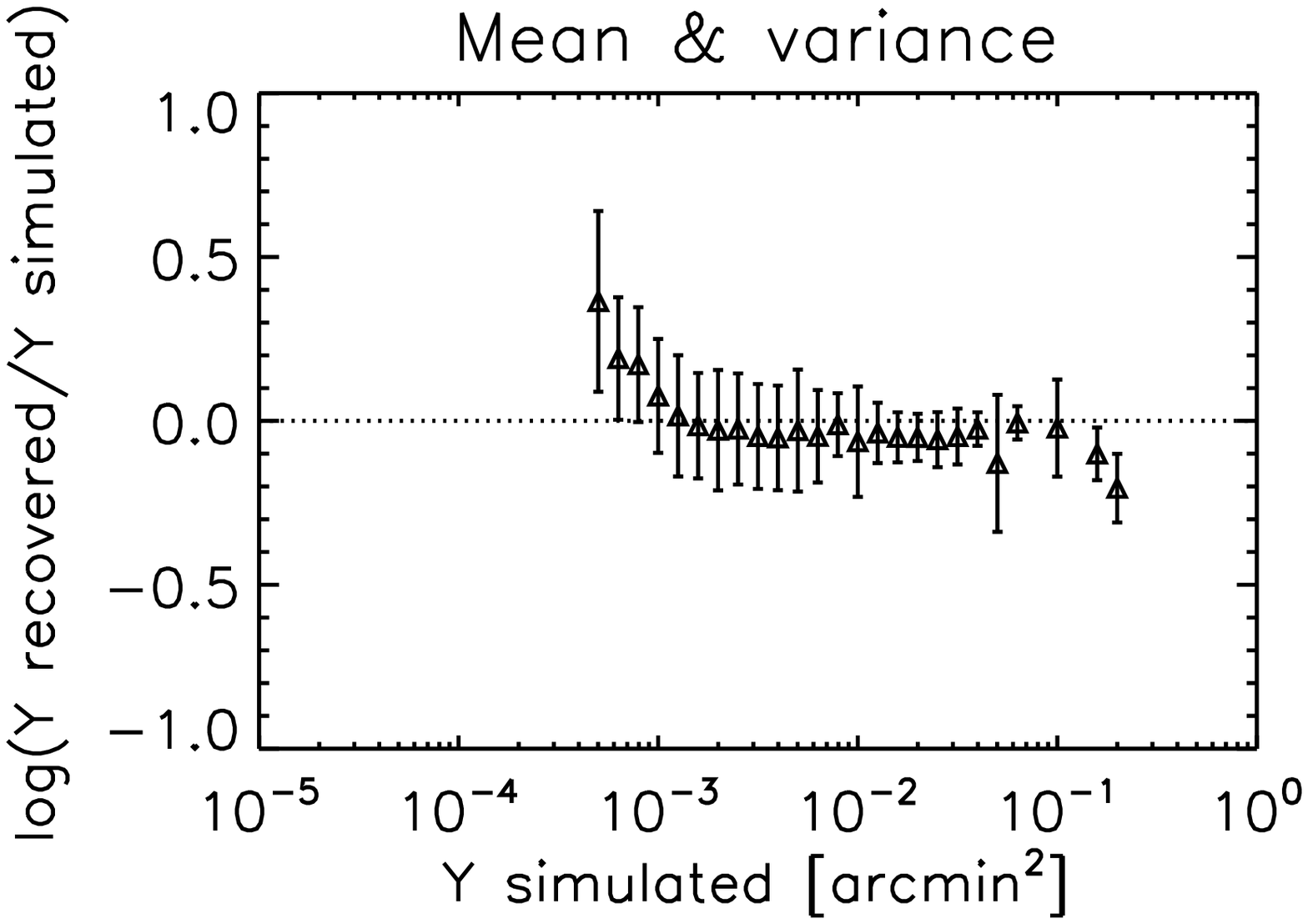}  &
\includegraphics[scale=0.45]{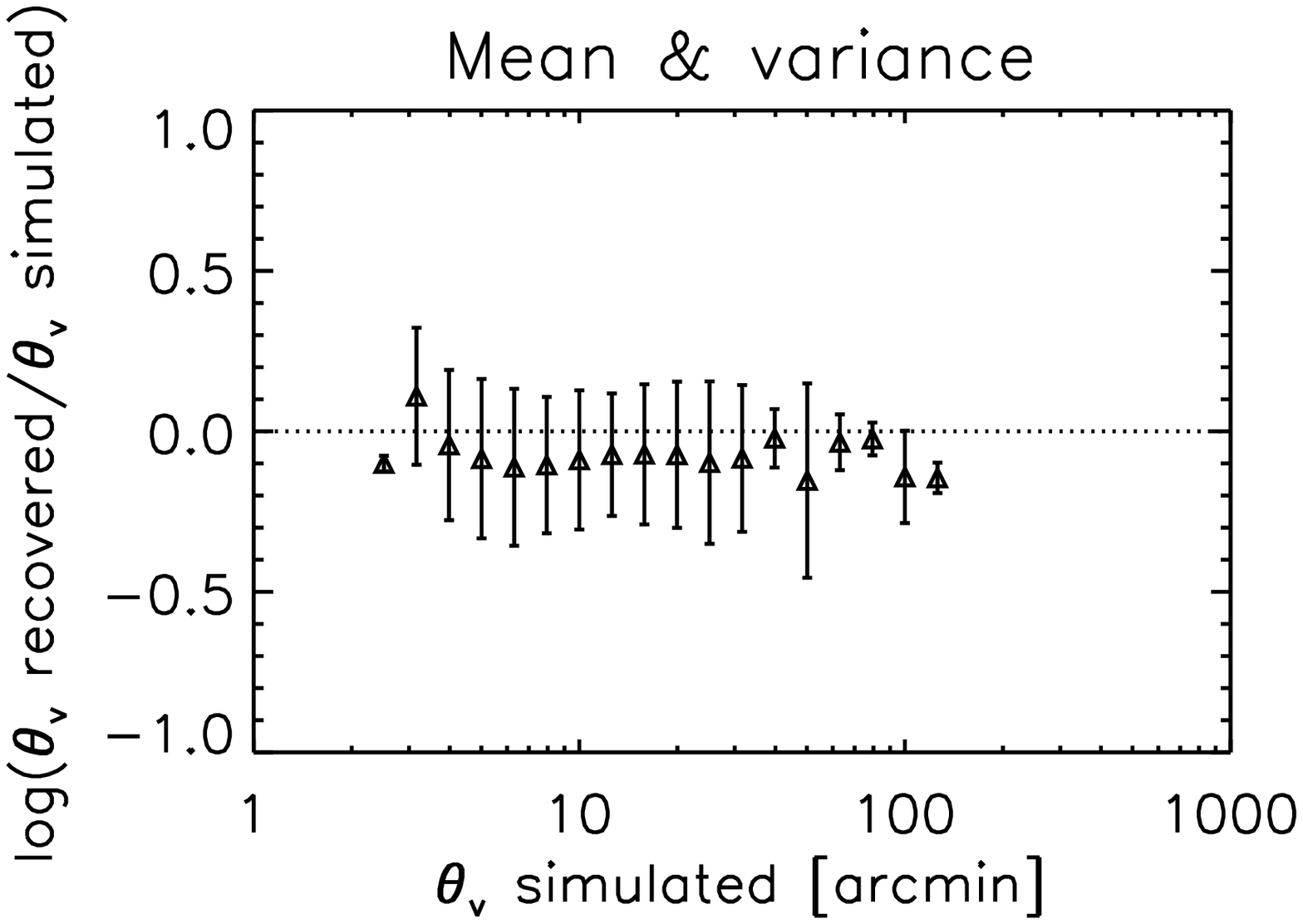} \\
\end{tabular}
\caption{{\bf ILC1}}
\end{center}
\end{table}

\clearpage

\begin{table}[htbp]
\begin{center}
\begin{tabular}{cc}
\includegraphics[scale=0.45]{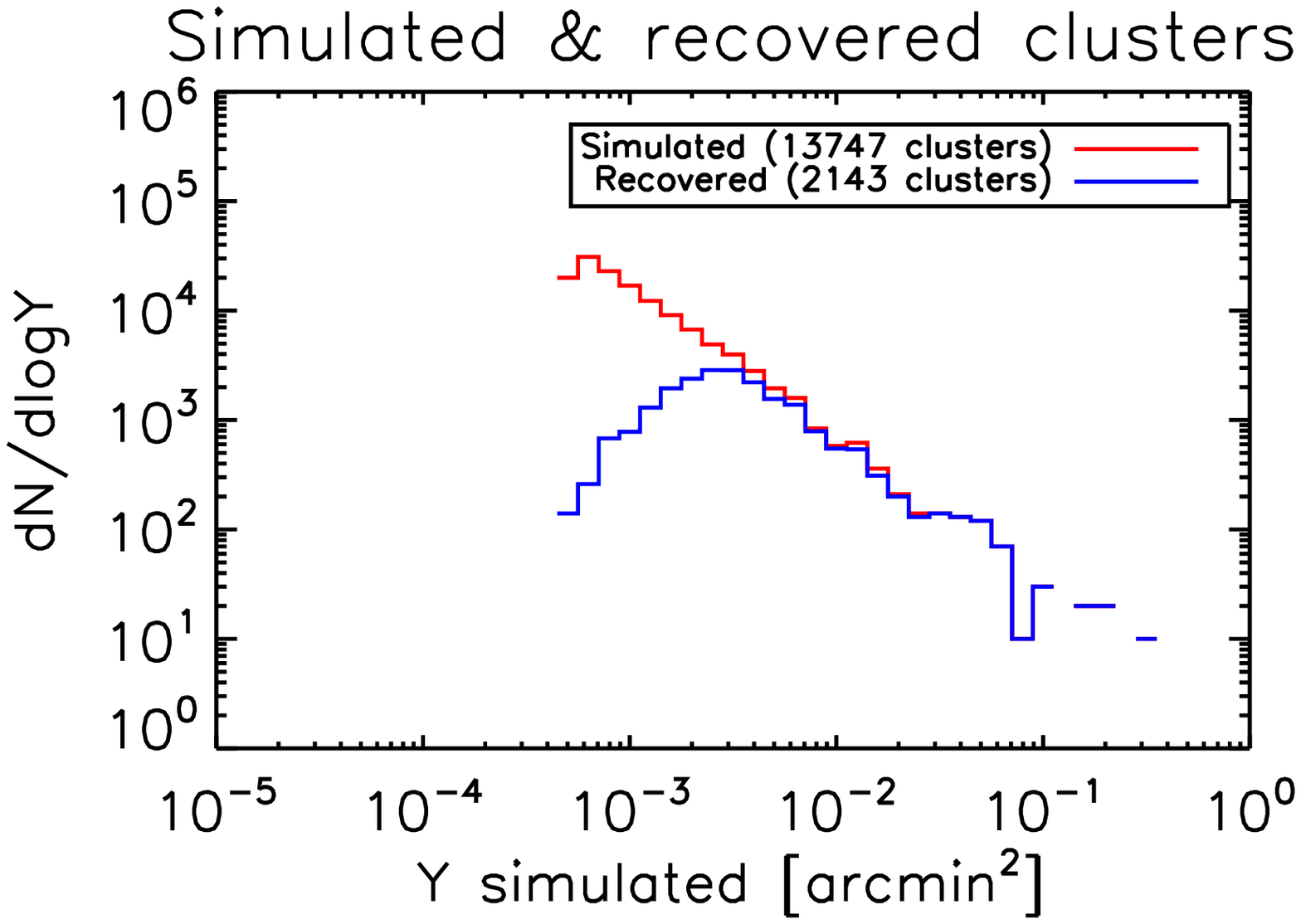}  &
\includegraphics[scale=0.45]{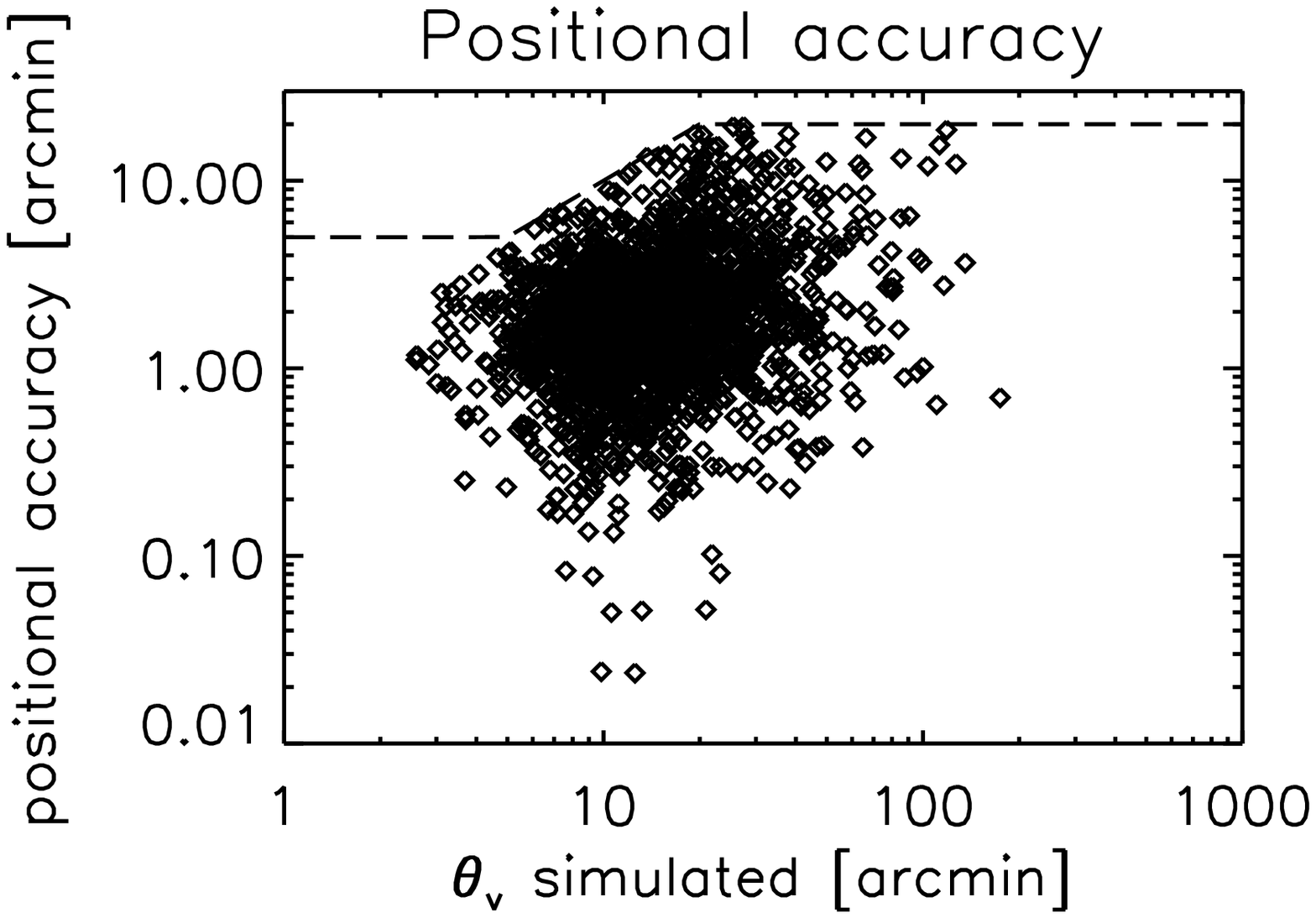} \\
\includegraphics[scale=0.45]{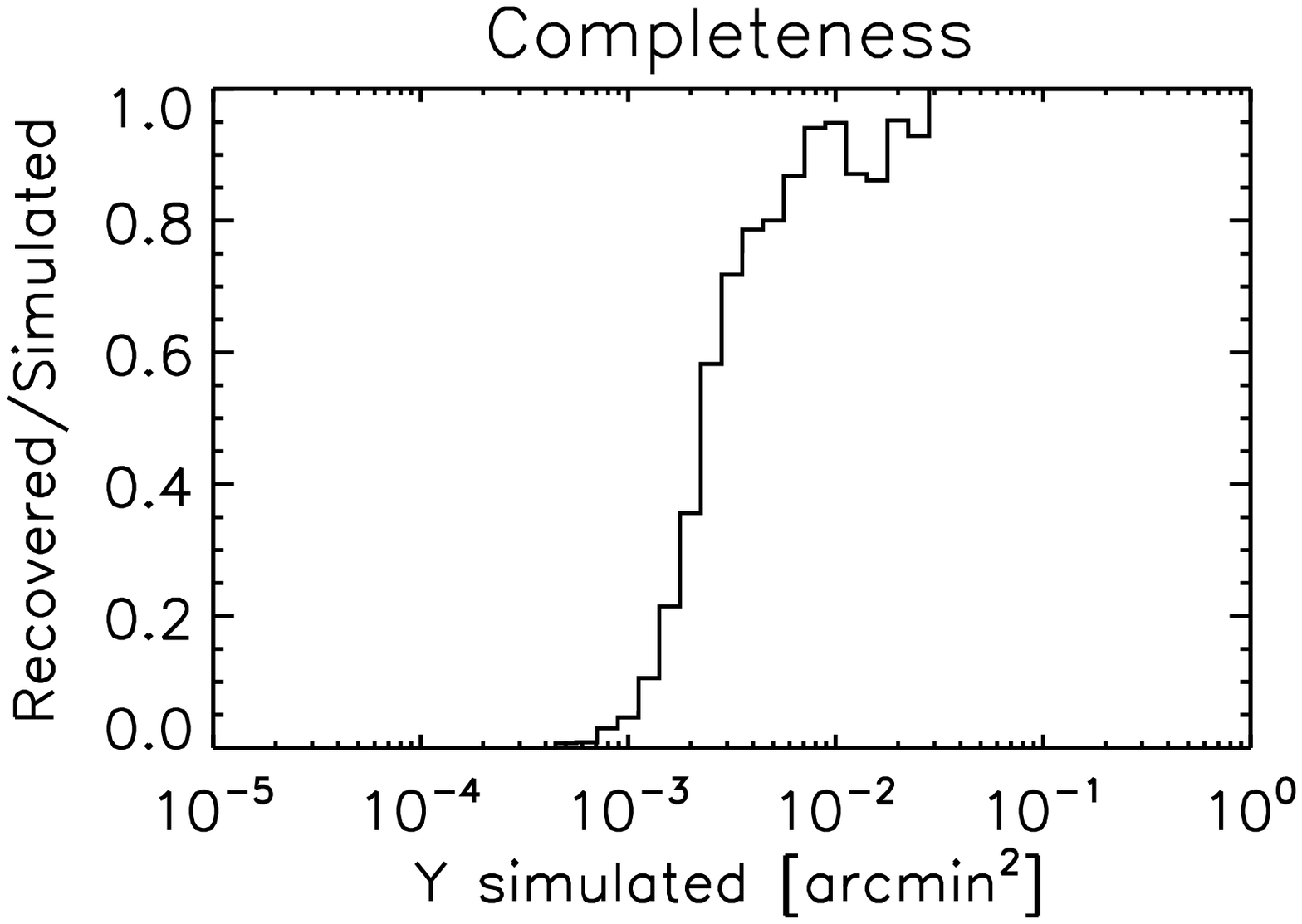}  &
\includegraphics[scale=0.45]{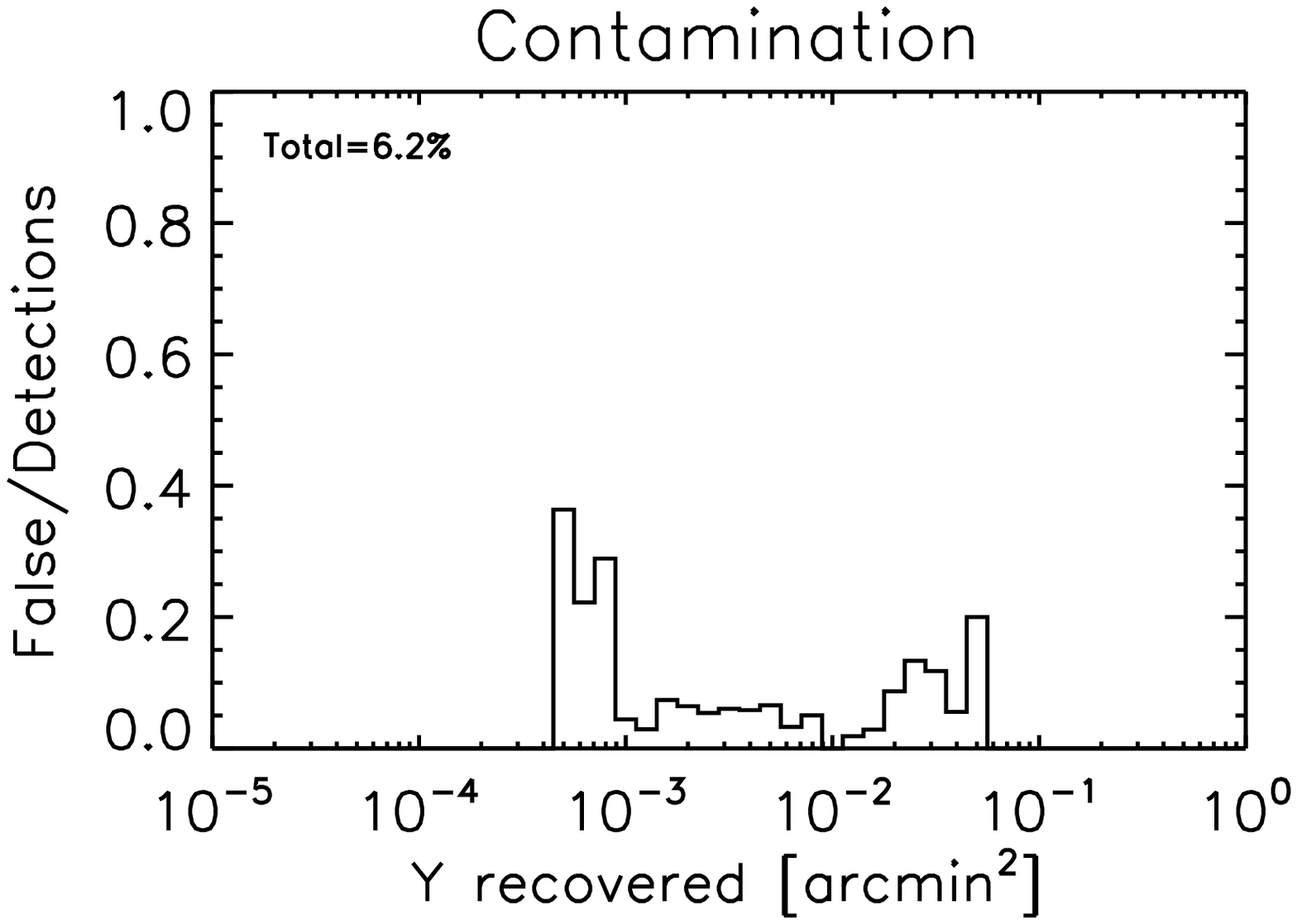} \\
\includegraphics[scale=0.45]{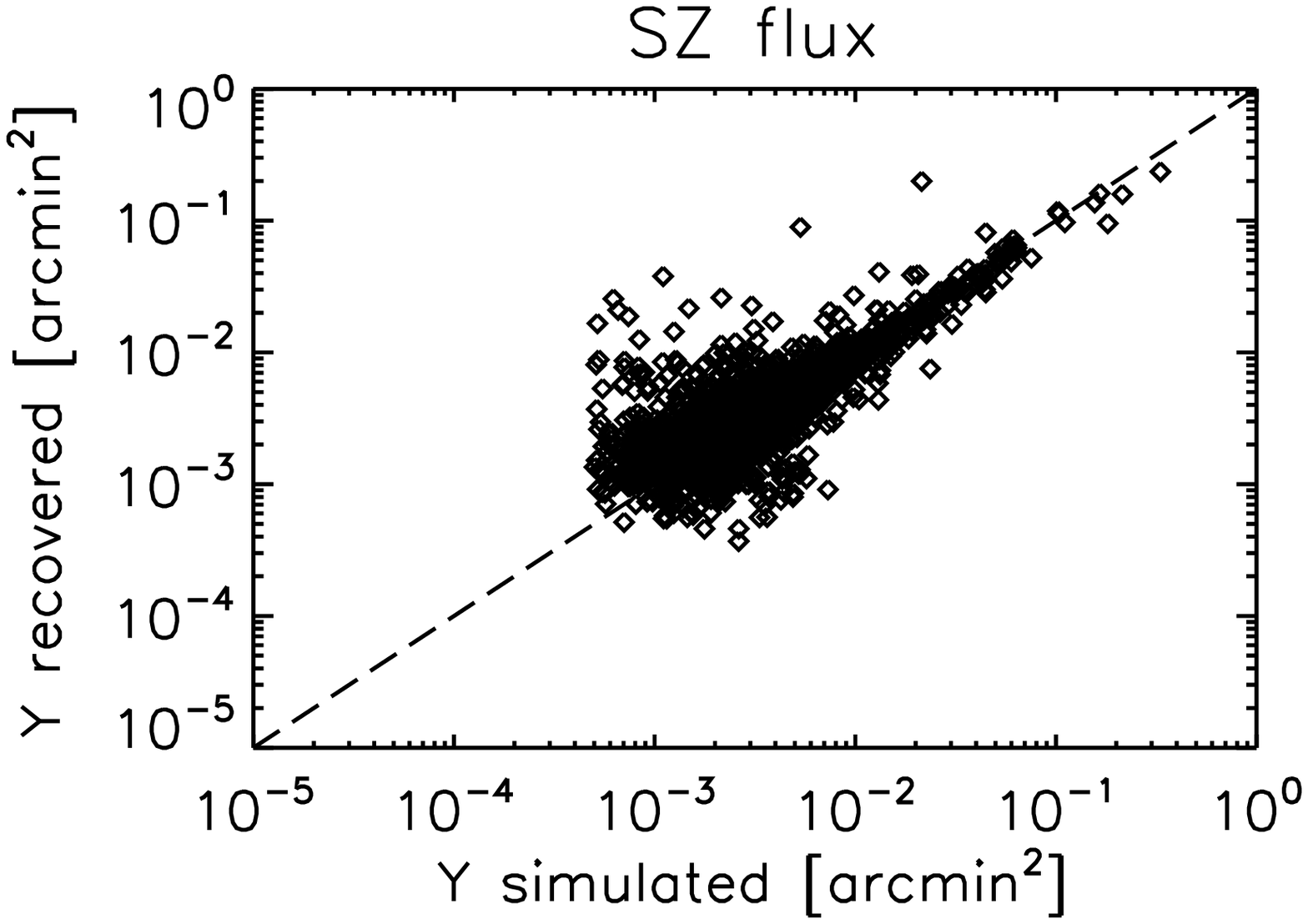}  &
\includegraphics[scale=0.45]{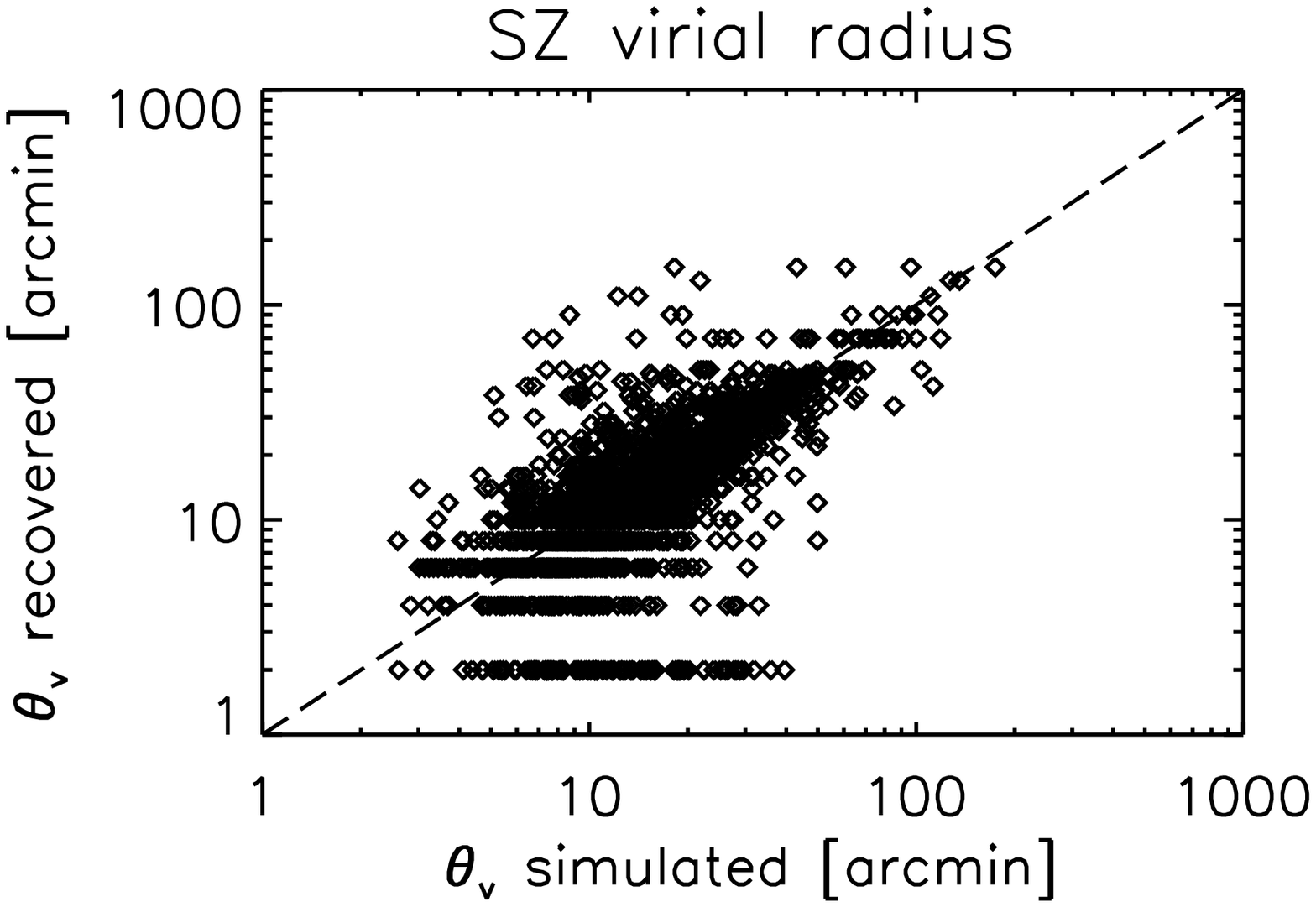}  \\
\includegraphics[scale=0.45]{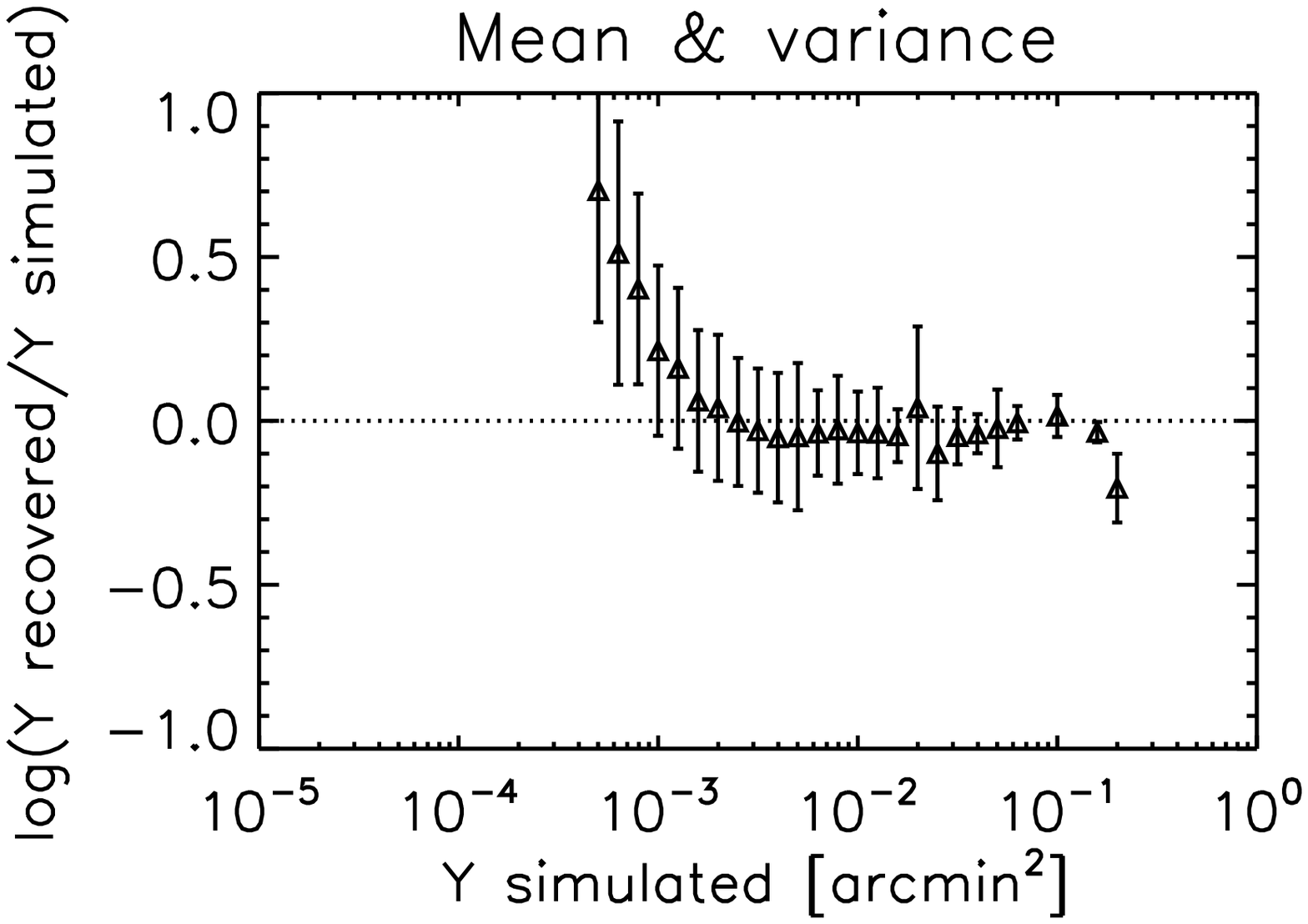} &
\includegraphics[scale=0.45]{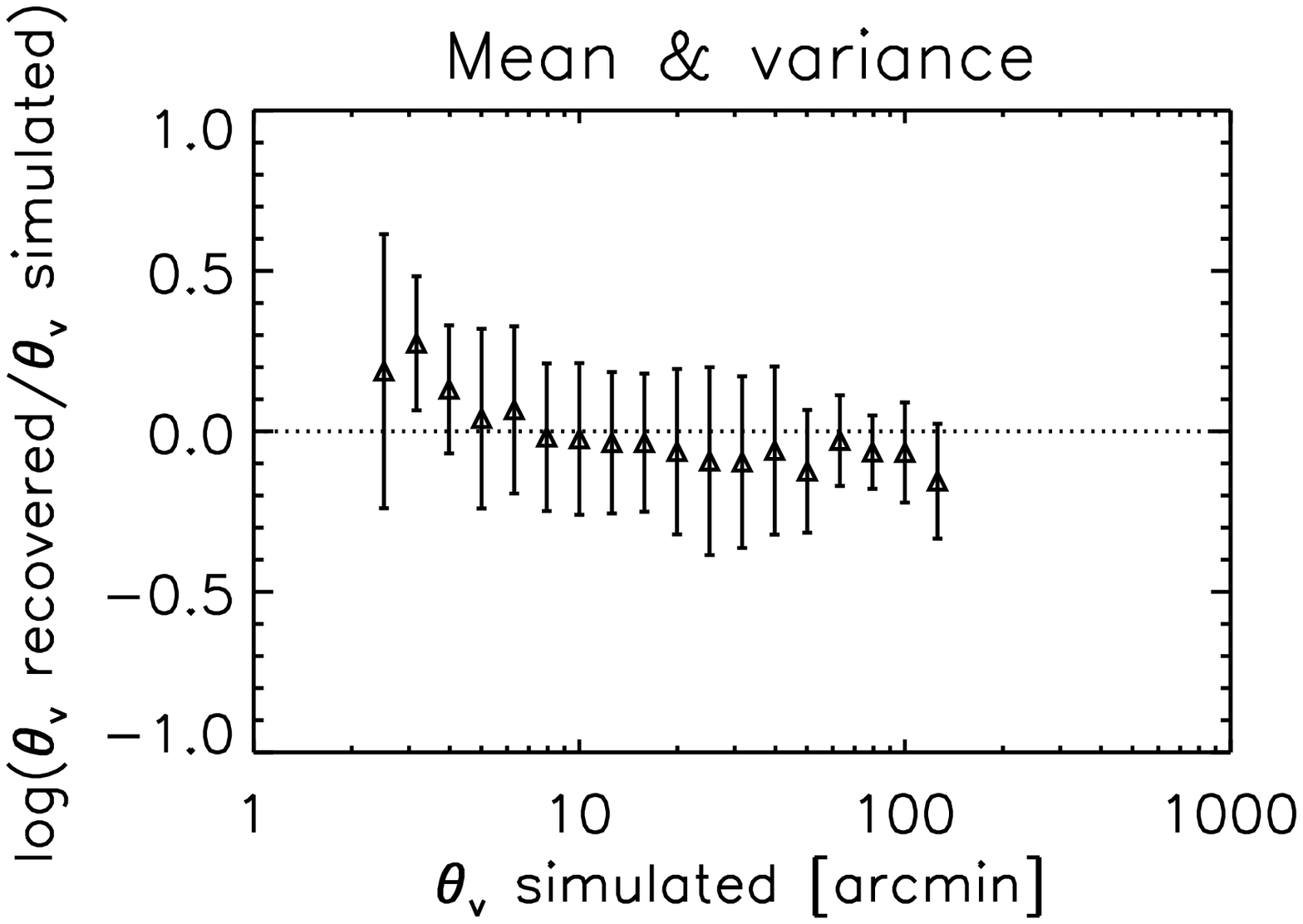} \\
\end{tabular}
\caption{{\bf ILC2}}
\end{center}
\end{table}

\clearpage

\begin{table}[htbp]
\begin{center}
\begin{tabular}{cc}
\includegraphics[scale=0.45]{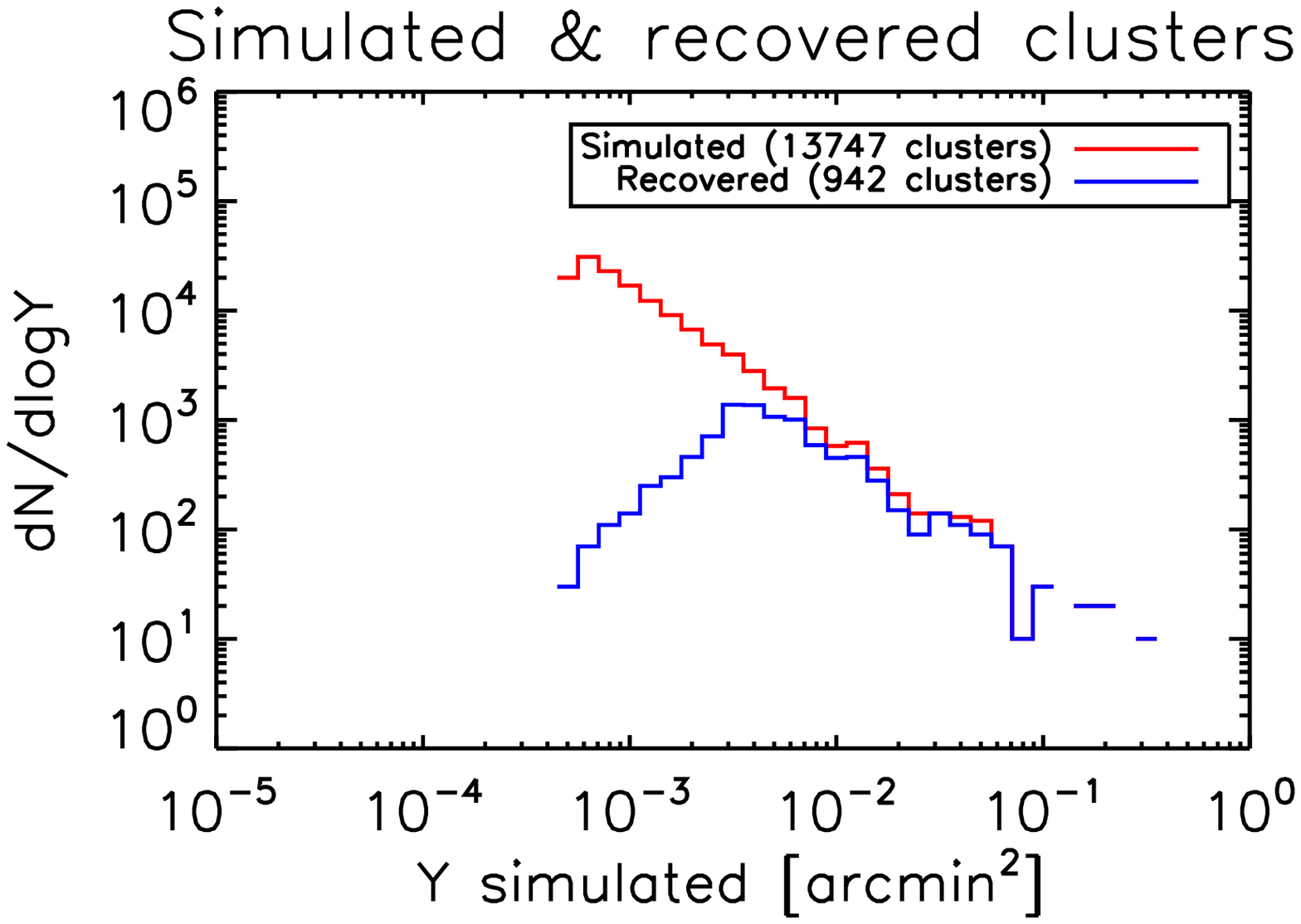}  &
\includegraphics[scale=0.45]{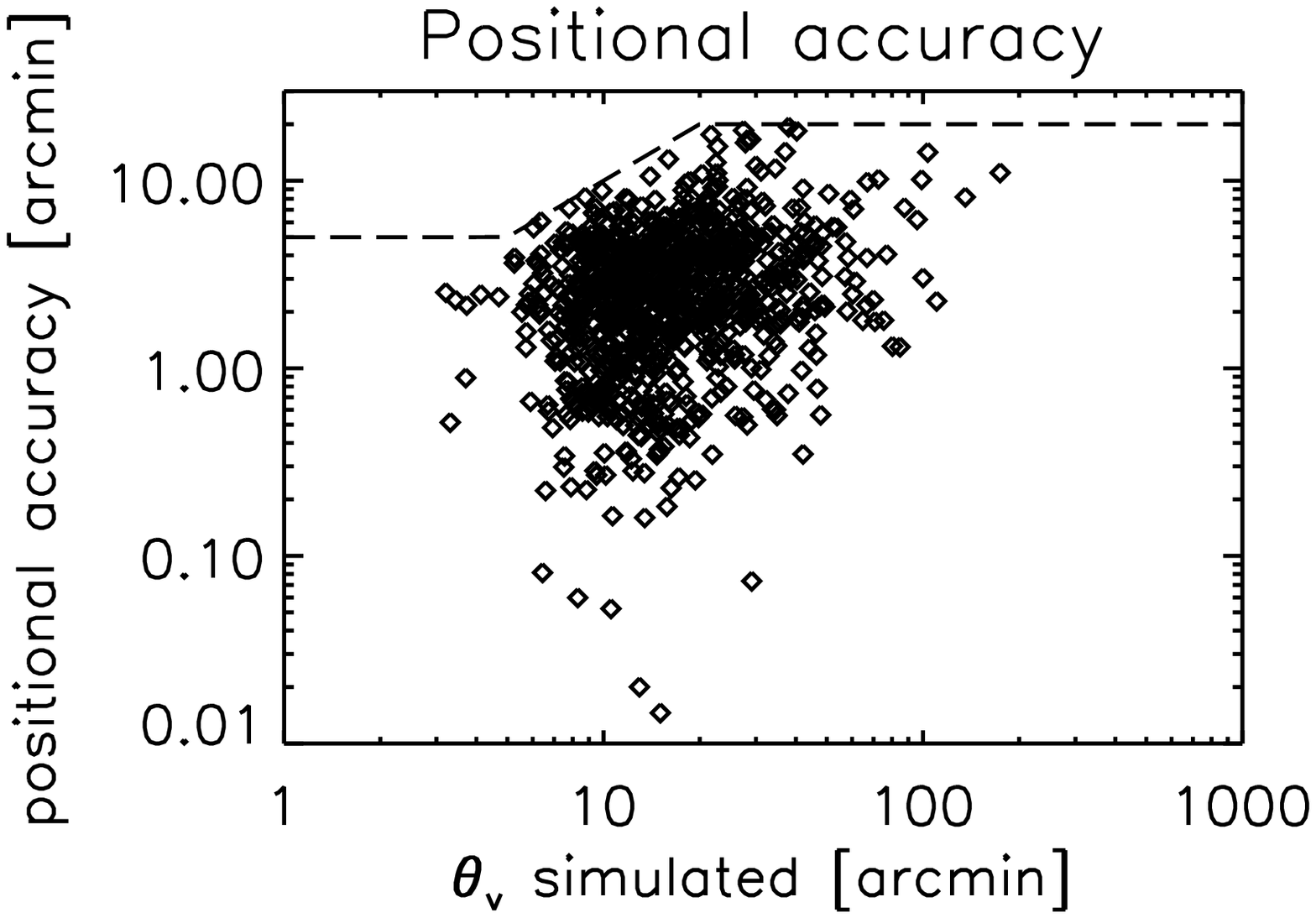} \\
\includegraphics[scale=0.45]{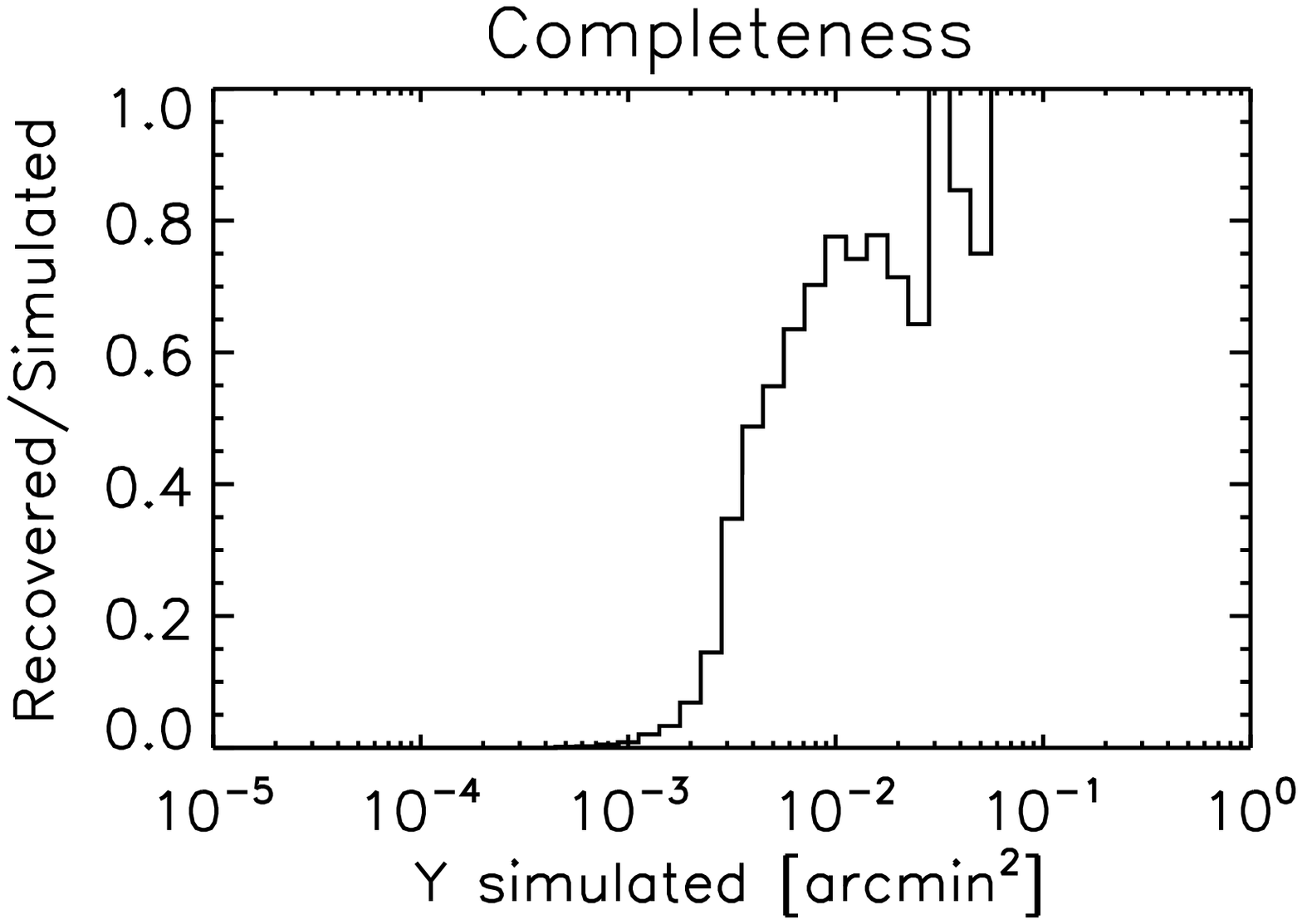}  &
\includegraphics[scale=0.45]{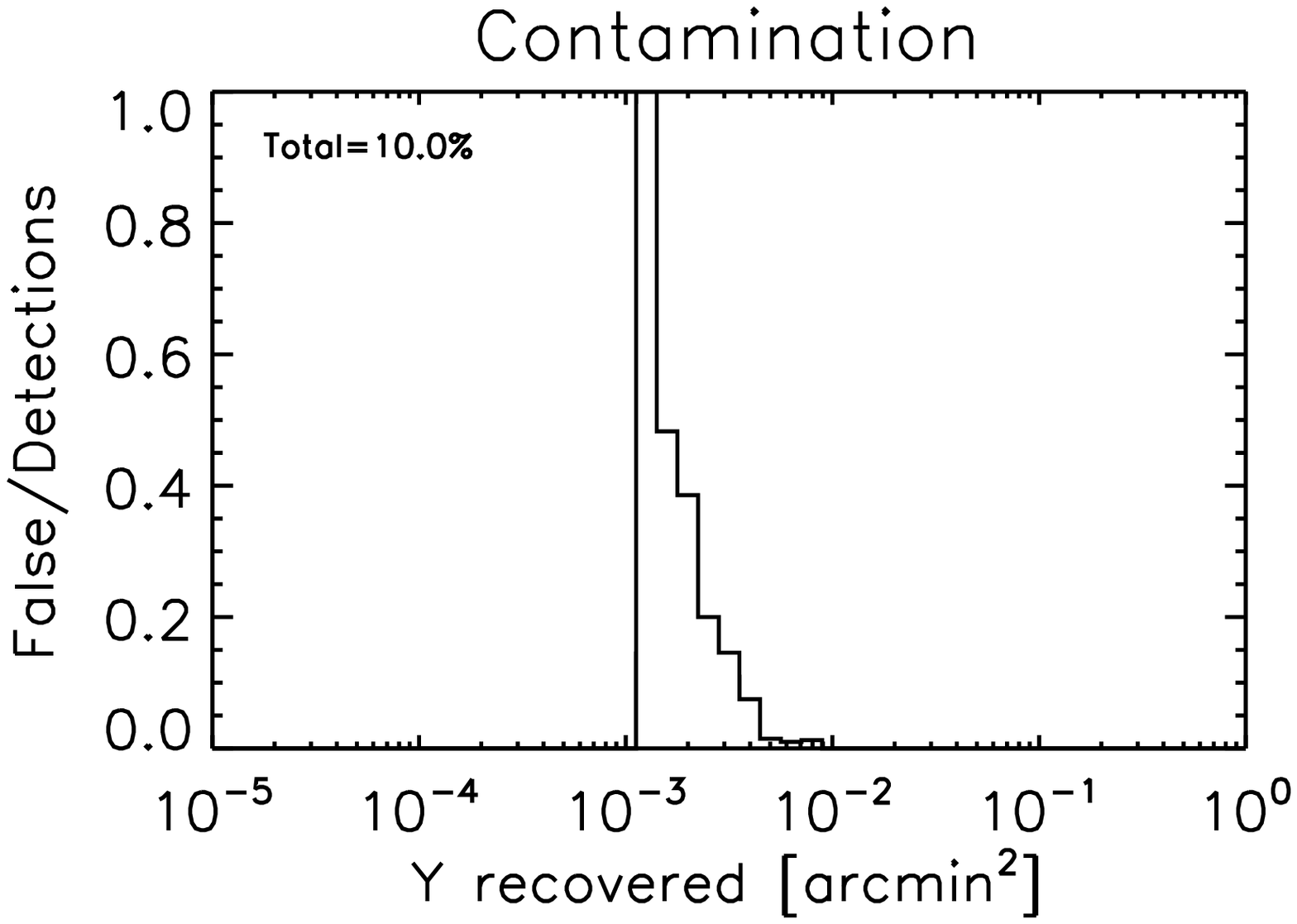} \\
\includegraphics[scale=0.45]{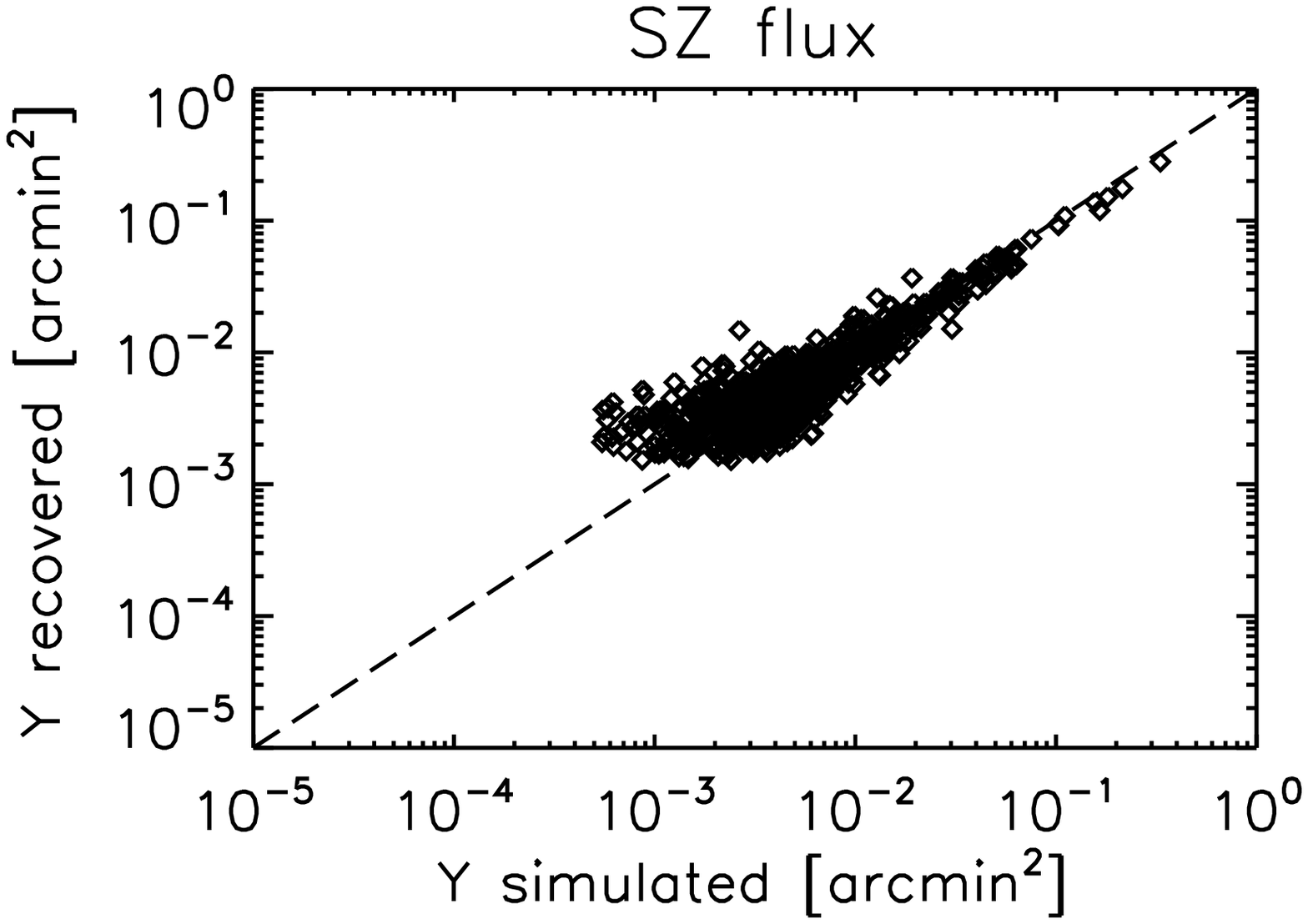}  &
\includegraphics[scale=0.45]{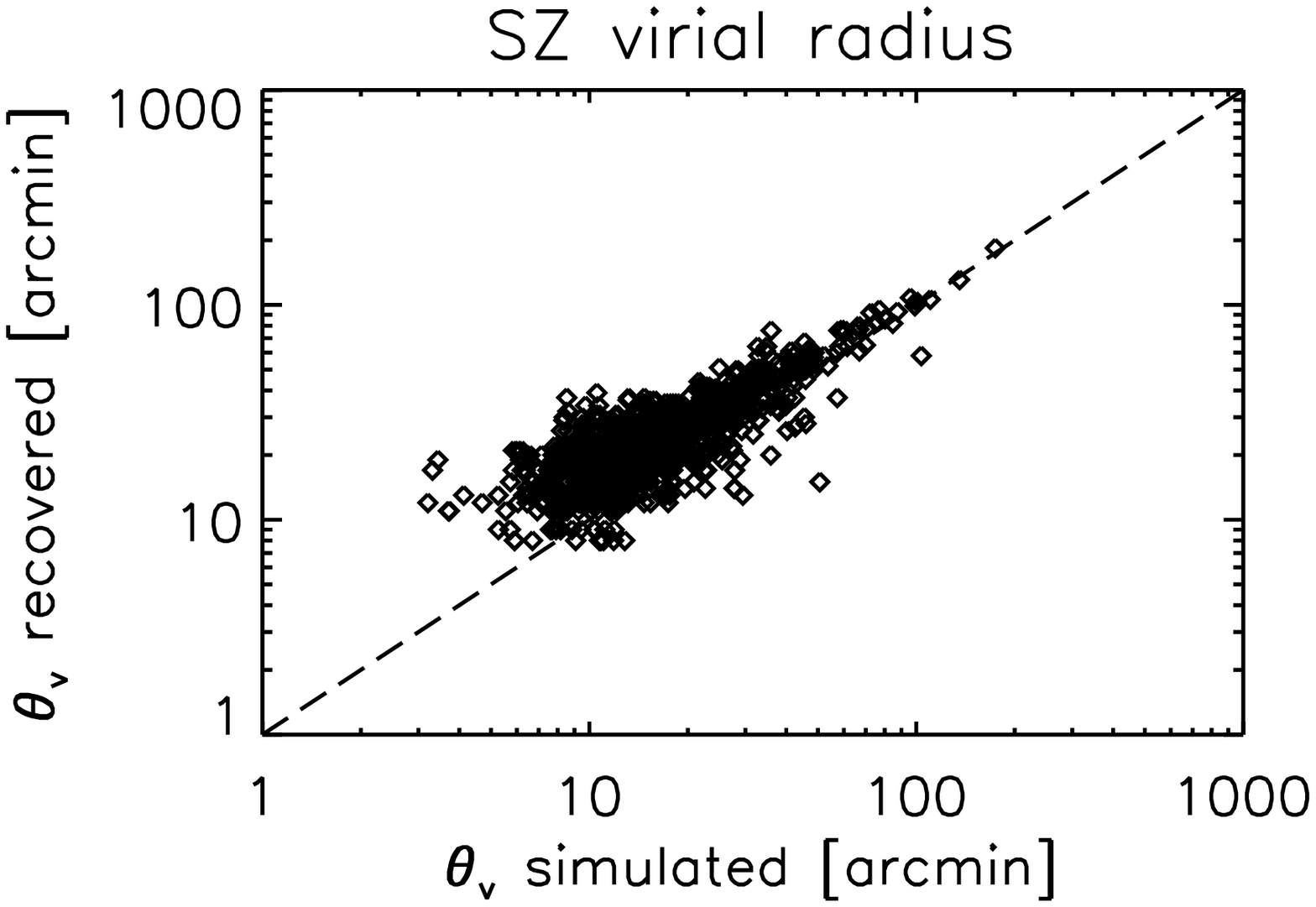}  \\
\includegraphics[scale=0.45]{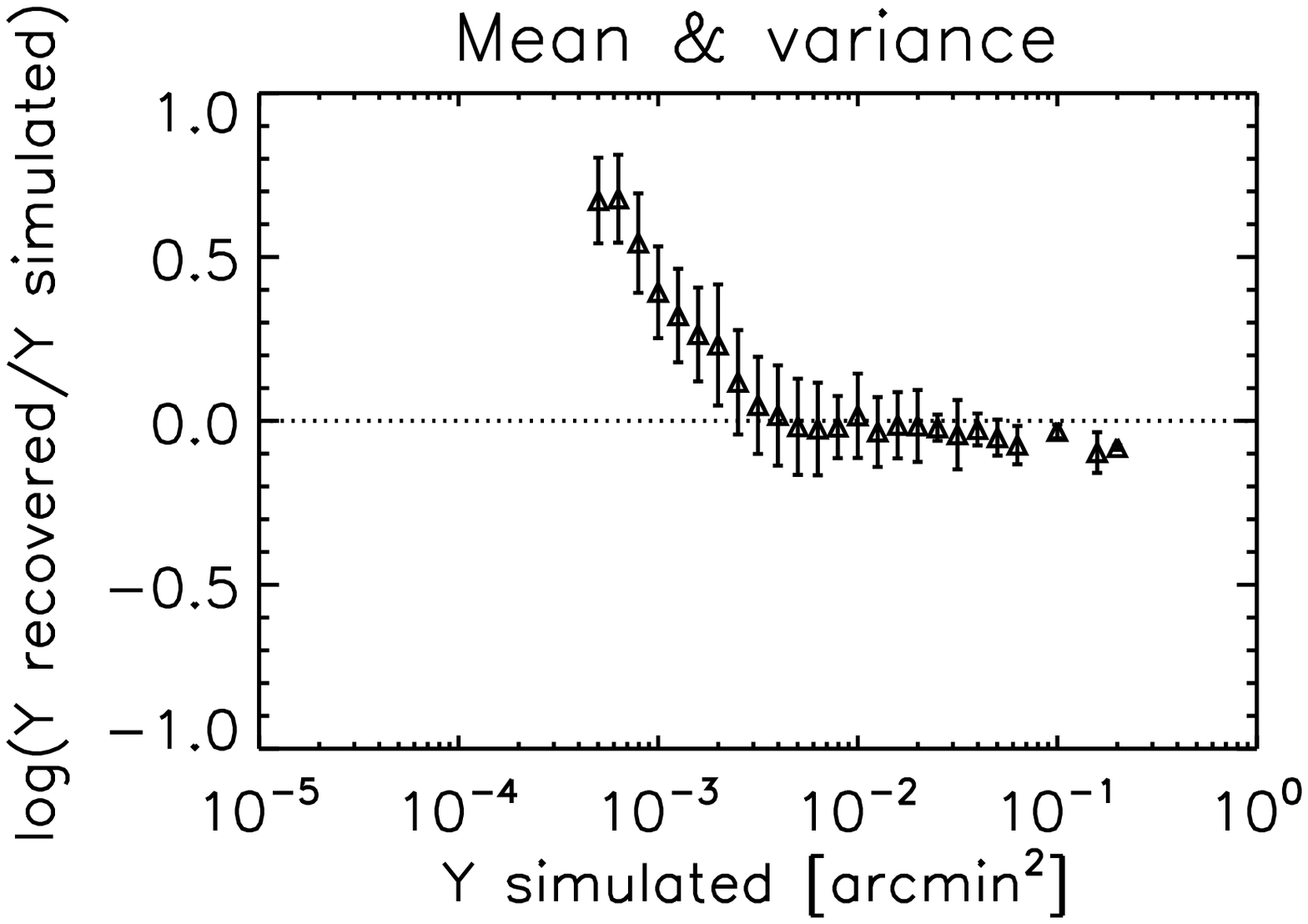} &
\includegraphics[scale=0.45]{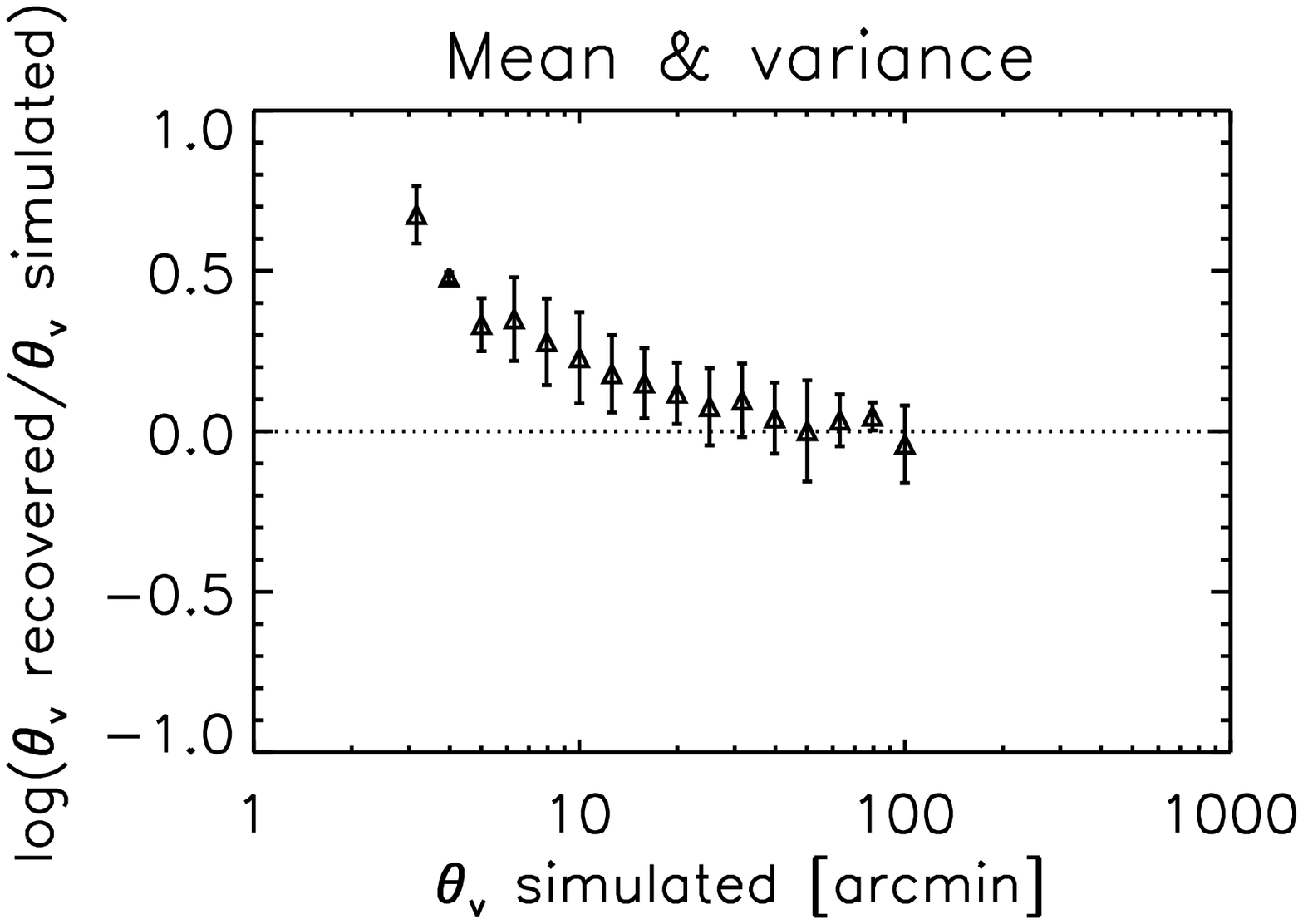} \\
\end{tabular}
\caption{{\bf ILC3}}
\end{center}
\end{table}

\clearpage

\begin{table}[htbp]
\begin{center}
\begin{tabular}{cc}
\includegraphics[scale=0.45]{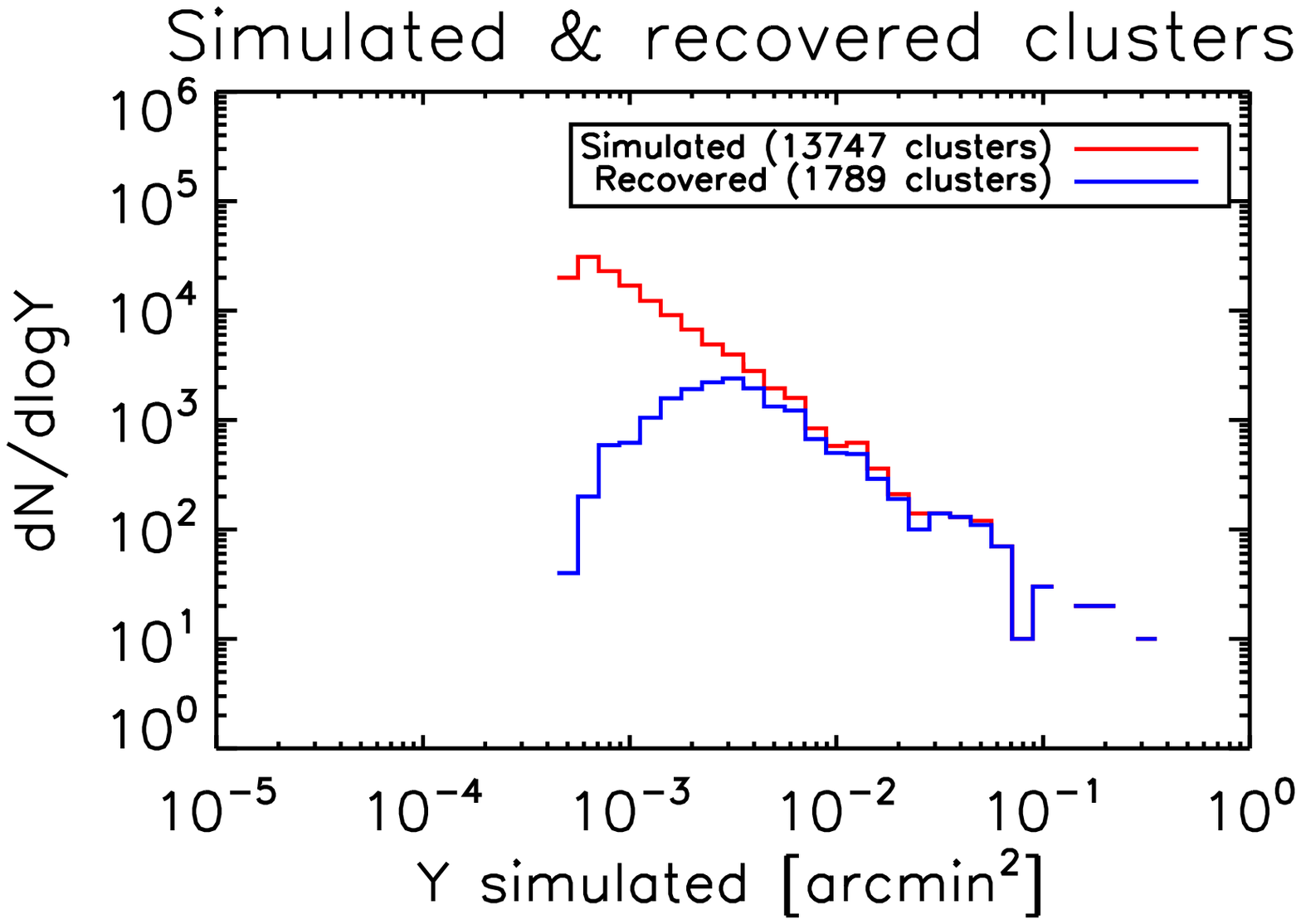}  &
\includegraphics[scale=0.45]{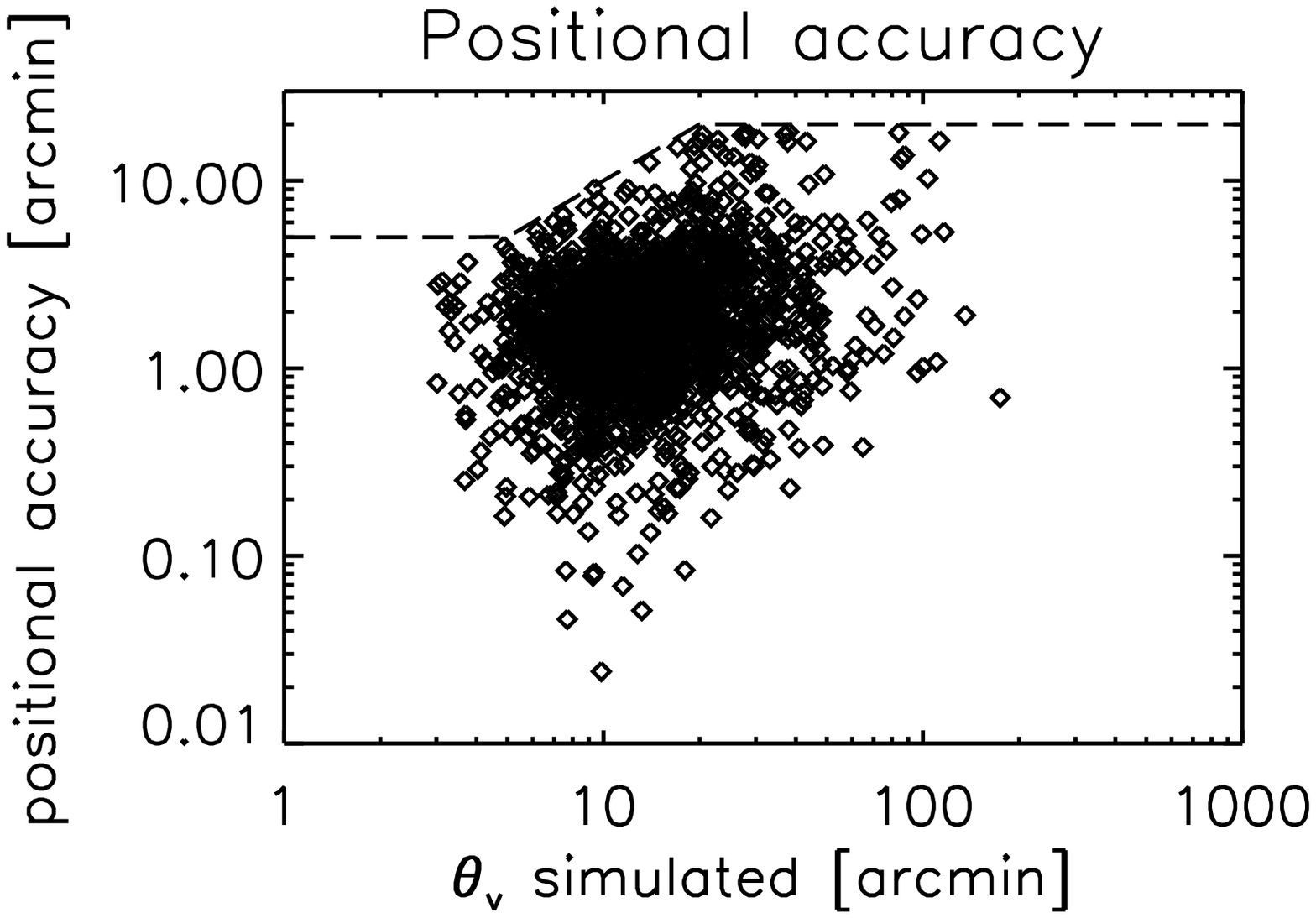} \\
\includegraphics[scale=0.45]{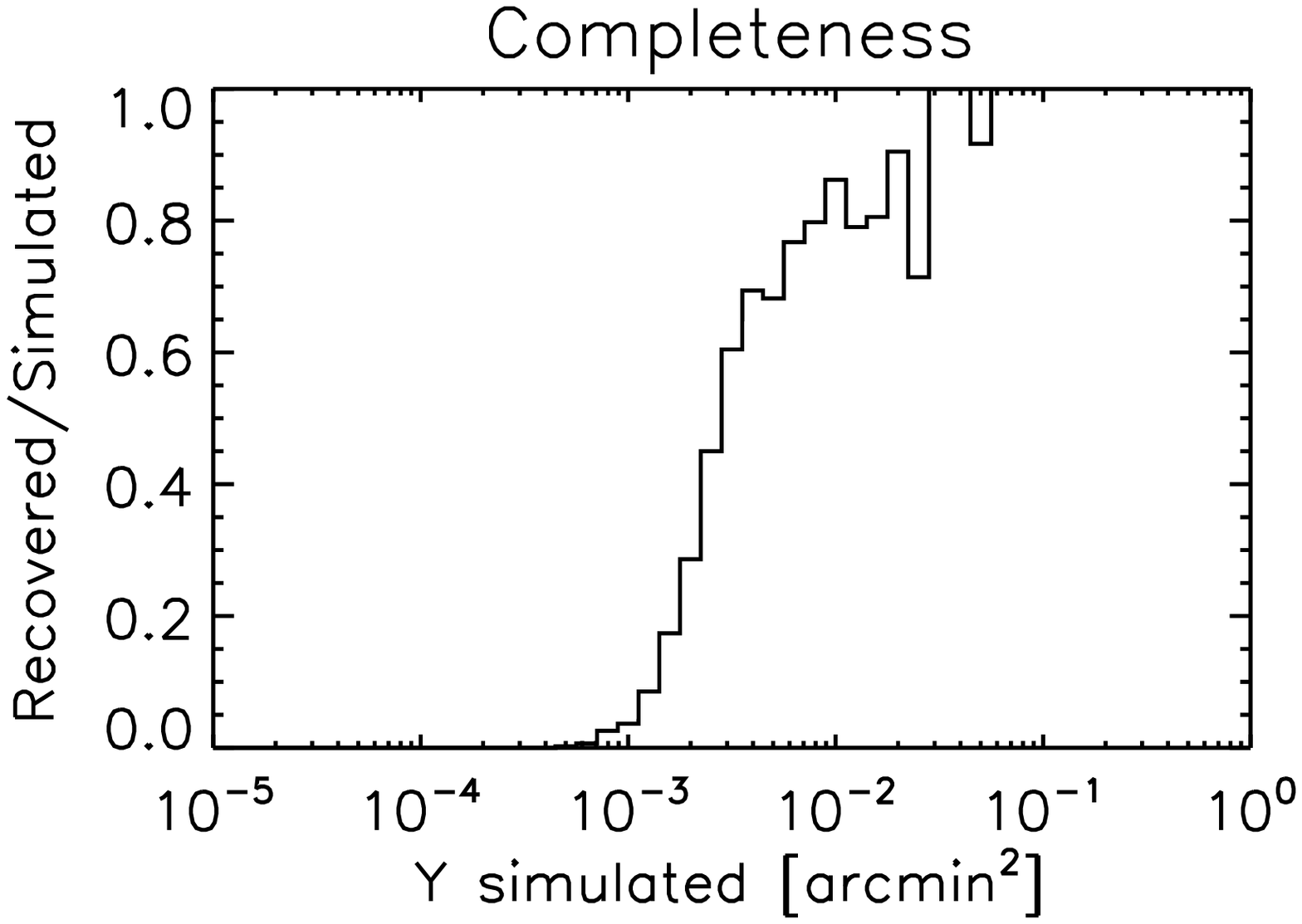}  &
\includegraphics[scale=0.45]{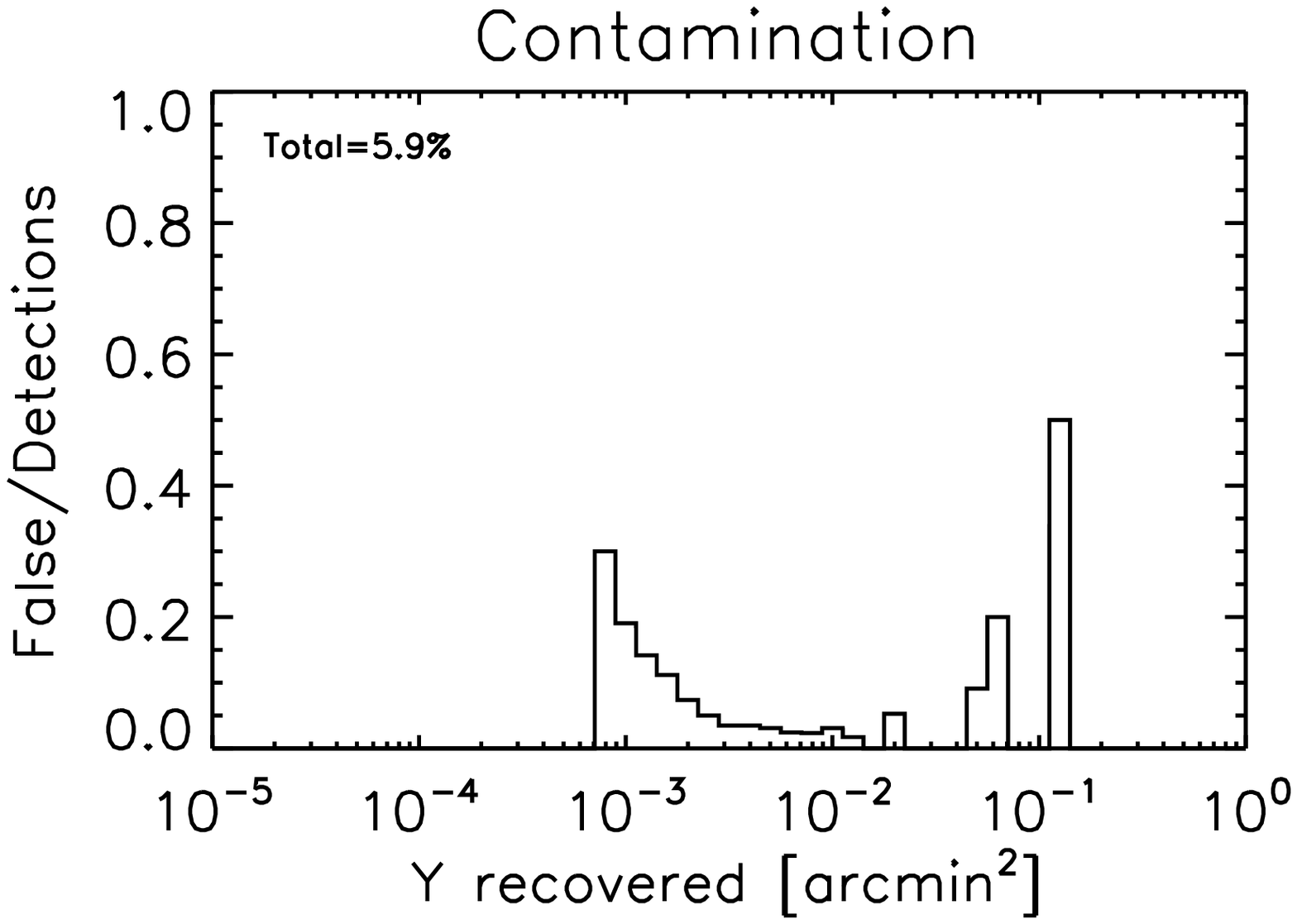} \\
\includegraphics[scale=0.45]{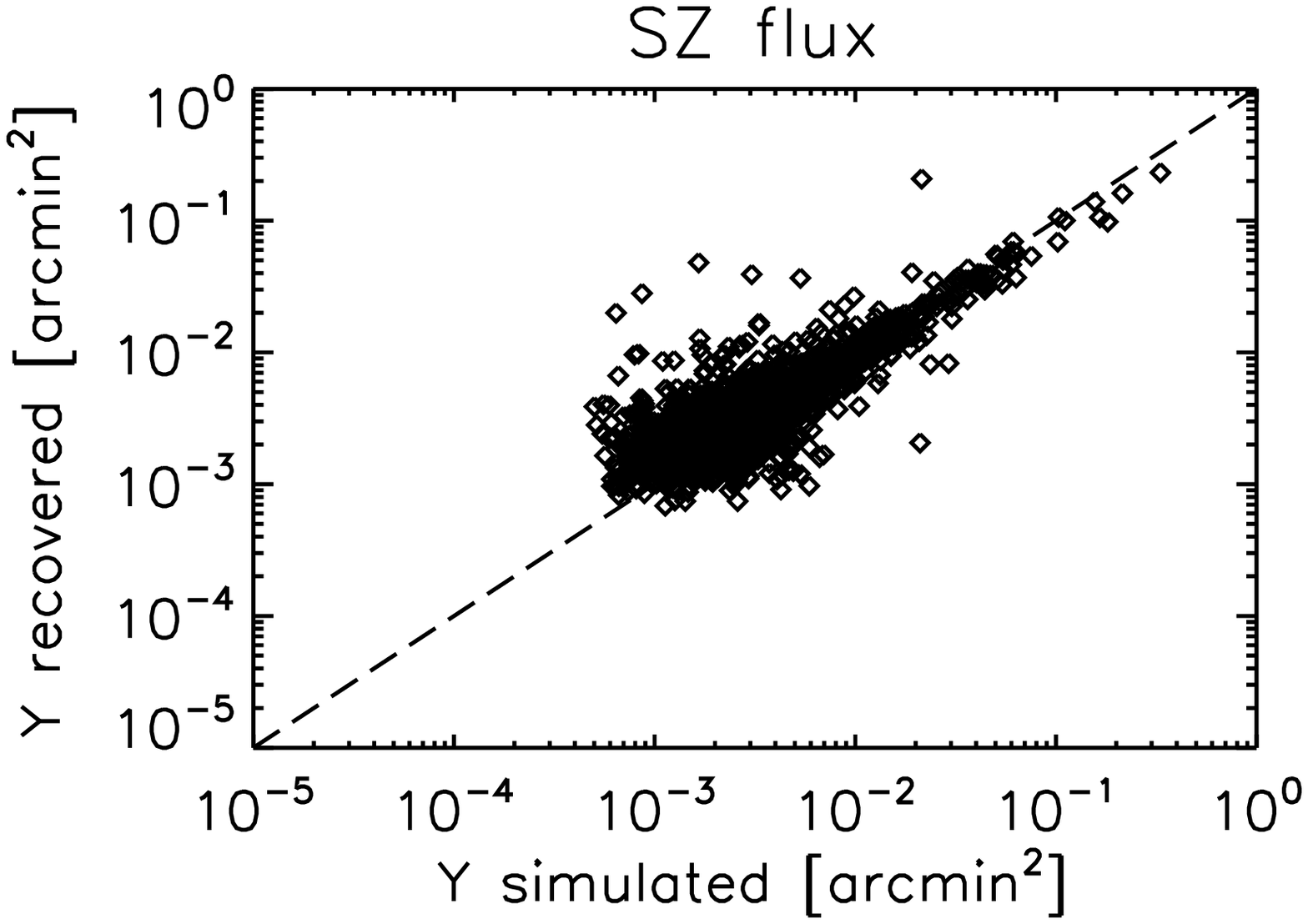}  &
\includegraphics[scale=0.45]{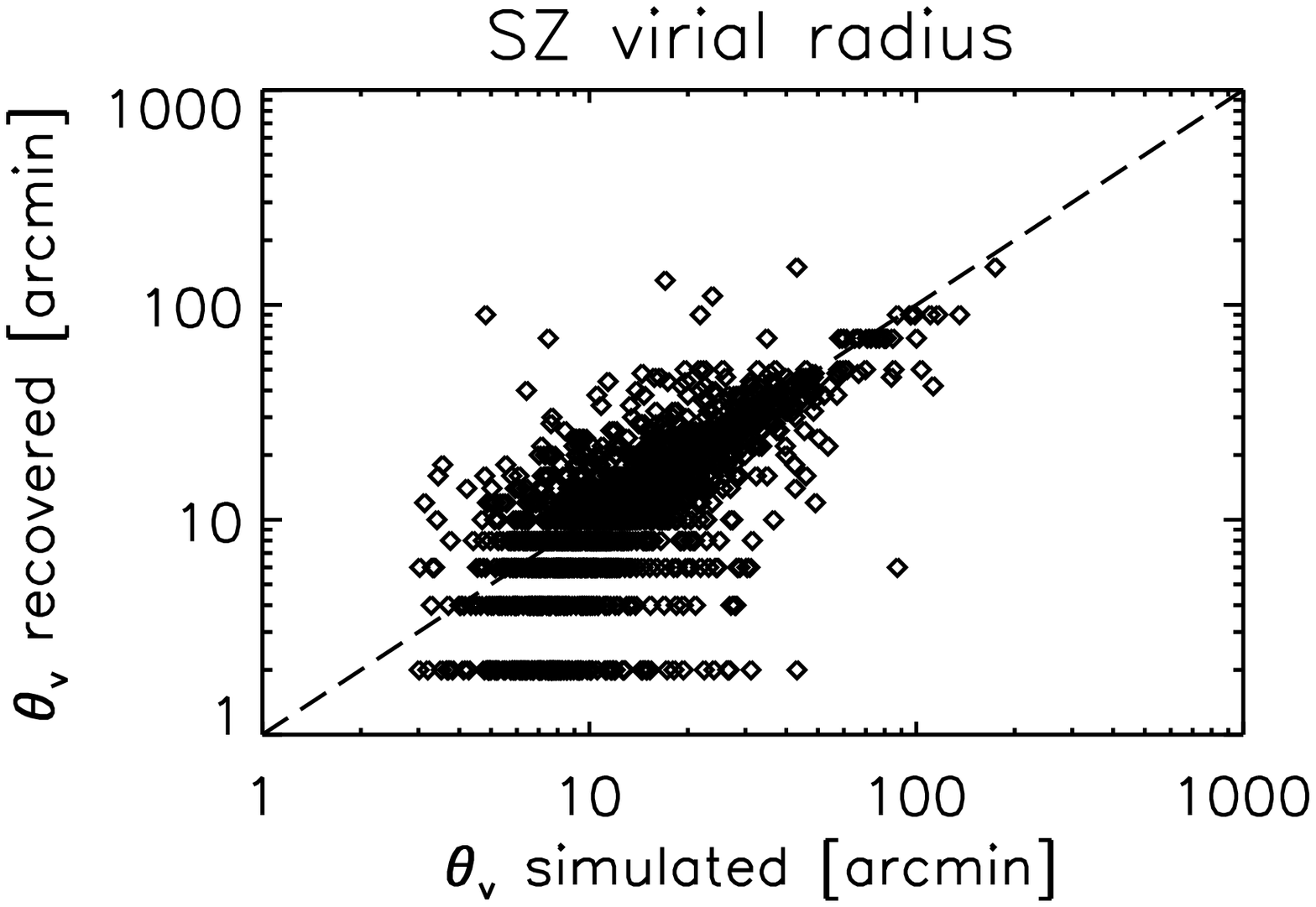} \\
\includegraphics[scale=0.45]{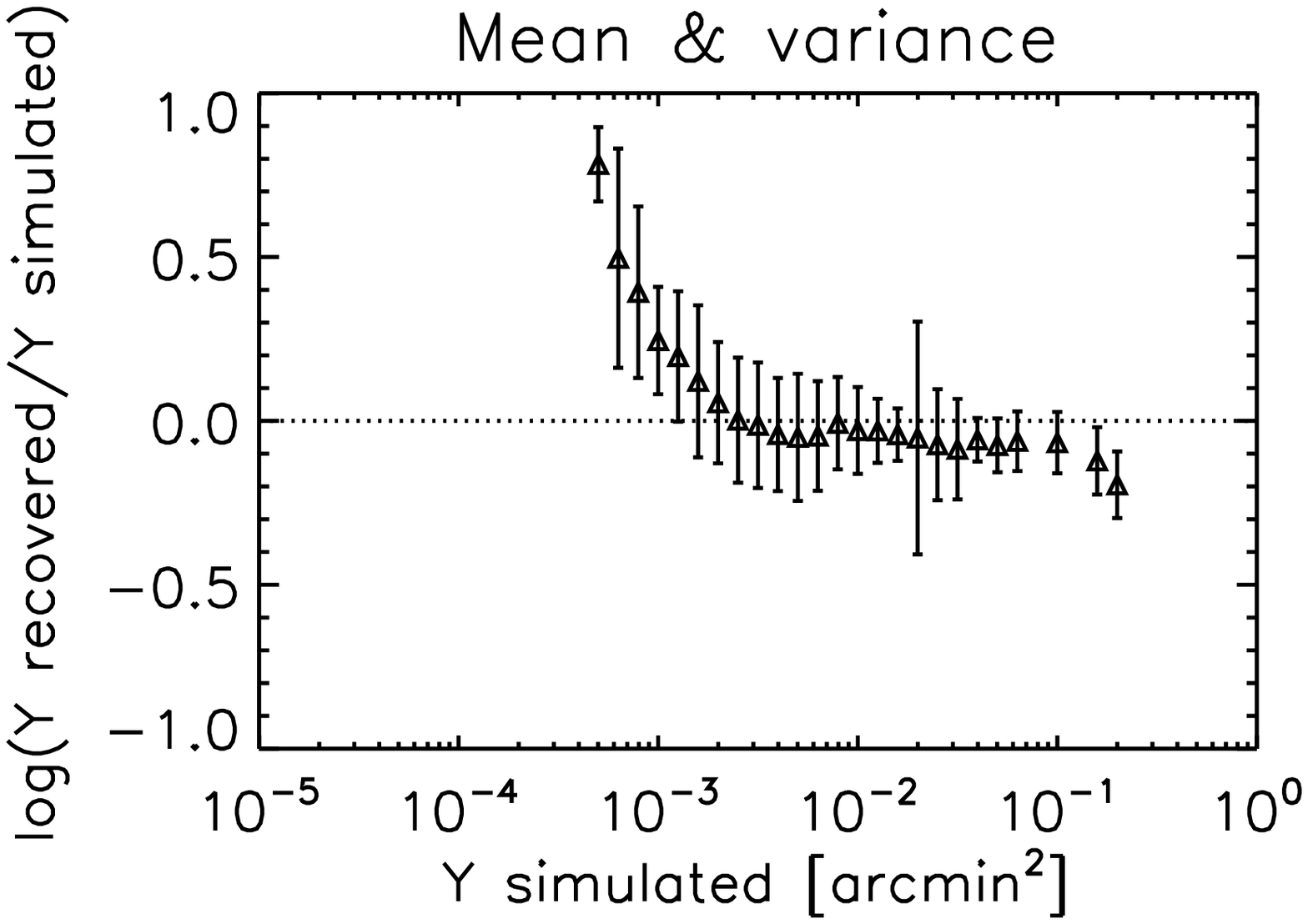}  &
\includegraphics[scale=0.45]{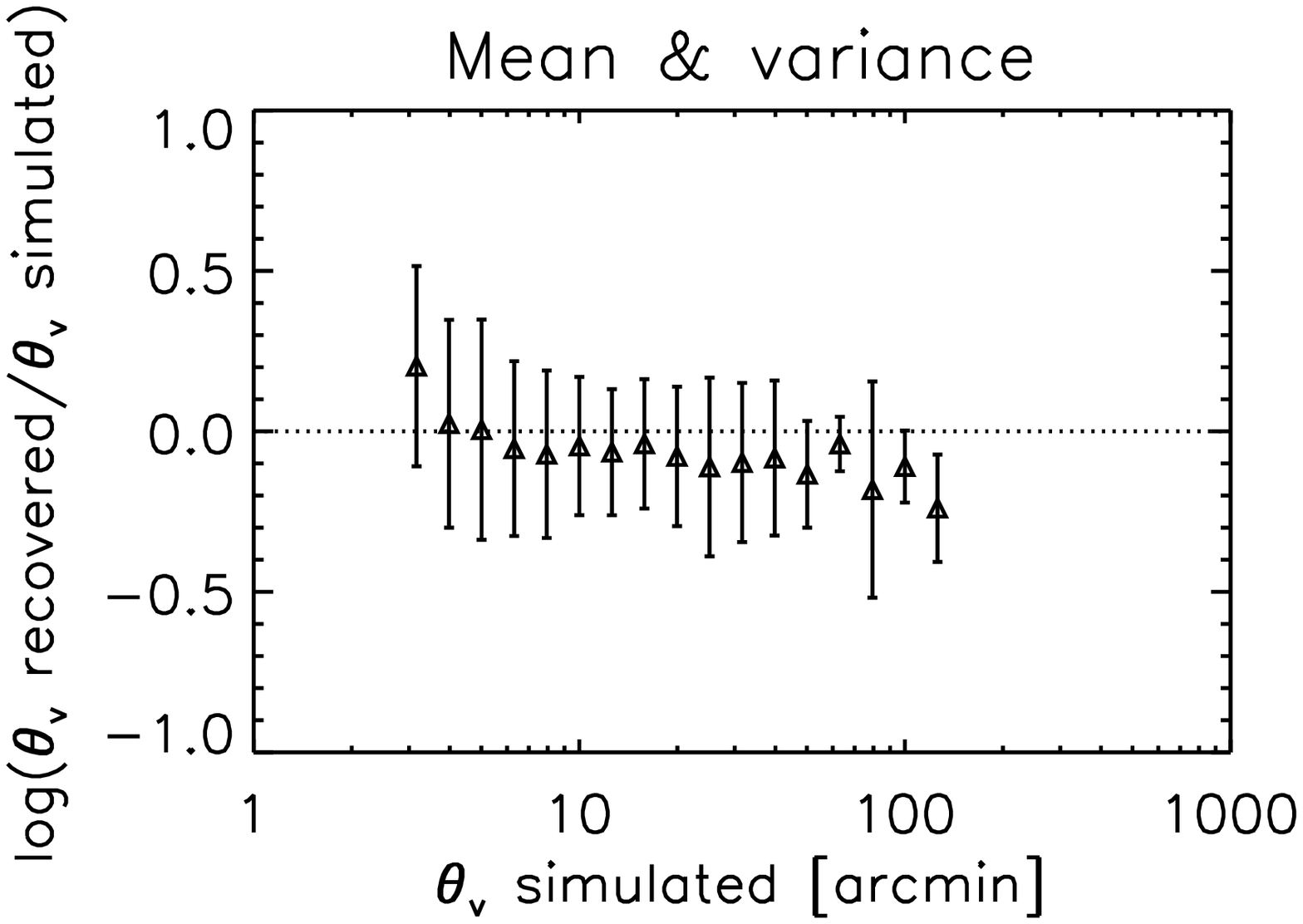} \\
\end{tabular}
\caption{{\bf ILC4}}
\end{center}
\end{table}

\clearpage

\begin{table}[htbp]
\begin{center}
\begin{tabular}{cc}
\includegraphics[scale=0.45]{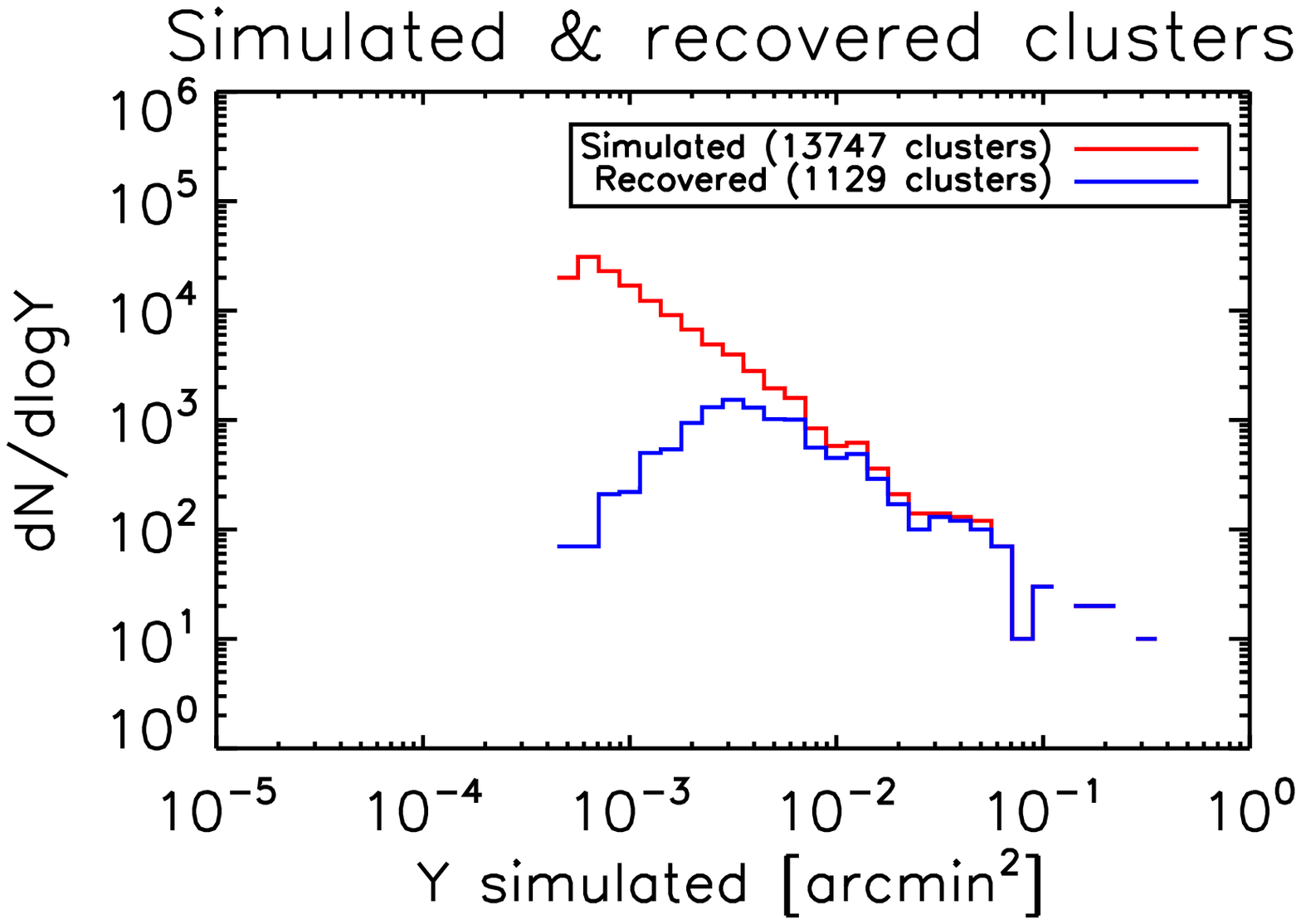}  &
\includegraphics[scale=0.45]{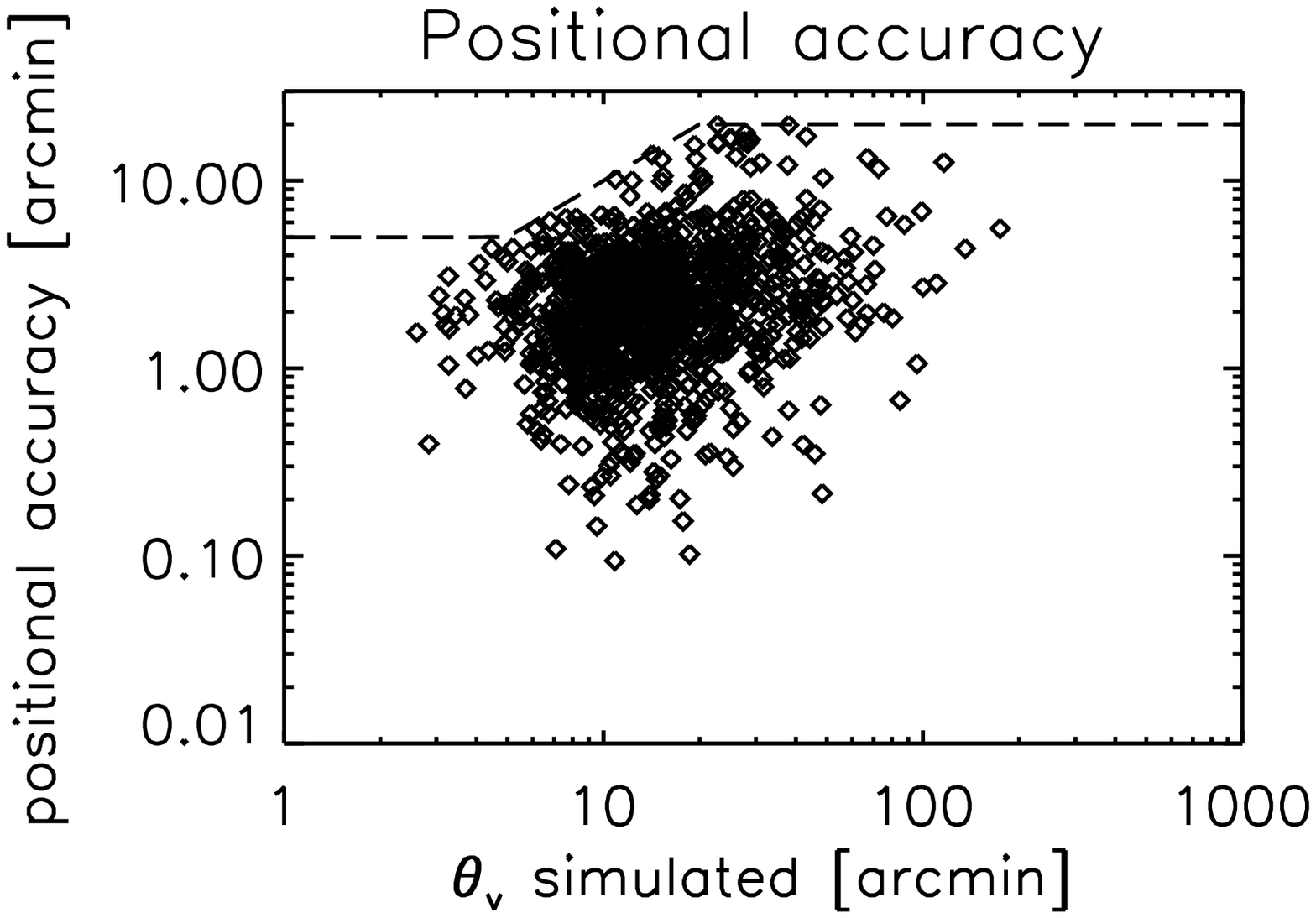} \\
\includegraphics[scale=0.45]{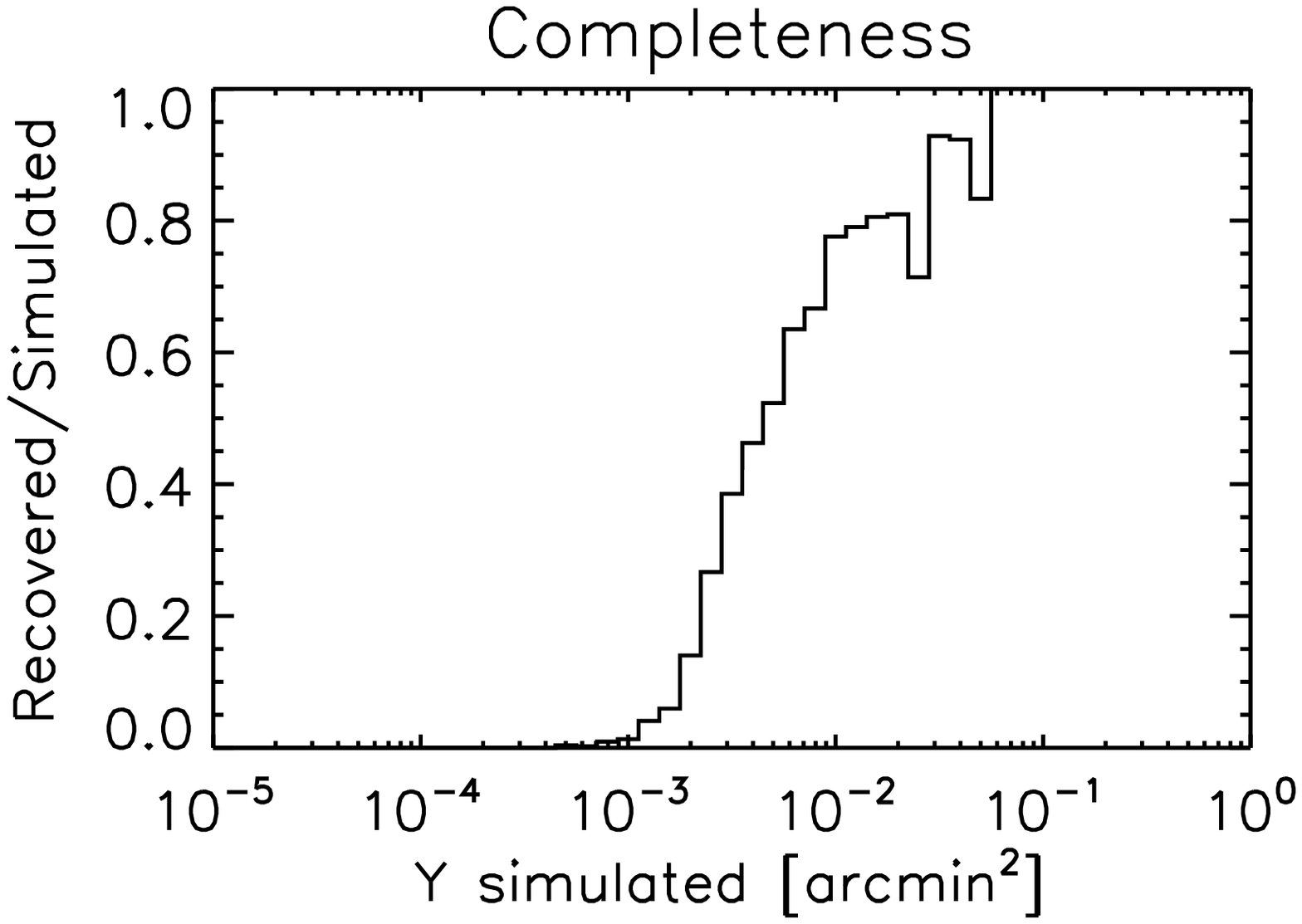}  &
\includegraphics[scale=0.45]{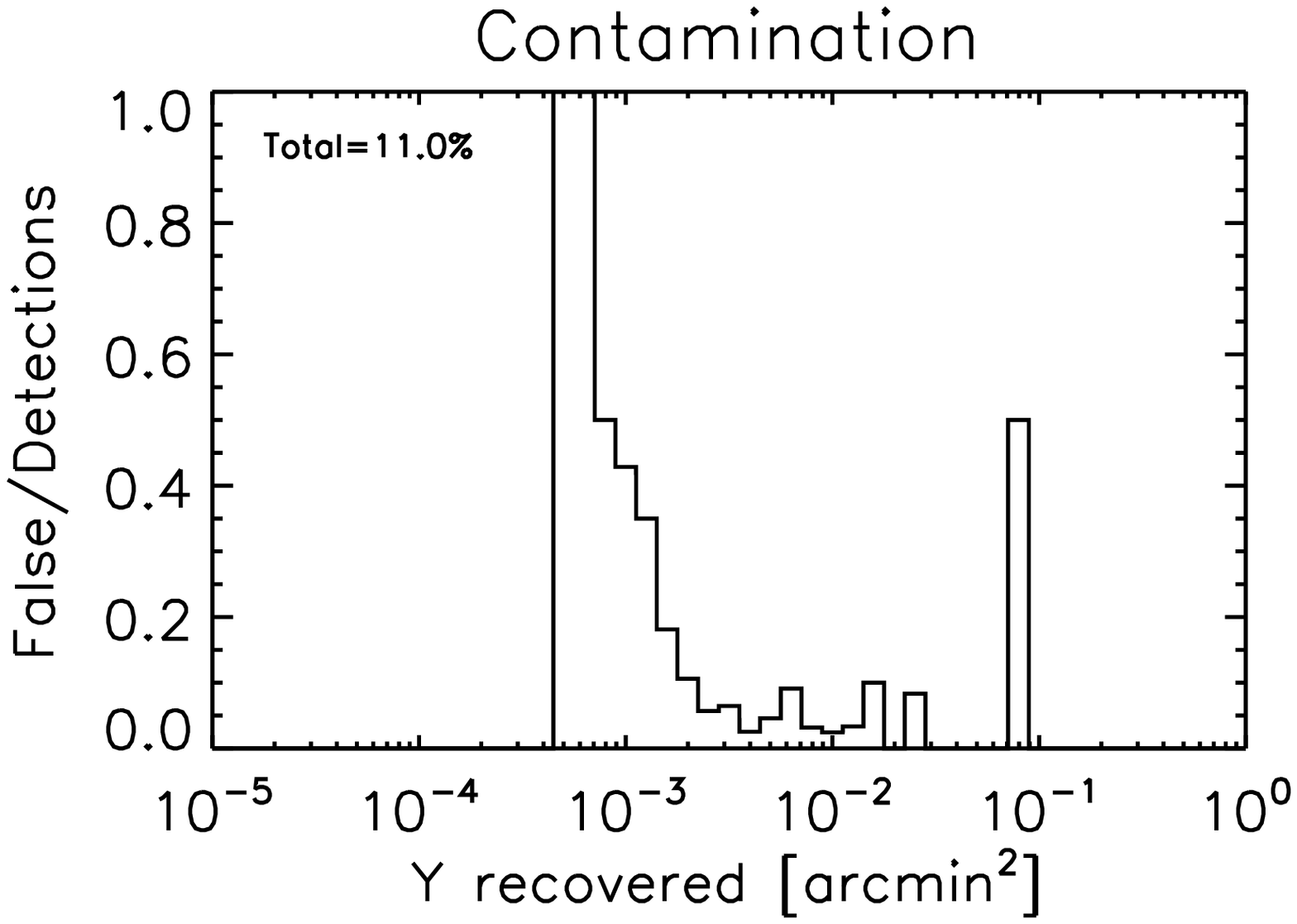} \\
\includegraphics[scale=0.45]{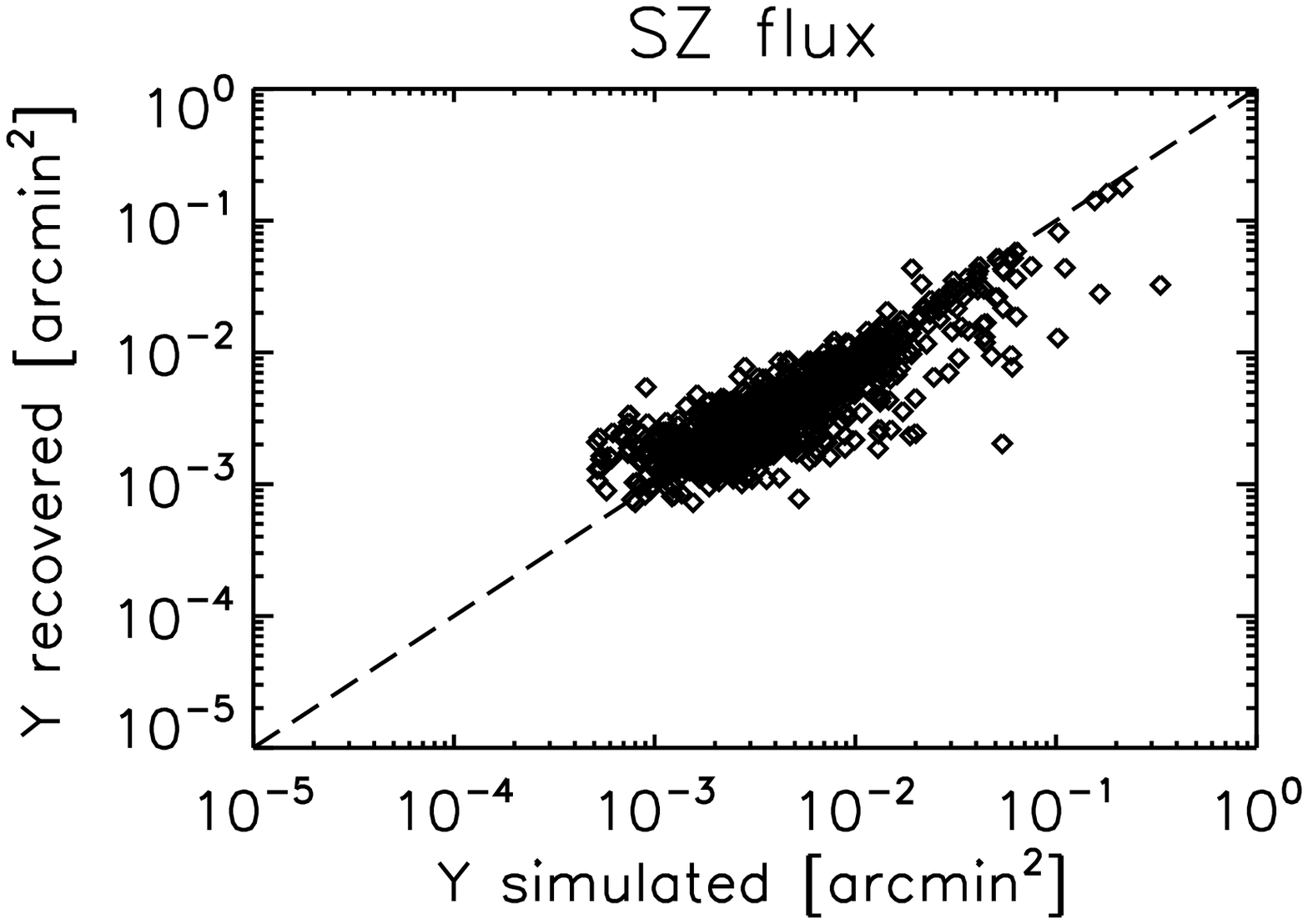}  &
\\ 
\includegraphics[scale=0.45]{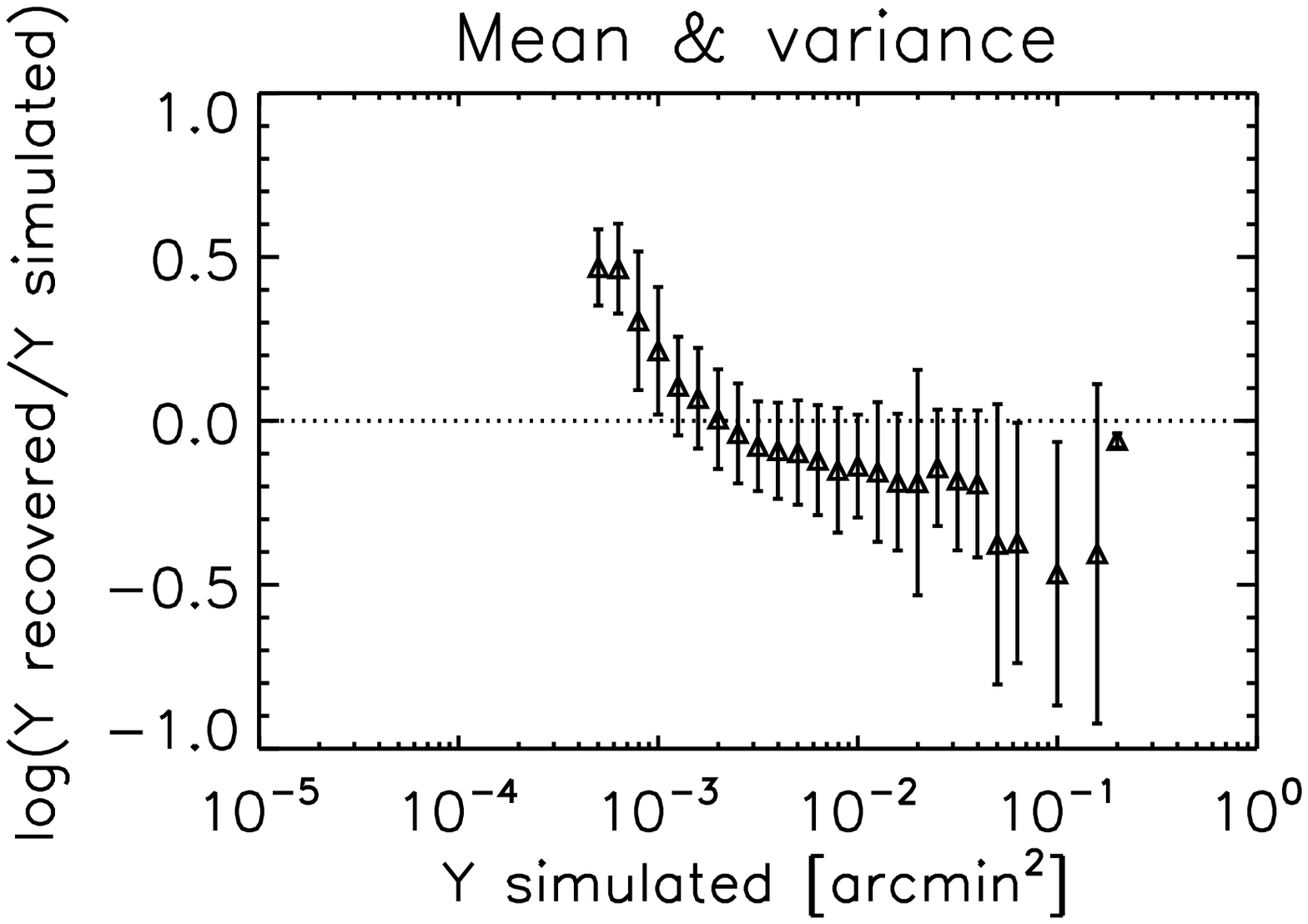}  &
\\ 
\end{tabular}
\caption{{\bf ILC5}}
\end{center}
\end{table}

\clearpage

\begin{table}[htbp]
\begin{center}
\begin{tabular}{cc}
\includegraphics[scale=0.45]{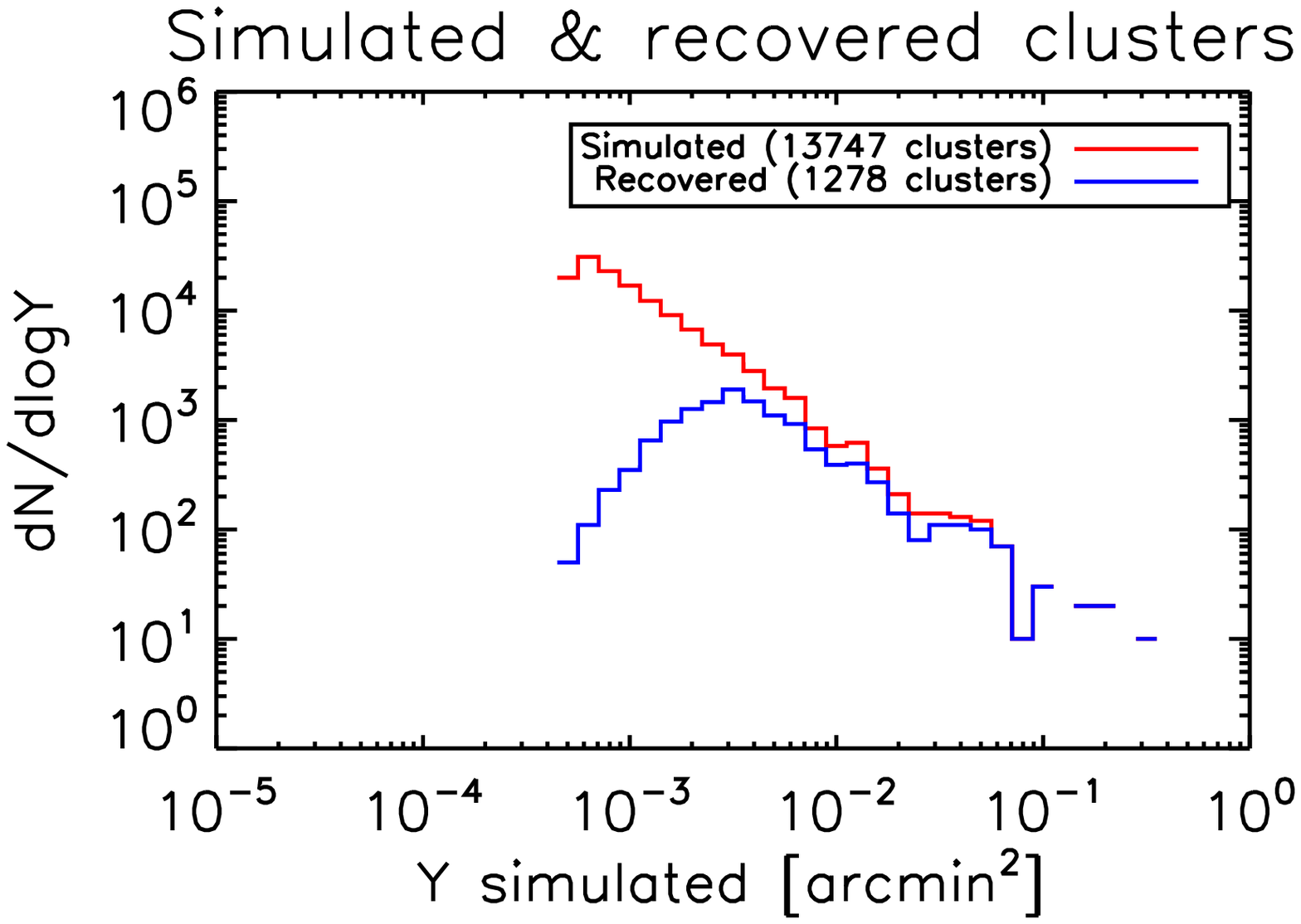}  &
\includegraphics[scale=0.45]{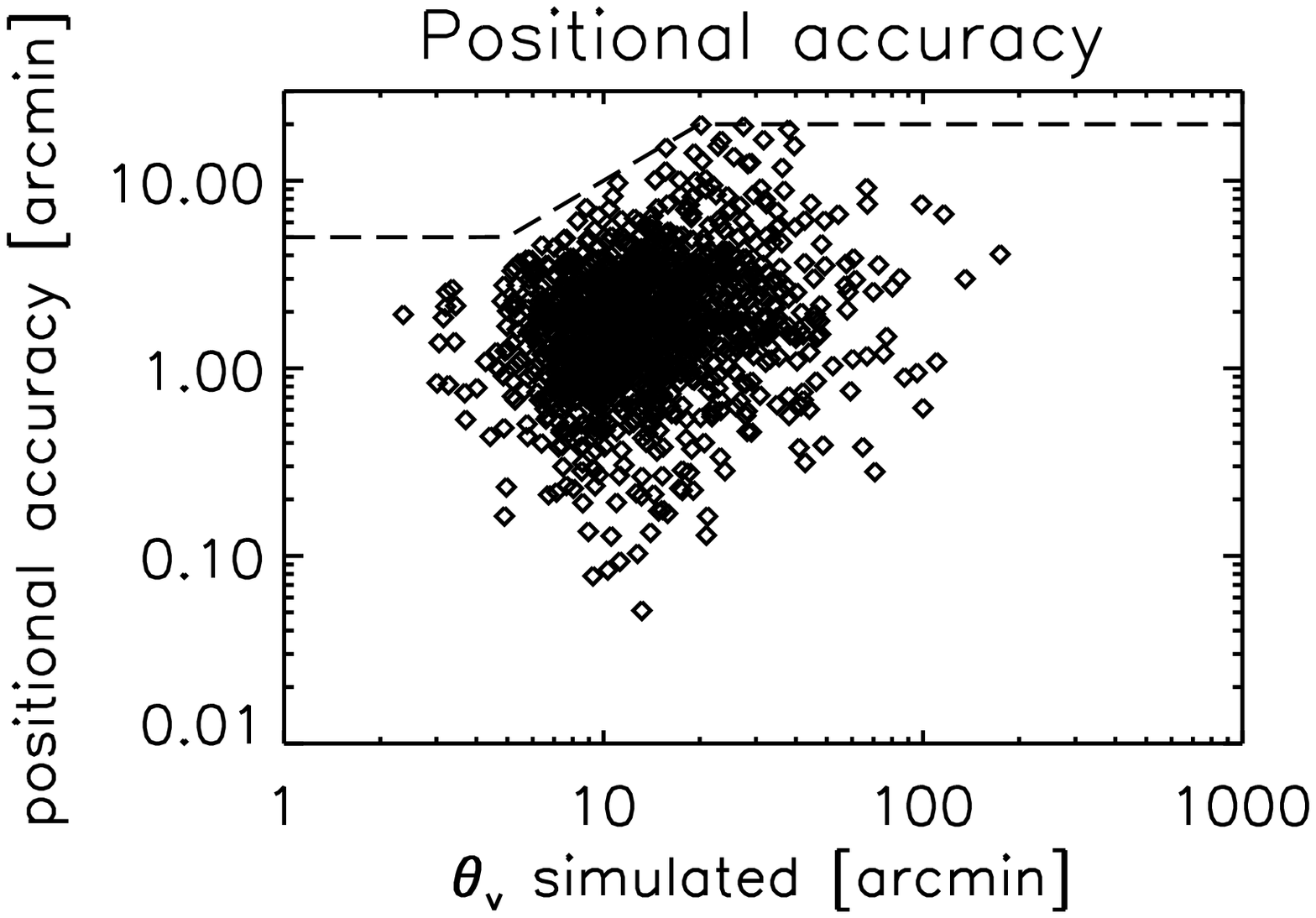} \\
\includegraphics[scale=0.45]{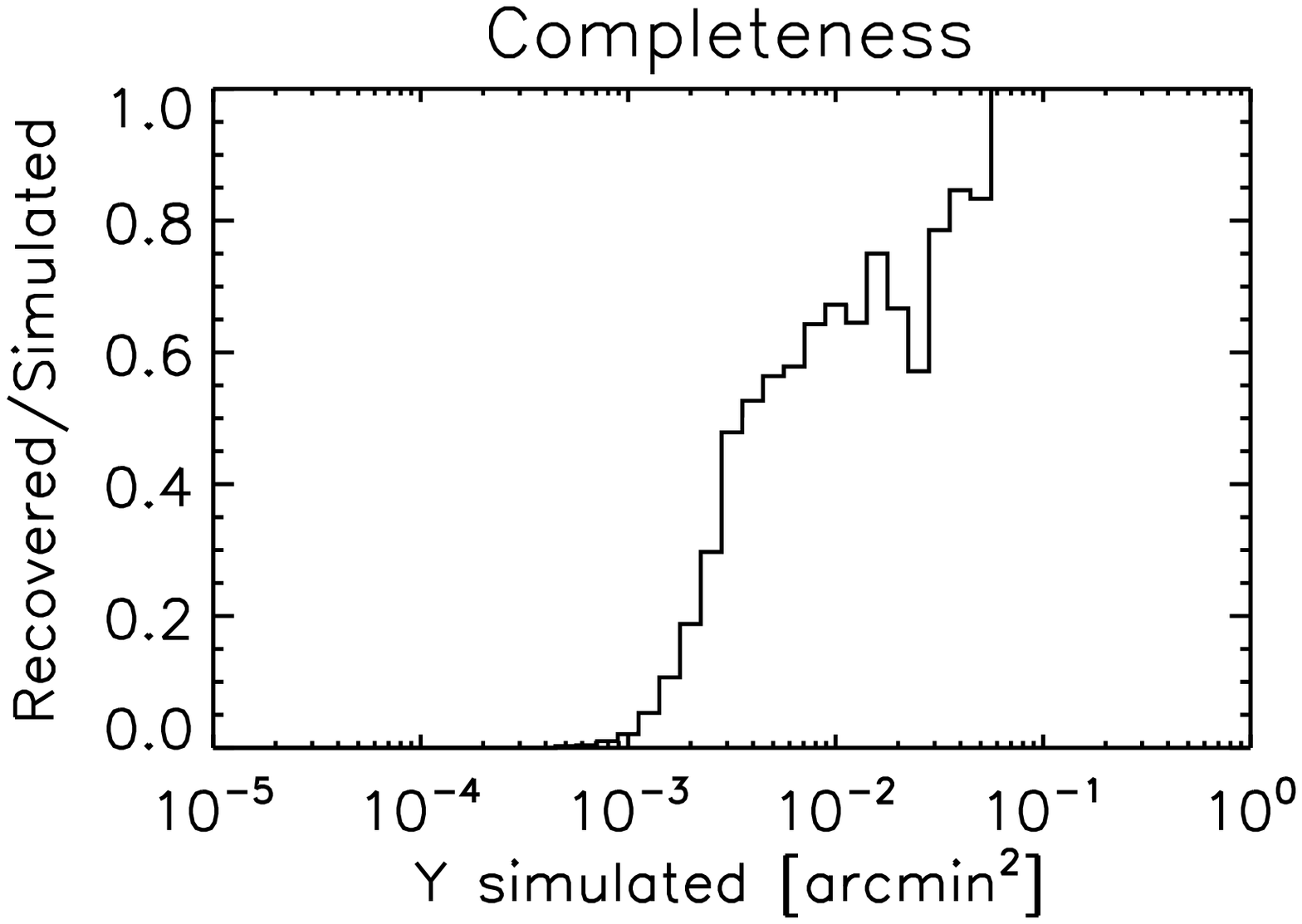}  &
\includegraphics[scale=0.45]{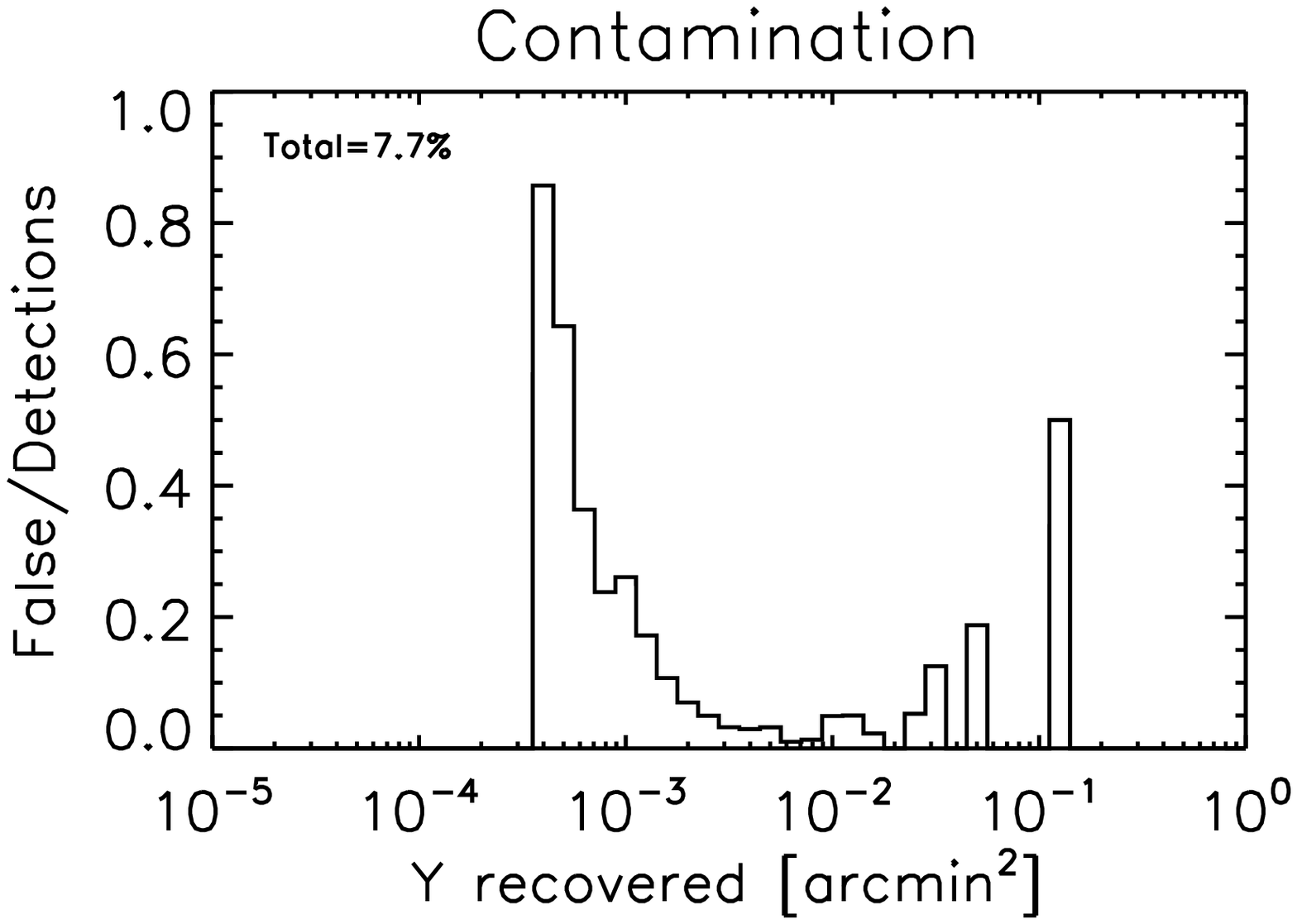} \\
\includegraphics[scale=0.45]{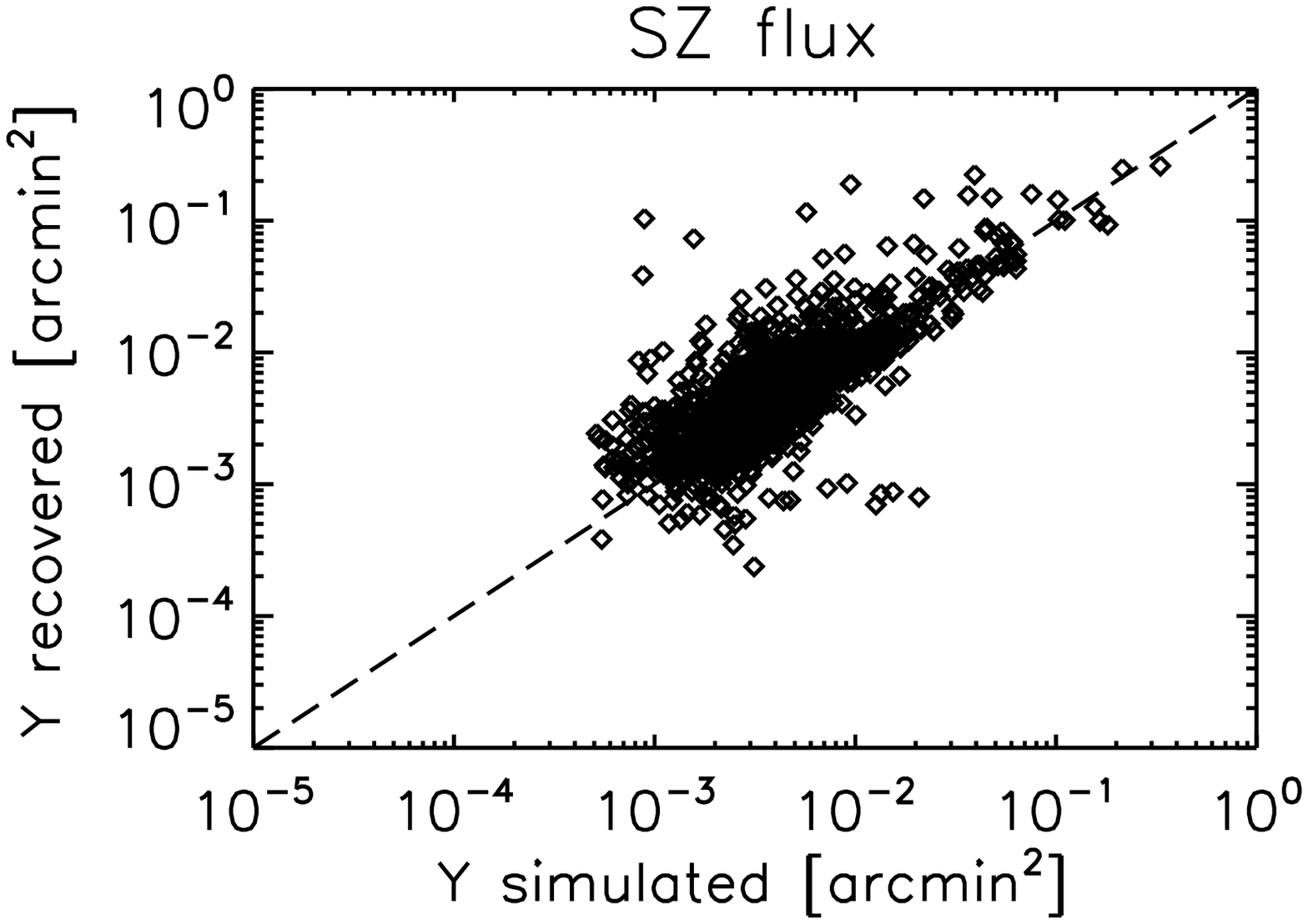}  &
\includegraphics[scale=0.45]{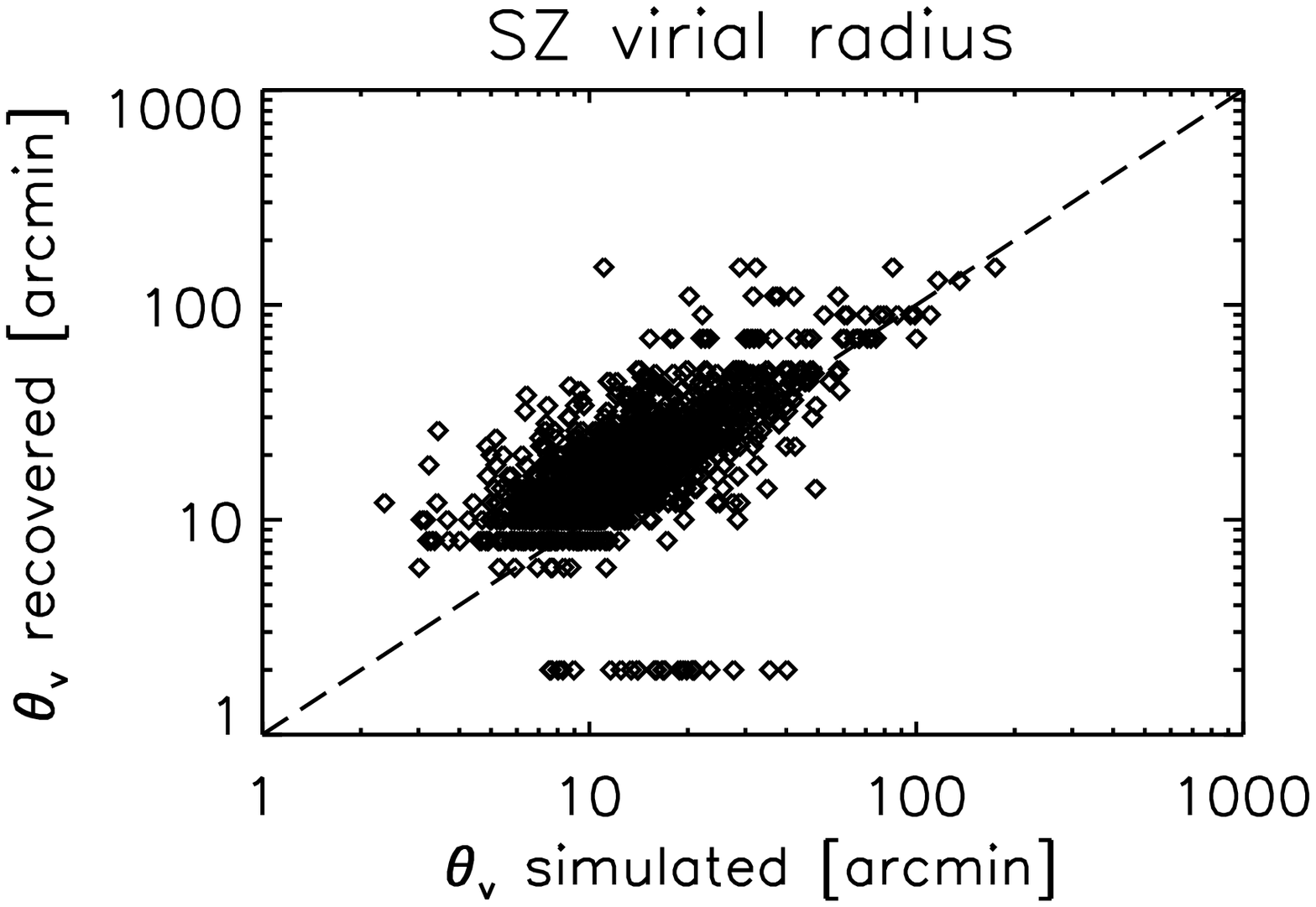} \\
\includegraphics[scale=0.45]{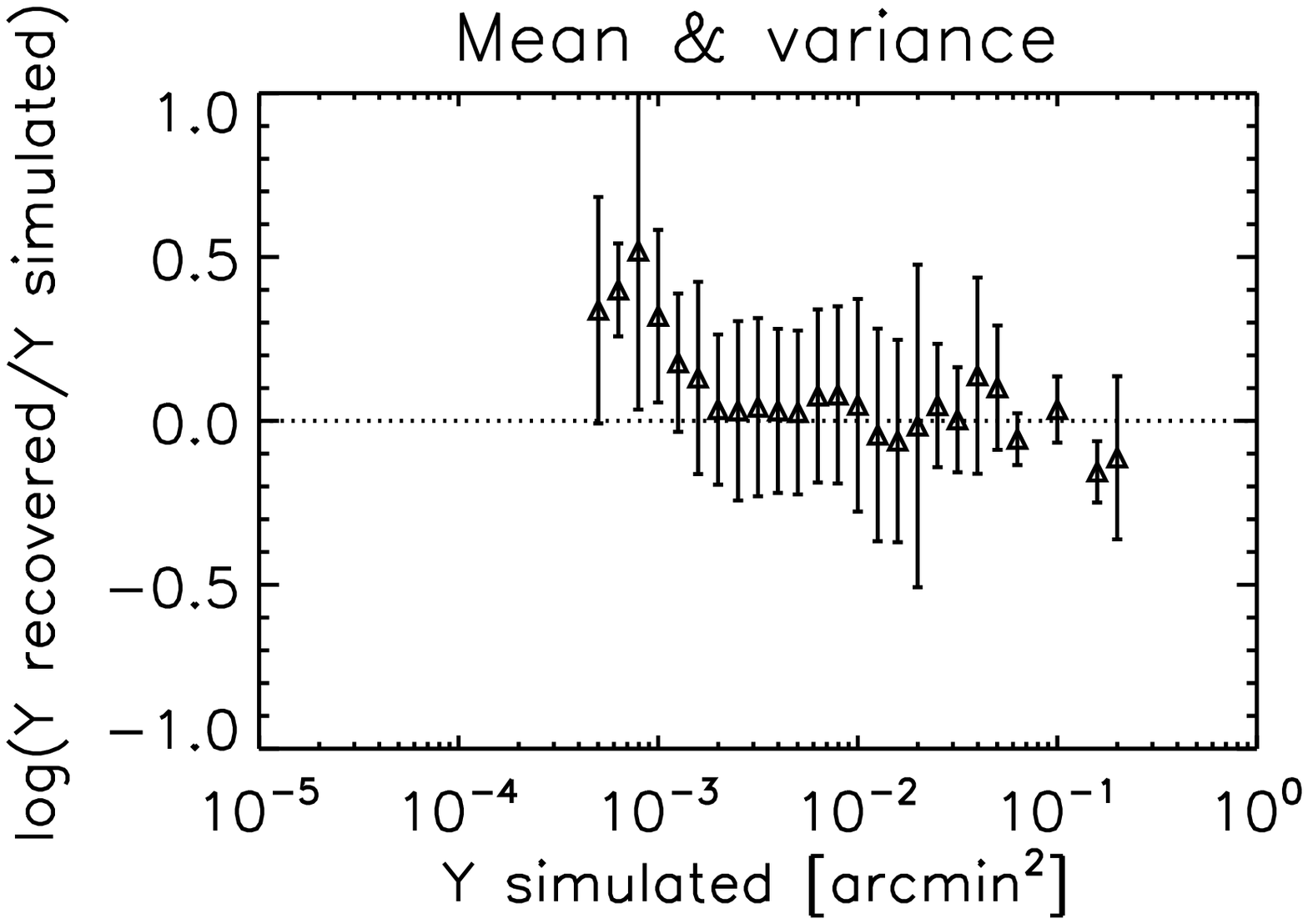}  &
\includegraphics[scale=0.45]{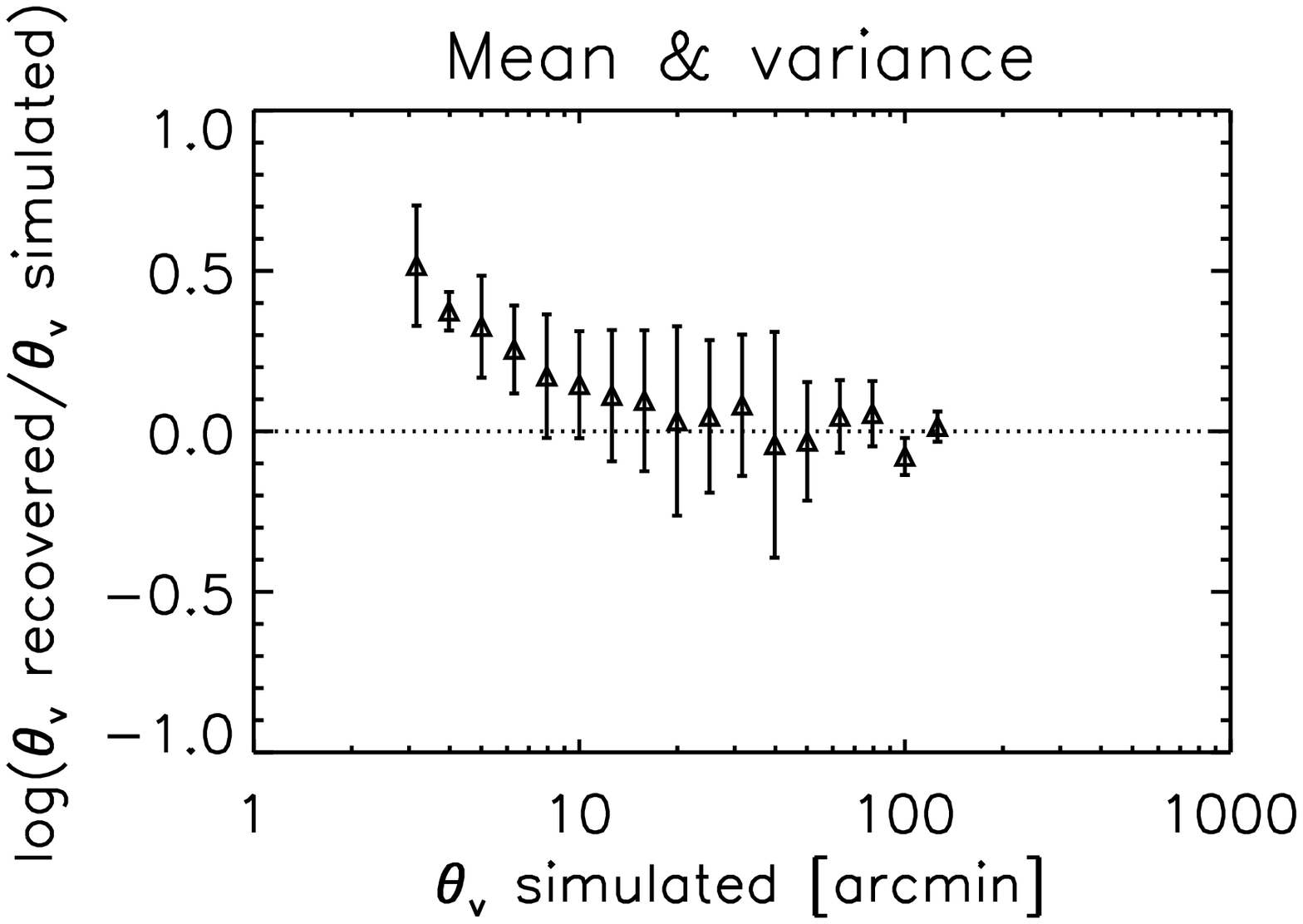} \\
\end{tabular}
\caption{{\bf GMCA}}
\end{center}
\end{table}

\clearpage

\begin{table}[htbp]
\begin{center}
\begin{tabular}{cc}
\includegraphics[scale=0.45]{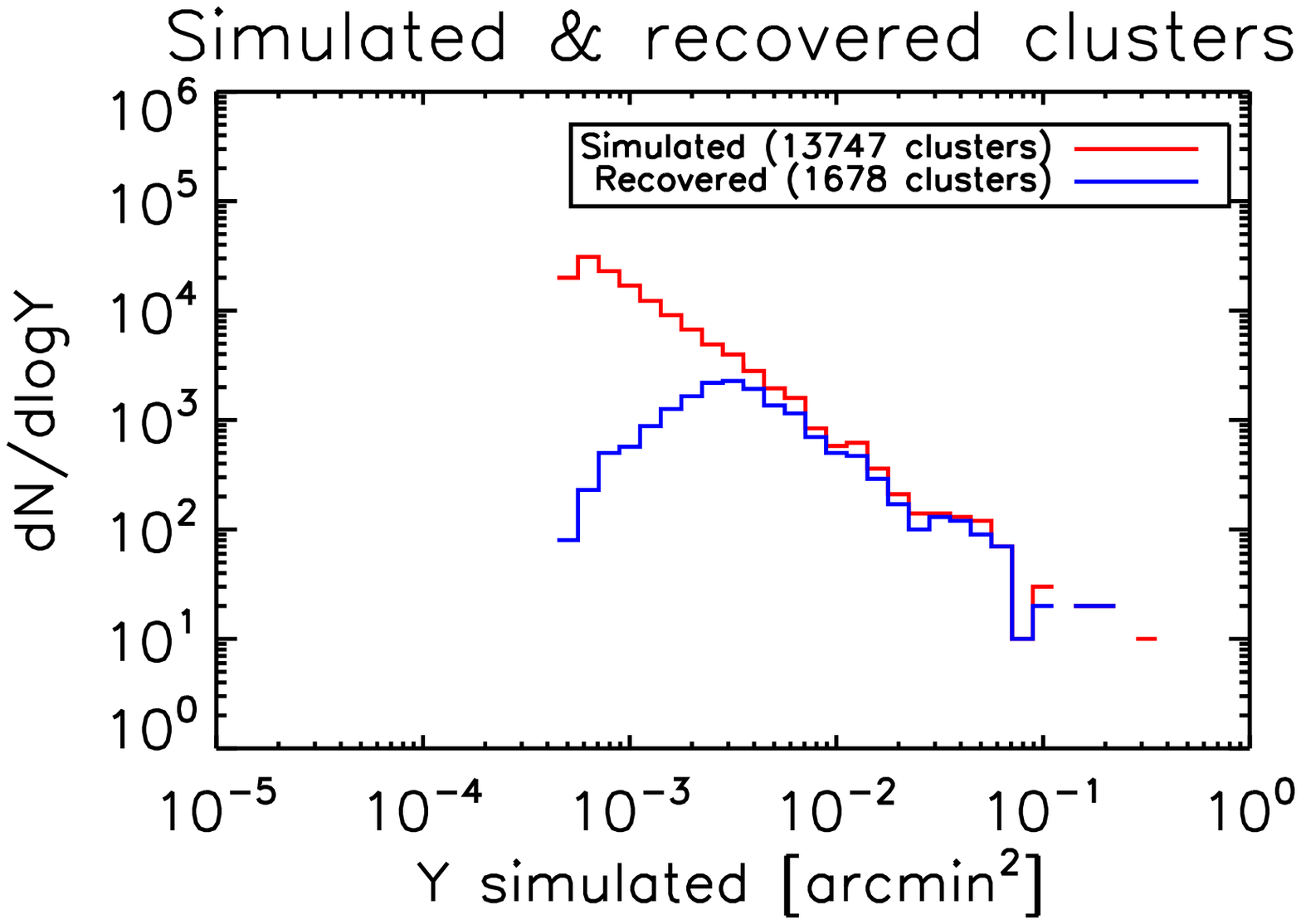}  &
\includegraphics[scale=0.45]{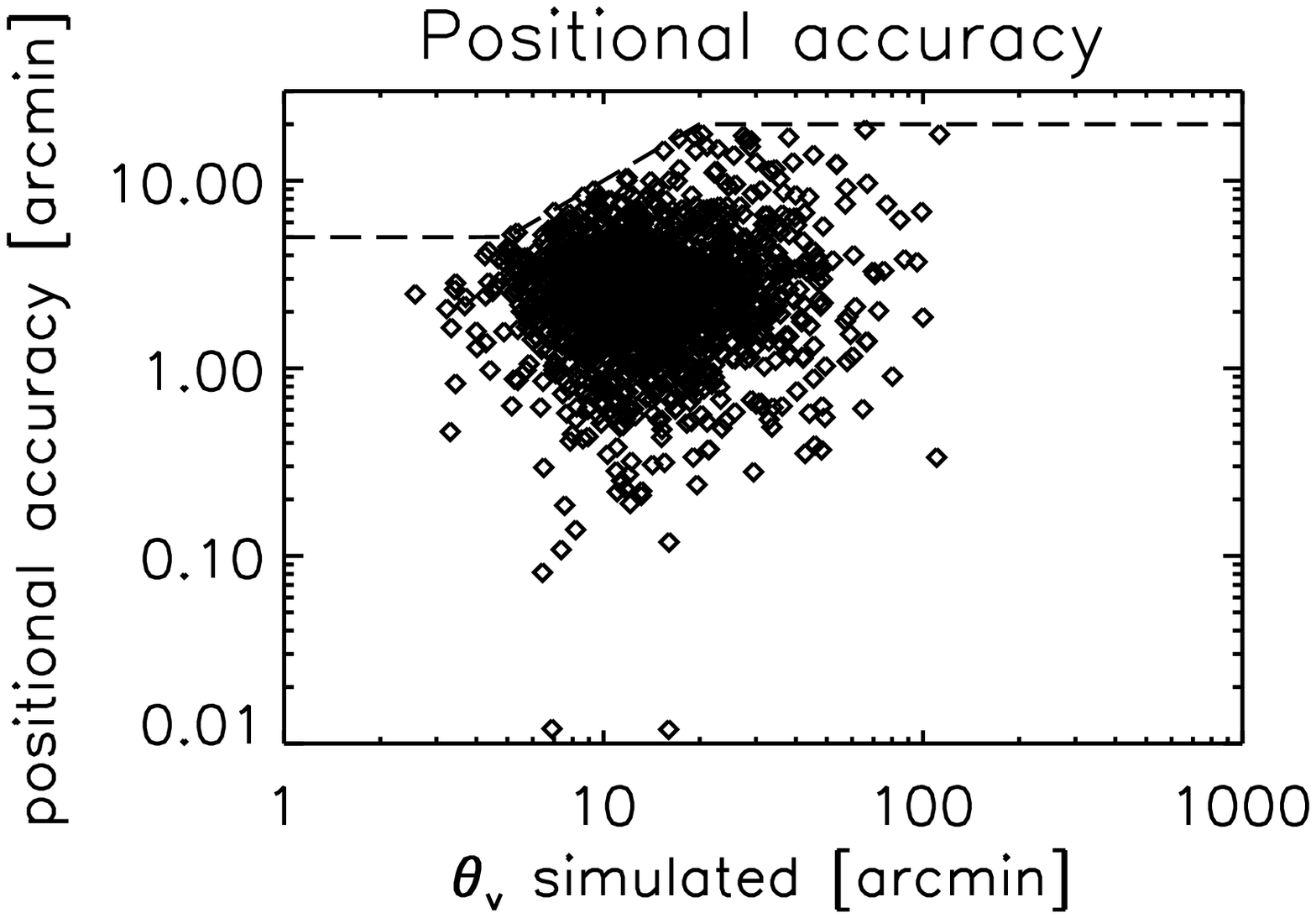} \\
\includegraphics[scale=0.45]{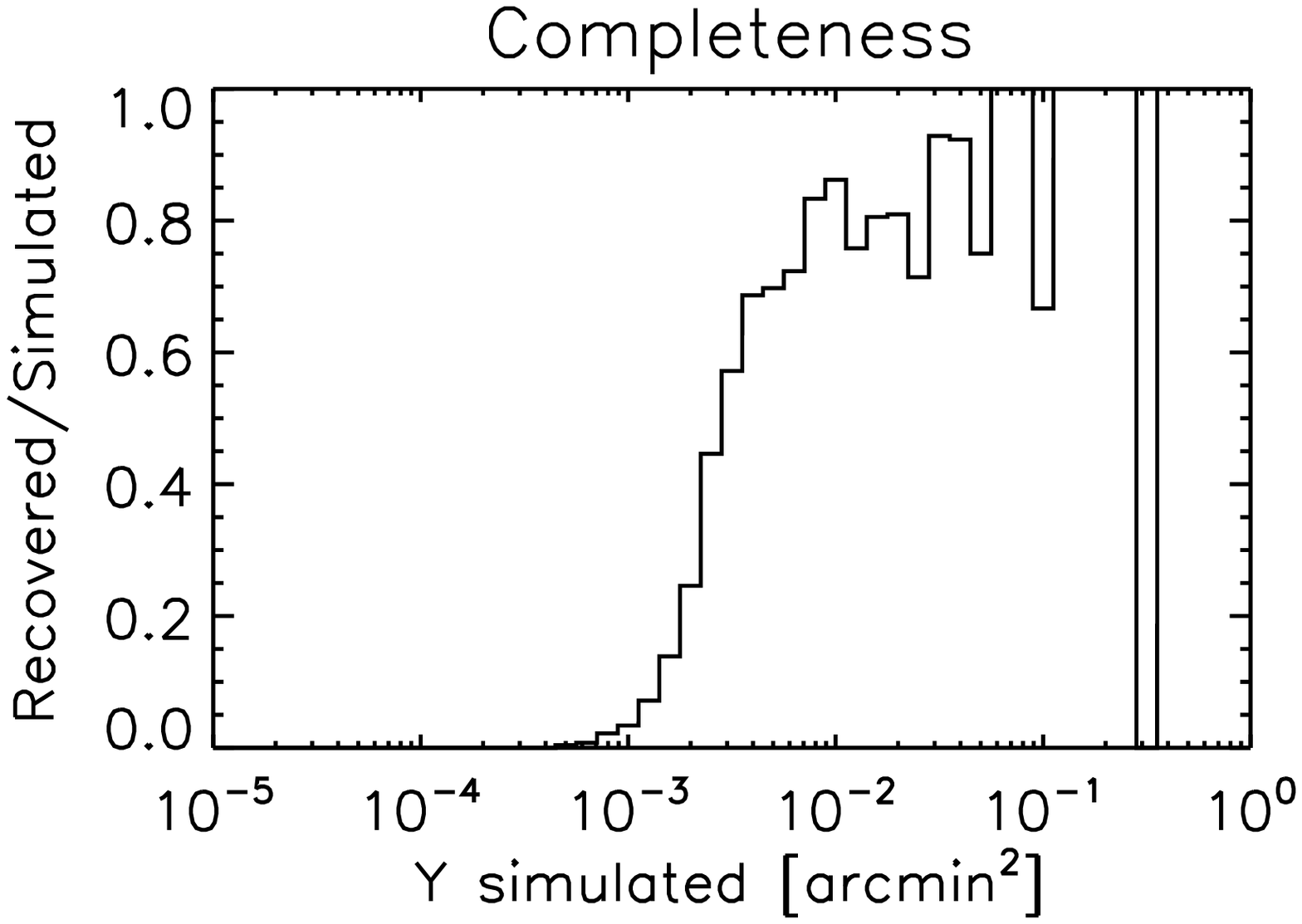}  &
\includegraphics[scale=0.45]{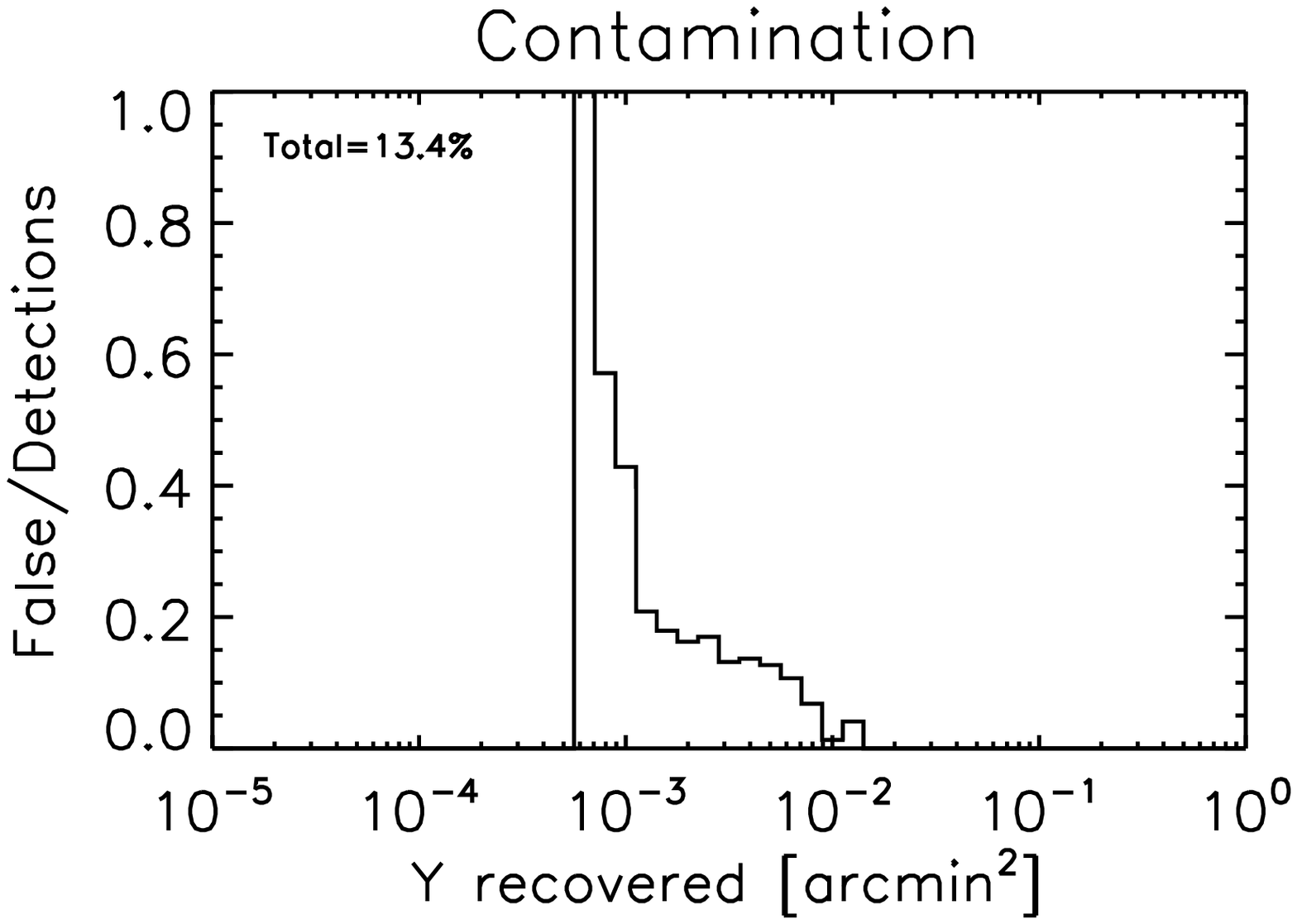} \\
\includegraphics[scale=0.45]{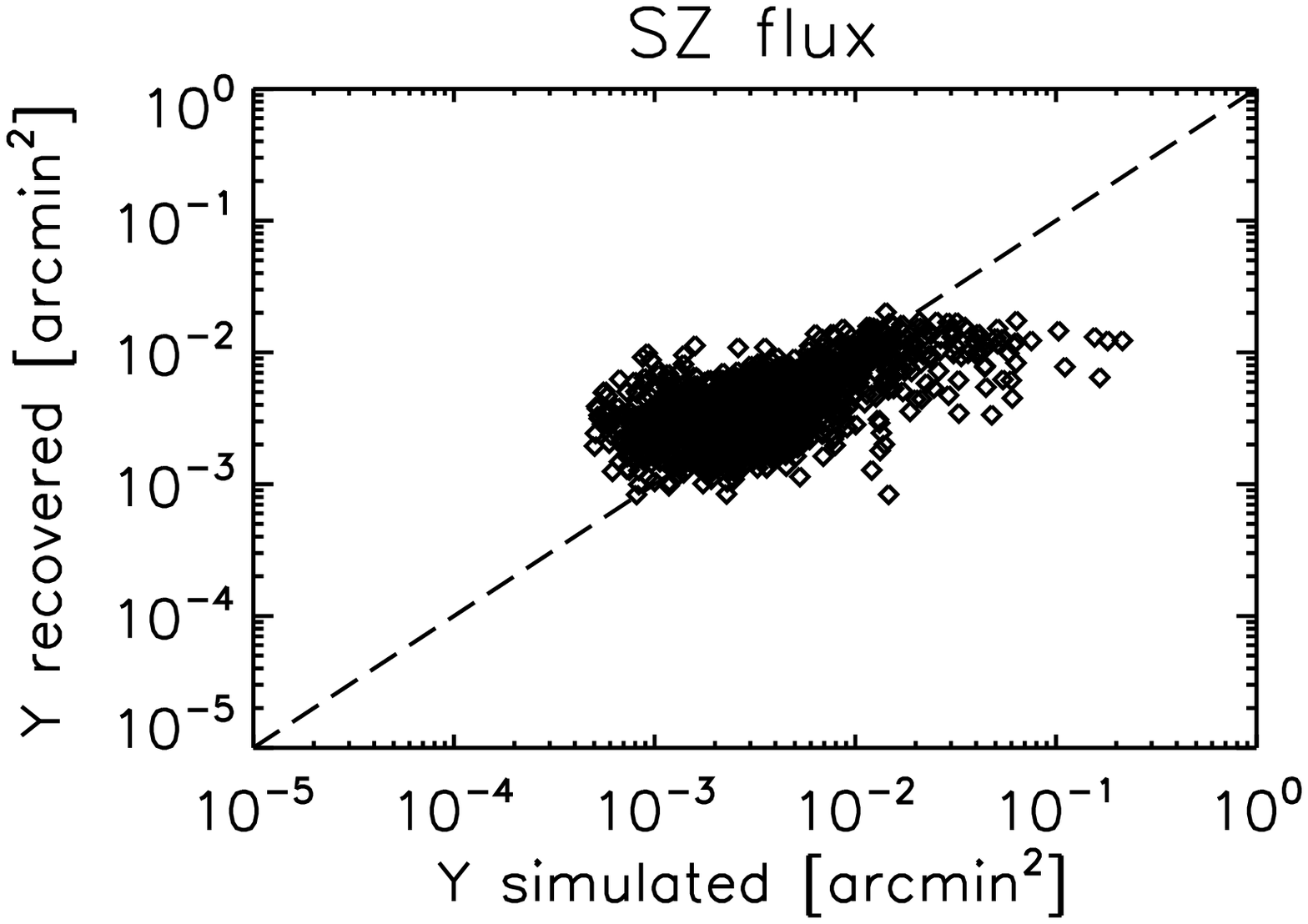}  &
\includegraphics[scale=0.45]{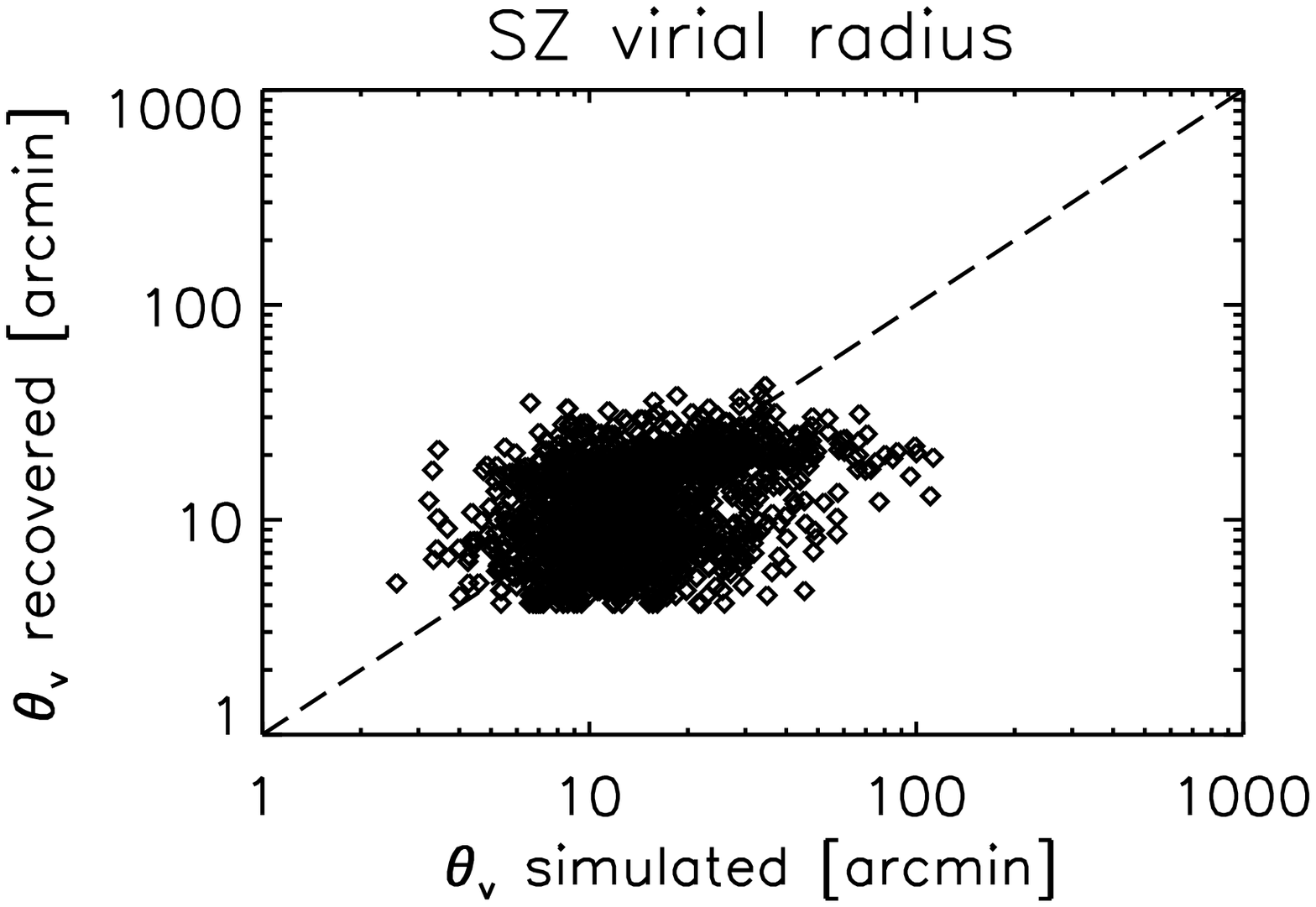} \\
\includegraphics[scale=0.45]{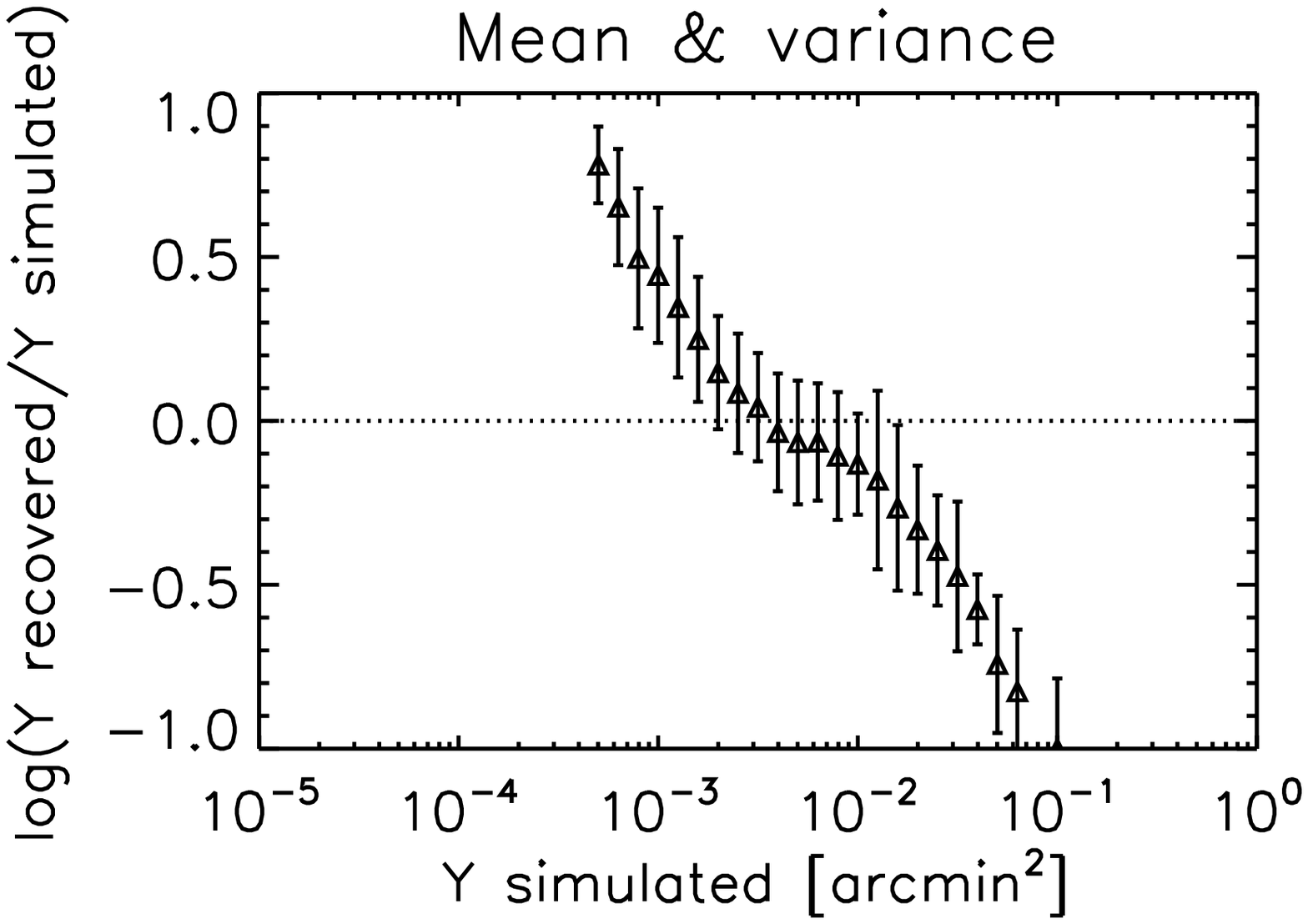} &
\includegraphics[scale=0.45]{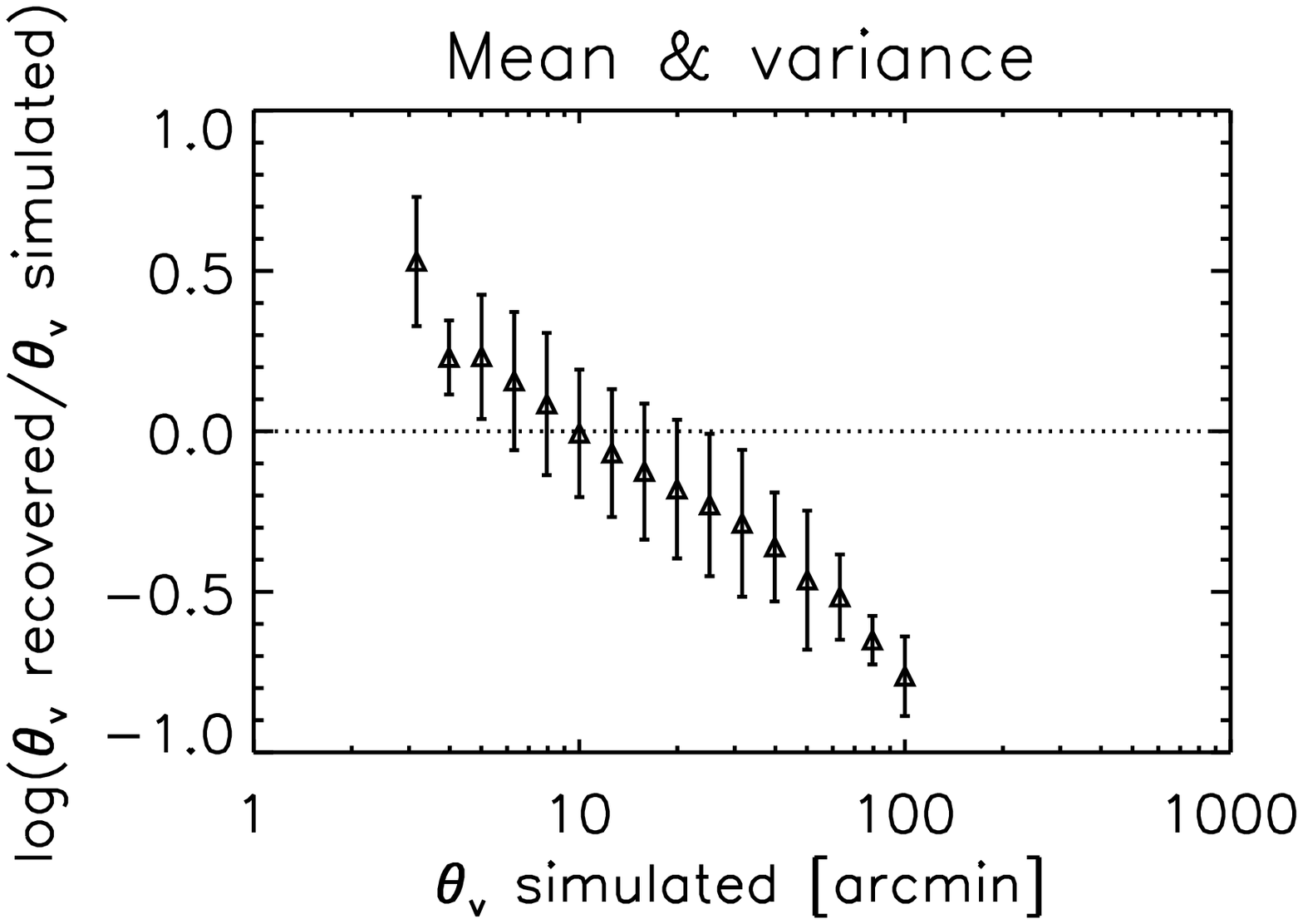} \\
\end{tabular}
\caption{{\bf BNP}}
\end{center}
\end{table}

\clearpage

\section{Comparison plots for Challenge SZ v2 (Run 2)}
\label{app:complotv2}

The plots are obtained at purity=0.9 (except when the catalog does not permit this value to be reached).

\clearpage

\begin{table}[htbp]
\begin{center}
\begin{tabular}{cc}
\includegraphics[scale=0.45]{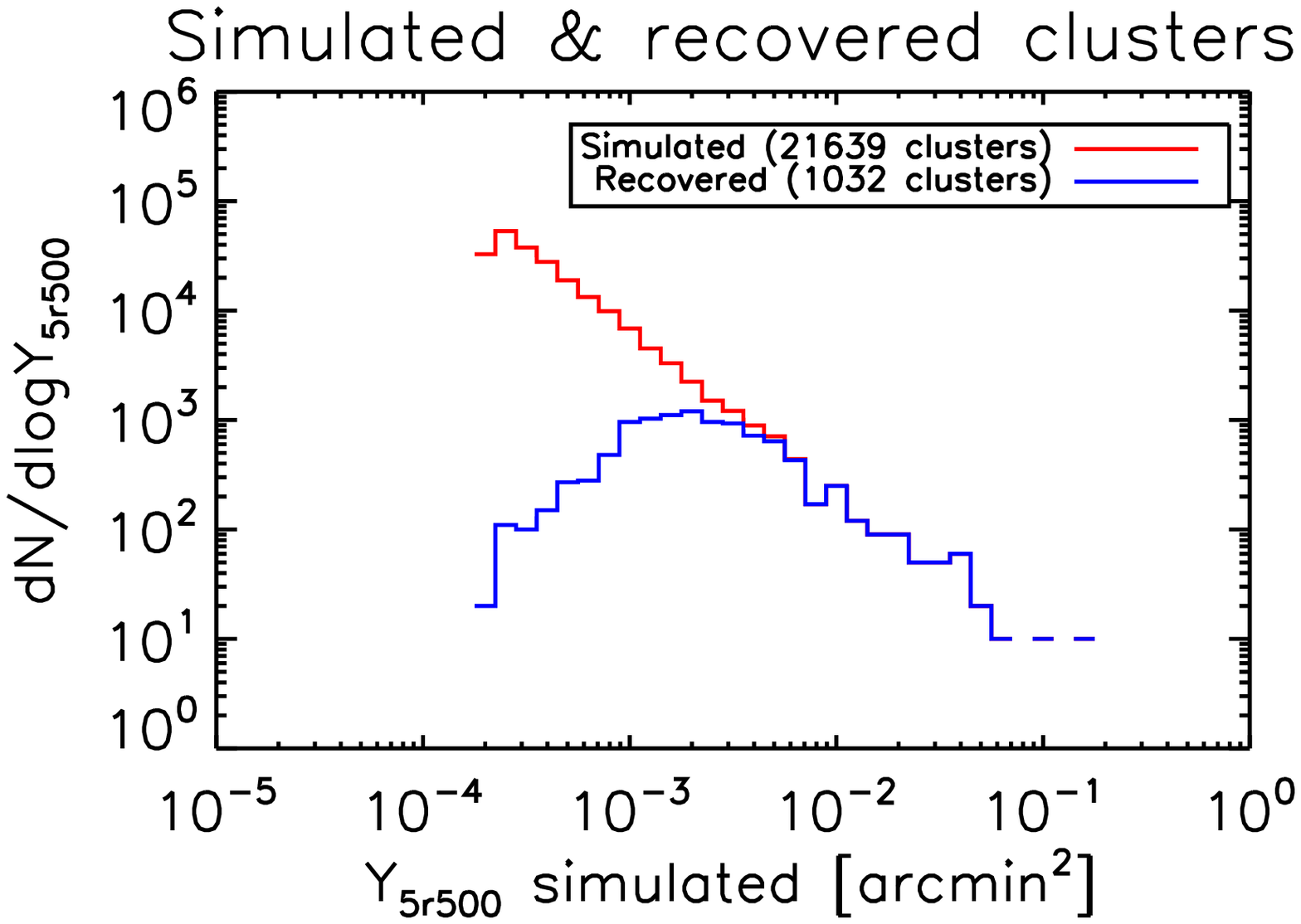}  &
\includegraphics[scale=0.45]{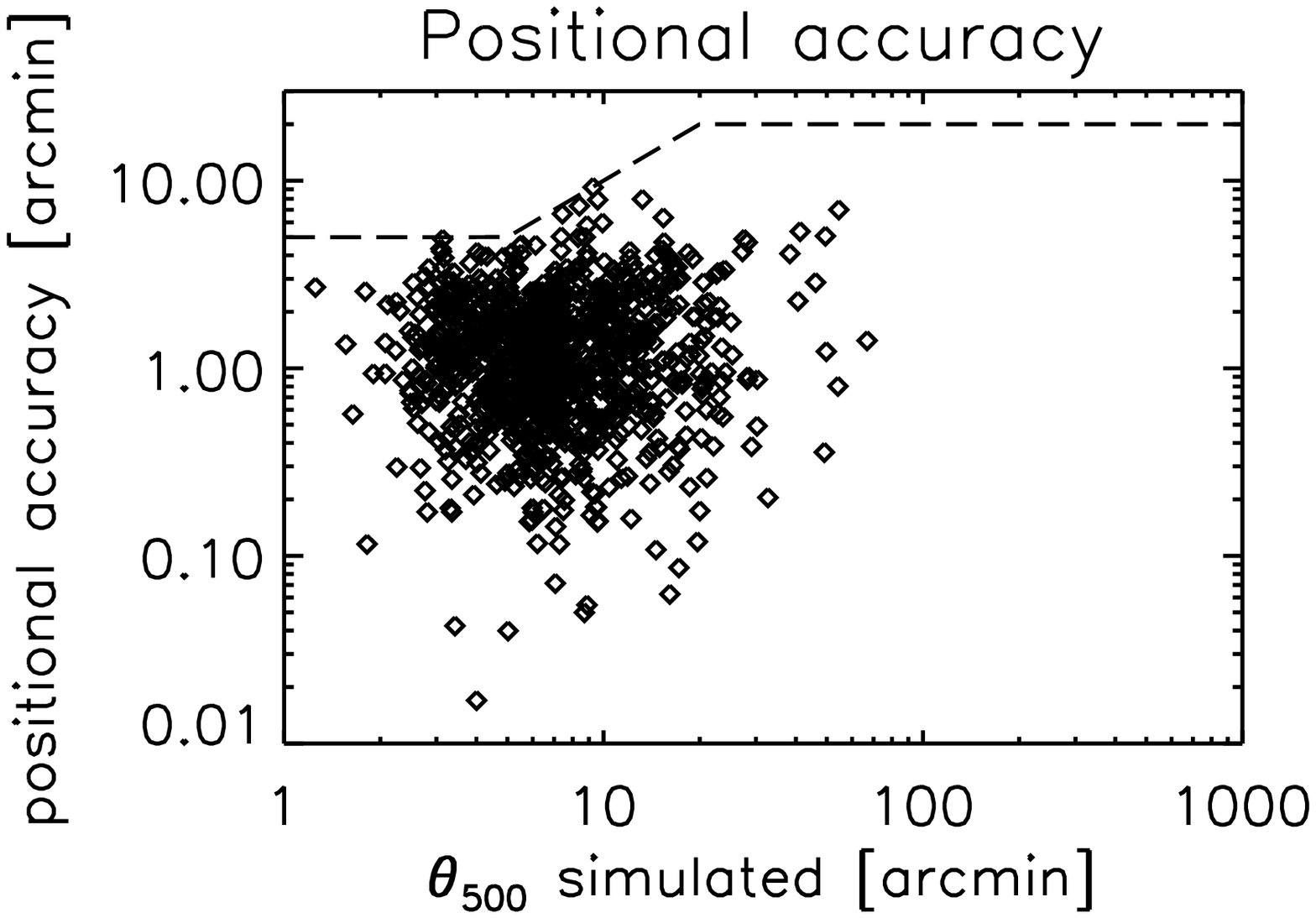} \\
\includegraphics[scale=0.45]{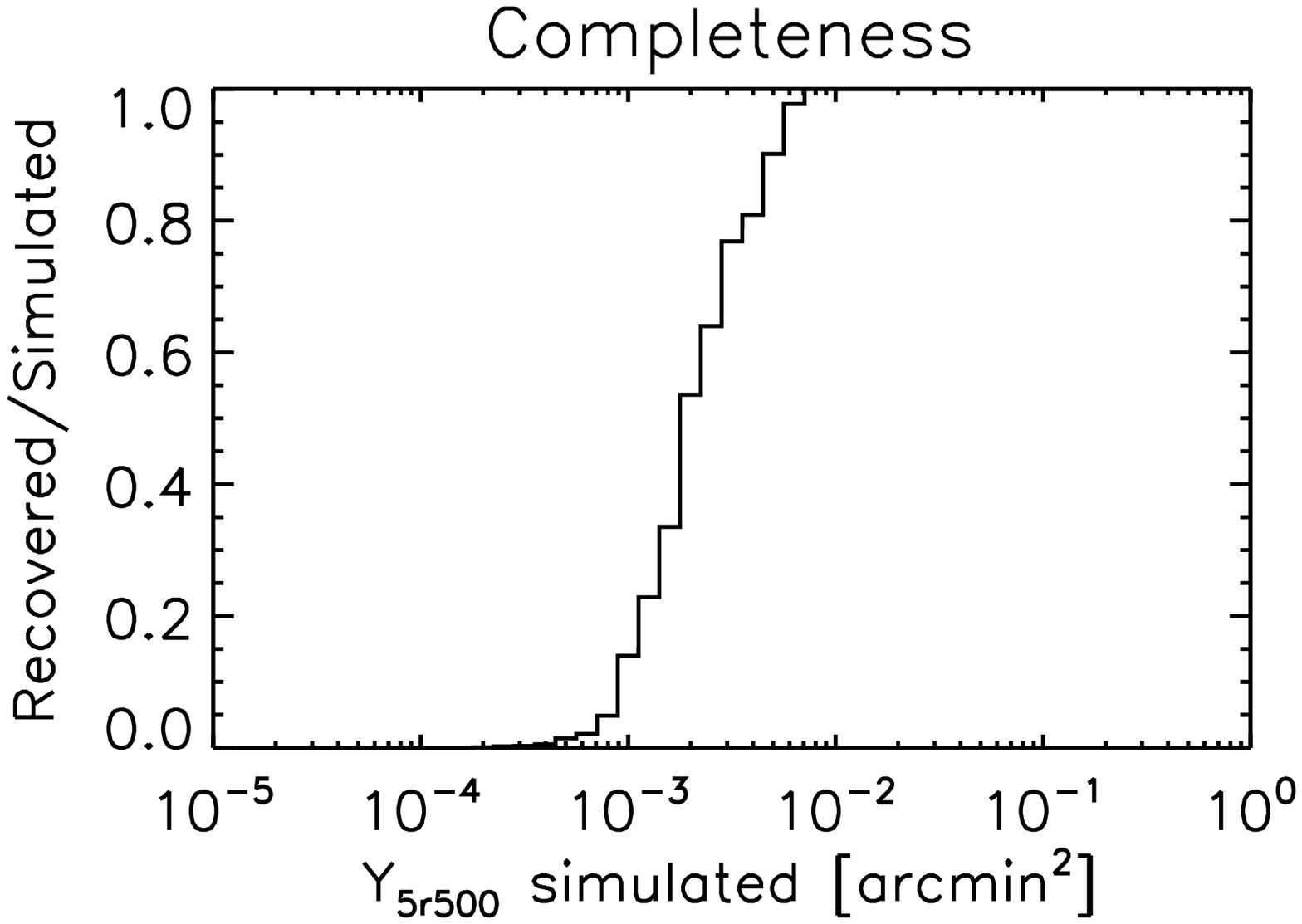}  &
\includegraphics[scale=0.45]{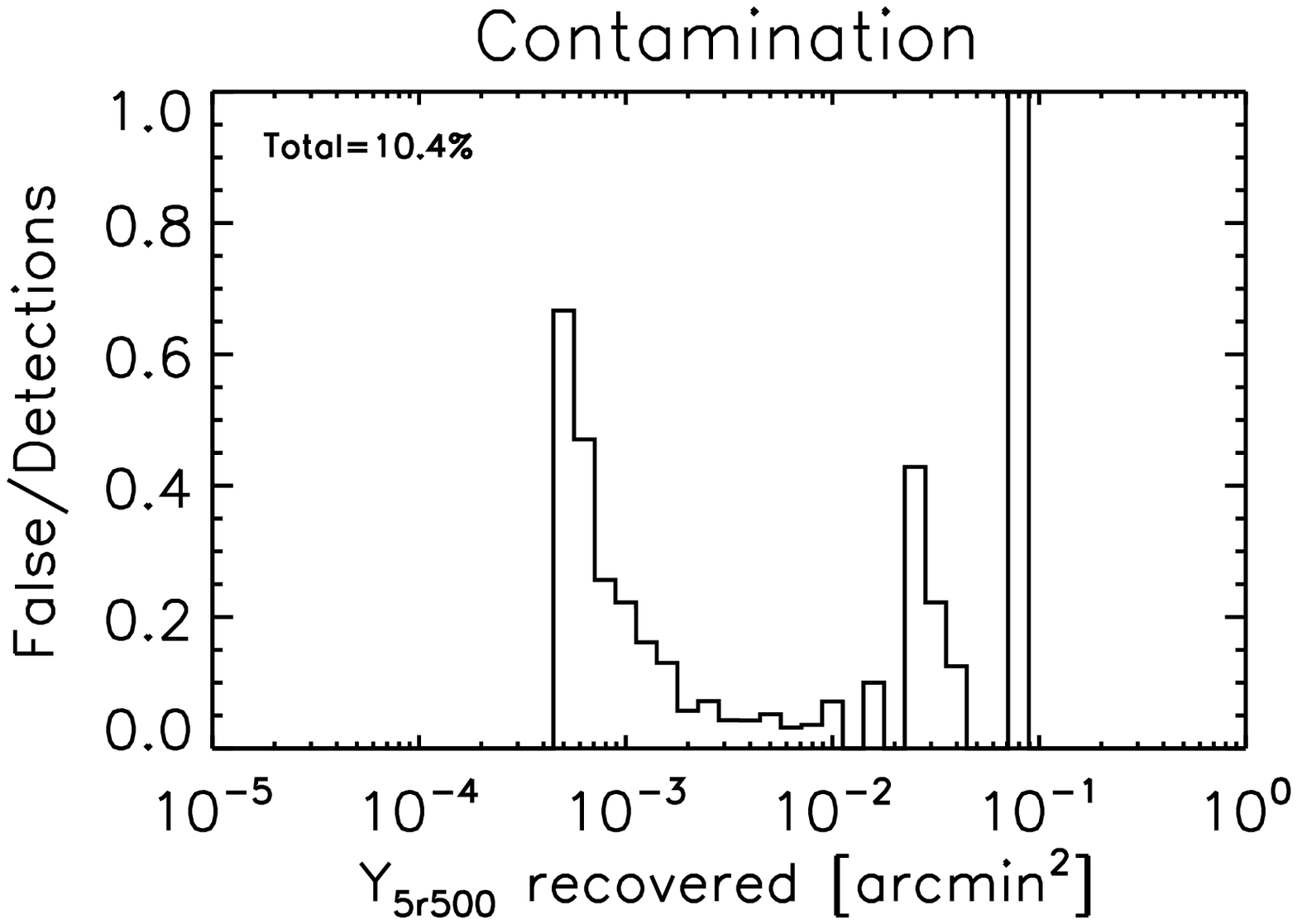} \\
\includegraphics[scale=0.45]{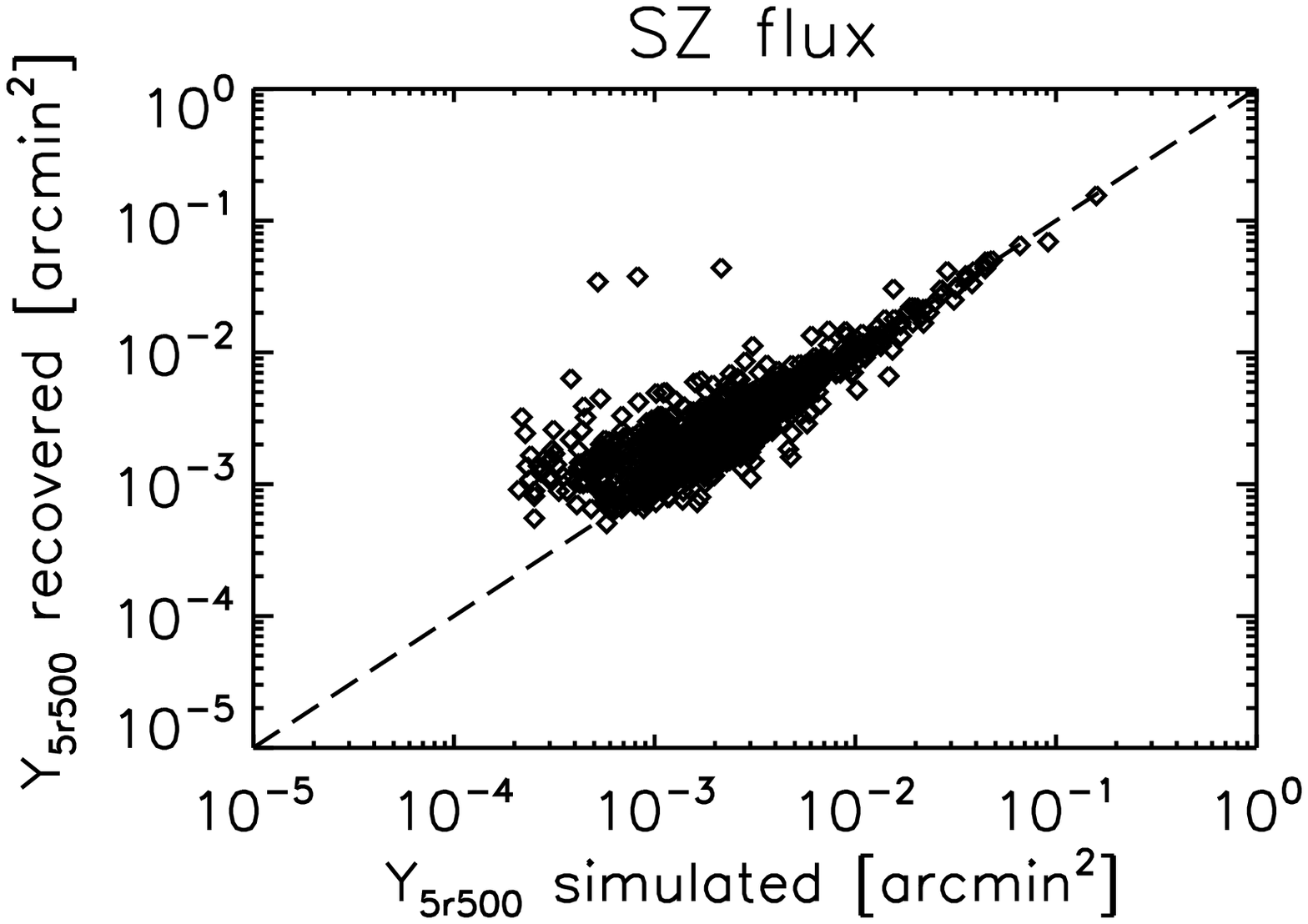}  &
\includegraphics[scale=0.45]{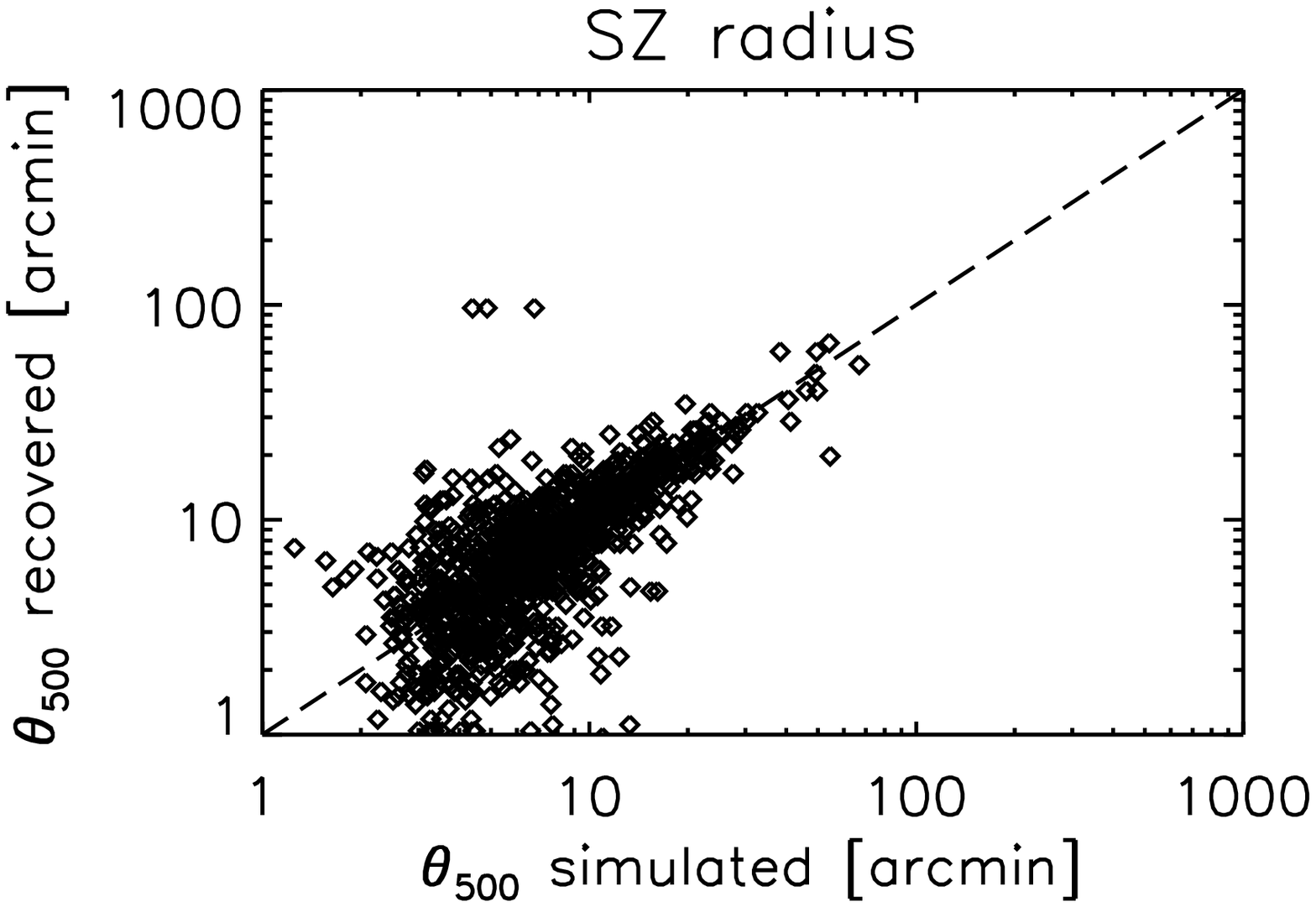} \\
\includegraphics[scale=0.45]{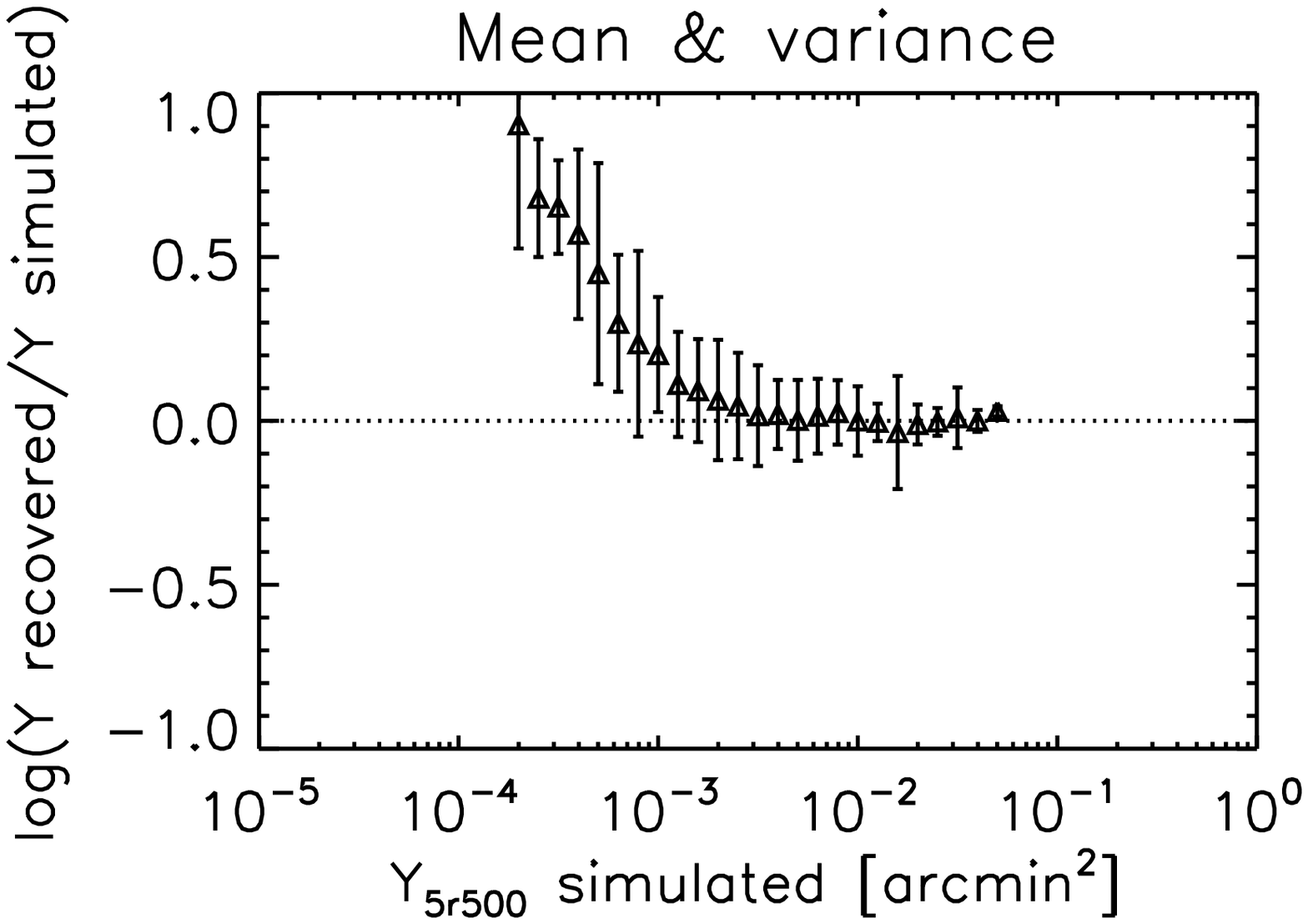} &
\includegraphics[scale=0.45]{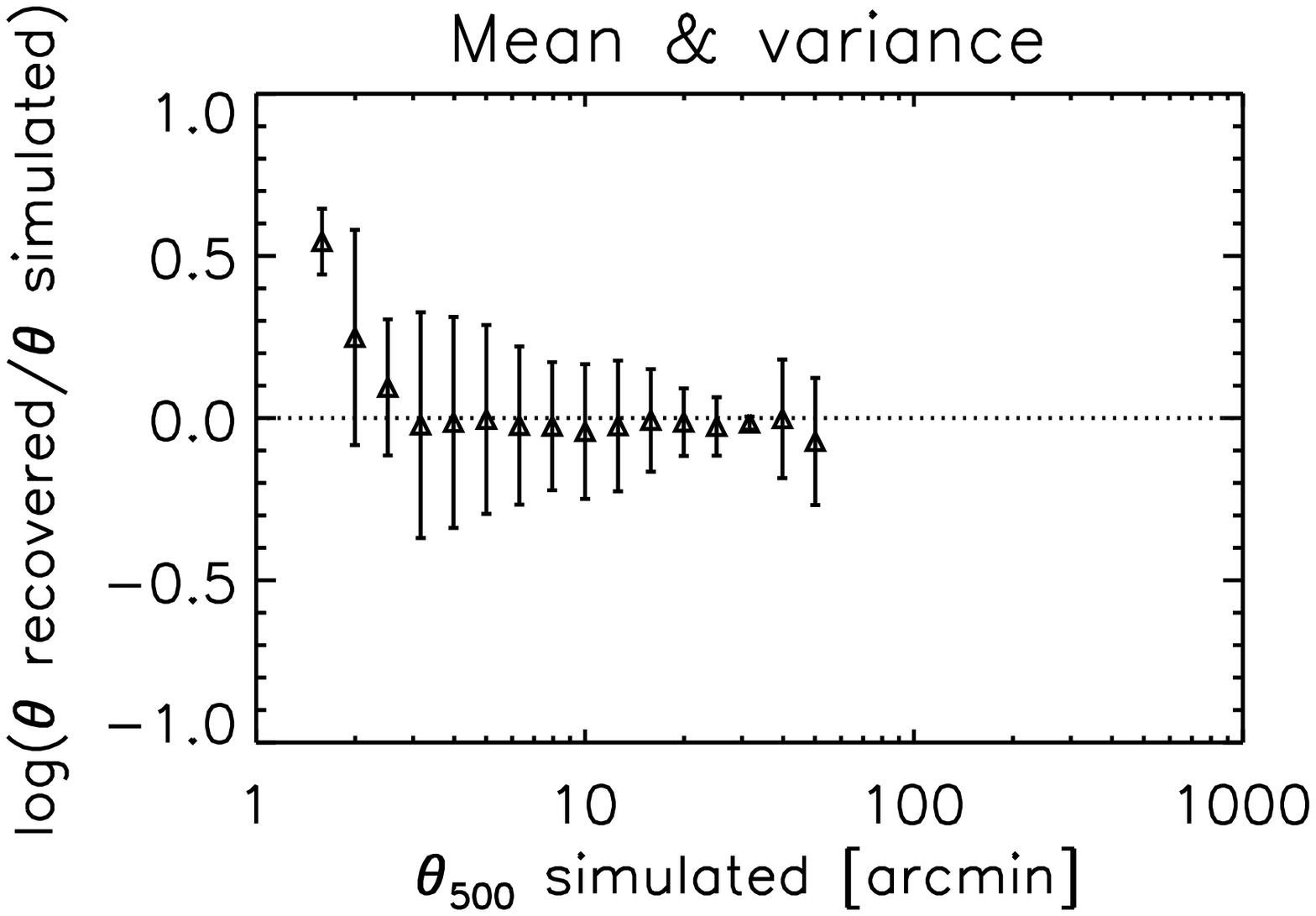} \\
\end{tabular}
\caption{{\bf MMF1}}
\end{center}
\end{table}

\clearpage

\begin{table}[htbp]
\begin{center}
\begin{tabular}{cc}
\includegraphics[scale=0.45]{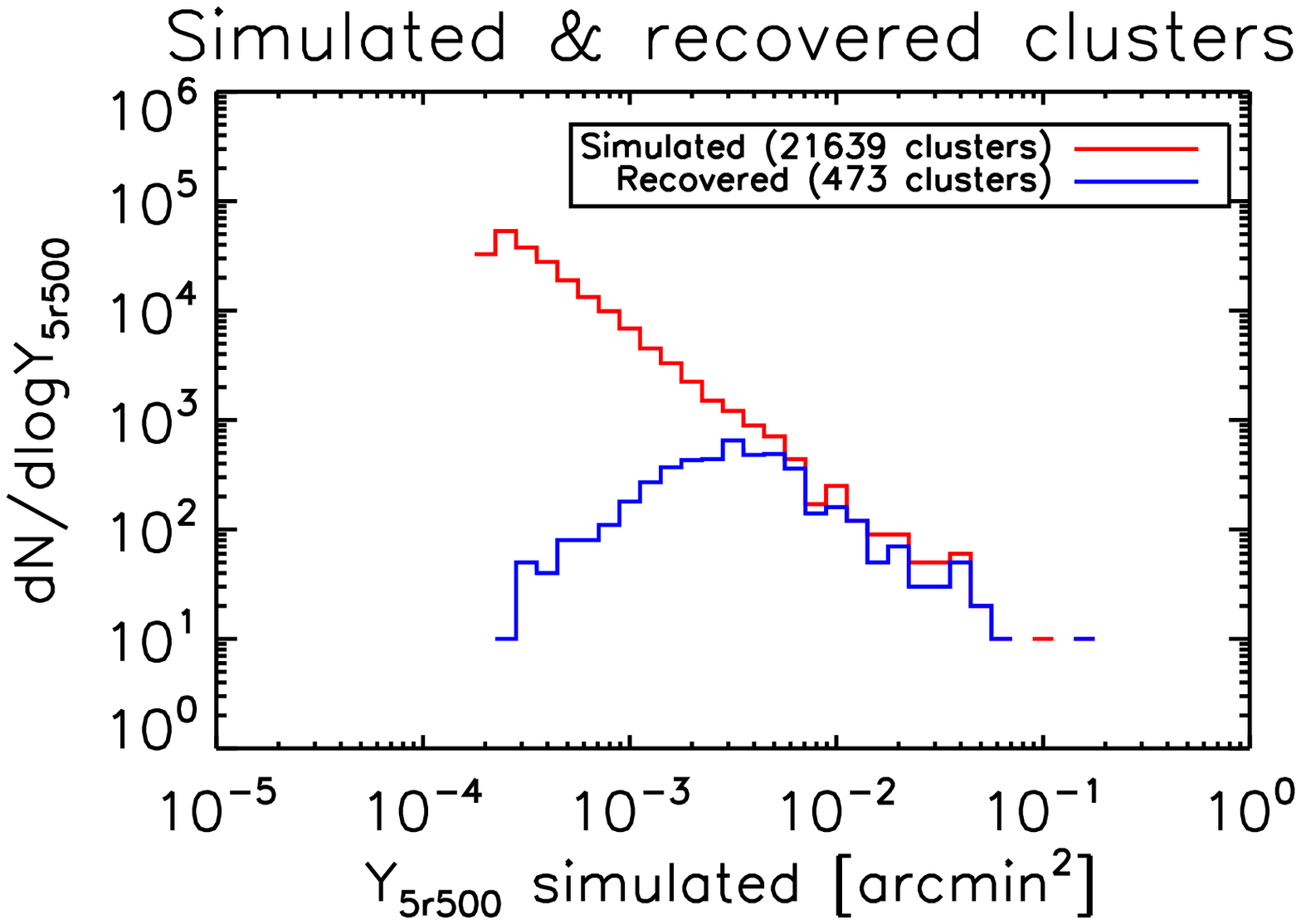}  &
\includegraphics[scale=0.45]{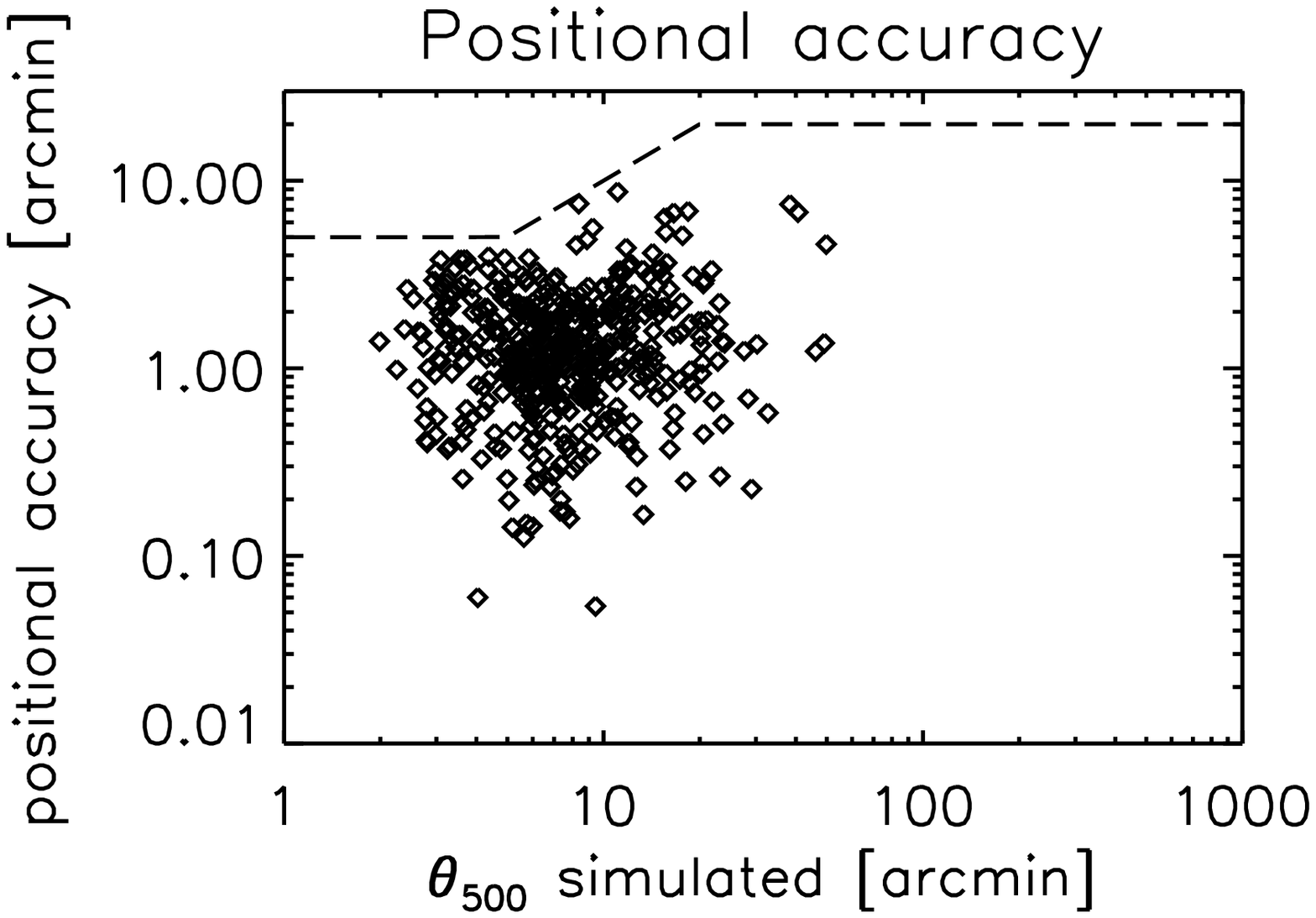} \\
\includegraphics[scale=0.45]{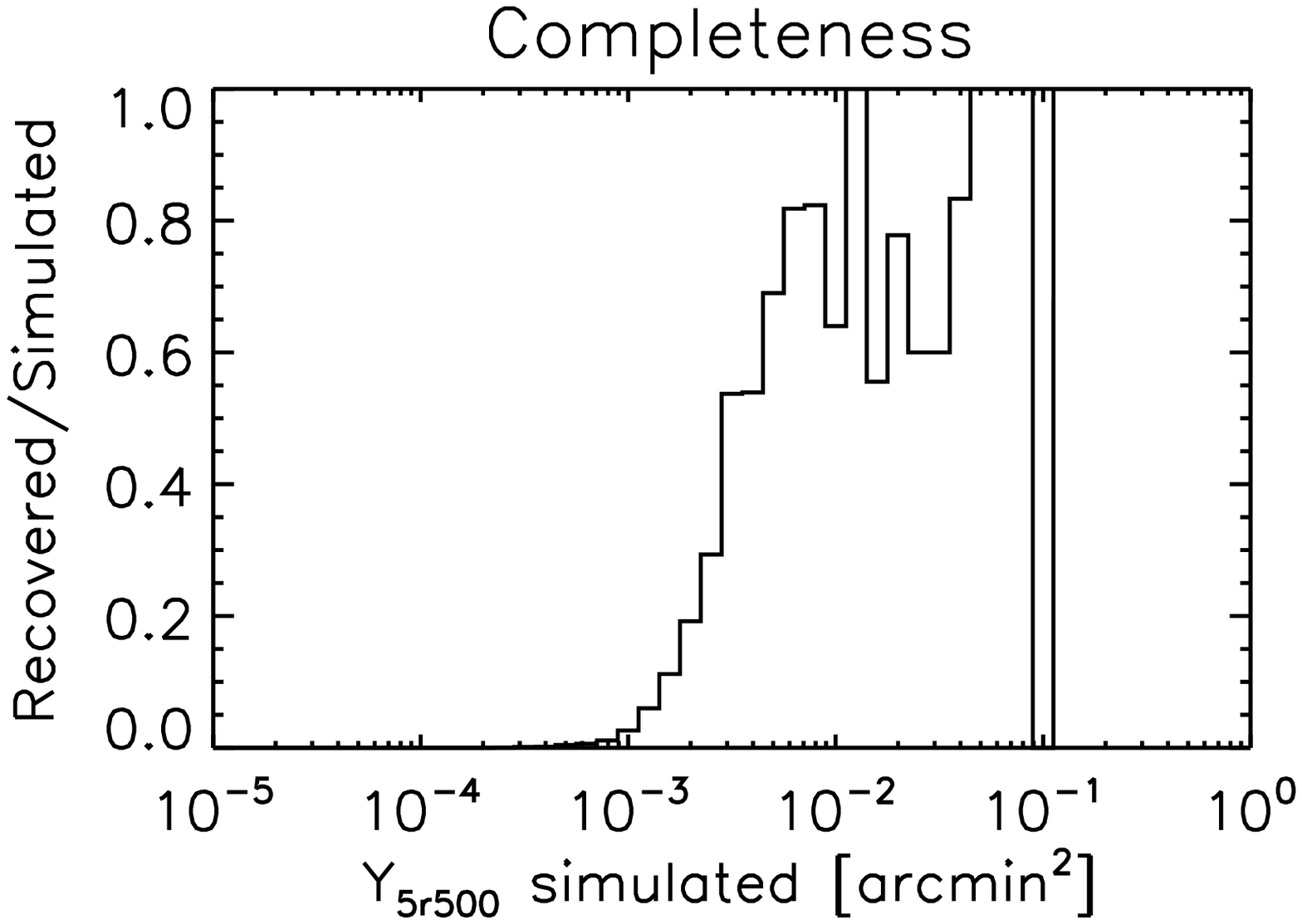}  &
\includegraphics[scale=0.45]{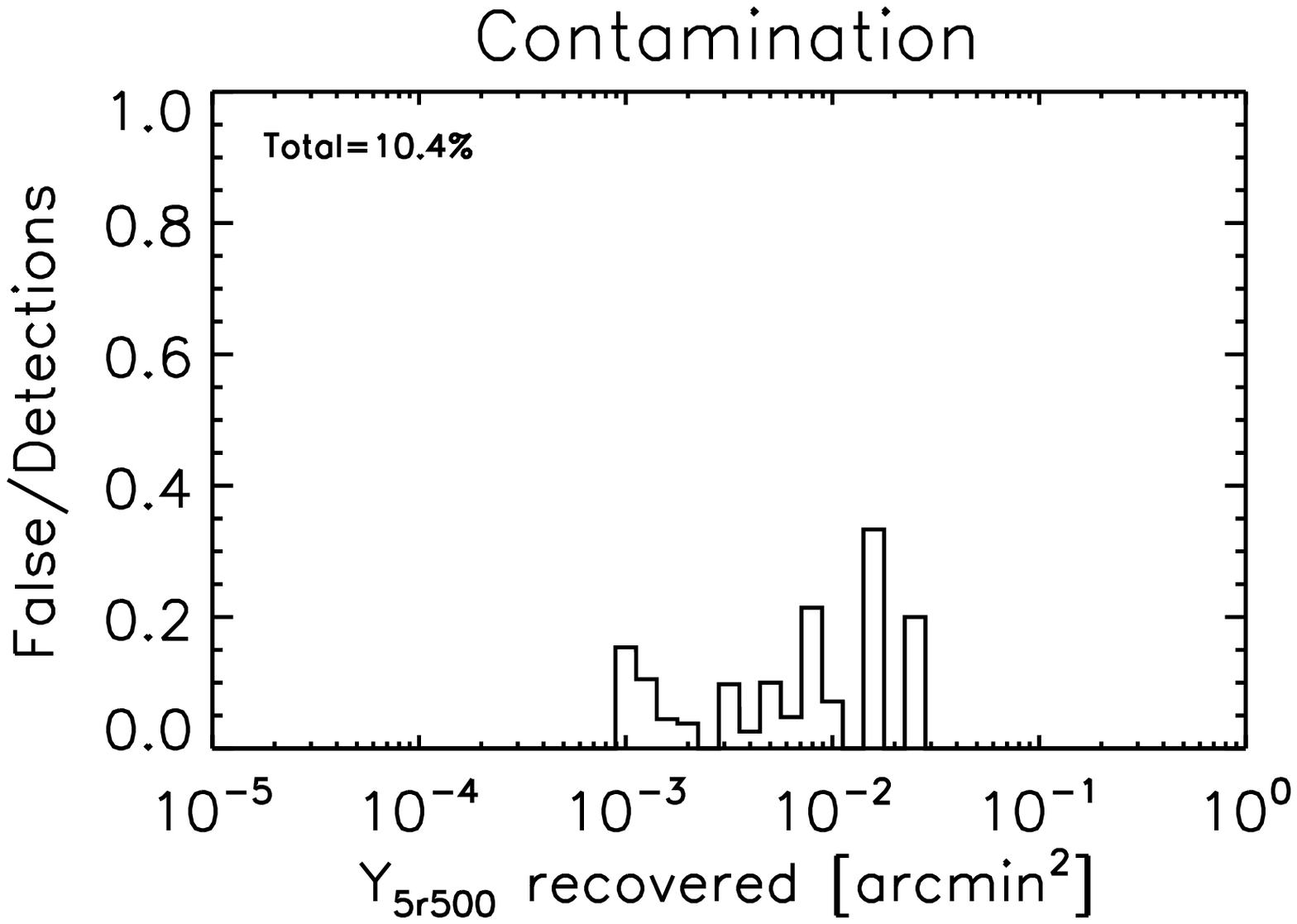} \\
\includegraphics[scale=0.45]{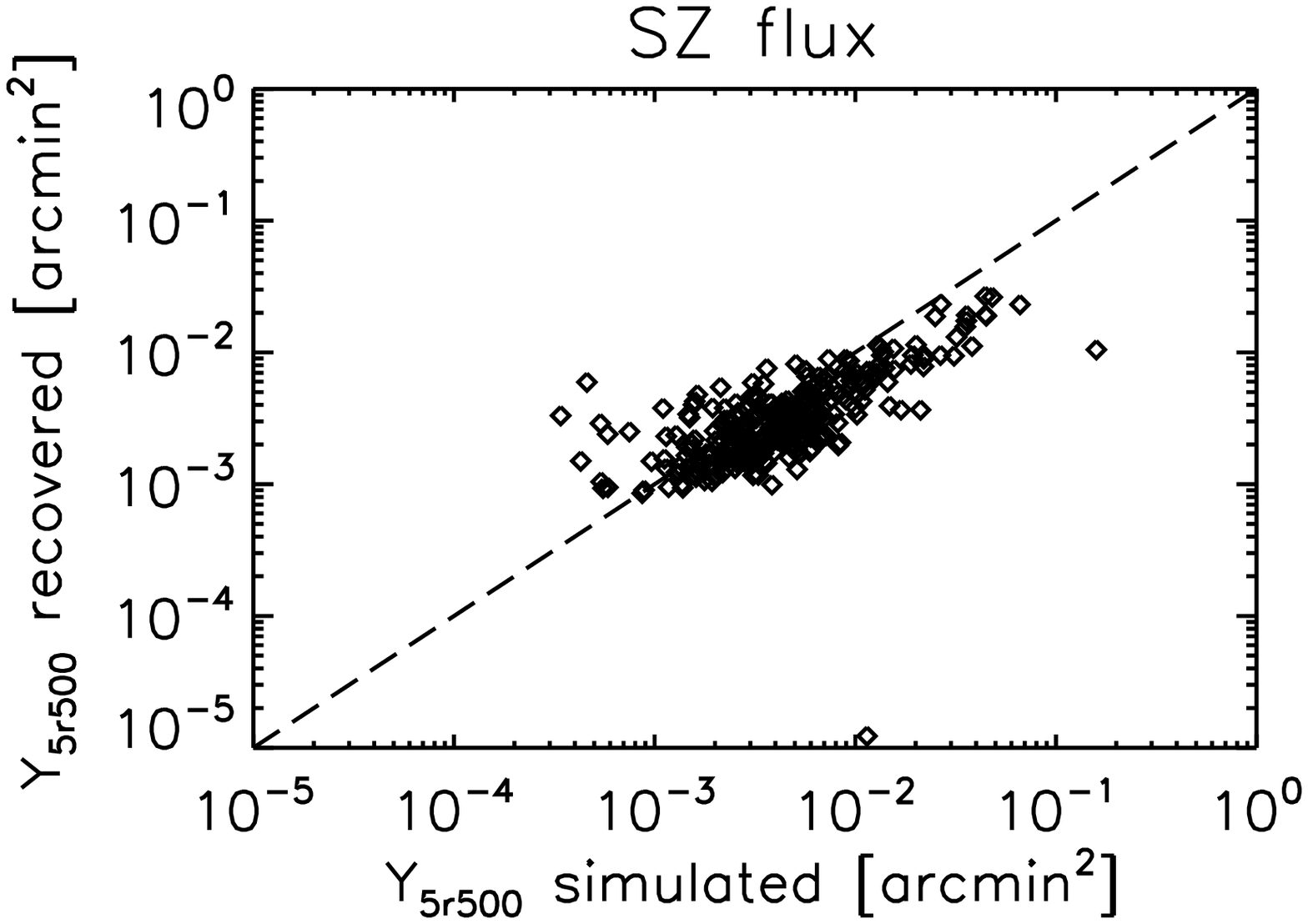}  &
\includegraphics[scale=0.45]{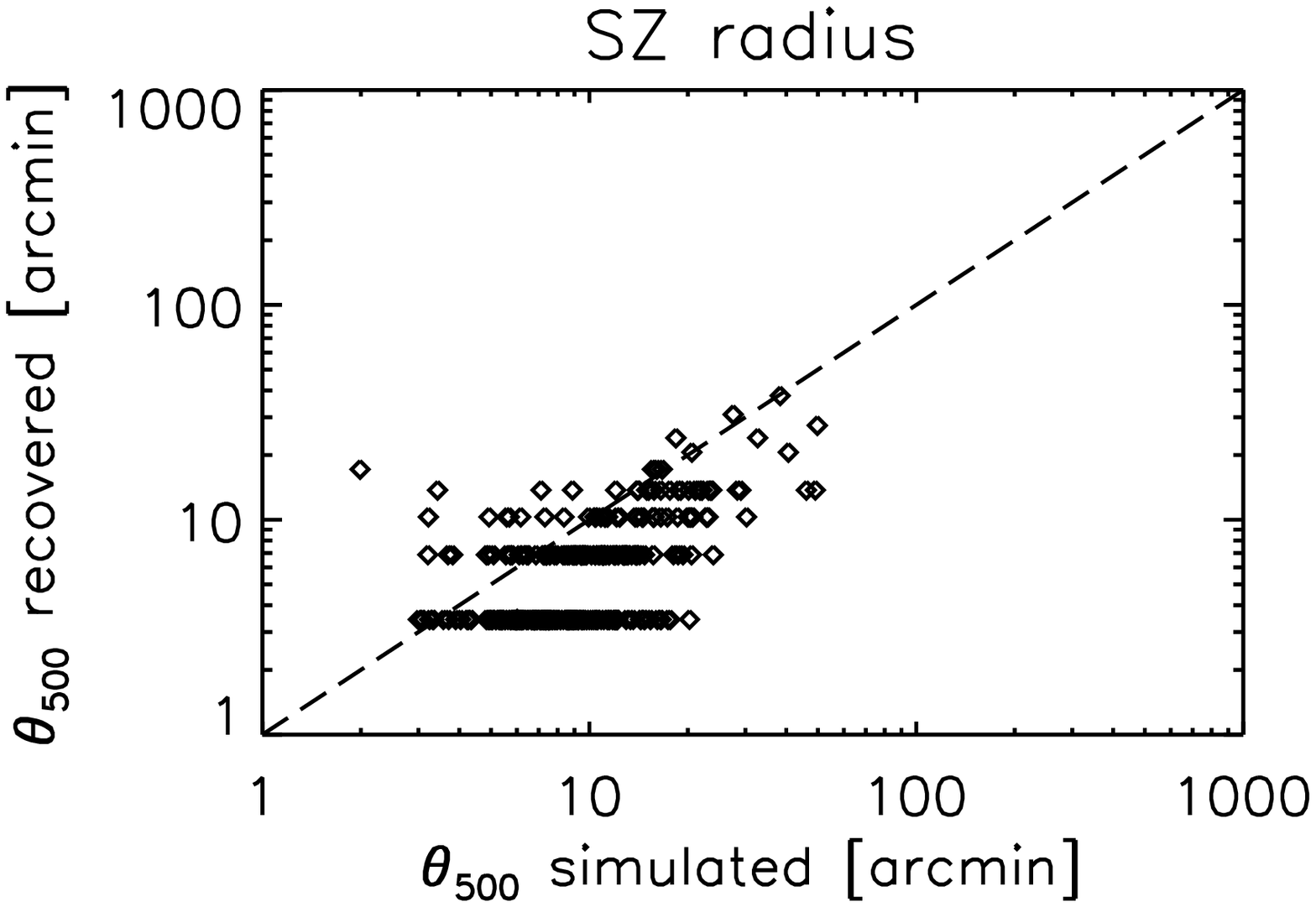} \\
\includegraphics[scale=0.45]{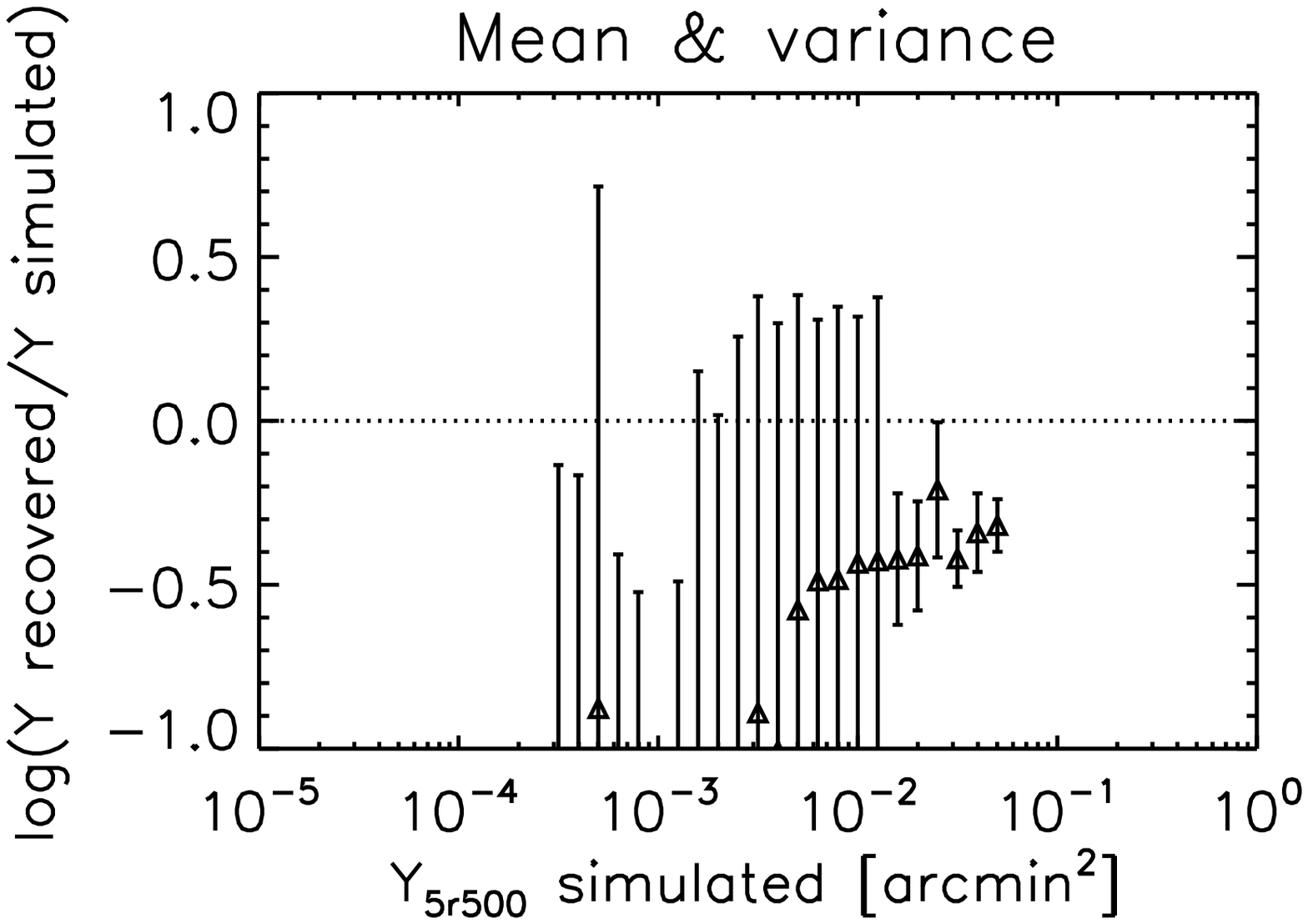} &
\includegraphics[scale=0.45]{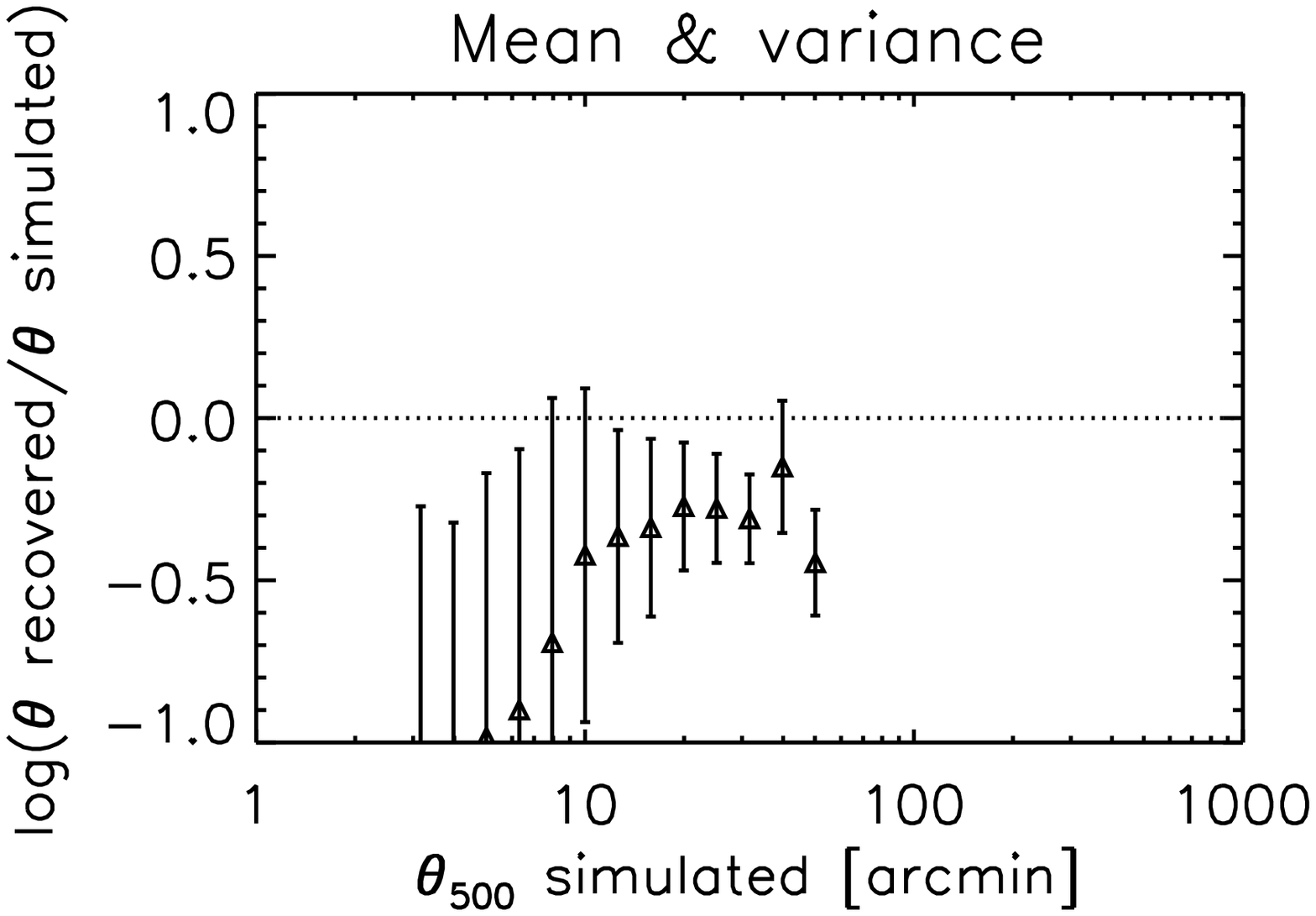} \\
\end{tabular}
\caption{{\bf MMF2}}
\end{center}
\end{table}

\clearpage

\begin{table}[htbp]
\begin{center}
\begin{tabular}{cc}
\includegraphics[scale=0.45]{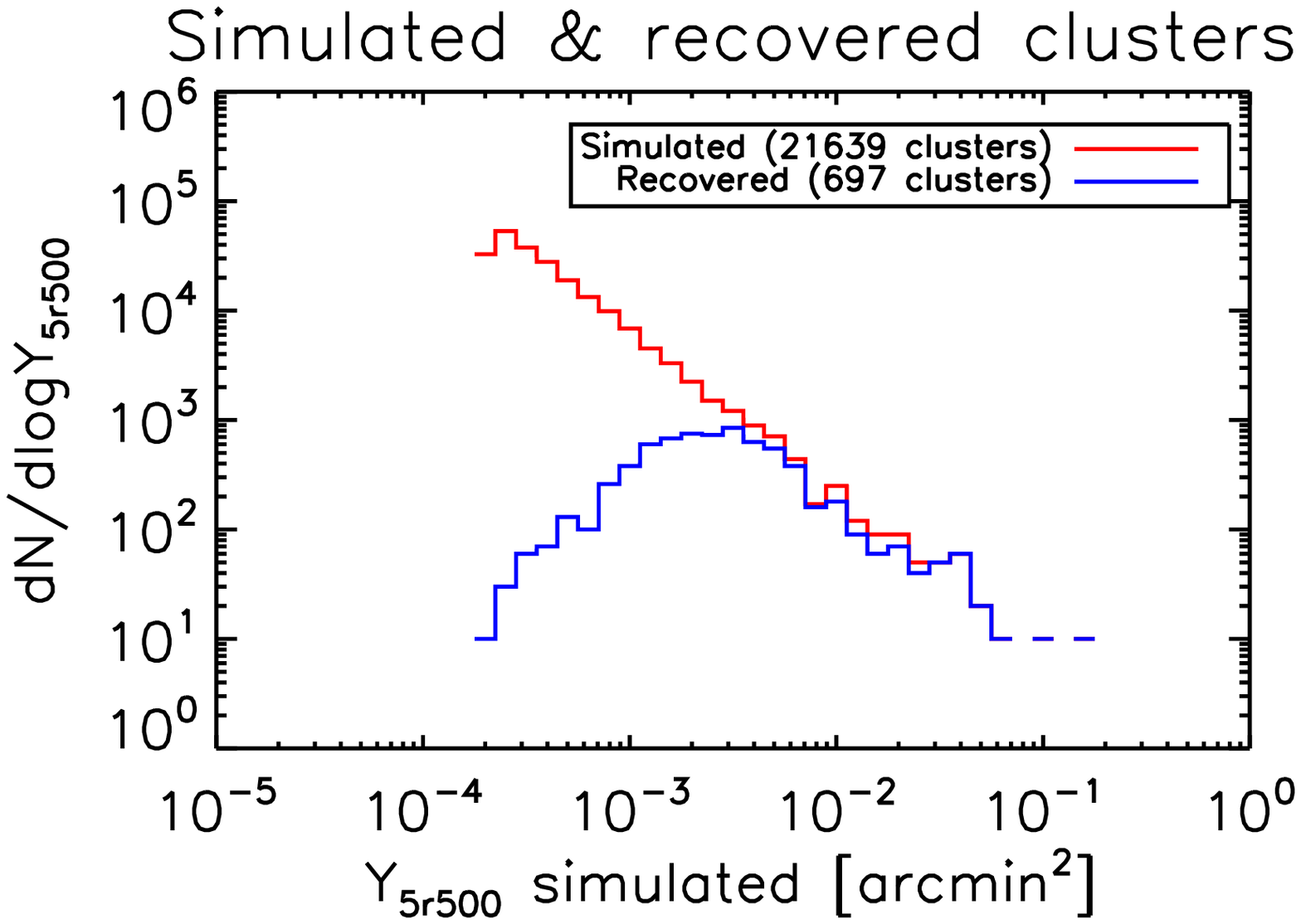}  &
\includegraphics[scale=0.45]{Figs/CompCSZv2/melin_run2_v2_positional_accuracy.eps} \\
\includegraphics[scale=0.45]{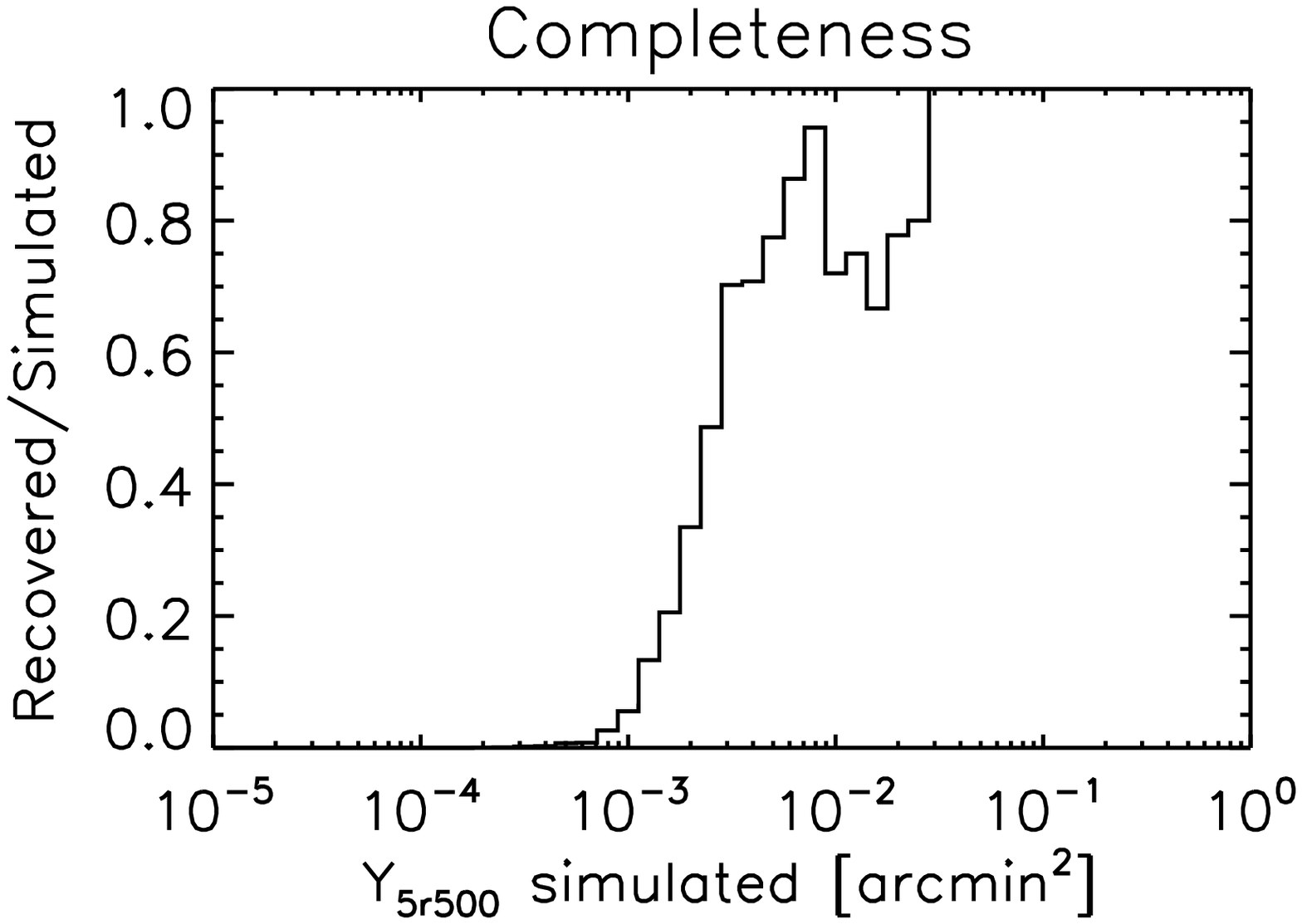}  &
\includegraphics[scale=0.45]{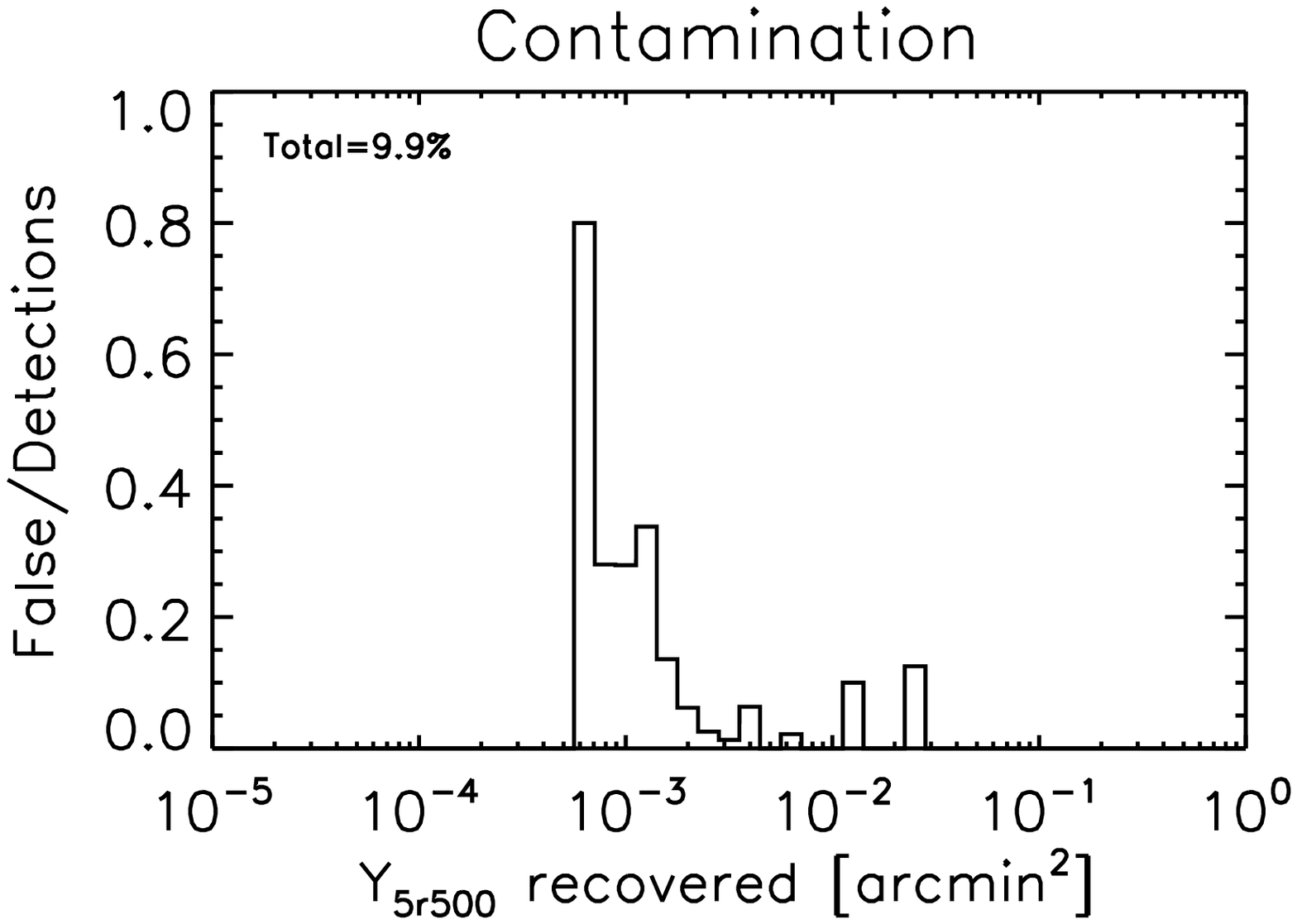} \\
\includegraphics[scale=0.45]{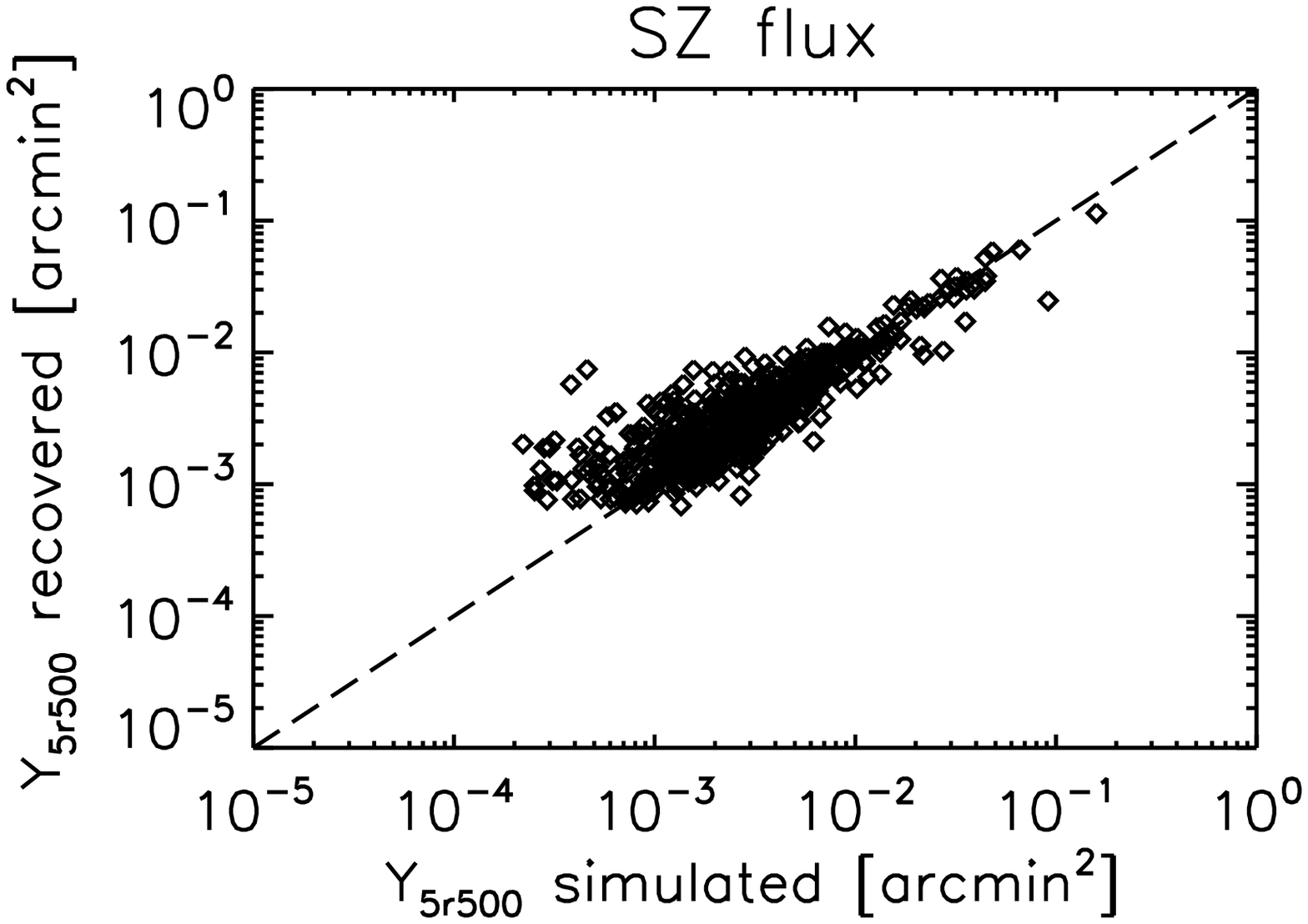}  &
\includegraphics[scale=0.45]{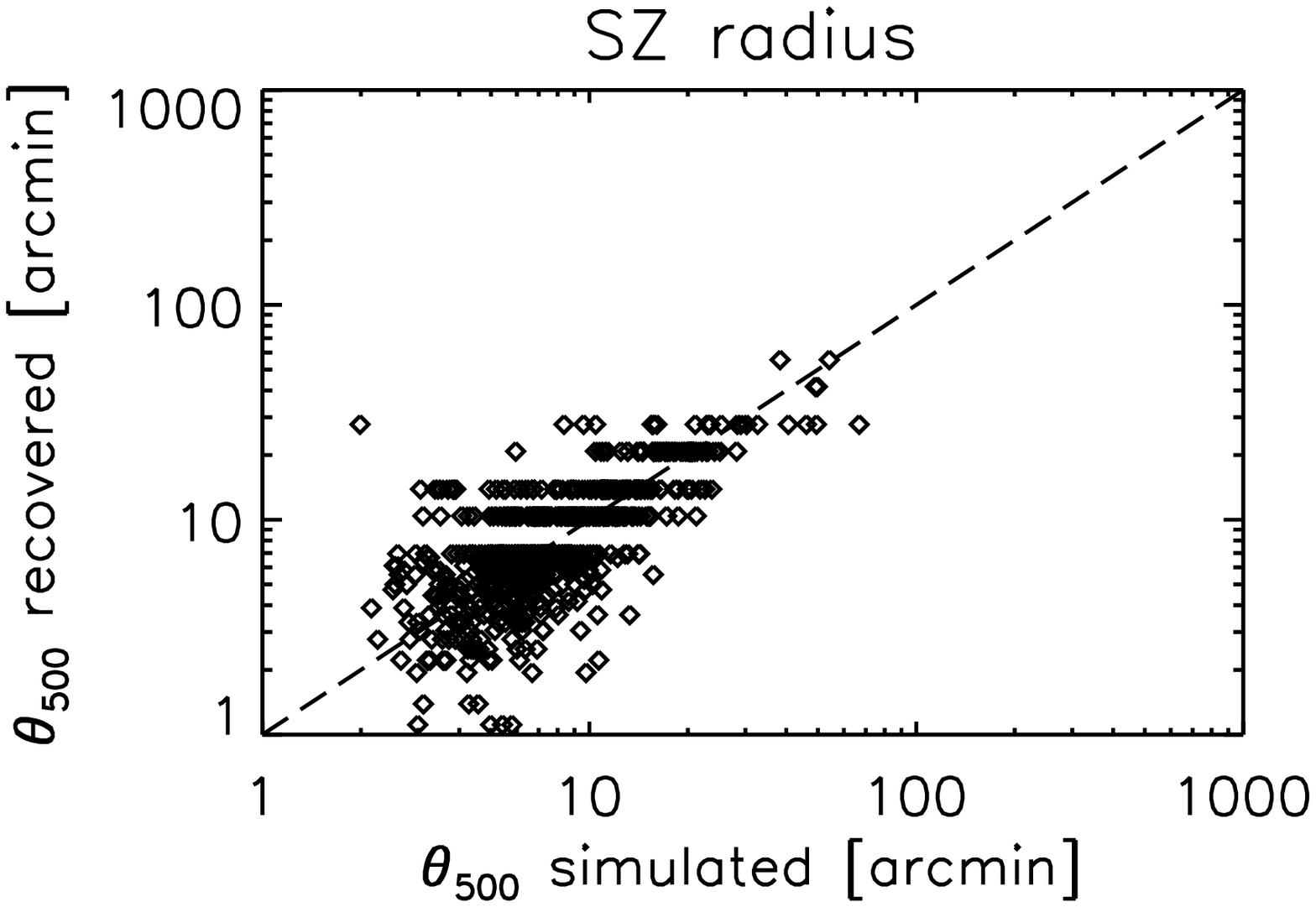} \\
\includegraphics[scale=0.45]{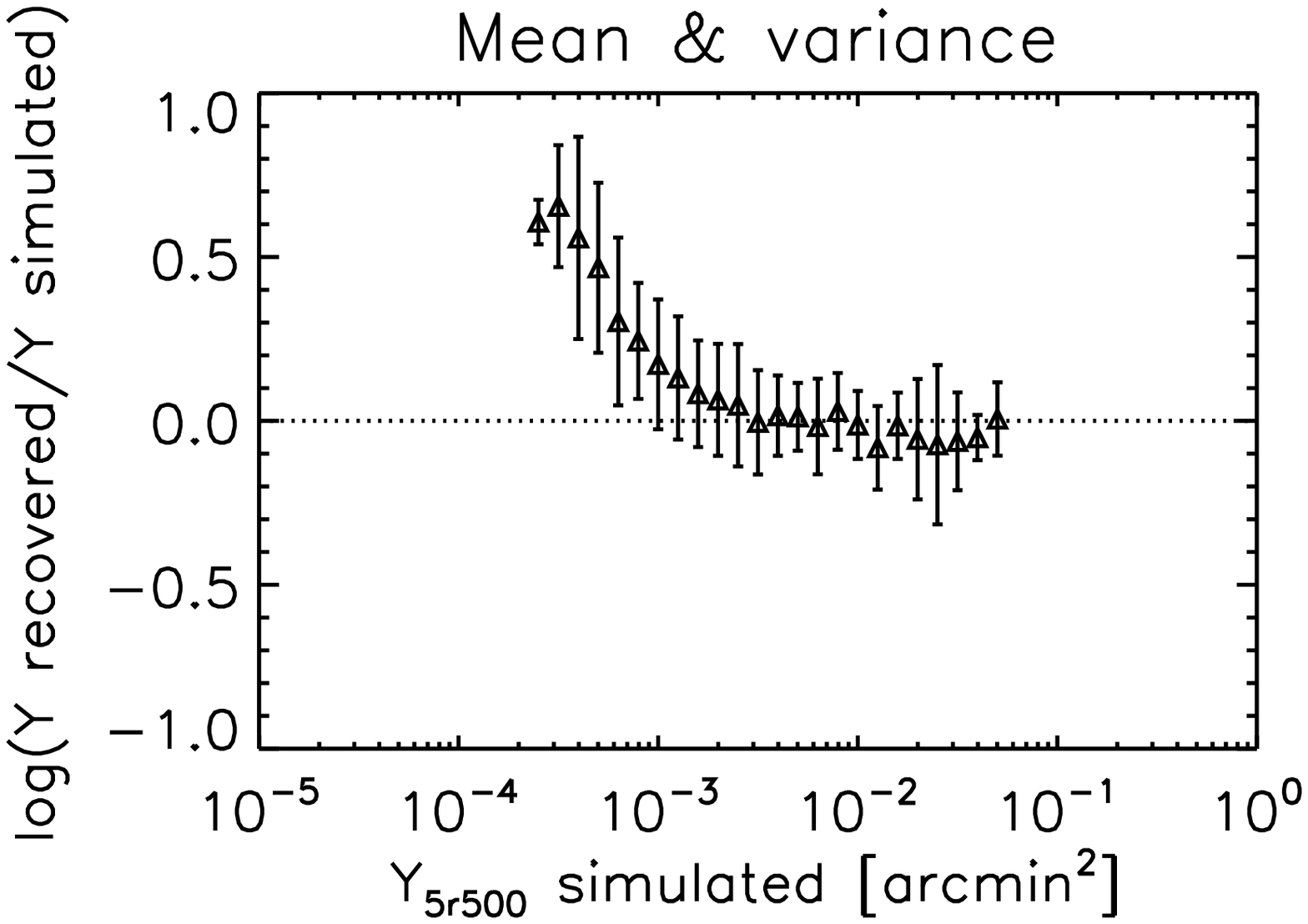}  &
\includegraphics[scale=0.45]{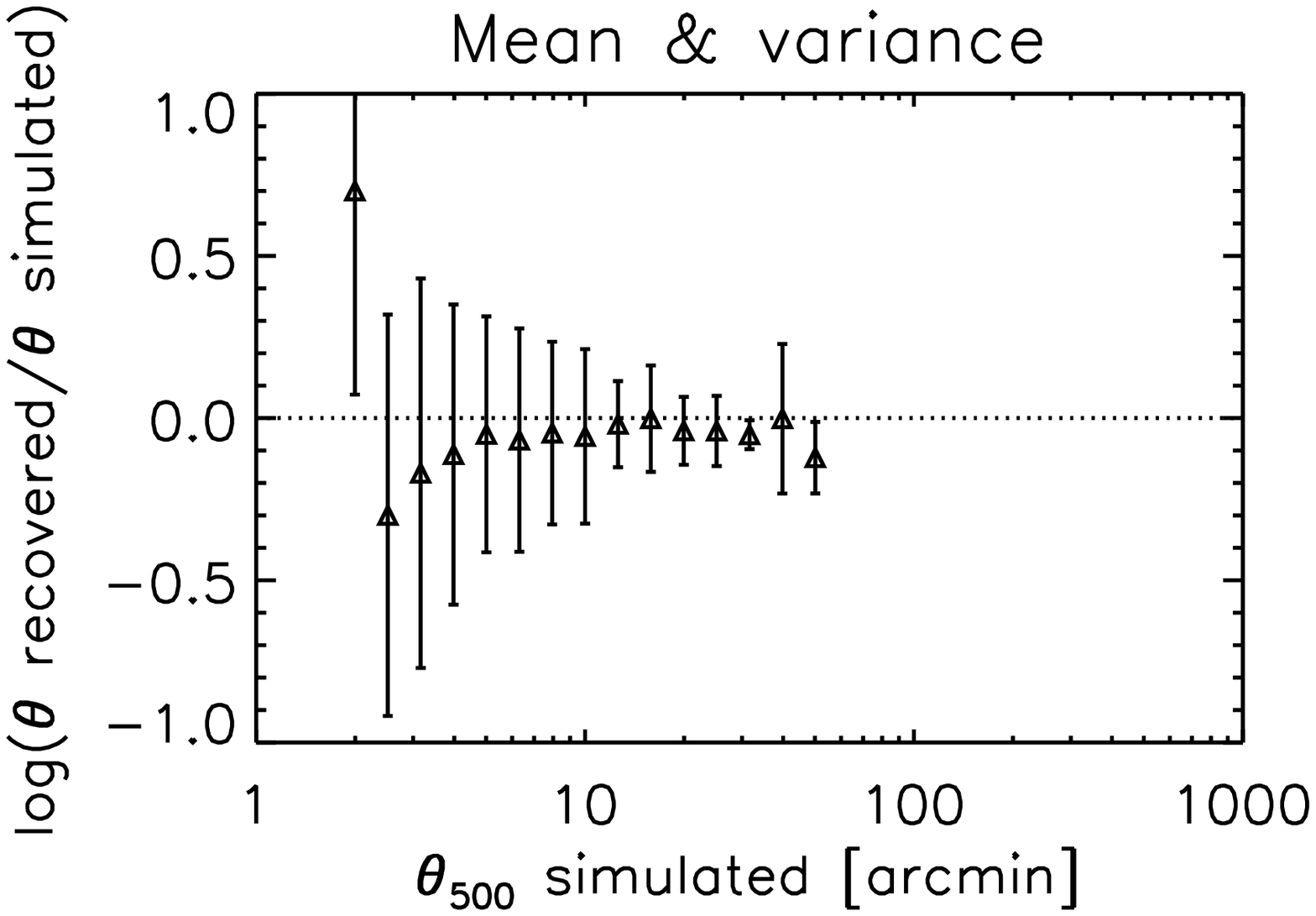} \\
\end{tabular}
\caption{{\bf MMF3}}
\end{center}
\end{table}

\clearpage

\begin{table}[htbp]
\begin{center}
\begin{tabular}{cc}
\includegraphics[scale=0.45]{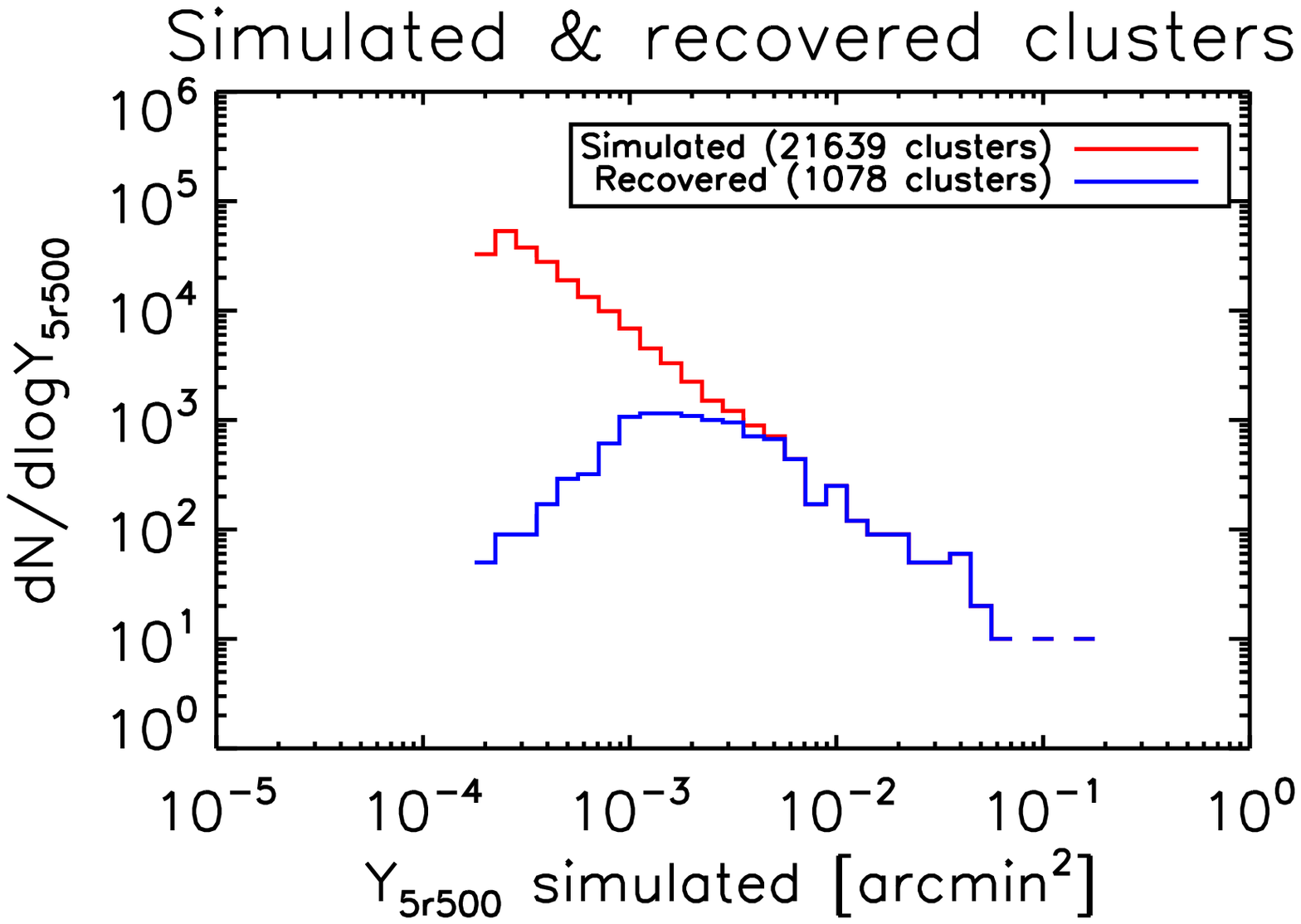}  &
\includegraphics[scale=0.45]{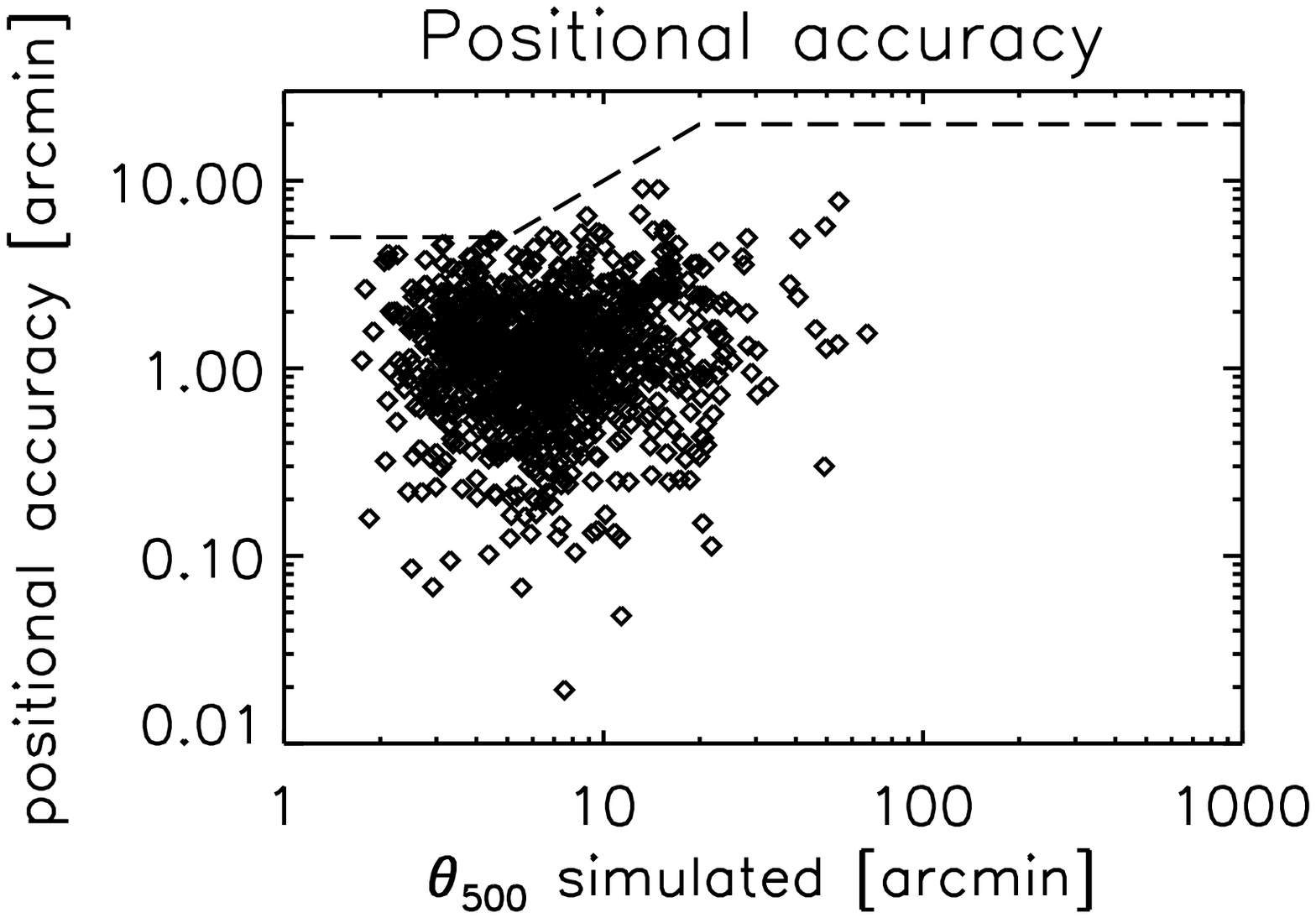} \\
\includegraphics[scale=0.45]{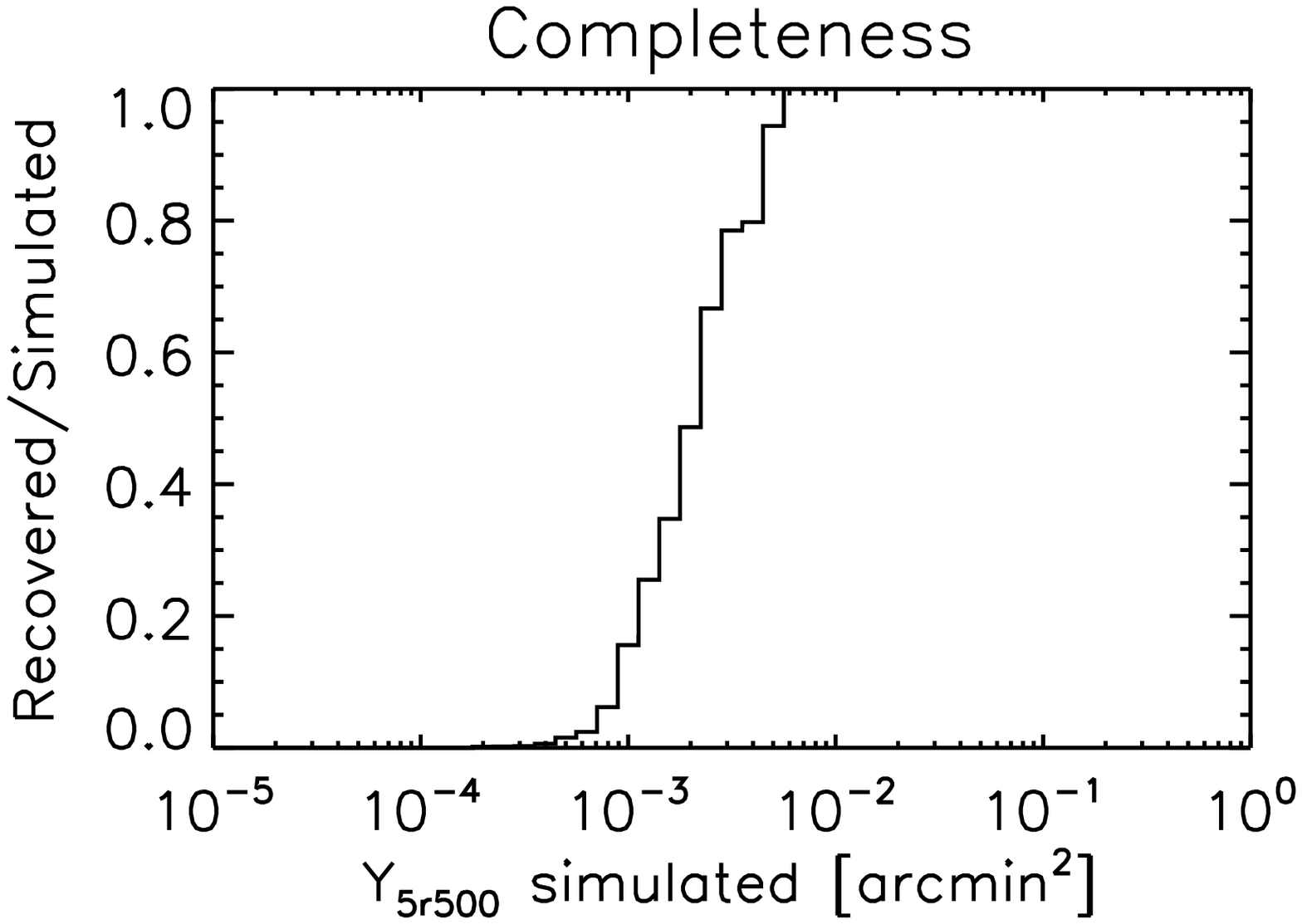}  &
\includegraphics[scale=0.45]{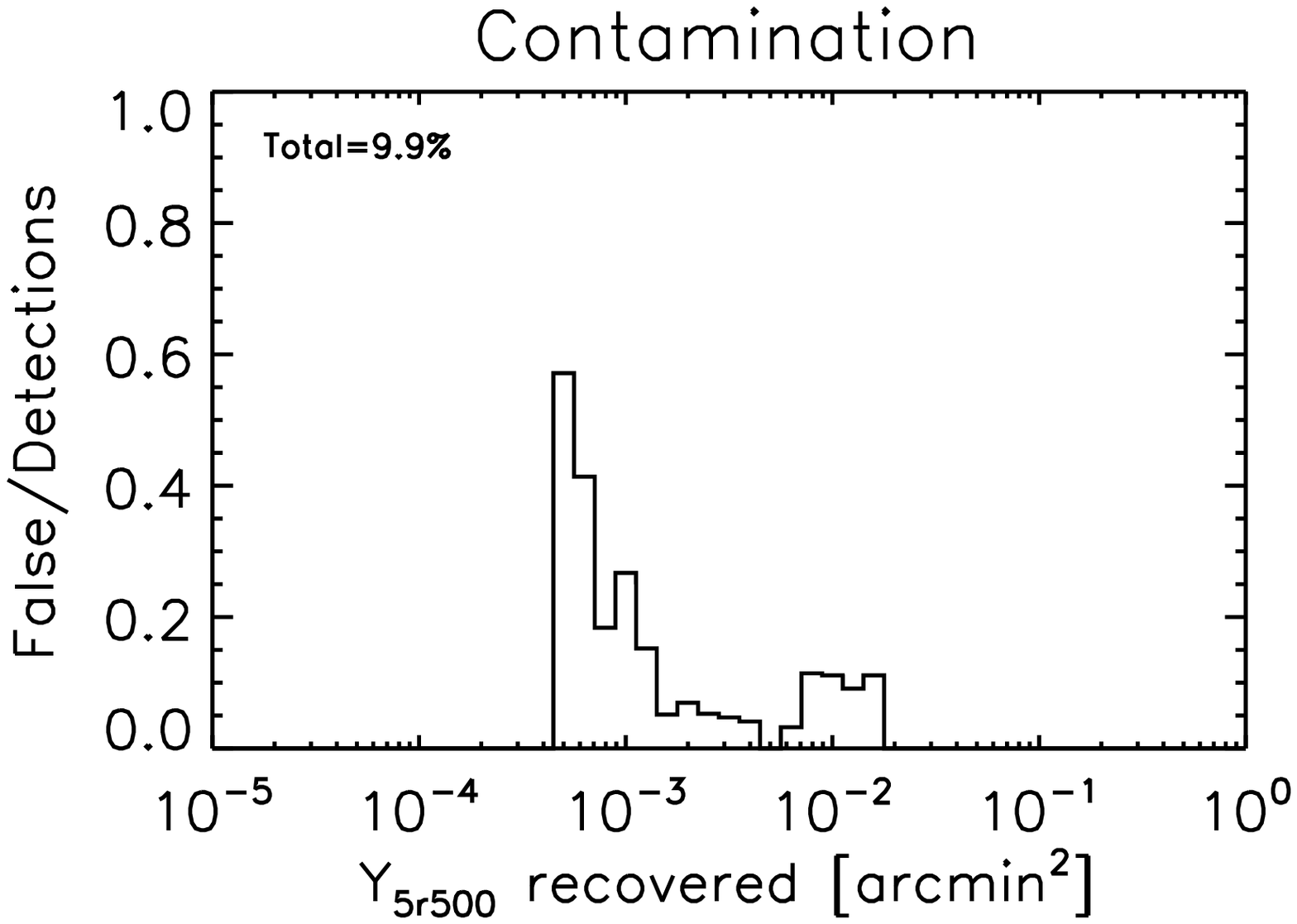} \\
\includegraphics[scale=0.45]{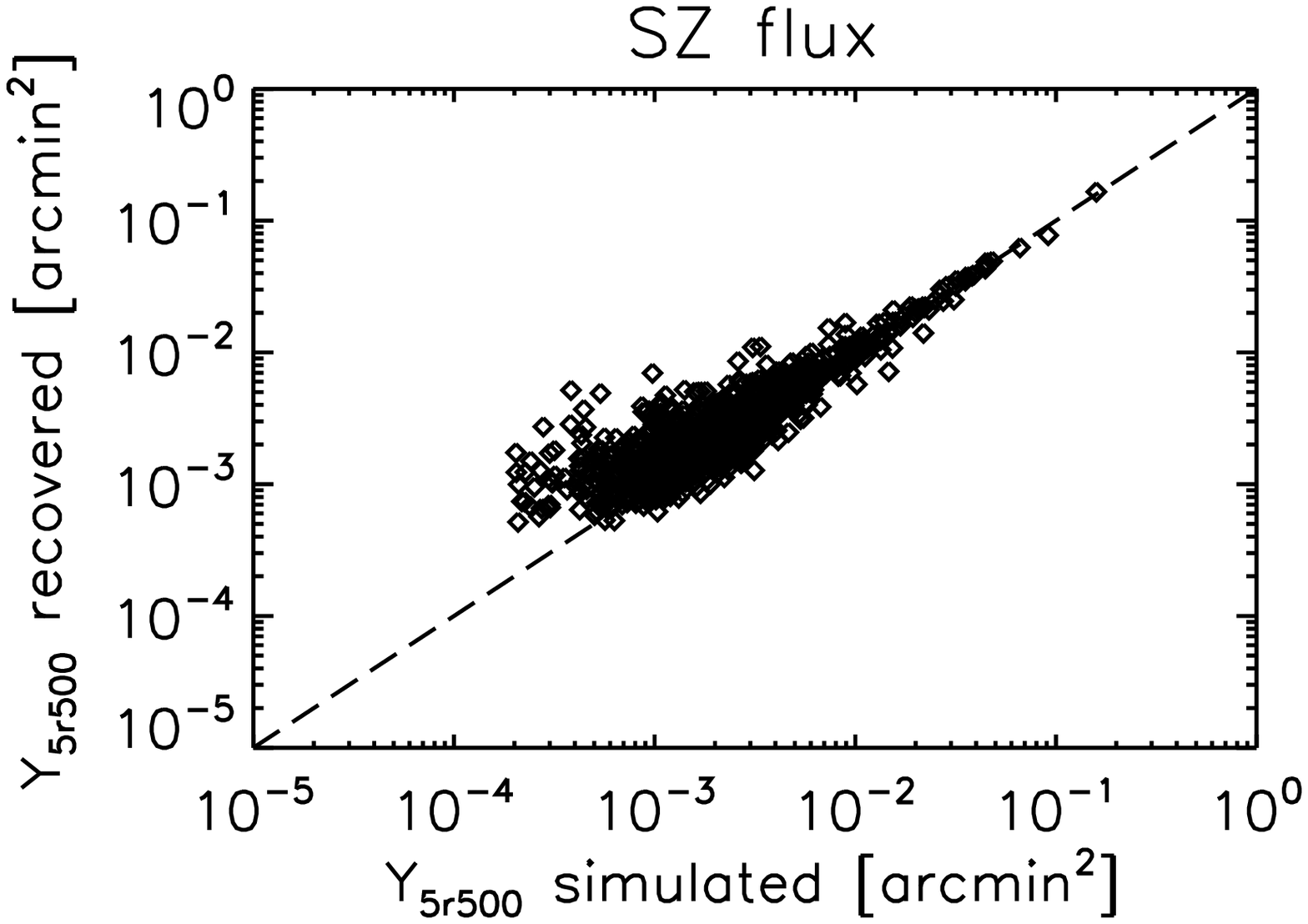}  &
\includegraphics[scale=0.45]{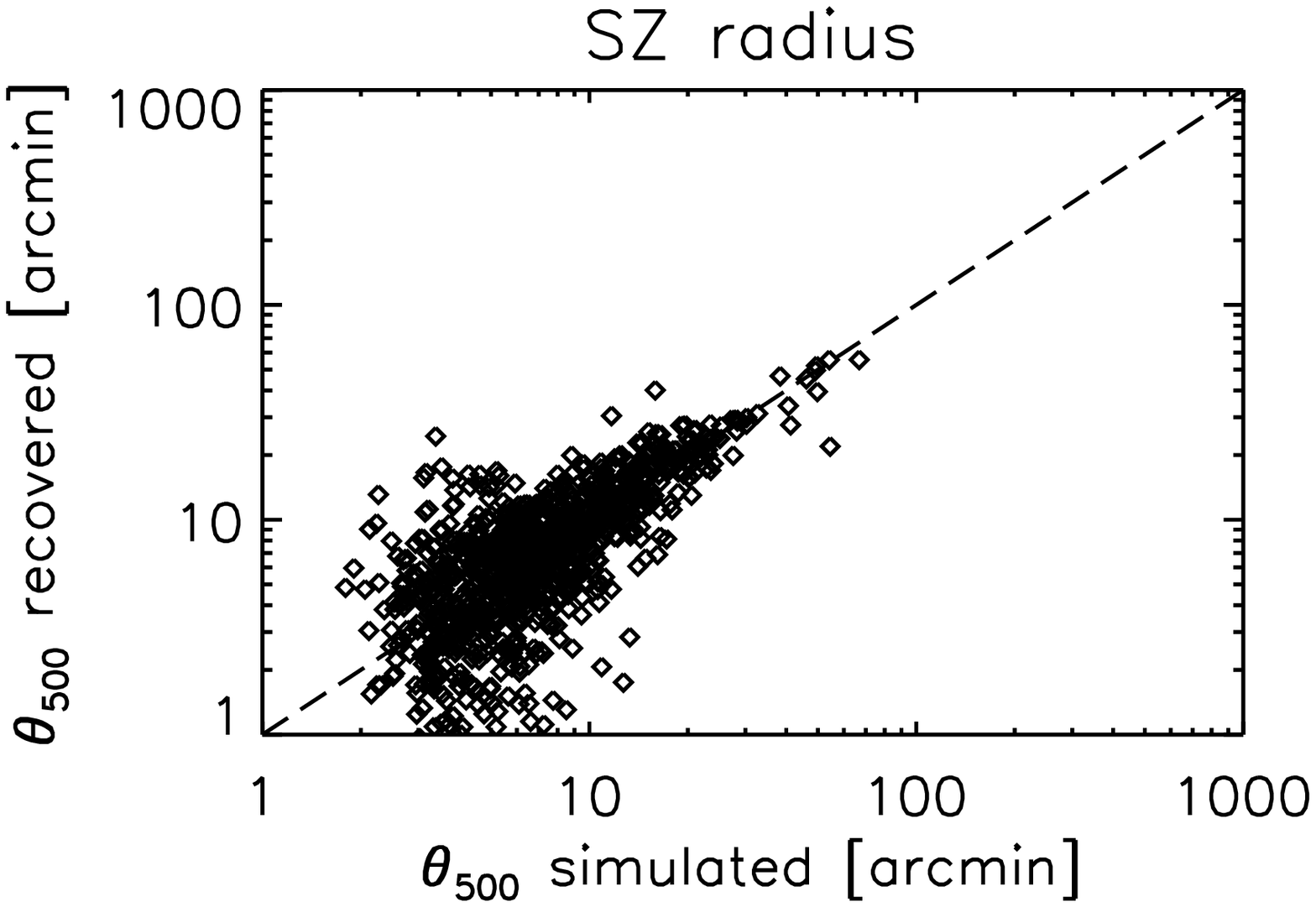} \\
\includegraphics[scale=0.45]{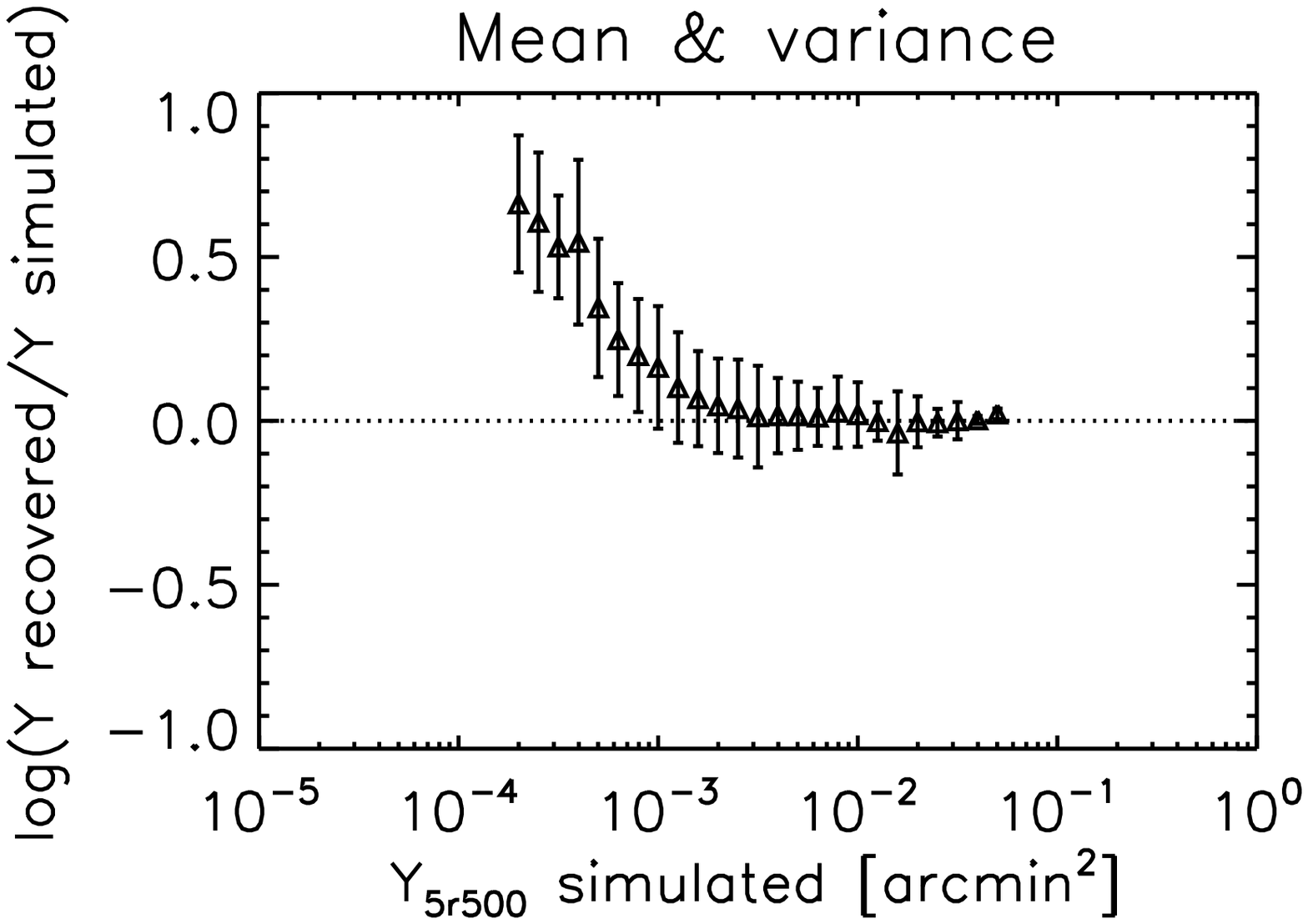}  &
\includegraphics[scale=0.45]{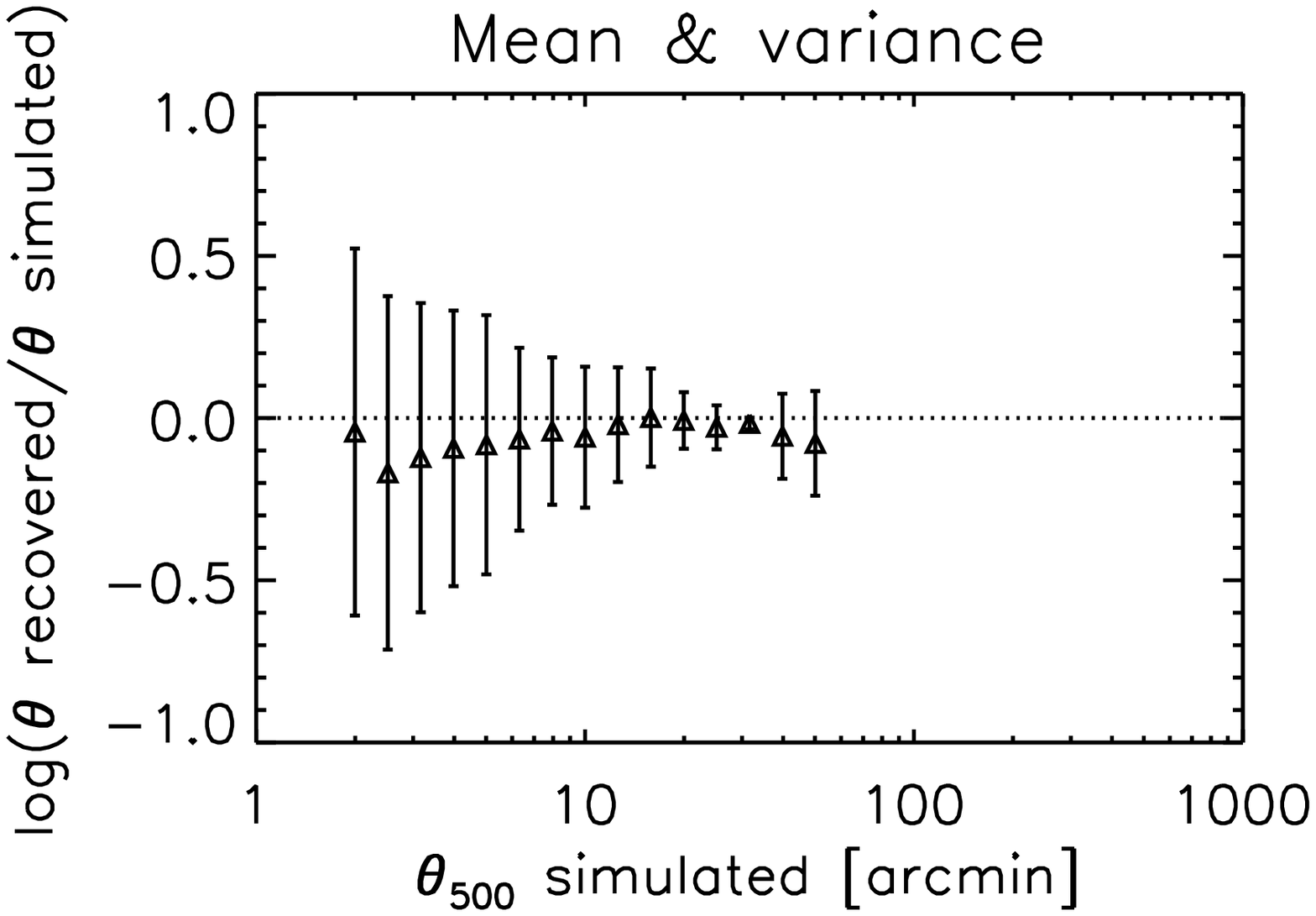} \\
\end{tabular}
\caption{{\bf PS}}
\end{center}
\end{table}

\clearpage

\begin{table}[htbp]
\begin{center}
\begin{tabular}{cc}
\includegraphics[scale=0.45]{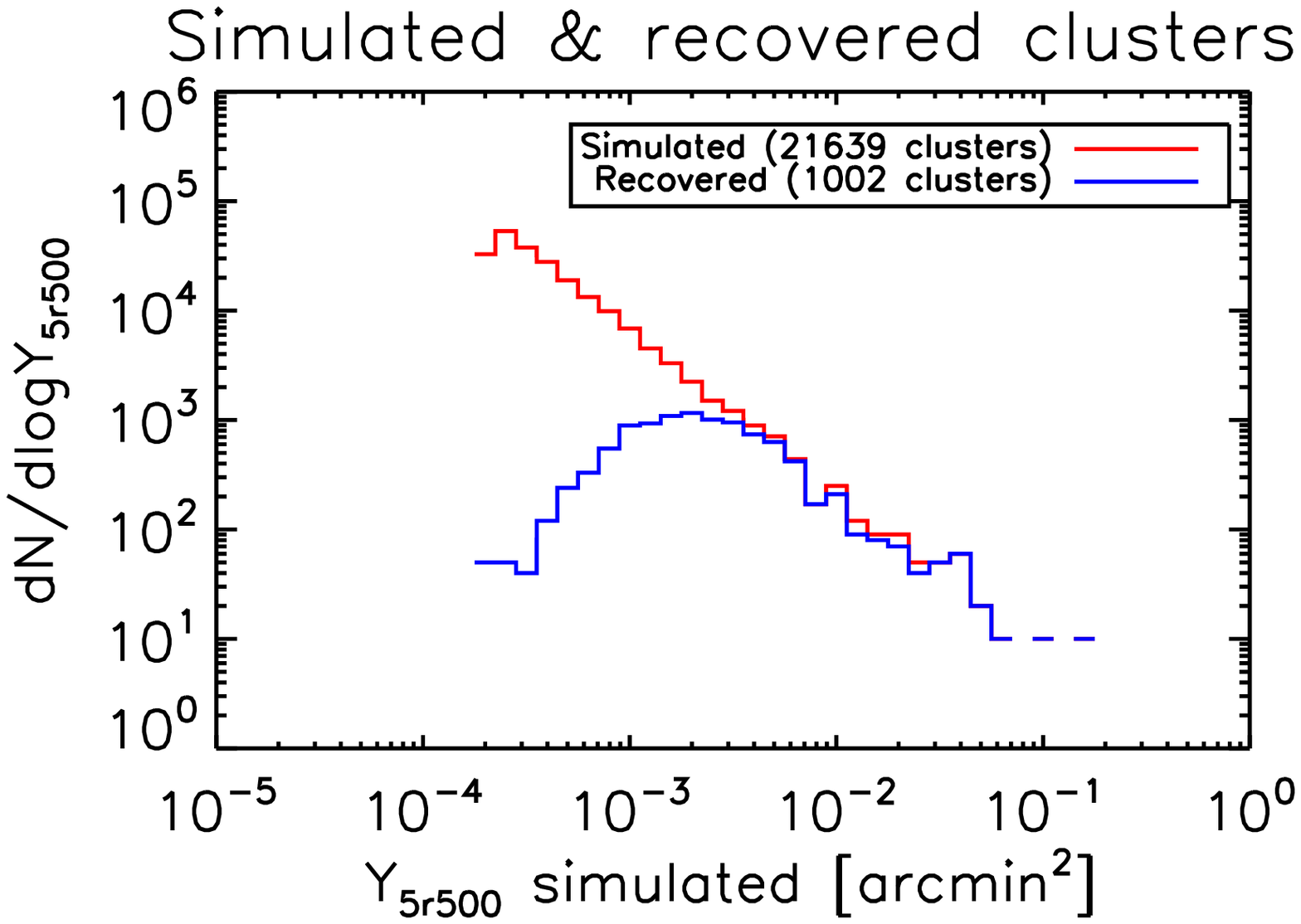}  &
\includegraphics[scale=0.45]{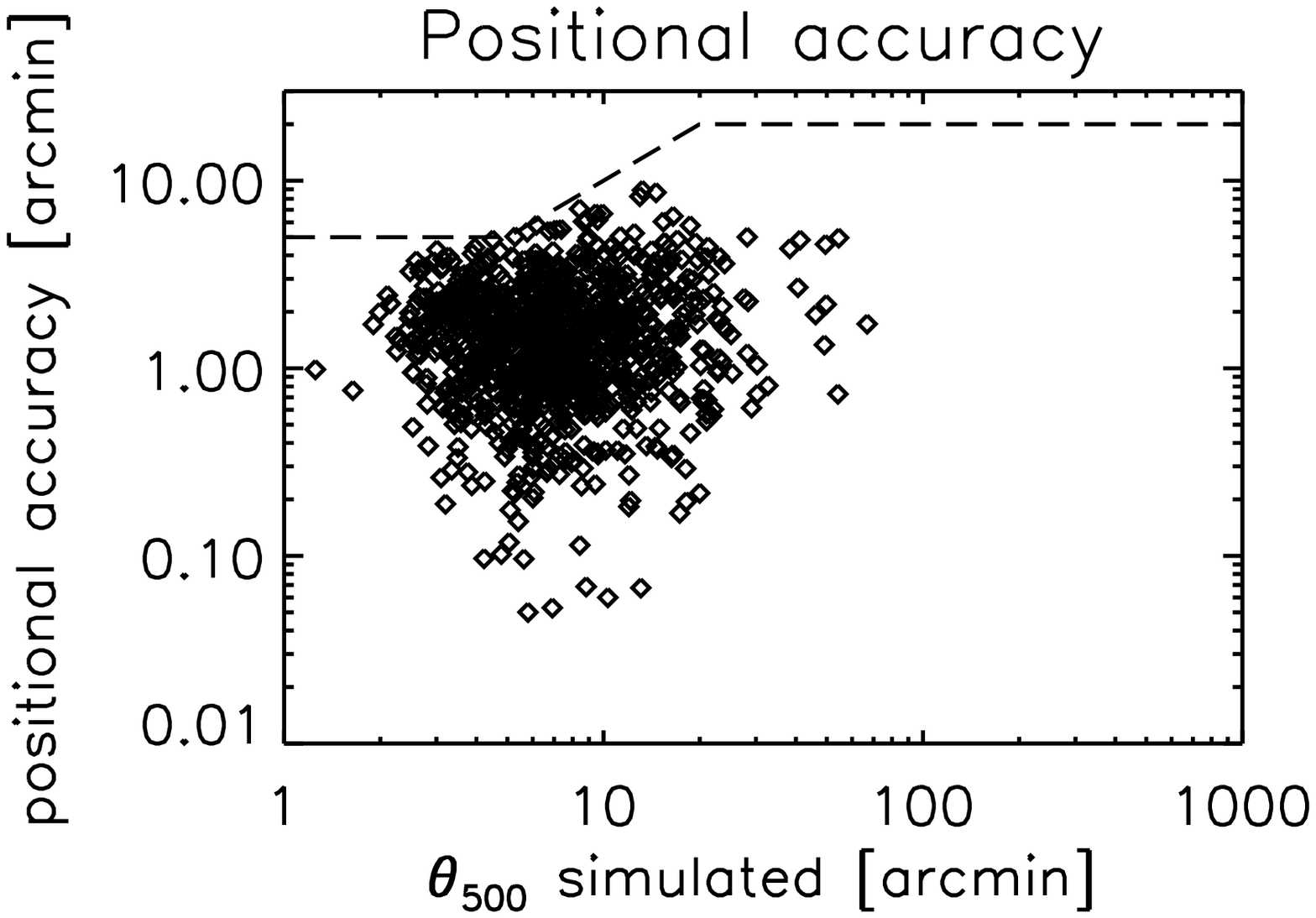} \\
\includegraphics[scale=0.45]{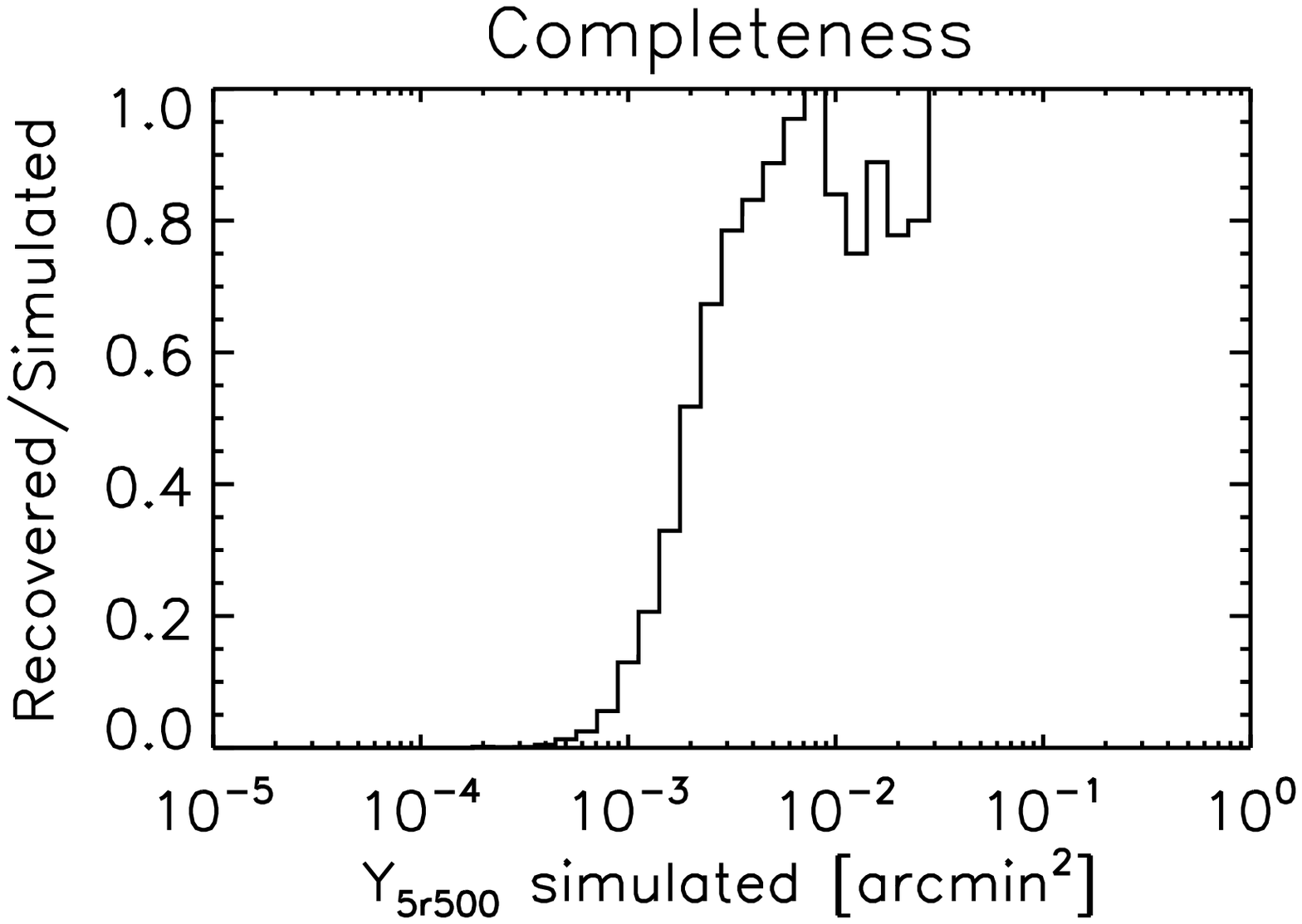}  &
\includegraphics[scale=0.45]{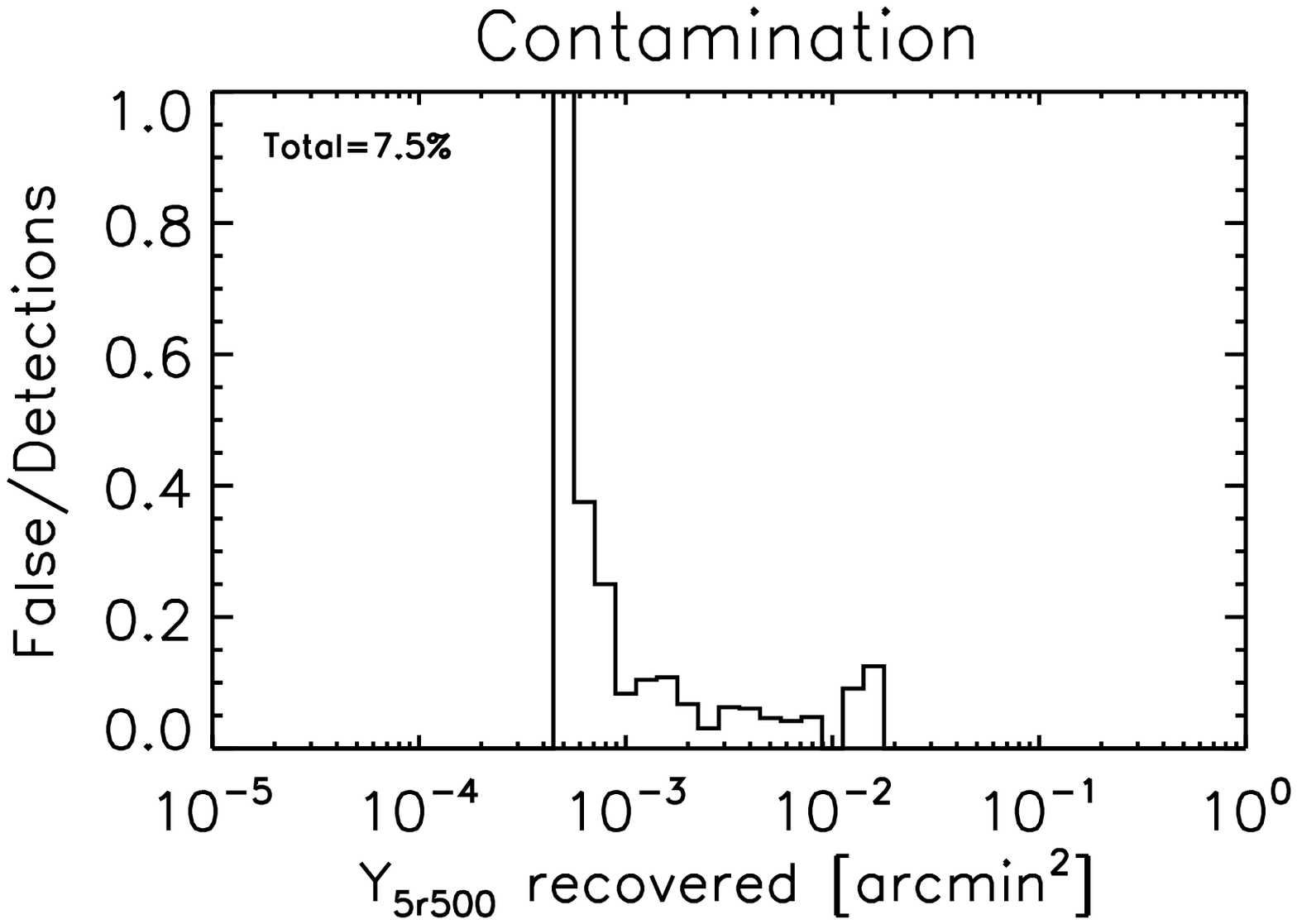} \\
\includegraphics[scale=0.45]{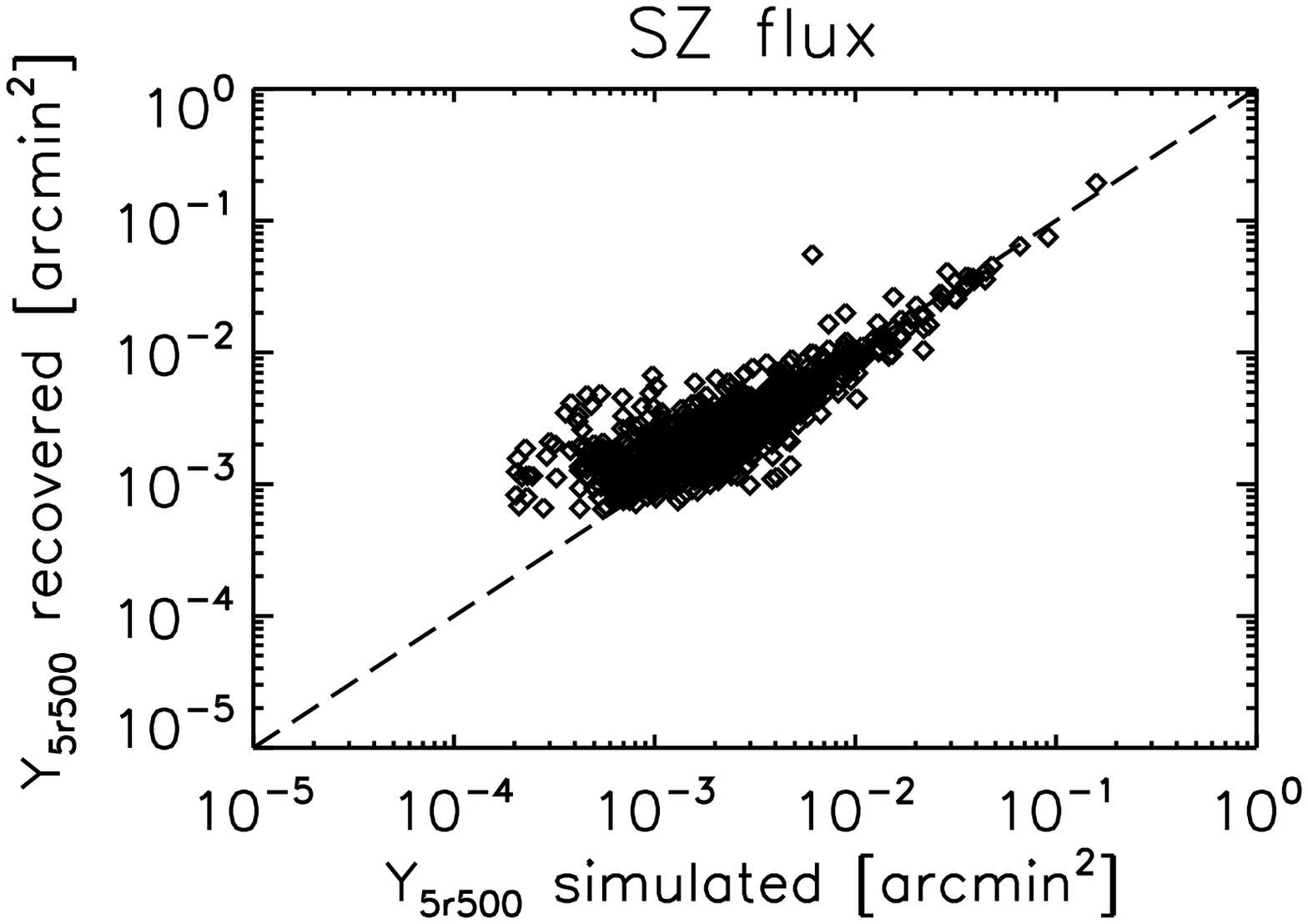}  &
\includegraphics[scale=0.45]{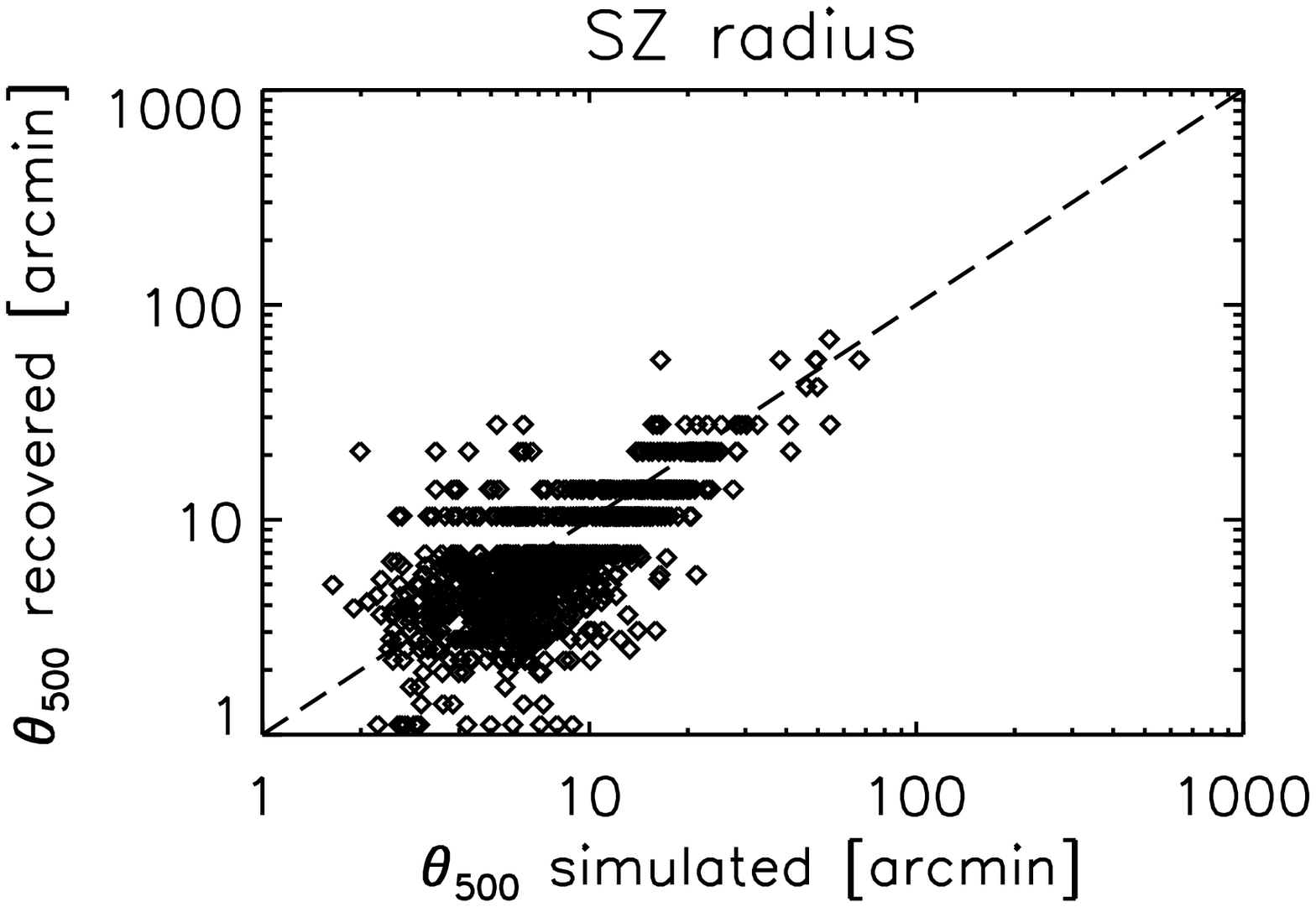} \\
\includegraphics[scale=0.45]{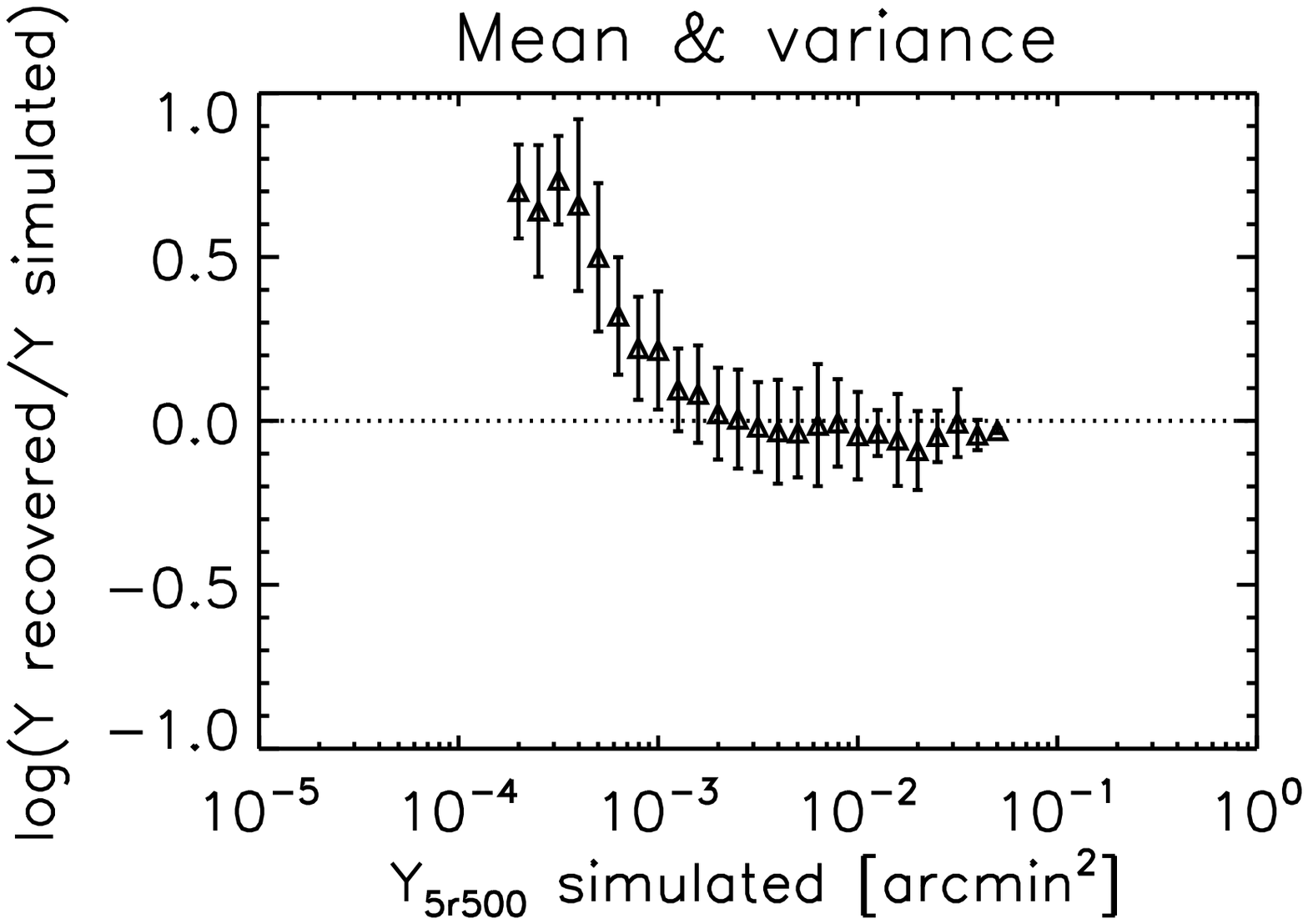}  &
\includegraphics[scale=0.45]{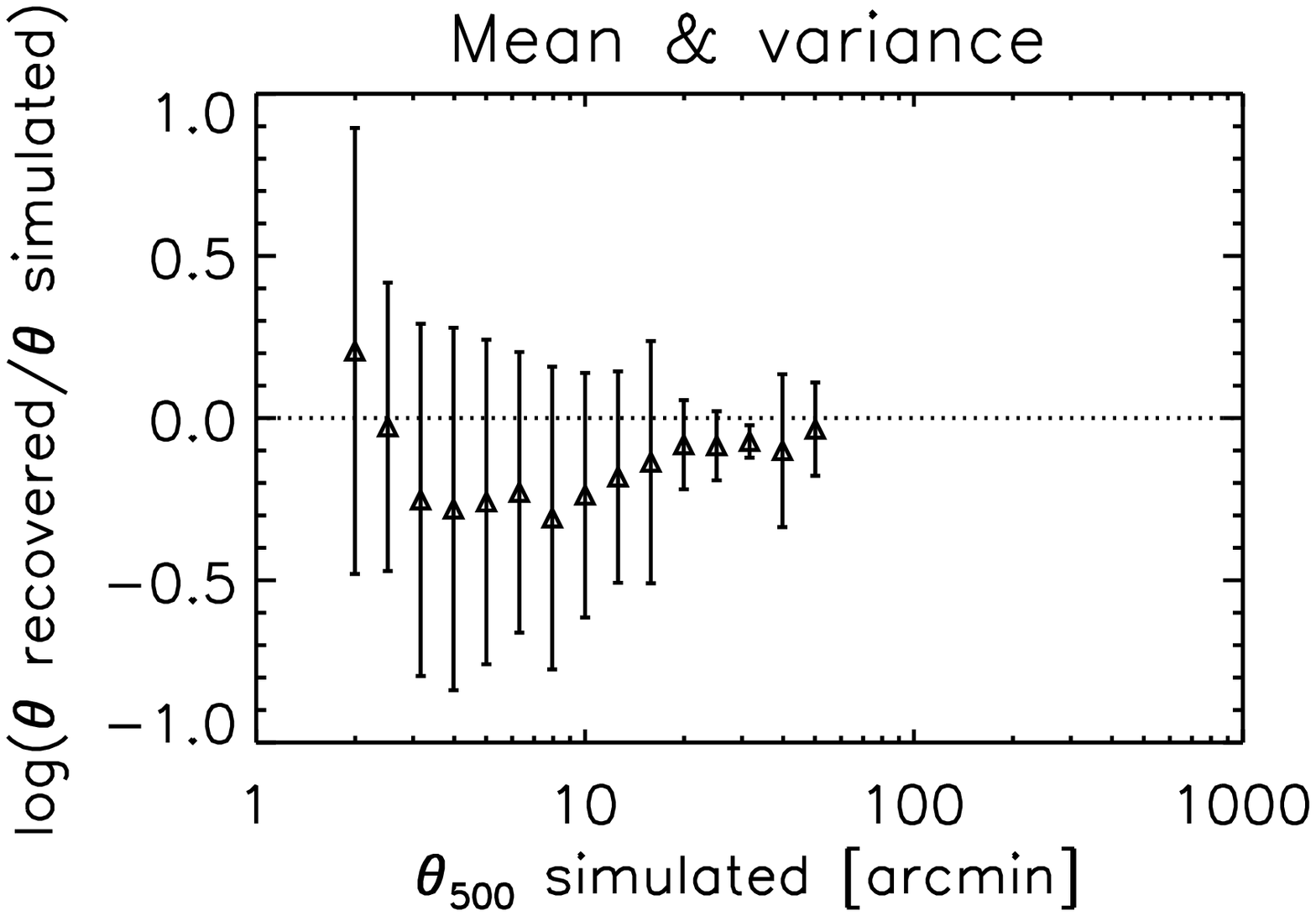} \\
\end{tabular}
\caption{{\bf ILC2}}
\end{center}
\end{table}

\clearpage

\begin{table}[htbp]
\begin{center}
\begin{tabular}{cc}
\includegraphics[scale=0.45]{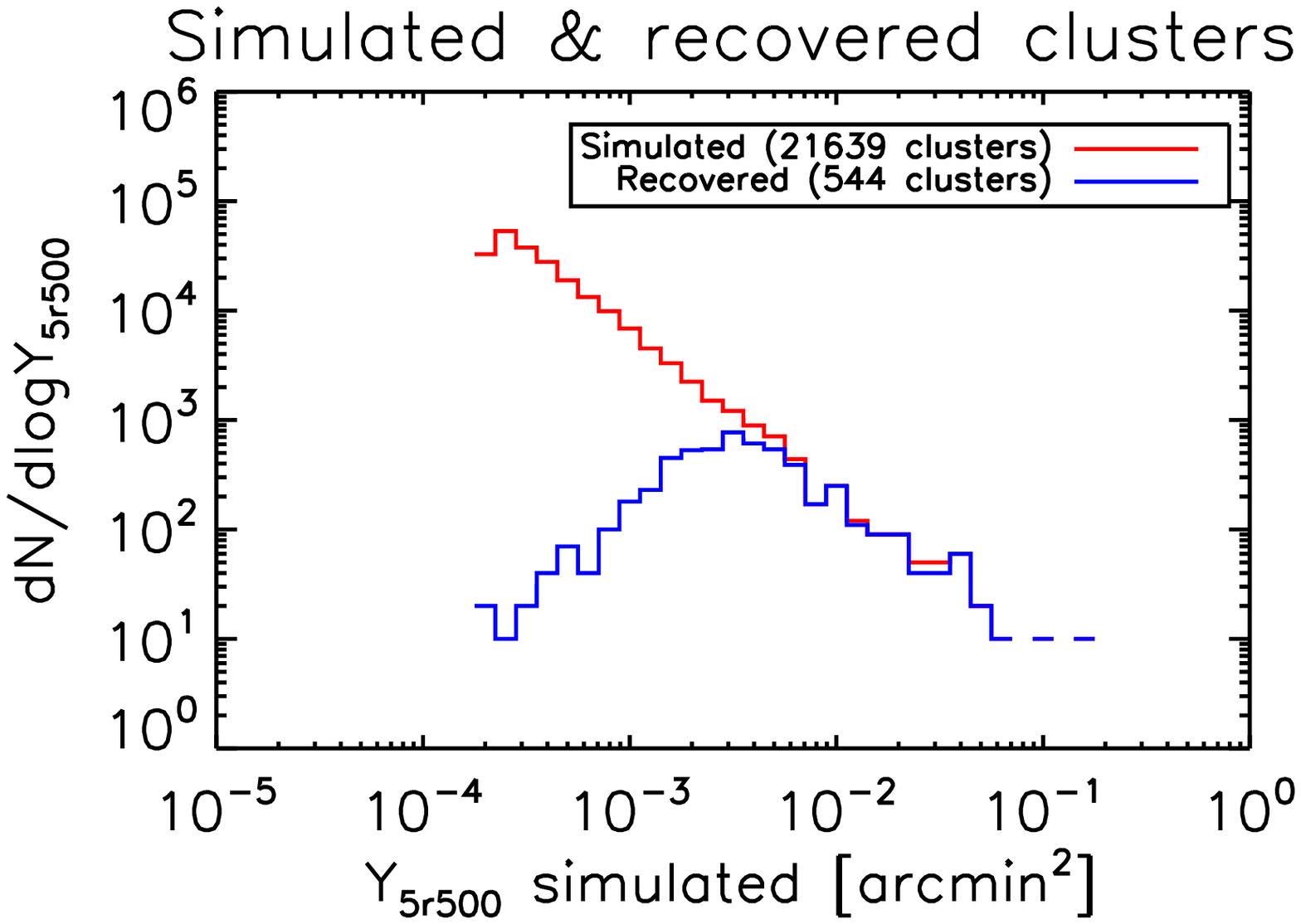}  &
\includegraphics[scale=0.45]{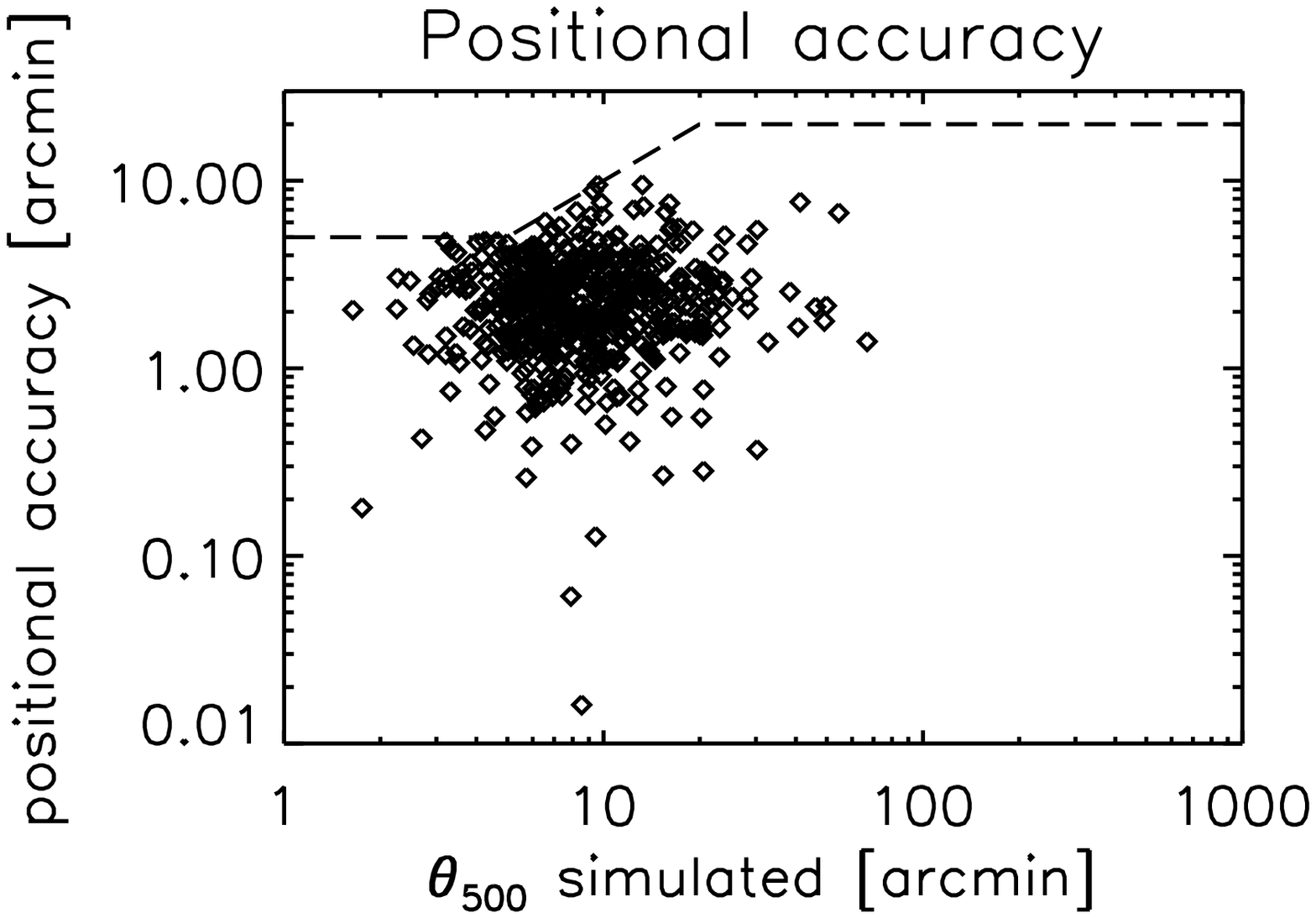} \\
\includegraphics[scale=0.45]{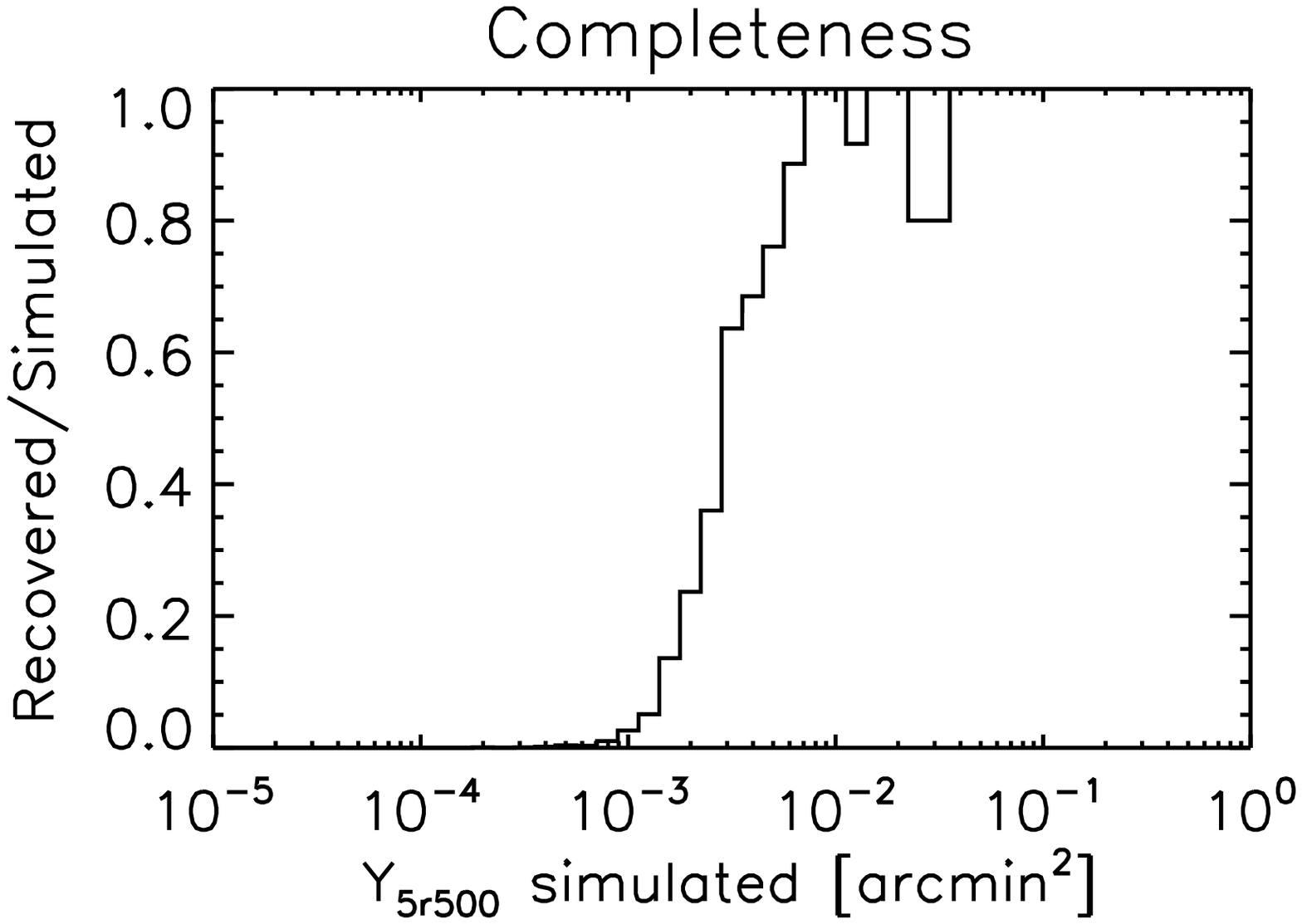}  &
\includegraphics[scale=0.45]{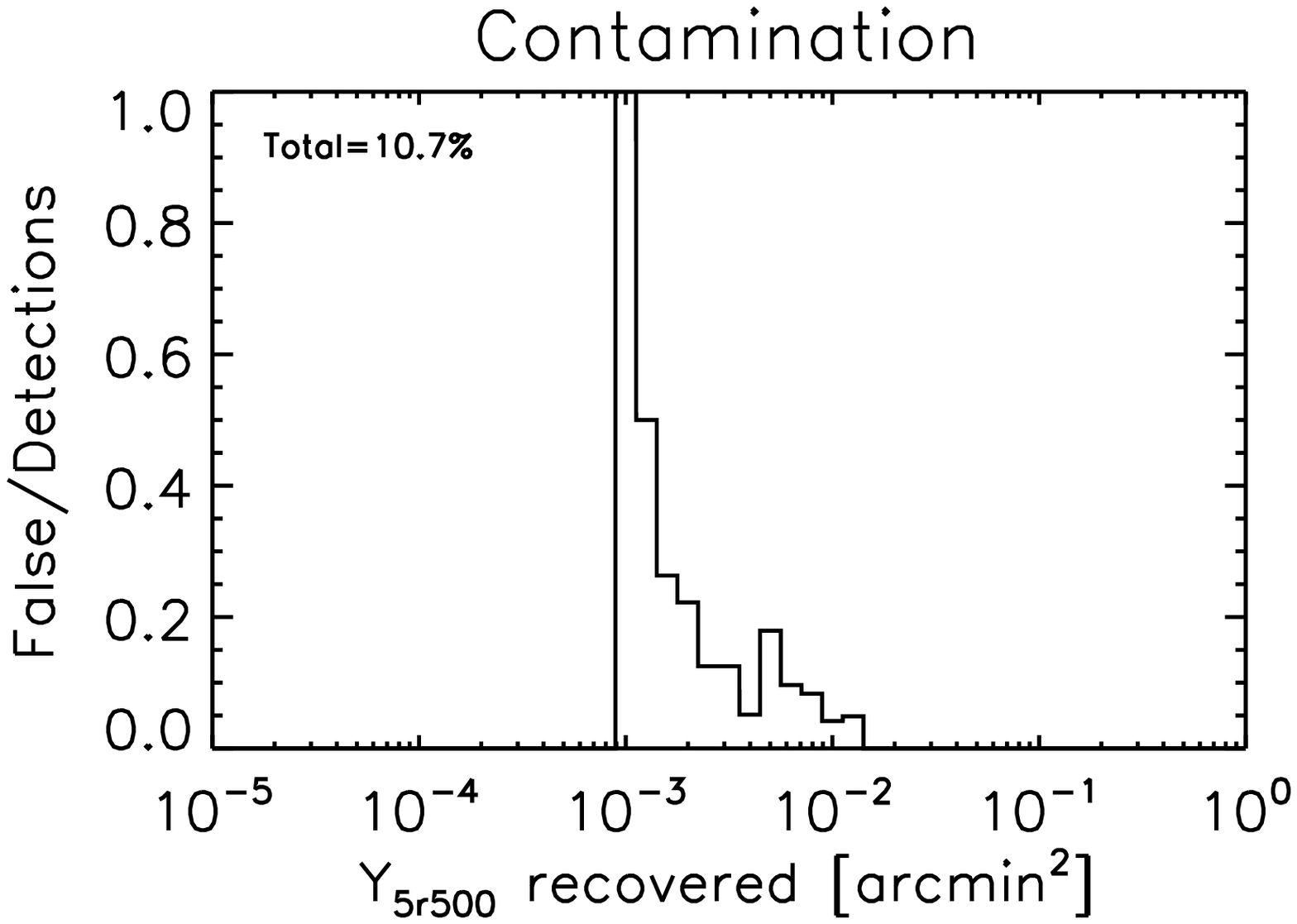} \\
\includegraphics[scale=0.45]{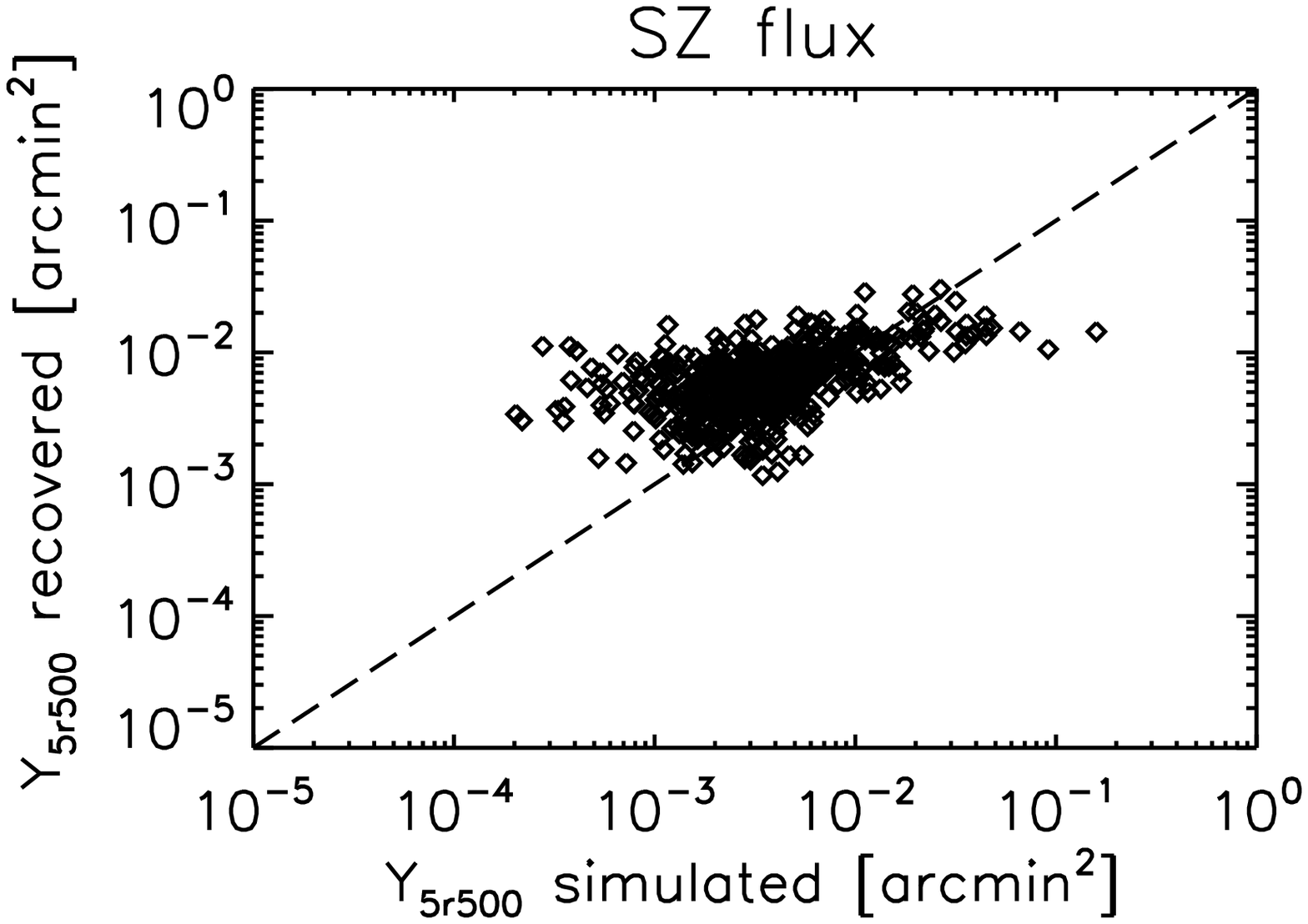}  &
\includegraphics[scale=0.45]{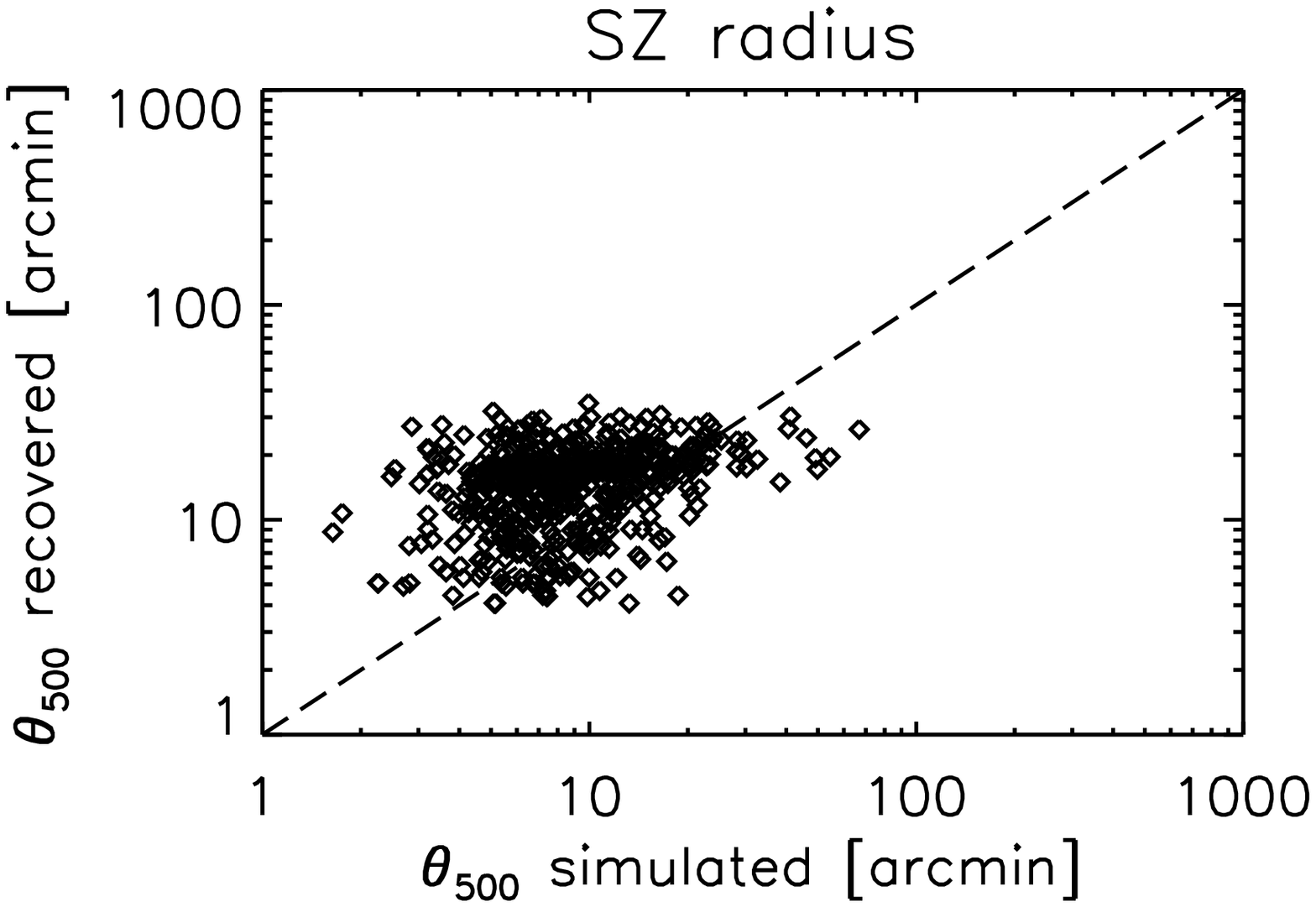} \\
\includegraphics[scale=0.45]{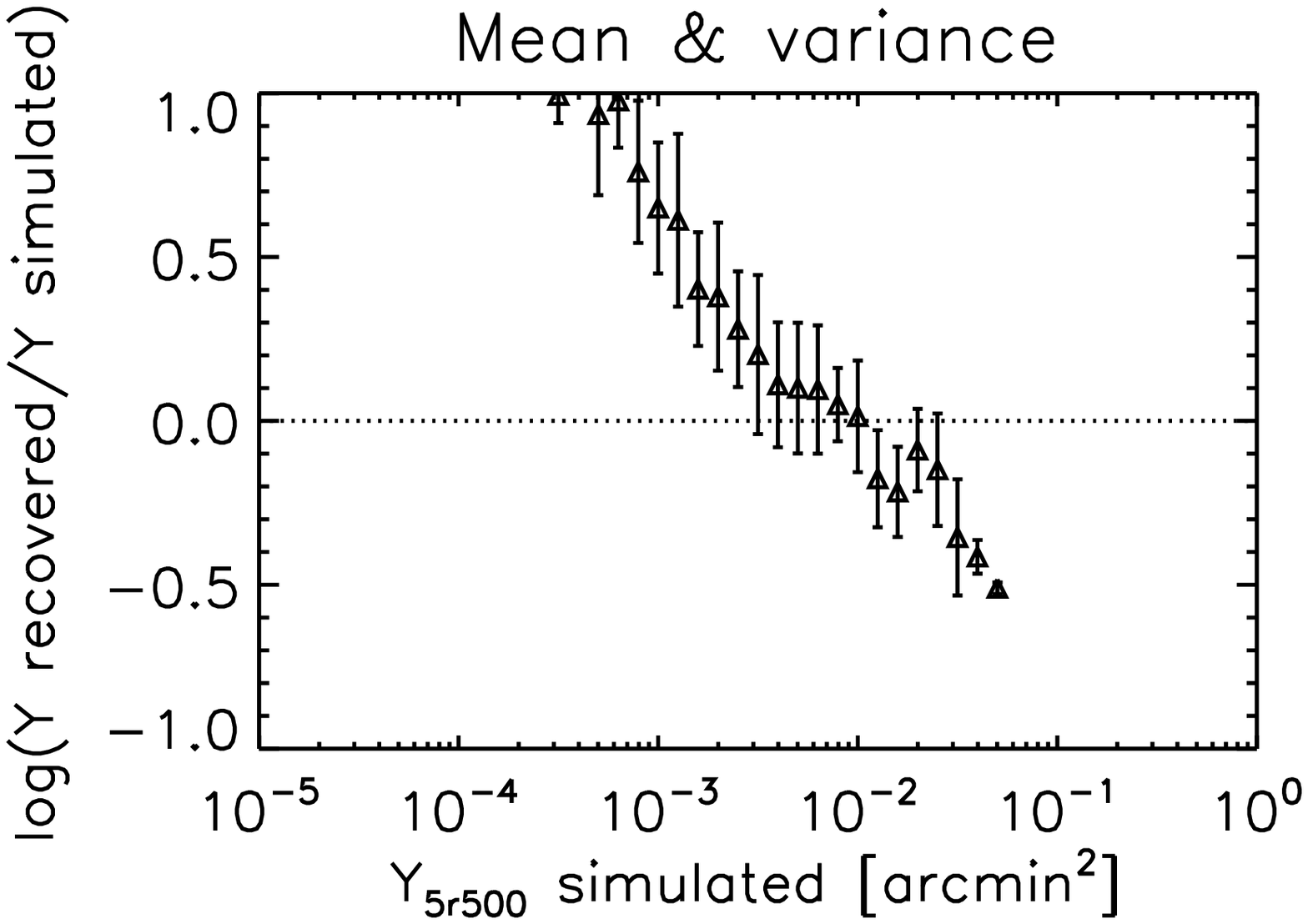} &
\includegraphics[scale=0.45]{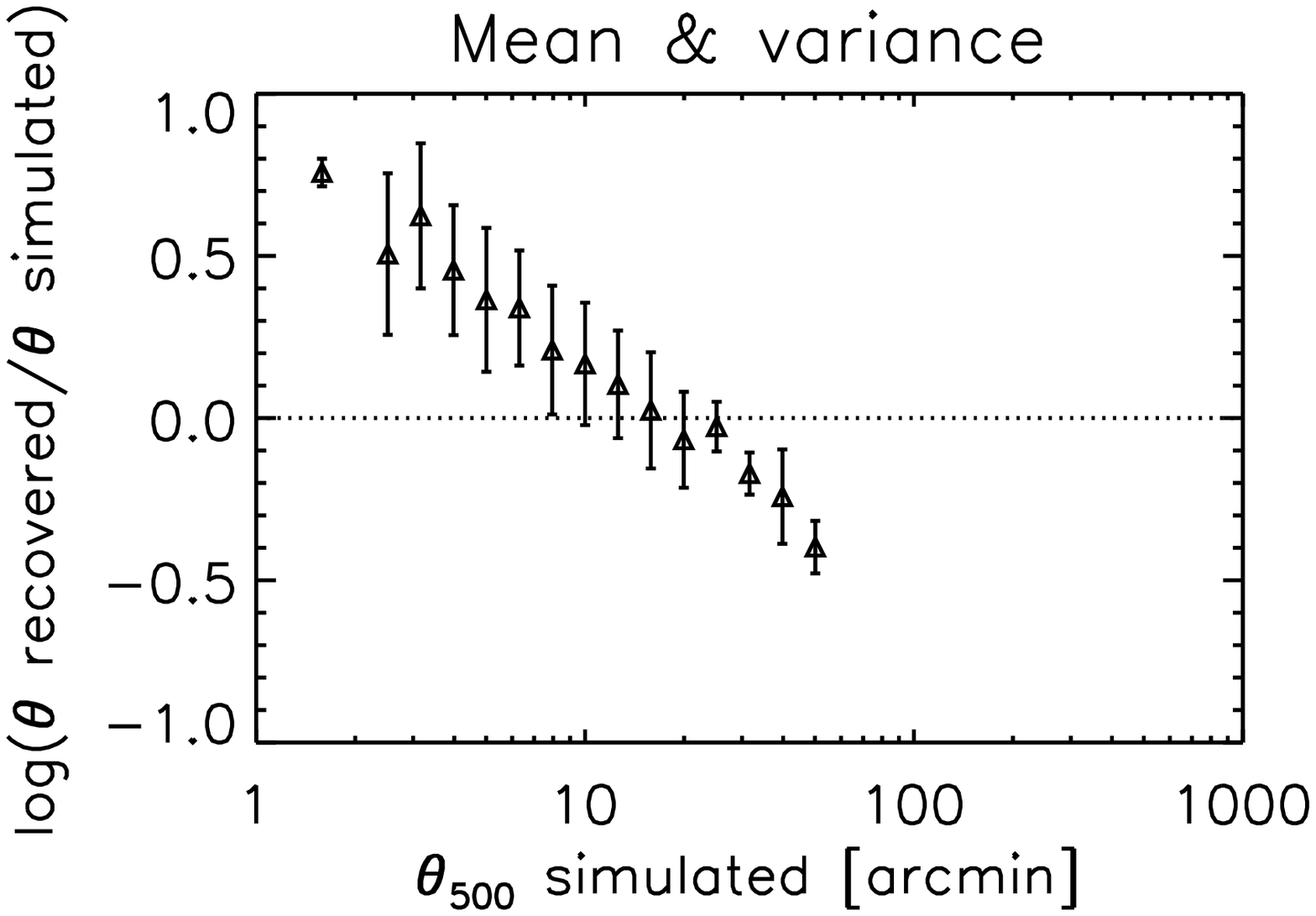} \\
\end{tabular}
\caption{{\bf BNP}}
\end{center}
\end{table}

\clearpage

\end{document}